# THÈSE DE DOCTORAT

de l'Université de recherche Paris Sciences et Lettres
PSL Research University

Préparée à
l'École Normale Supérieure

*Spectral Inference Methods on Sparse Graphs*: *Theory and Applications*

## Méthodes spectrales d'inférence sur les graphes parcimonieux
Théorie et applications

École doctorale n° 564 *Physique en Île-de-France*
Spécialité *Physique théorique*

### Composition du jury

M. Jean-Philippe Bouchaud
CFM et École Polytechnique
Rapporteur

M. Laurent Massoulié
INRIA et Microsoft
Rapporteur

M. Marc Mézard
ENS
Membre du jury

M. David Saad
Aston University
Membre du jury

Mme Lenka Zdeborová
CNRS et CEA
Membre du jury

M. Florent Krzakala
ENS et Paris 6
Directeur de thèse

Soutenue par **Alaa Saade**
le 3 octobre 2016

Dirigée par **Florent Krzakala**

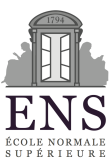
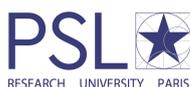

Explain ! Explain !! Explain !!!

— *The Daleks*

# ABSTRACT

In an era of unprecedented deluge of (mostly unstructured) data, graphs are proving more and more useful, across the sciences, as a flexible abstraction to capture complex relationships between complex objects. One of the main challenges arising in the study of such networks is the inference of macroscopic, large-scale properties affecting a large number of objects, based solely on the microscopic interactions between their elementary constituents. Statistical physics, precisely created to recover the macroscopic laws of thermodynamics from an idealized model of interacting particles, provides significant insight to tackle such complex networks.

In this dissertation, we use methods derived from the statistical physics of disordered systems to design and study new algorithms for inference on graphs. Our focus is on spectral methods, based on certain eigenvectors of carefully chosen matrices, and sparse graphs, containing only a small amount of information. We develop an original theory of spectral inference based on a relaxation of various mean-field free energy optimizations. Our approach is therefore fully probabilistic, and contrasts with more traditional motivations based on the optimization of a cost function. We illustrate the efficiency of our approach on various problems, including community detection, randomized similarity-based clustering, and matrix completion.

# RÉSUMÉ

Face au déluge actuel de données principalement non structurées, les graphes ont démontré, dans une variété de domaines scientifiques, leur importance croissante comme language abstrait pour décrire des interactions complexes entre des objets complexes. L'un des principaux défis posés par l'étude de ces réseaux est l'inférence de propriétés macroscopiques à grande échelle, affectant un grand nombre d'objets ou d'agents, sur la seule base des interactions microscopiques qu'entretiennent leurs constituants élémentaires. La physique statistique, créée précisément dans le but d'obtenir les lois macroscopiques de la thermodynamique à partir d'un modèle idéal de particules en interaction, fournit une intuition décisive dans l'étude des réseaux complexes.

Dans cette thèse, nous utilisons des méthodes issues de la physique statistique des systèmes désordonnés pour mettre au point et analyser de nouveaux algorithmes d'inférence sur les graphes. Nous nous concentrons sur les méthodes spectrales, utilisant certains vecteurs propres de matrices bien choisies, et sur les graphes parcimonieux, qui contiennent une faible quantité d'information. Nous développons une théorie originale de l'inférence spectrale, fondée sur une relaxation de l'optimisation de certaines énergies libres en champ moyen. Notre approche est donc entièrement probabiliste, et diffère considérablement des motivations plus classiques fondées sur l'optimisation d'une fonction de coût. Nous illustrons l'efficacité de notre approche sur différents problèmes, dont la détection de communautés, la classification non supervisée à partir de similarités mesurées aléatoirement, et la complétion de matrices.

**Mots clés :** opérateur non retraçant, Hessienne de Bethe, méthodes spectrales, détection de communautés, partitionnement spectral, complétion de matrices, modèles graphiques, inférence bayésienne, approximations de champ moyen, systèmes désordonnés, propagation des convictions, algorithmes de passage de messages.

*They danced down the streets like dingledodies, and I shambled after as I've been doing all my life after people who interest me, because the only people for me are the mad ones, the ones who are mad to live, mad to talk, mad to be saved, desirous of everything at the same time, the ones who never yawn or say a commonplace thing, but burn, burn, burn like fabulous yellow roman candles exploding like spiders across the stars and in the middle you see the blue centerlight pop and everybody goes "Awww!".*

— Jack Kerouac

## REMERCIEMENTS

This dissertation covers part of my work as a Ph.D. student from September 2013 to September 2016 in the Laboratoire de Physique Statistique at the Ecole Normale Supérieure in Paris, under the supervision of Florent Krzakala. Some of the ideas and figures presented here have previously appeared in the following publications, some of which have been considerably reworked to fit naturally in the general framework developed in this dissertation.

### PUBLICATIONS RELATED TO THIS DISSERTATION

1. Alaa Saade, Florent Krzakala, and Lenka Zdeborová. « Spectral density of the non-backtracking operator on random graphs. » In: *EPL (Europhysics Letters)* 107.5 (2014), p. 50005.

We compute the spectral density of the non-backtracking operator on sparse random graphs using the non-rigorous cavity method. We show the existence of a phase transition between a region in the complex plane where the spectral density is finite, and another one where it strictly vanishes. This fact provides a physical explanation of the superiority of this operator in the clustering of sparse graphs. This work is presented in section 3.1.4.

2. Alaa Saade, Florent Krzakala, and Lenka Zdeborová. « Spectral Clustering of graphs with the Bethe Hessian. » In: *Advances in Neural Information Processing Systems (NIPS)*. 2014, pp. 406–414.

This paper introduces an efficient spectral algorithm for community detection in sparse networks, based on the Bethe Hessian. It is up to now the simplest and most efficient optimal spectral method on the popular stochastic block model. Our approach is based on the Bethe Hessian, and is reviewed in chapter 5, which however differs substantially from the paper, and offers in particular more statistical physics insight into the community detection problem.

3. Alaa Saade, Florent Krzakala, Marc Lelarge, and Lenka Zdeborová. « Spectral detection in the censored block model. » In: *2015 IEEE International Symposium on Information Theory (ISIT)*. IEEE. 2015, pp. 1184–1188.

The censored block model, or planted spin glass (see section 1.6.5) is a simple inference problem in which we try to recover hidden binary variables from a small number of noisy pairwise comparisons between them. An information-theoretic lower bound on the number of comparisons necessary to partially recover the hidden variables was known. We prove rigorously that when above this bound, efficient recovery is possible, and introduce two optimal algorithms for this task. An augmented version of this work is presented in chapter 4, where the relation to the phase diagram of the planted spin glass is highlighted.

4. Alaa Saade, Florent Krzakala, and Lenka Zdeborová. « Matrix Completion from Fewer Entries: Spectral Detectability and Rank Estimation. » In: *Advances in Neural Information Processing Systems (NIPS).* 2015, pp. 1261–1269.

We consider the problem of reconstructing a low rank matrix from a minimal number of revealed entries. We address two aspects of this problem: *rank inference*, i. e. the ability to reliably estimate the rank of the matrix from as few revealed entries as possible, and *reconstruction accuracy*, i. e. the ability to achieve a small reconstruction error on the missing entries. We propose a spectral algorithm based on the Bethe Hessian and analyze its performance for both tasks. In a random matrix setting, we compute analytically the number of revealed entries required to infer the rank. We also evaluate empirically the reconstruction error, and show that our algorithm compares favorably to other existing methods. This work is presented in chapter 8, where the discussion of the connection of our approach with the Hopfield model is extended with respect to the original paper. Additional numerical results are also presented.

5. Alaa Saade, Marc Lelarge, Florent Krzakala, and Lenka Zdeborová. « Clustering from Sparse Pairwise Measurements. » In: *2016 IEEE International Symposium on Information Theory (ISIT).* IEEE. 2016, To appear.

In similarity-based clustering, a similarity graph involving n items is constructed after computing (typically) all the pairwise similarities between the items. In this paper, We consider instead the problem of grouping items into clusters based on few random pairwise comparisons between the items. We introduce three closely related algorithms for this task: a belief propagation algorithm approximating the Bayes optimal solution, and two spectral algorithms based on the non-backtracking and Bethe Hessian operators. For the case of two symmetric clusters, we conjecture that these algorithms are asymptotically optimal in that they detect the clusters as soon as it is information theoretically possible to do so on a model. We substantiate this claim with rigorous arguments for the spectral approach based on the non-backtracking operator. This work is presented in chapter 6, where the emphasis is on connecting our results to the general theory developed in the second part of this dissertation.

6. Alaa Saade, Florent Krzakala, Marc Lelarge, and Lenka Zdeborová. « Fast Randomized Semi-Supervised Clustering. » In: *arXiv preprint arXiv:1605.06422* (2016).

We investigate a semi-supervised variant of the previous problem, and introduce a *local* and highly efficient algorithm, based on the non-backtracking operator, that allows to cluster large datasets to excellent accuracy, using a small number of pairwise similarity measures between the items. On a simple but reasonable model, we prove rigorous guarantees on the performance of our algorithm, and show in particular that its error decays exponentially with the number of measured similarities. Numerical experiments on real-world datasets show that our approach outperforms a popular alternative by a large margin. This work is presented in chapter 7, where we emphasize the connections with the general theory developed in this dissertation, and also include additional real-world experiments. This paper is currently under review.

7. Alaa Saade, Florent Krzakala, and Lenka Zdeborová. « Spectral Bounds for the Ising Ferromagnet on an Arbitrary Given Graph. » In preparation. 2016.

This paper, to be submitted shortly, is based on chapter 9 of this dissertation. We derive rigorous upper bounds on the partition function, the magnetizations and the correlations in the ferromagnetic Ising model, on arbitrary graphs. Our bounds hold in a certain high-temperature region, specified by a condition on the spectral radius of the non-backtracking operator. Our approach extends previous bounds on the high temperature expansion of the Ising model, for which an explicit expression was known only in very special cases. We compute these bounds explicitly on arbitrary graphs, in terms of the non-backtracking and Bethe Hessian operators, and show that they can be computed efficiently. As a by-product, we prove that in the setting we consider, susceptibility propagation converges to a solution that admits a closed form expression, and that it yields an upper bound on the correlations.

### SCOPE OF THIS DISSERTATION

In this dissertation, I have tried to develop an original and general theory of spectral inference methods based on relaxations of mean-field approximations. In particular, chapters 2 and 3 consider a generic setting that encompasses most of the applications considered in the previously listed papers. These two chapters, written with a focus on intuition, present general results that are then illustrated, starting from chapter 4, on the various problems considered in the papers. In writing these last chapters, I have tried to precise some of the intuitive arguments of chapters 2 and 3. In particular, some of these chapters, although based on previously published work, may be substantially different from their corresponding paper.

The following work was completed during my Ph.D. and is not addressed in this dissertation.

8. Jie Lin, Alaa Saade, Edan Lerner, Alberto Rosso, and Matthieu Wyart. « On the density of shear transformations in amorphous solids. » In: *EPL (Europhysics Letters)* 105.2 (2014), p. 26003.

This paper is based on work started during my research internship at New York University from April to July 2012, in the group of Matthieu Wyart. This work studies the stability of amorphous solids, focusing on the distribution $\mathbb{P}(x)$ of the local stress increase $x$ that would lead to an instability. We argue that this distribution behaves as $\mathbb{P}(x) \sim x^\theta$, where the exponent $\theta$ is larger than zero if the elastic interaction between rearranging regions is non-monotonic, and increases with the interaction range. For a class of finite-dimensional models we show that stability implies a lower bound on $\theta$, which is found to lie near saturation. For quadrupolar interactions these models yield $\theta \approx 0.6$ for $d = 2$ and $\theta \approx 0.4$ in $d = 3$ where $d$ is the spatial dimension, accurately capturing previously unresolved observations in atomistic models, both in quasi-static flow and after a fast quench. In addition, we compute the Herschel-Buckley exponent in these models and show that it depends on a subtle choice of dynamical rules, whereas the exponent $\theta$ does not.

9. Alaa Saade, Francesco Caltagirone, Igor Carron, Laurent Daudet, Angélique Drémeau, Sylvain Gigan, and Florent Krzakala. « Random Projections through multiple optical scattering: Approximating kernels at the speed of light. » In: *2016 IEEE International Conference on Acoustics, Speech and Signal Processing (ICASSP).* IEEE. 2016, pp. 6215–6219.

This contribution, involving a sizable amount of experimental work, results from a collaboration with Sylvain Gigan (LKB-ENS), and aims at designing new optical devices able to speed up some bottleneck computations in machine learning applications. This paper describes an apparatus that performs random projections at the speed of light using the scattering properties of a random medium, such as a microscope glass slide covered with white paint pigments. We use this device on a simple classification task with a kernel machine, and show that, on the MNIST dataset of handwritten digits, our experimental results closely match the theoretical performance of the corresponding kernel. We filed a patent describing this device in 2015.

10. Aurélien Decelle, Janina Hüttel, Alaa Saade, and Cristopher Moore. « Computational Complexity, Phase Transitions, and Message-Passing for Community Detection. » In: *Statistical Physics, Optimization, Inference, and Message-Passing Algorithms.* Ed. by Florent Krzakala, Federico Ricci-Tersenghi, Lenka Zdeborová,

This book chapter is based on the lectures given by Cristopher Moore at the Les Houches School of Physics in October 2013. It covers topics in computational complexity (P, NP, and NP-completeness) as well as phase transitions in random graphs, satisfiability, and community detection.

# CONTENTS

# NOTATIONS

| | |
|---|---|
| $[n]$ | Set of integers larger or equal to 1 and lower or equal to $n$ |
| $G = (V, E)$ | An undirected graph with vertex set $V$ and edge set $E \subset V^2$ |
| $(i \to j)$ | The directed edge between vertices $i$ and $j$ (for $(ij) \in E$) |
| $\vec{E}$ | The set of directed edges of a graph $G = (V, E)$ |
| $\partial i$ | The set of neighbors of node $i$ in the graph $G$ |
| $\sigma$ | A collection of $n$ random variables $\sigma_i$ for $i \in n$ |
| $\mathcal{X}$ | Alphabet, or range of possible values of each variable $\sigma_i$ |
| $\mathbb{P}_i(\sigma_i)$ | Marginal distribution of the variable $\sigma_i$ in the JPD $\mathbb{P}(\sigma)$ |
| $\mathbb{P}_{ij}(\sigma_i, \sigma_j)$ | Joint marginal distribution of the variables $\sigma_i, \sigma_j$ in the JPD $\mathbb{P}(\sigma)$ |
| $\lvert S \rvert$ | Cardinal of the set $S$ |
| $\propto$ | Proportional to |
| $\mathbf{1}(P)$ | Indicator function, returns 1 if the proposition $P$ is true, 0 otherwise |
| $\mathcal{G}(n, p)$ | Erdős-Rényi random graph with $n$ vertices and edge probability $p$ |
| w.h.p | With high probability, i. e. with probability tending to 1 as $n \to \infty$ |
| $O(v_n)$ | $u_n = O(v_n)$ if there exists a constant $C > 0$ such that $\lvert u_n \rvert \leqslant C \lvert v_n \rvert, \forall n$ |
| $\rho(M)$ | The spectral radius of $M$, i.e the modulus of its largest eigenvalue |
| $\overset{\mathcal{D}}{=}$ | Equality in distribution between two random variables |

# ACRONYMS

| | |
|---|---|
| BP | belief propagation |
| MRF | Markov random field |
| JPD | joint probability distribution |
| i.i.d | independent and identically distributed |

| w.h.p | with high probability |
|---|---|
| SBM | stochastic block model |
| nMF | naive mean-field |
| TAP | Thouless-Anderson-Palmer |
| lSBM | labeled stochastic block model |
| RMSE | root mean square error |
| KL | Kullback–Leibler |
| SVD | singular value decomposition |
| RBM | restricted Boltzmann Machine |

## LIST OF FIGURES

Graphs, sometimes called networks, provide a flexible language to describe complex systems, composed of many items maintaining complicated interactions. Examples of such systems abound in all branches of science, ranging from biological or artificial neural networks, to social networks, the Internet and the World-Wide Web, electrical power grids, ecological networks or telecommunication networks... This list could go on without significantly scratching the surface of the ubiquity of graphs.

In this dissertation, we are interested in inference problems defined on sparse graphs. While precise definitions will be given in the upcoming first chapter, these problems can be intuitively thought of as prediction or estimation tasks based on scarce data, that can be conveniently encoded in the form of a graph with "few" edges. A typical example, studied in this dissertation, is community detection, where the aim is to identify homogeneous groups of items based on their connections. An illustration of this problem is shown on figure 0.1, where we have represented a network of political blogs during the 2004 U.S. presidential election. By using an algorithm developed in chapter 5, we are able to distinguish the conservative blogs from the liberal ones based solely on their hyperlinking pattern, represented as a graph.

Our focus is on *spectral* methods, which encode the graph in a carefully chosen matrix, and use certain eigenvectors of this matrix to make predictions. In the context of graph clustering, spectral methods are often introduced as relaxations of certain optimization problems [98], where a cost function is introduced that promotes grouping together strongly connected vertices. In contrast with this motivation, we adopt here a fully probabilistic approach. More precisely, we introduce spectral algorithms approximating the marginals of certain probability distributions, encoding the similarity between the vertices.

This dissertation is organized into three parts, the last two of which present original contributions. The first part is composed of chapter 1, and introduces the general language and framework of inference on graphs. We review generalities about graphical probabilistic models, and gradually specialize to the particular models which will be useful in the subsequent applications. We also review some standard results on sparse random graphs, mean-field approximations, as well as several variations of the Ising model of statistical physics, which is the cornerstone of the following analyses.

The second part introduces the original theory underlying our main contributions. In chapter 2, we identify a class of probabilistic mod-els that exhibit a phase transition, and give necessary conditions for non-trivial inference to be possible. These conditions are expressed in terms of two related matrices, one called the non-backtracking operator, and the other called the Bethe Hessian. We give an intuitive interpretation of their interesting eigenvectors, and propose two general spectral algorithms for approximate inference on sparse graphs. In chapter 3, we study in detail, using the cavity method of statistical physics, the spectral properties of these two matrices on weighted random graphs, and further explore their mutual relationship. We illustrate our general theory on the simple but enlightening example of the planted spin glass in chapter 4, for which we sketch a rigorous treatment.

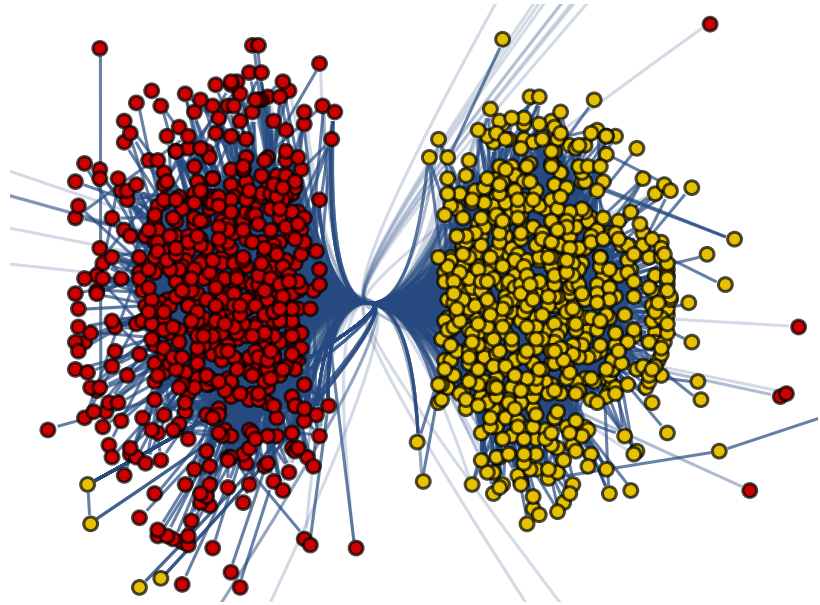

Figure 0.1 – Network of political blogs during the 2004 U.S. presidential election. Each vertex in this network represents a blog, and each edge represents a hyperlink between two blogs. This visualization clearly separates the Democrats (in red) from the Republicans (in yellow). We show in chapter 5 how to recover this partition with a 93% accuracy. Data collected by [8].

The third part of this dissertation is devoted to applications of the previous general theory. In chapter 5, we introduce an efficient and optimal algorithm for community detection. In chapter 6, we consider the problem of classifying items based on a minimal number of comparisons between them. We introduce several related algorithms and illustrate their performance on a model and on toy datasets. Then, in chapter 7, we consider a semi-supervised version of the same problem, and introduce a highly efficient algorithm with guaranteed performance on a model. On real world data, we show that our approach outperforms a state-of-the-art algorithm by a large margin. In chap-ter 8, we take a look at the matrix completion problem, and introduce an efficient algorithm able to infer the missing entries of a matrix from a small number of revealed entries. Coming back to statistical physics, we show in chapter 9 that the matrices introduced in this dissertation allow to derive rigorous upper bounds on many quantities of interest in the ferromagnetic Ising model, on arbitrary graphs. We conclude in chapter 10 by outlining a possible direction for future applications of the present ideas in the context of unsupervised learning.

This work applies statistical physics methods and intuition to statistical inference and machine learning problems. This is an awkward task, as each of these fields has its own folklore. In section 1.6, we review at a high level the phenomenology of the Ising model in its various flavors. We have tried to include in this section most of the statistical physics background necessary for the discussion of the second part of this dissertation. Many of these statistical physics results are "well known" without being fully rigorously proven yet. We have tried to clearly separate non-rigorous results from rigorous ones, both when referring to previous work and in the present contributions. Throughout the dissertation, we use graffiti in the margin to highlight important concepts or quantities and facilitate browsing through the chapters. We use footnotes to make additional comments.

*On the use of graffiti in this dissertation*

## Part I

## INFERENCE ON GRAPHS

This first part introduces the general background for the methods we develop in the following. We start by introducing the general formalism of (undirected) graphical models, gradually restricting to pairwise models. We review some standard results on sparse random graphs, mean-field theory and the Ising model, and outline the general approach we adopt in the subsequent parts.

# MARKOV RANDOM FIELDS

Graphical models are a powerful paradigm for multivariate statistical modeling. They allow to encode information about the conditional dependencies of a large number of interacting variables in a compact way, and provide a unified view of inference and learning problems in areas as diverse as statistical physics, computer vision, coding theory or machine learning (see *e. g.* [86, 151] for reviews of applications). In this first chapter, we motivate and introduce in section 1.1 the formalism of undirected graphical models, often called Markov random fields (MRFs). We then focus in section 1.3 on the particular case of pairwise MRFs which will be particularly important in the following. The analyses of this dissertation will apply to models drawn from certain random graph ensembles, and section 1.4 is devoted to a review of some of their basic properties. Computing the marginals of pairwise MRFs is computationally hard, and we will systematically resort to mean-field approximations, reviewed in section 1.5. We next turn to a presentation of the Ising model and its variants, and give in section 1.6 a high level description of their phase diagram, which will guide our intuition in the following. We finally outline in section 1.7 the general strategy we will apply to tackle the various problems in this dissertation.

We note that while MRFs are particularly well-suited to represent non-causal probabilistic interactions between variables, they are not the only example of graphical models. In particular, we will not address Bayesian networks [1], which are defined on directed acyclic graphs. The reader is referred to [86] for a comprehensive introduction to graphical models.

## 1.1 MOTIVATION

One of the first goals of a graphical model is to give a high-level, compact representation of a joint probability distribution (JPD). Far from being of purely academic concern, this problem has a considerable practical importance when dealing with the distribution of even a moderate number of correlated random variables. Imagine for instance having to represent the JPD of 300 binary variables. In general, such a JPD is specified by $2^{300}$ numbers — the probabilities of the different assignments of all the variables. This already requires more bits than the total number of atoms in the observable universe [2]. In practical applications, it is common to face millions or billions of variables, with hundreds of possible values for each of them. For instance, in computer vision, the probabilistic analysis of a single megapixel image in grayscale requires dealing with the JPD of more than a million variables (the pixels), with 256 values for each one of them (their intensities).

---

[1]. As shown in [161], Bayesian networks can in fact be transformed into the MRFs on which we focus.

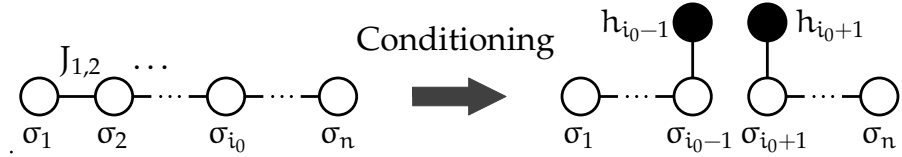

Figure 1.1 – Graphical model associated with the Ising chain 1.1. After conditioning on the value of the spin $\sigma_{i_0}$, the JPD splits into two independent parts, with the addition of local fields $h_{i_0-1}$ and $h_{i_0+1}$ acting on the neighbors of $\sigma_{i_0}$.

Fortunately, it is often the case that there is some kind of structure in the distribution of the variables, which can be stated in terms of (conditional) independence properties. As a simple example, consider the case of a one-dimensional Ising model on a chain of length $n \in \mathbb{N}$, defined by the JPD

*Ising chain*

$$\mathbb{P}(\sigma) = \frac{1}{\mathcal{Z}} \exp \sum_{i=1}^{n-1} J_{i,i+1} \sigma_i \sigma_{i+1} , \qquad (1.1)$$

where $\sigma = (\sigma_i)_{i \in [n]} \in \{\pm 1\}^n$ is a set of binary *spins*, and the $J_{i,i+1} \in \mathbb{R}$ for $i \in [n-1]$ are called *couplings*. $\mathcal{Z}$ is a normalization called the *partition function*. Let us consider the effect of conditioning on a given spin $\sigma_{i_0}$. We define two *fields* $h_{i_0-1} = J_{i_0-1,i_0} \sigma_{i_0}$ and $h_{i_0+1} = J_{i_0,i_0+1} \sigma_{i_0}$ which summarize the influence of $\sigma_{i_0}$ on its neighbors. We can then write

$$\mathbb{P}(\sigma \mid \sigma_{i_0}) = \mathbb{P}(\sigma_1, \dots, \sigma_{i_0-1} \mid \sigma_{i_0}) \mathbb{P}(\sigma_{i_0+1}, \dots, \sigma_n \mid \sigma_{i_0}) , \qquad (1.2)$$

where

$$\mathbb{P}(\sigma_1, \dots, \sigma_{i_0-1} \mid \sigma_{i_0}) \propto \exp \sum_{i=1}^{i_0-2} J_{i,i+1} \sigma_i \sigma_{i+1} + h_{i_0-1} \sigma_{i_0-1} , \quad (1.3)$$

$$\mathbb{P}(\sigma_{i_0+1}, \dots, \sigma_n \mid \sigma_{i_0}) \propto \exp \sum_{i=i_0+1}^{n-1} J_{i,i+1} \sigma_i \sigma_{i+1} + h_{i_0+1} \sigma_{i_0+1} . \qquad (1.4)$$

This property, depicted on figure 1.1, means that the spins that are *left* of $\sigma_{i_0}$ in the chain are independent of the spins that are *right* of $\sigma_{i_0}$, when conditioned on $\sigma_{i_0}$. It is a special case of a more general *global Markov property*, which we state in the next section.

---

2. The current estimate of the number of atoms in the observable universe is $10^{80} \approx 2^{266}$.

Markov random fields exploit such conditional independence properties of a JPD to provide a compact, graphical representation of the distribution. In the following section, we introduce MRFs as a generalization of the Ising chain example, and state their independence properties.

## 1.2 GENERAL MARKOV RANDOM FIELDS

Throughout the dissertation, we let $\sigma = (\sigma_i)_{i \in [n]} \in \mathcal{X}^n$ denote $n$ random variables, jointly distributed according to the JPD $\mathbb{P}$. We will assume that the *alphabet* $\mathcal{X}$ is finite, although many of the results of the next section can be naturally extended to real random variables. Our first aim is to find a compact representation of $\mathbb{P}$ that highlights the conditional dependencies between the $n$ variables.

*We consider $n$ random variables $(\sigma_i)_{i \in [n]}$ over a finite alphabet $\mathcal{X}$*

### 1.2.1 *Definition and Markov property*

In the Ising chain example of the previous section, global correlations between spins arise as a consequence of *local* interactions between neighboring spins on the chain. Such a structural property can be generalized, and more complex interactions can be modeled by replacing the chain with a general graph $G = (V, E)$, where $V = [n]$ is the set of vertices, representing the random variables $\sigma_i$ for $i \in [n]$, and the set of edges $E$ represents *direct interactions* between the random variables. We allow for interactions involving more than two variables by defining *compatibility functions* on the *cliques* of the graph. A clique of $G$ is a fully connected subgraph of $G$, i.e. a subset $C \subset V$ of the vertices such that for any $i, j \in C$, $(ij) \in E$. We associate with each clique $C$ a compatibility function, or *potential*, $\psi_C : \mathcal{X}^{|C|} \to \mathbb{R}_+$, where $|C|$ is the number of vertices in the clique $C$. Generalizing example (1.1), we define a *Markov random field* (MRF), also called *undirected graphical model*, by the following factorization [3]

*Cliques*

*Potentials*

*Definition of a MRF*

$$\mathbb{P}(\sigma) = \frac{1}{\mathcal{Z}} \prod_{C \in \mathcal{C}} \psi_C \left( (\sigma_i)_{i \in C} \right) , \tag{1.5}$$

where $\mathcal{C}$ is the set of cliques of the graph $G$, and the partition function $\mathcal{Z}$ ensures that the distribution is normalized. Note that we recover the Ising chain example of the previous section by taking $G$ to be the chain graph with edge set $E = (i, i+1)_{i \in [n-1]}$, and noticing that the cliques of $G$ consist of pairs of neighboring spins. The corresponding compatibility functions are, for $i \in [n-1]$,

*Partition function*

$$\psi_{i,i+1}(\sigma_i, \sigma_{i+1}) = \exp J_{i,i+1} \sigma_i \sigma_{i+1} . \tag{1.6}$$

---

3. Note that it is equivalent to require the factorization (1.5) to hold on the maximal cliques of $G$, i.e. the cliques that are contained by no other clique.

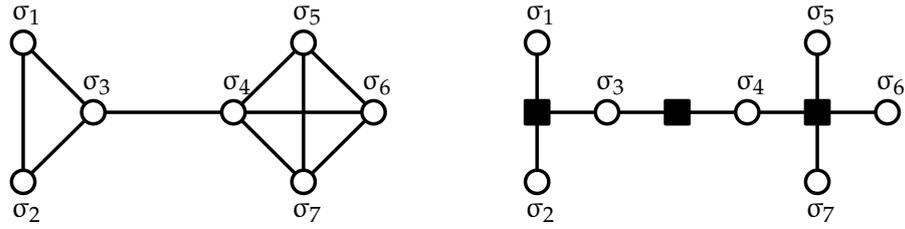

Figure 1.2 – On the left is an example of a MRF encoding the factorization $\mathbb{P}(\sigma) \propto \psi_{123}(\sigma_1, \sigma_2, \sigma_3)\psi_{34}(\sigma_3, \sigma_4)\psi_{4567}(\sigma_4, \sigma_5, \sigma_6, \sigma_7)$, with its factor graph representation on the right.

*Global Markov property*

The definition (1.5) is justified by the following *global Markov property*, that generalizes the conditional independence property (1.2) of the Ising chain. Let $A, B, S \subset V$ be three disjoint sets of nodes of the graph G. Assume that $S$ *separates* A and B, in the sense that any path in the graph G linking a node in A and a node in B must include a node in S. Then any MRF, i.e. any distribution of the form (1.5) verifies

$$\mathbb{P}\big((\sigma_i)_{i \in A \cup B} \mid (\sigma_i)_{i \in S}\big) = \mathbb{P}\big((\sigma_i)_{i \in A} \mid (\sigma_i)_{i \in S}\big)\mathbb{P}\big((\sigma_i)_{i \in B} \mid (\sigma_i)_{i \in S}\big). \tag{1.7}$$

The factorization (1.5) can therefore efficiently encode complex interactions and independence properties between a large number of interactions. In fact, under a small additional assumption, it is possible to show that a JPD that verifies the global Markov property with respect to a graph G *must* factorize in the form (1.5). This is the content of the *Hammersley-Clifford* theorem which we recall here.

*Hammersley-Clifford theorem*

**Theorem 1.2.1.** *(Hammersley and Clifford, 1971) Let $\mathbb{P}$ be a distribution over the set $\mathcal{X}^n$ verifying the global Markov property (1.7) with respect to a graph G. Assume that $\mathbb{P}(\sigma) > 0$ for any $\sigma \in \mathcal{X}^n$. Then there exist compatibility functions $\psi_C$ indexed by the cliques of the graph G such that $\mathbb{P}$ factorizes into the form (1.5).*

A consequence of this theorem is that, as an alternative to the definition (1.5), (positive) MRFs can equivalently be defined as distributions verifying the global Markov property, a definition often chosen in the literature. In fact, the Hammersley-Clifford theorem also implies the equivalence between the global Markov property and so-called local Markov properties ([86]), any of which can be used as a definition for MRFs.

### 1.2.2 *Factor graph representation*

One of the main advantages of MRFs is that they can be conveniently represented graphically (see left panel of figure 1.2 for an example). However, to recover the factorization associated with a given graphical representation, one has to look for the cliques of the graph, which can be cumbersome, especially when many variables are involved. An alternative visual representation of the same factorization property is given by *factor graphs*, which explicitly depict the interactions between variables. A factor graph is formally defined as a bipartite graph $G = (V, F, E)$ where $V$ is the set of variable nodes, $F$ is the set of factor nodes, representing the potentials, and $E$ is the set of edges. To represent a MRF of the form (1.5) as a factor graph, we associate with each variable $\sigma_i$ ($i \in [n]$) a variable node (represented as a circle on figure 1.2), and with each potential $\psi_C$ a factor node (represented as a square on figure 1.2). We then link each variable to the factors that depend on it.

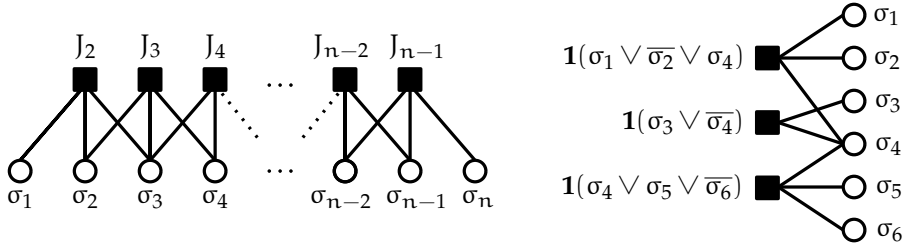

Figure 1.3 – Factor graphs associated with the 3-spin model (1.8) (left) and the SAT formula (1.9) (right).

*Factor graph representation*

### 1.2.3  *First examples*

Factor graphs give an explicit and convenient representation of the conditional independencies of a MRF, and are extensively used in many fields. Let us illustrate their use on two examples, represented graphically on figure 1.3. The first example is drawn from statistical physics, and is called the p-*spin model*. It is a generalization of the Ising model (1.1) with p-body interactions ($p > 2$) between the spins, instead of only pairwise interactions. The probability of a spin configuration $\sigma \in \{\pm 1\}^n$ in the 3-spin model on a chain is given by

*p-spin model*

$$\mathbb{P}(\sigma) = \frac{1}{\mathcal{Z}} \exp \sum_{i=2}^{n-1} J_i \sigma_{i-1} \sigma_i \sigma_{i+1} \,. \tag{1.8}$$

The factor graph associated with this distribution is depicted on figure 1.3.

A second example, from the field of theoretical computer science, is that of *satisfiability*. A satisfiability problem (or SAT problem for short) is specified by $n$ boolean variables $(\sigma_i)_{i \in [n]} \in \{\text{true}, \text{false}\}^n$, and a *formula*, defined as the logical AND of a certain number of *clauses* that the variables must satisfy. An example of a SAT formula over $n = 6$ variables is

*SAT problem*

$$(\sigma_1 \vee \overline{\sigma_2} \vee \sigma_4) \wedge (\sigma_3 \vee \overline{\sigma_4}) \wedge (\sigma_4 \vee \sigma_5 \vee \overline{\sigma_6}) \,, \tag{1.9}$$

where $\overline{\sigma_i}$ denotes the negation of the variable $\sigma_i$. The problem of finding an assignment $\sigma$ of the variables that satisfies a given SAT formula is equivalent to finding the configurations with non-vanishing probability in an associated MRF. For instance, the MRF associated with the SAT formula (1.9) is

$$\mathbb{P}(\sigma) = \frac{1}{\mathcal{Z}}\, \mathbf{1}(\sigma_1 \vee \overline{\sigma_2} \vee \sigma_4)\mathbf{1}(\sigma_3 \vee \overline{\sigma_4})\mathbf{1}(\sigma_4 \vee \sigma_5 \vee \overline{\sigma_6})\,. \qquad (1.10)$$

The associated factor graph is shown on figure 1.3.

*Factor graphs and inference*

These two examples barely scratch the surface of the extensive use of factor graphs in many problems. Whenever the problem at hand requires performing *inference*, e.g. computing the marginals or the modes of a high dimensional distribution, the fist step in finding a solution is usually to draw a factor graph representation of the problem, and to use it to derive approximate inference algorithms, e.g. belief propagation (see section 1.5). This approach is standard in problems as diverse as error-correcting coding [129], compressed sensing [106], computer vision [119], and many more [86, 151].

In the following, we will be mainly interested in the special case of *pairwise* MRFs, both in the interest of simplicity and because the original contributions of this dissertation are naturally expressed in terms of pairwise models. In the next section, we motivate the use of pairwise models, and show that they can in fact be used to represent any MRF.

## 1.3    PAIRWISE MARKOV RANDOM FIELDS

### 1.3.1    *Definition and examples*

An interesting particular case of the general MRF (1.5) is obtained by setting trivial compatibility functions $\psi_C((\sigma_i)_{i \in C}) = 1$, for all cliques with size $|C| > 2$. On a given graph $G = ([n], E)$, the resulting model, called a *pairwise* MRF, reads

*Pairwise MRF*

$$\mathbb{P}(\sigma) = \frac{1}{\mathcal{Z}} \prod_{(ij) \in E} \psi_{ij}(\sigma_i, \sigma_j) \prod_{i=1}^{n} \phi_i(\sigma_i)\,, \qquad (1.11)$$

where we have separated the single-variable potentials $(\phi_i)_{i \in [n]}$ associated with cliques of size $|C| = 1$, from the pairwise potentials $(\psi_{ij})_{(ij) \in E}$ associated with cliques of size $|C| = 2$. Note that for convenience, we are slightly abusing our notations by indexing the pairwise potentials, defined on each edge $e \in E$, by indices $i, j \in [n]$ such that $e = (ij)$. Since the edge $(ij)$ is the same as the edge $(ji)$, for equation (1.11) to be consistent regardless of the labeling of the edges, the pairwise potentials must verify the following symmetry

*Symmetry of the potentials*

$$\psi_{ij}(\sigma_i, \sigma_j) = \psi_{ji}(\sigma_j, \sigma_i) \qquad \forall (ij) \in E, \sigma_i, \sigma_j \in \mathcal{X}\,. \qquad (1.12)$$

When the distribution (1.11) is strictly positive for any assignment $\sigma$, it admits the following *exponential representation*

$$\mathbb{P}(\sigma) = \frac{1}{Z} \exp\left( -\sum_{(ij) \in E} \epsilon_{ij}(\sigma_i, \sigma_j) - \sum_{i=1}^{n} \epsilon_i(\sigma_i) \right), \qquad (1.13)$$

*Exponential representation*

where the *energies* $\epsilon_i, \epsilon_{ij}$ for $i \in [n]$ and $(ij) \in E$ are defined as

$$\epsilon_i(\sigma_i) = -\log \phi_i(\sigma_i), \qquad \epsilon_{ij}(\sigma_i, \sigma_j) = -\log \psi_{ij}(\sigma_i, \sigma_j). \qquad (1.14)$$

*Energies*

The choice of the word "energy" highlights the connection with statistical physics. Indeed, expression (1.13) is nothing but the *Boltzmann distribution* of the so-called Potts model of statistical physics [159].

We will represent pairwise models of the form (1.11) by simply drawing the graph $G = ([n], E)$, in which each node (pictured as a white circle) represents a variable, and each edge $(ij) \in E$ carries a pairwise potential $\psi_{ij}$. The single-variable potentials $\phi_i$ for $i \in [n]$ are pictured as black circles connected to a single variable node. We stress that while the graph $G$ may contain cliques of size larger than 2, they are not associated with potentials in a pairwise MRF model. For instance, the triangle graph in the margin represents the factorization

*Representation of pairwise models*

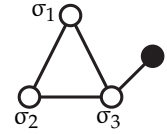

$$\mathbb{P}(\sigma) \propto \psi_{12}(\sigma_1, \sigma_2)\psi_{13}(\sigma_1, \sigma_3)\psi_{23}(\sigma_2, \sigma_3)\phi_3(\sigma_3), \qquad (1.15)$$

which does not include a term of the form $\psi_{123}(\sigma_1, \sigma_2, \sigma_3)$.

The Potts model (1.13) is popular in computer vision [21, 22, 54], where it is used to denoise a partially observed image by encouraging nearby pixels to have similar colors. In this case, the graph $G$ is taken to be a regular lattice where each node, representing a pixel, is connected with its four adjacent pixels in the image. The single-variable energies $\epsilon_i(\sigma_i)$ are chosen to be minimized when $\sigma_i = \hat{\sigma}_i$, where $\hat{\sigma}_i$ is the observed value of the pixel. One simple example is

*Computer vision example*

$$\epsilon_i(\sigma_i) = -h\,\mathbf{1}(\sigma_i = \hat{\sigma}_i), \qquad (1.16)$$

where $h > 0$ can be interpreted physically as a local field, quantifying the confidence that we have in the observed value of the pixel. The pairwise energies $\epsilon_{i,j}(\sigma_i, \sigma_j)$ act as smoothing terms, and are chosen to be minimized when $\sigma_i = \sigma_j$, e.g.

$$\epsilon_{ij}(\sigma_i, \sigma_j) = -J\,\mathbf{1}(\sigma_i = \sigma_j), \qquad (1.17)$$

where $J > 0$ is a coupling controlling the strength of the smoothing. The resulting graphical model is depicted on figure 1.4.

Pairwise models of the form (1.11) provide a principled approach to the probabilistic analysis of data in which we expect correlations to arise as a consequence of two-body interactions between random variables. An interesting application in biology is the so-called *direct coupling analysis* [109] of co-evolution in protein sequences. The primary

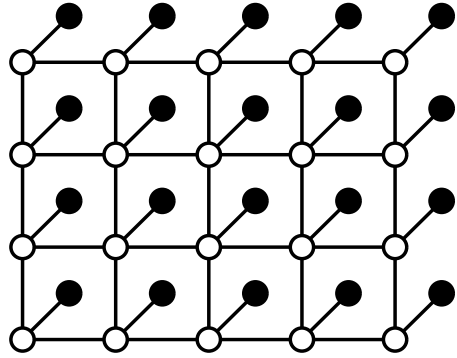

Figure 1.4 – Pairwise MRF for computer vision. Each variable node (white) represents a pixel, and is attached to a single-variable potential (black) containing the observed value of pixel. Pairwise potentials promote smoothness by correlating nearby pixels in the image.

*An example in biology*  structure of a protein can be regarded as a sequence of $n$ amino acids $(\sigma_i)_{i \in [n]}$ taking values in an alphabet $\mathcal{X}$ with 21 possible values (20 amino acids plus possible "gaps" arising from insertion or deletion mutations). The function of the protein is (mainly) determined by its three-dimensional structure, which, in turn, is determined by a pairwise *contact map*, encoding which pairs of amino acids are in contact in the folded protein. Although such contacts are the result of complex physical and chemical interactions, they can in fact be inferred from purely statistical considerations. The key observation is that evolution strongly constrains these pairs of interacting amino acids. Indeed, if two amino acids $i$ and $j$ are in contact after the folding of the protein, then a mutation affecting $\sigma_i$ must come with a mutation affecting $\sigma_j$ accordingly, otherwise the protein will be denatured, and the mutation will not be selected. These two-body interactions result in strong correlations in the sequences of homologous proteins, i. e. proteins sharing the same function. By sequencing homologous proteins in various living organisms, it is possible to compute these correlations, and to learn a model of the form (1.13) that is able to reproduce these correlations. Using this approach, [109] showed that the learned pairwise potentials allow to recover, with an impressive accuracy, the contact map of the protein.

*Learning*  The previous example is an illustration of an *inverse*, or *learning* problem, where we have access to some statistics which we believe can be explained by a model of the form (1.11), and we try to learn a graph $G = ([n], E)$ and potentials $(\psi_{ij})_{(ij) \in E}$, $(\phi_i)_{i \in [n]}$ that can reproduce the observed data. In the converse direction, it may happen that we have access to some data in the form of pairwise constraints on the variables. In this case, we may *postulate* a model of the form (1.11) encoding the available data, as well as any belief we may have on the structure of the data, into the potentials $(\psi_{ij})_{(ij) \in E}$, $(\phi_i)_{i \in [n]}$ of a cer-tain graph $G = ([n], E)$, and use this model to make predictions. For instance, to predict the most probable state of a particular variable $\sigma_i$ for some $i \in [n]$, one may compute its marginal distribution $\mathbb{P}_i(\sigma_i)$ and look for its mode. This approach is usually called *inference*, and will be the main subject of the original contributions of this dissertation. This strategy is exemplified by the computer vision example represented on figure 1.4. Here, the available observed pixels are encoded into single-variable potentials, and the pairwise potentials translate our belief that neighboring pixels in an image usually have similar colors. Similarly, in the following, we will consider graph clustering problems, in which the pairwise potentials will encode the similarity (or dissimilarity) between two objects, and the marginals of model (1.11) will serve to label the objects.

Another motivation for considering pairwise MRFs stems for the maximum entropy principle, and is the subject of the next section.

### 1.3.2 *Maximum entropy principle*

Generalizing the biological example of the previous section, assume that we wish to learn the multivariate probability distribution $\mathbb{P}$ of a random variable $\sigma = (\sigma_i)_{i \in [n]} \in \mathcal{X}^n$, given some realizations of this random variable. More precisely, assume that we have access to $p$ independent and identically distributed (i.i.d) samples $\sigma^{(1)}, \ldots, \sigma^{(p)}$ from the distribution $\mathbb{P}$, and that we wish to find a distribution $\mathbb{Q}$ such that, for all $\alpha$ in some index set $\mathcal{I}$,

$$\mathbb{E}_{\mathbb{Q}}\left[f_\alpha(\sigma)\right] = \hat{\mu}_\alpha \, , \qquad (1.18)$$

*Moment matching conditions*

where $\sigma$ has distribution $\mathbb{Q}$, each $f_\alpha : \mathcal{X}^n \to \mathbb{R}$ is a fixed function, and the $(\hat{\mu}_\alpha)_{\alpha \in \mathcal{I}}$ are the empirical averages of the functions $(f_\alpha)_{\alpha \in \mathcal{I}}$ on the data we are given, i.e.

$$\hat{\mu}_\alpha = \frac{1}{p} \sum_{j=1}^{p} f_\alpha\left(\sigma^{(j)}\right) \, . \qquad (1.19)$$

*Empirical moments*

In words, we are looking for a probability distribution $\mathbb{Q}$ such that the expected value of each function $f_\alpha$ for $\alpha \in \mathcal{I}$ is equal to the empirical average of this function on the data. By the law of large numbers, in the limit where the number of samples $p \to \infty$, we have $\hat{\mu}_\alpha \to \mathbb{E}_{\mathbb{P}}(f_\alpha(\sigma))$, so that $\mathbb{Q} = \mathbb{P}$ is a solution. However, the system of equations (1.18) is strongly under-determined, and there exists in general an infinite number of solutions. As usual in physics and science in general, we favor the "least constrained" solution[4], i.e. the solution that assumes the least additional structure in the distribution $\mathbb{Q}$.

---

4. This general principle of parsimony is known as Occam's razor, and is vividly criticized, as is the maximum entropy principle.

To quantify what we mean by "least constrained", we introduce the (Gibbs-Shannon) *entropy* of the distribution $\mathbb{Q}$, defined as

$$S(\mathbb{Q}) = -\sum_{\sigma \in \mathcal{X}^n} \mathbb{Q}(\sigma) \log \mathbb{Q}(\sigma) \,. \tag{1.20}$$

The entropy of the distribution $\mathbb{Q}$ quantifies the *unpredictability* of a variable $\sigma$ distributed according to $\mathbb{Q}$. The entropy of a deterministic distribution (which support is a single element of $\mathcal{X}^n$) vanishes — the value of $\sigma$ can be exactly predicted without doing an experiment. On the other hand, the uniform distribution on $\mathcal{X}^n$ has maximum entropy, as the uncertainty on the value of a realization of $\sigma$ is maximal in this case. The principle of maximum entropy states that among all the distributions $\mathbb{Q}$ that verify the constraints (1.11), we should favor the distribution that has maximum entropy. Therefore $\mathbb{Q}$ is the solution of a constrained optimization problem which Lagrangian is given by

$$\mathcal{L} = -\sum_{\sigma \in \mathcal{X}^n} \mathbb{Q}(\sigma) \log \mathbb{Q}(\sigma) + \sum_{\alpha \in \mathcal{I}} \lambda_\alpha \left( \sum_{\sigma \in \mathcal{X}^n} \mathbb{Q}(\sigma) f_\alpha(\sigma) - \hat{\mu}_\alpha \right) \,, \tag{1.21}$$

where the $\lambda_\alpha$ for $\alpha \in \mathcal{I}$ are Lagrange multipliers. For any given $\sigma \in \mathcal{X}^n$, we get by differentiating

$$\frac{\partial \mathcal{L}}{\partial \mathbb{Q}(\sigma)} = -1 - \log \mathbb{Q}(\sigma) + \sum_{\alpha \in \mathcal{I}} \lambda_\alpha f_\alpha(\sigma) \,, \tag{1.22}$$

so that $\mathbb{Q}$ takes the form of an exponential (or Boltzmann) distribution

$$\mathbb{Q}(\sigma) = \frac{1}{\mathcal{Z}} \exp \left( -\sum_{\alpha \in \mathcal{I}} \lambda_\alpha f_\alpha(\sigma) \right) \,, \tag{1.23}$$

for some normalization $\mathcal{Z}$. The general form (1.23) reduces to the pairwise exponential representation of (1.13) for a certain graph $G = ([n], E)$ when $\mathcal{I} = ([n] \times \mathcal{X}) \cup (E \times \mathcal{X}^2)$, and for any $\sigma \in \mathcal{X}^n$

$$f_{i,a}(\sigma) = \mathbf{1}(\sigma_i = a), \qquad\qquad \forall i \in [n], a \in \mathcal{X}, \tag{1.24}$$

$$f_{ij,a,b}(\sigma) = \mathbf{1}(\sigma_i = a)\,\mathbf{1}(\sigma_j = b), \qquad \forall (ij) \in E, (a,b) \in \mathcal{X}^2. \tag{1.25}$$

The energies $(\epsilon_i)_{i \in [n]}, (\epsilon_{ij})_{(ij) \in E}$ of (1.13) are related to the Lagrange multipliers of (1.23) by the identities

$$\epsilon_i(a) = \lambda_{i,a}, \qquad\qquad \forall i \in [n], a \in \mathcal{X}, \tag{1.26}$$

$$\epsilon_{ij}(a, b) = \lambda_{ij,a,b}, \qquad\qquad \forall (ij) \in E, (a,b) \in \mathcal{X}^2 \,. \tag{1.27}$$

The constraints (1.18) then correspond to prescribing the one-variable and two-variables marginals of $\mathbb{Q}$, which, in the limit of large number of samples $p$, verify for all $i \in [n], (ij) \in E, a, b \in \mathcal{X}$

$$\mathbb{Q}_i(\sigma_i = a) = \hat{\mu}_{i,a} \xrightarrow[p \to \infty]{} \mathbb{P}_i(\sigma_i = a) \,, \tag{1.28}$$

$$\mathbb{Q}_{ij}(\sigma_i = a, \sigma_j = b) = \hat{\mu}_{ij,a,b} \xrightarrow[p \to \infty]{} \mathbb{P}_{ij}(\sigma_i = a, \sigma_j = b) \,. \tag{1.29}$$

To sum up, among all the distributions over $\mathfrak{X}^n$ with prescribed one-variable and two-variables marginals, the Potts model of equation (1.13) is the one with maximum entropy.

### 1.3.3 *From general to pairwise Markov random fields*

As stated previously, our main motivations for focusing on the pairwise model (1.11) instead of the more general MRF (1.5) are both simplicity, and the fact that the contributions we describe in following are naturally expressed in terms of pairwise models. For completeness however, we show here that any MRF of the general form (1.5) may in fact be rewritten in the pairwise form (1.11), at the expense of increasing the size of the dictionary $\mathfrak{X}$ [160].

We start from the factor graph representation of a general MRF. As sketched in the margin, the potentials corresponding to cliques of size 1 and 2 are already of the pairwise form (1.11) and are readily translated. To deal with factors corresponding to larger cliques, we introduce a new variable node with a corresponding single-variable potential, and tune the potentials adequately.

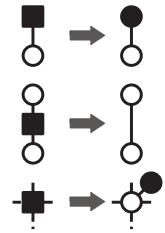

To be definite, consider the MRF depicted on figure 1.5, which is of the form

$$\mathbb{P}(\sigma_1, \sigma_2, \sigma_3, \sigma_4) \propto \psi_{123}(\sigma_1, \sigma_2, \sigma_3)\psi_{34}(\sigma_3, \sigma_4)\psi_4(\sigma_4) \,. \qquad (1.30)$$

We replace the $\psi_{123}$ potential with a new variable node $\sigma_5 \in \mathfrak{X}^3$ (note that we have enlarged the alphabet), attached to a single-variable potential $\phi_5$. We write the components of $\sigma_5$ as $\sigma_5 = \left(\sigma_5^{(1)}, \sigma_5^{(2)}, \sigma_5^{(3)}\right)$. The new pairwise model on the right of figure 1.5 factorizes as

$$\begin{aligned} \mathbb{P}'(\sigma_1, \sigma_2, \sigma_3, \sigma_4, \sigma_5) \propto {} & \phi_5(\sigma_5)\psi_{15}(\sigma_1, \sigma_5)\psi_{25}(\sigma_2, \sigma_5) \\ & \psi_{35}(\sigma_3, \sigma_5)\psi_{34}(\sigma_3, \sigma_4)\phi_4(\sigma_4) \,. \end{aligned} \qquad (1.31)$$

We tune the newly defined potentials $\psi_{i5}$ for $i \in [3]$ in the following way

$$\psi_{i5}(\sigma_i, \sigma_5) = \mathbb{1}\left(\sigma_i = \sigma_5^{(i)}\right) \,, \qquad (1.32)$$

so that, in the pairwise MRF, the probability of a configuration vanishes unless $\sigma_5 = (\sigma_1, \sigma_2, \sigma_3)$. It only remains to fix the single-variable potential $\phi_5$ to equal the potential $\psi_{123}$ that we removed

$$\phi_5(x_5) = \psi_{123}\left(\sigma_5^{(1)}, \sigma_5^{(2)}, \sigma_5^{(3)}\right) \,. \qquad (1.33)$$

With these choices, the distribution represented by the pairwise model is related to the distribution represented by the original factor graph by

$$\mathbb{P}'(\sigma_1, \sigma_2, \sigma_3, \sigma_4, \sigma_5) = \mathbb{1}(\sigma_5 = (\sigma_1, \sigma_2, \sigma_3))\mathbb{P}(\sigma_1, \sigma_2, \sigma_3, \sigma_4) \,, \qquad (1.34)$$

so that the marginals of $\mathbb{P}'$ are trivially related to those of $\mathbb{P}$. The procedure we have described on this example is straightforwardly adapted to any clique potential, and allows to turn any factor graph into a pairwise MRF. Note that this is achieved at the expense of an increase in the size of the alphabet, which goes from $\mathcal{X}$ to $\mathcal{X}^{|C|_{\max}}$, where $|C|_{\max}$ is the size of the largest clique in the original MRF. This cost is however not completely superfluous, in the sense that an inference algorithm such as belief propagation (see section 1.5) has a time complexity of $O(\mathcal{X}^{|C|_{\max}})$ on general MRFs, while it can be made to run in time $O(\mathcal{X}^2)$ on pairwise models. On the other hand, using pairwise models simplifies the formulation of belief propagation which is defined using only one type of messages, whereas in general factor graphs, there are two types of messages (factor to node and node to factor). As the previous procedure shows, we can choose without loss of generality to restrict the presentation to pairwise models, as in the classical paper [160].

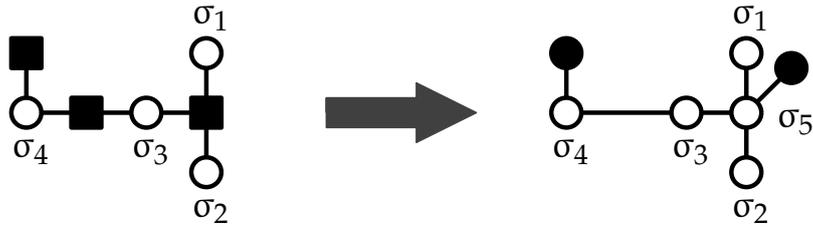

Figure 1.5 – Conversion from a factor graph with a 3-body interaction (left) to a pairwise MRF (right).

## 1.4    RANDOM GRAPH ENSEMBLES

In the applications of the following chapters, we will consider pairwise models defined on random graphs. We define here the main ensembles of random graphs we will be considering, and list some of their prominent properties that will important to our study.

### 1.4.1    *The Erdős-Rényi model*

*Sampling from $\mathcal{G}(n, p)$*

Perhaps the best known ensemble of random graphs is the so-called *Erdős-Rényi model* $\mathcal{G}(n, p)$ [41]. In the $\mathcal{G}(n, p)$ model, a graph $G = ([n], E)$ with $n$ vertices (or nodes) is generated by including in the edge set $E$ each pair $(ij)$ for $1 \leqslant i < j \leqslant n$ with probability $p$, independently for each pair. The *degree* $d_i$ of vertex $i$ in $G$ is the number of neighbors of $i$ in $G$. We write $\partial i$ for the set of neighbors of $i$, so that $d_i = |\partial i|$. In $\mathcal{G}(n, p)$, the degree of any node follows the same binomial distribution, for any $i \in [n]$

*The degree distribution is binomial*

$$\mathbb{P}(d_i = d) = \binom{k}{n-1} p^d (1-d)^{n-1-d} . \tag{1.35}$$

We call $\mathbb{E}[d_i]$ the *average connectivity* (or simply average degree) of G. In $\mathcal{G}(n, p)$, the average connectivity is $p(n-1)$. In any graph (deterministic or random), the number of edges $|E|$ is related to the degrees by the so-called *handshake lemma*

*Average connectivity*

$$2\,|E| = \sum_{i=1}^{n} d_i \,,\qquad(1.36)$$

*Handshake lemma*

so that $|E|$ is a binomial random variable with expectation $\mathbb{E}\,[|E|] = p\,n(n-1)/2$. In the following, we will be mostly interested in the limit $n \to \infty$, with the scaling

$$p = \frac{\alpha}{n}\,,\qquad(1.37)$$

*Sparse regime*

where $\alpha > 0$ is therefore identified with the average connectivity. When $\alpha$ is fixed, independently of $n$, we say that the resulting graph is *sparse*, as the expected number of edges scales linearly with the number of vertices

$$\mathbb{E}\,[|E|] \underset{n\to\infty}{\sim} \frac{\alpha n}{2}\,.\qquad(1.38)$$

In this regime, the degree of a given node $i$ converges in probability to a Poisson random variable with mean $\alpha$

*Poissonian degree in the sparse case*

$$\mathbb{P}(d_i = d) \underset{n\to\infty}{\longrightarrow} e^{-\alpha}\frac{\alpha^d}{d!}\,.\qquad(1.39)$$

We say that a property holds with high probability (w.h.p) if it holds with a probability tending to 1 as $n \to \infty$. The Erdős-Rényi model undergoes several phase transitions when $n \to \infty$, as the average connectivity $\alpha$ is varied [41].

— For $\alpha < 1$, a typical realization G is w.h.p composed of a large number of small disconnected trees, each containing of the order of $\log n$ vertices.
— For $\alpha > 1$, G has w.h.p a unique *giant connected component* containing a finite fraction of the vertices.
— For $\alpha < \log(n)$, G contains w.h.p isolated vertices of degree 0.
— For $\alpha > \log(n)$, G is w.h.p connected.

*Phase transitions in the Erdős-Rényi model*

These phase transitions give us a clue about what to expect from an inference algorithm on a random graph G. As a generic example, anticipating on the following, assume that we are trying to infer latent labels $(\sigma_i)_{i\in[n]}$ carried by the nodes of the graph G, based on information (e. g. comparisons or similarities) encoded on the edges of G. If two vertices $i$ and $j$ belong to different connected components of the graph G, then there is no way to decide whether $\sigma_i = \sigma_j$ or not, since we do not have access to any (even indirect) comparison between the two vertices. As a consequence, we expect two different regimes.

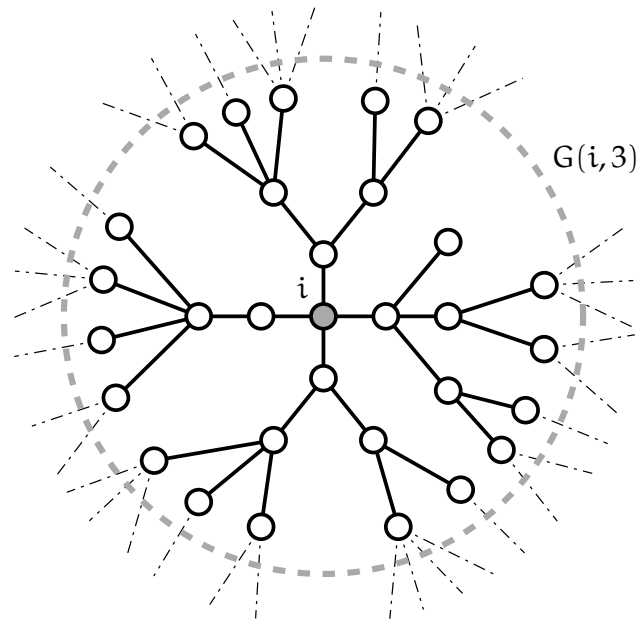

Figure 1.6 – Illustration of the locally tree-like property of a sparse random graph. The gray dashed circle contains the subgraph $G(i,3)$ of all nodes that are at distance at most 3 from $i$. $G(i,3)$ contains no cycle.

— $\alpha = O(1)$ is the regime of so-called *detectability*, in which we may hope to correctly infer the labels of a finite fraction of the vertices, but not all of them.

— $\alpha \sim \kappa \log n$ for some constant $\kappa > 1$ is the regime of so-called *perfect recovery*, in which we may hope to correctly infer the labels of all the vertices.

*Detectability vs. perfect recovery*

This hand-waving argument will be made more precise on the specific problems we consider in the following. The contributions of this dissertation are interested in the detectability regime.

One of the most interesting properties of random graphs with small average connectivity $\alpha = O(\log n)$ (including both detectability and perfect recovery regimes) is that they are *locally tree like*. More precisely, for a fixed $k \in \mathbb{N}$, and for some vertex $i$ chosen uniformly at random in $[n]$, let $G_{i,k}$ denote the subgraph of $G$ consisting of all the vertices and edges of $G$ that are at most $k$ steps away from $i$. Then for any fixed $k$, $G(i,k)$ w.h.p a tree. This property is depicted graphically on figure 1.6 and stems from the following heuristic exploration process. Starting from node $i$, we discover its neighbors by drawing (on average) $\alpha$ vertices in $[n]$ uniformly at random. Then we discover its neighbors' neighbors by repeating the same procedure for each neighbor of $i$. Assuming that each new neighbor has not been discovered before, we have discovered after $k$ iterations of the previous scheme $\alpha^k$ vertices on average. When $\alpha = O(\log n)$, the number of

*Locally tree-like property*

*j is k steps away from i if the shortest path connecting i and j in G is of length k*

discovered vertices is $o(n)$, so that w.h.p, they are all distinct, and there is no cycle in $G(i, k)$. This argument can be made precise, and a full proof can be found *e. g.* in [19]. An equivalent popular statement of the locally-tree like property is that there is w.h.p no cycle of length smaller than $\log_\alpha n$.

We now turn to generalizations of the Erdős-Rényi model that will allow us to encode information in the topology of the graph, or to assign labels to the edges of the graph. We stress that these generalized ensembles will inherit the properties of the Erdős-Rényi model of same average degree, including the thresholds for the emergence of the giant component and for the connectedness of the graph, as well as the locally-tree like property.

### 1.4.2   *The stochastic block model*

The stochastic block model (SBM) is a simple generalization of the Erdős-Rényi model in which the probability of presence of an edge between two vertices is not uniform, but depends on latent variables carried by the vertices. The stochastic block model (SBM) was first introduced in the context of mathematical sociology in [65] and has played a major role in the theoretical understanding of the performance of community detection algorithms [32]. Indeed, unlike the Erdős-Rényi model, the SBM allows to generate graphs with non-uniform density of edges, *e. g.* with communities of nodes more densely connected than the average. Specifically, an instance of the SBM is specified by

| | |
|---|---|
| $n$ | the number of vertices |
| $q$ | the number of communities |
| $(f_\sigma)_{\sigma \in [q]}$ | the relative sizes of the communities |
| $(p_{\sigma, \sigma'})_{\sigma, \sigma' \in [q]}$ | a $q \times q$ matrix of edge probabilities. |

A graph $G = ([n], E)$ is generated from the SBM in the following way. First we assign each vertex $i \in [n]$ to a community $\sigma_i \in [q]$ independently with probability $\mathbb{P}(\sigma_i = \sigma) = f_\sigma$, for $\sigma \in [q]$. Then for each pair of vertices $(ij)$ such that $1 \leqslant i < j \leqslant n$, we include the edge $(ij)$ in $E$ with probability $\mathbb{P}\left((ij) \in E \mid \sigma_i, \sigma_j\right) = p_{\sigma_i, \sigma_j}$. The aim is usually, given the graph, to recover the community memberships $(\sigma_i)_{i \in [n]}$ of the nodes.

We will be mainly interested in the sparse case, i. e. for $\sigma, \sigma' \in [q]$

$$p_{\sigma, \sigma'} = \frac{\alpha_{\sigma, \sigma'}}{n},$$ (1.40)

with $\alpha_{\sigma, \sigma'}$ fixed independently of $n$, although we will occasionally mention some results in the perfect recovery regime $\alpha_{\sigma, \sigma'} \sim \kappa \log n$. In the sparse case, the average connectivity of the graph is given by

$$\alpha = \sum_{\sigma, \sigma'=1}^{q} f_\sigma f_{\sigma'} \alpha_{\sigma, \sigma'} = O(1).$$ (1.41)

In a social network, $\alpha$ can be interpreted as the average number of friends a person in the network has[5]. While the sparse regime is more challenging, as the graph contains less information, it is also arguably the most relevant in real world applications. Indeed, if the total number of users of Facebook doubles overnight, we do not expect that the number of friends of a given user will also double. Said differently, the number of friends that a person can have is constrained by time, location and cognitive skills[6]. It is therefore reasonable to assume this number to be independent of the total number of people in the world.

*Relevance of the sparse assumption*

The properties of the Erdős-Rényi model $\mathcal{G}(n, \alpha/n)$ listed in the previous section also apply in the SBM with average connectivity $\alpha$. In particular, as argued in the previous section, we cannot hope to perfectly recover the community memberships of all the nodes in the sparse case. Instead, we may ask what conditions must verify the parameters $(f_\sigma)_{\sigma \in [q]}$ and $(\alpha_{\sigma,\sigma'})_{\sigma,\sigma' \in [q]}$ to *detect* the communities, i.e. to recover the community memberships $(\sigma_i)_{i \in [n]}$ better than by guessing randomly. Remarkably, based on arguments originating in statistical physics, [32] conjectured a very precise answer to this question. To state the result, we consider a particular, *symmetric* instance of the SBM, with equally sized communities ($f_\sigma = 1/q$, $\forall \sigma \in [q]$), and

*Symmetric SBM*

$$\alpha_{\sigma,\sigma'} = \alpha_{\text{in}} \, \mathbf{1}\left(\sigma = \sigma'\right) + \alpha_{\text{out}} \, \mathbf{1}\left(\sigma \neq \sigma'\right),\tag{1.42}$$

for $\sigma, \sigma' \in [q]$. In words, the probability of two nodes being connected takes only two different values, depending on whether the nodes belong in the same community or not. Therefore, $\alpha_{\text{in}}$ is interpreted as the average connectivity inside each community, while $\alpha_{\text{out}}$ is the average connectivity between different communities. The average connectivity in the whole graph is

$$\alpha = \frac{\alpha_{\text{in}} + (q-1)\alpha_{\text{out}}}{q}.\tag{1.43}$$

*Assortative and disassortative*

The case $\alpha_{\text{in}} > \alpha_{\text{out}}$ is termed *assortative*, and corresponds to the case where the nodes are more densely connected inside a community than the average. The opposite case $\alpha_{\text{in}} < \alpha_{\text{out}}$ is termed *disassortative*, and is useful to describe e. g. predator-prey, or mating networks. Examples of both cases are pictured on figure 1.7.

---

5. The SBM (sparse or not) has however many other limitations when it comes to modeling reality, e. g. the fact that it cannot account for the heavy-tailed degree distributions observed in real world networks.

6. In the 90s, the British anthropologist Robin Dunbar attempted to compute the maximum number of "stable" friends a person can have. He found a correlation between the size of a primate's brain and the average size of its social group [39]. By extrapolation, Dunbar suggested a cognitive limit of 150 as the maximum number of close friends a human being can keep. Later studies based on the exchange of Christmas cards in contemporary Western societies proved in remarkable agreement with this prediction [63].

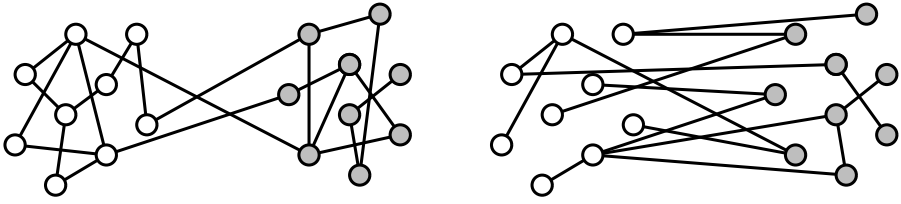

Figure 1.7 – Examples of assortative (left) and disassortative (right) communities.

The conjecture of [32] states that there exists a phase transition in detectability, in that a polynomial time algorithm can detect the communities in the limit $n \to \infty$ if and only if

$$|\alpha_{in} - \alpha_{out}| > q\sqrt{\alpha}. \qquad (1.44)$$

<div style="float:right; font-style:italic; width:20%">Detectability threshold in the SBM</div>

More precisely, if $q \leqslant 4$ in the assortative case (or $q \leqslant 3$ in the disassortative case), the authors of [89] claim that it is information-theoretically impossible to detect the communities unless (1.44) holds. For larger $q$, they argue that when below the transition (1.44), exponential algorithms such as exhaustive search can detect the communities, but no polynomial time algorithm can. A large part of this conjecture has now been proved rigorously [7]. More precisely, for the case of $q = 2$ communities, the impossibility result below the transition (1.44) was first proved in [110], and the detectability result was proved shortly later in [100, 111]. For larger $q$, [4] proved that a polynomial time algorithm (close to belief propagation, see section 1.5) succeeds in detecting the communities when (1.44) holds, and also supported the existence of an information-computation gap [8] for $q \geqslant 5$, by showing the existence of exponential time algorithms that can detect the communities when (1.44) does not hold.

<div style="float:right; font-style:italic; width:20%">Rigorous results on detectability in the SBM</div>

Finally, we mention for completeness that in the exact recovery regime $\alpha \sim \kappa \log n$, the situation is simpler, as there is no information-computation gap in the limit $n \to \infty$. More precisely, [3] showed that there exists a fundamental limit, expressed in terms of a certain divergence function, such that no algorithm can perfectly recover the communities below this limit, while a polynomial time algorithm can do so when above this limit.

<div style="float:right; font-style:italic; width:20%">Rigorous results on perfect recovery in the SBM</div>

### 1.4.3  *The labeled stochastic block model*

The labeled stochastic block model (lSBM), first introduced in [60], is a generalization of the SBM that allows to generate labeled graphs. A graph generated from the labeled stochastic block model (lSBM)

---

7. Note that a rigorous proof of the full conjecture implies showing that $P \neq NP$.

8. An information-computation gap is a mismatch between what is information-theoretically possible and what (known) polynomial time algorithms can achieve.

contains two sources of information. The first one is encoded in the topology of the graph, i. e. in the presence or absence of edges, just like in the original SBM. The second one stems from the values of the labels carried by the edges, which are assumed to be random variables with a distribution that depends only on the labels of the endpoints of the edges. More precisely, an instance of the lSBM is specified by

*Parameters of the lSBM*

| | |
|---|---|
| $n$ | the number of vertices |
| $q$ | the number of communities |
| $(f_\sigma)_{\sigma \in [q]}$ | the relative sizes of the communities |
| $\left(p_{\sigma,\sigma'}^{(e)}\right)_{\sigma,\sigma' \in [q]}$ | a $q \times q$ matrix of edge probabilities |
| $\mathcal{L}$ | a set of labels |
| $\left(p_{\sigma,\sigma'}^{(\ell)}\right)_{\sigma,\sigma' \in [q]}$ | a $q \times q$ matrix of label probability distributions |

The first four parameters are the same as the SBM — note that we now denote the edge probabilities $p^{(e)}$ to differentiate them from the label probabilities $p^{(\ell)}$. We use these parameters to assign each vertex $i \in [n]$ to a community $\sigma_i$, and to generate a graph $G = ([n], E)$ from the SBM (see section 1.4.2). Then, for each edge $(ij) \in E$, we draw a label $\ell_{ij}$ with probability $\mathbb{P}(\ell_{ij} = \ell \mid \sigma_i, \sigma_j) = p_{\sigma_i,\sigma_j}^{(\ell)}(\ell)$, for $\ell$ in the label set $\mathcal{L}$. Like in the SBM, the aim is, given the thus constructed labeled graph, to recover the community memberships $\sigma = (\sigma_i)_{i \in [n]}$ of the nodes. Again, we will be interested in the following scaling, for $\sigma, \sigma' \in [q]$,

$$p_{\sigma,\sigma'}^{(e)} = \frac{\alpha_{\sigma,\sigma'}}{n} \,, \tag{1.45}$$

with $\alpha_{\sigma,\sigma'} = O(1)$ (detectability), and only mention some results in the regime $\alpha_{\sigma,\sigma'} \sim \kappa \log n$ (perfect recovery). Like the SBM, this model is thought to exhibit phase transitions, both for detectability and perfect recovery, although it has been comparatively less studied than the SBM.

In the detectability regime, the different conjectures and results about this model mostly concern the *symmetric* lSBM defined for $\sigma, \sigma' \in [q], \ell \in \mathcal{L}$ by

*Symmetric lSBM*

$$f_\sigma = \frac{1}{q}$$
$$\alpha_{\sigma,\sigma'} = \alpha_{\text{in}} \mathbf{1}\left(\sigma = \sigma'\right) + \alpha_{\text{out}} \mathbf{1}\left(\sigma \neq \sigma'\right) \,, \tag{1.46}$$
$$\forall \ell \in \mathcal{L}, \quad p_{\sigma,\sigma'}^{(\ell)}(\ell) = p_{\text{in}}(\ell) \mathbf{1}\left(\sigma = \sigma'\right) + p_{\text{out}}(\ell) \mathbf{1}\left(\sigma \neq \sigma'\right) \,.$$

In the symmetric lSBM with $q = 2$ clusters and discrete edge set $\mathcal{L}$, [60] first conjectured that detectability is possible if and only if

*Conjectured detectability threshold in the lSBM*

$$\tau = \frac{1}{2} \sum_{\ell \in \mathcal{L}} \frac{(\alpha_{\text{in}} \, p_{\text{in}}(\ell) - \alpha_{\text{out}} \, p_{\text{out}}(\ell))^2}{\alpha_{\text{in}} \, p_{\text{in}}(\ell) + \alpha_{\text{out}} \, p_{\text{out}}(\ell)} > 1 \,. \tag{1.47}$$

The negative part of this conjecture (impossible detectability when $\tau < 1$) was proved in [94], who also exhibited three algorithms that can detect the hidden assignment $\sigma$ when $\tau$ is larger than some constant, strictly larger than 1.

In the following, we will be mostly interested in the special case where $\alpha_{\sigma,\sigma'} = \alpha, \forall \sigma, \sigma' \in [q]$, so that there is no information in the topology of the graph, only in the labels carried by the edges. For the case of $q = 2$ communities, we will introduce provably optimal algorithms that can detect $\sigma$ as soon as $\tau > 1$, and also extend the conjecture (1.47) to the case of real labels, and $q > 2$ communities. This model includes as a special case the planted spin glass introduced in section 1.6.5, called *censored block model* in the information theory community. In the censored block model, $\mathcal{L} = \{\pm 1\}$, and $p_{in}(+1) = p_{out}(-1) = 1 - \epsilon$, for some $\epsilon > 0$. This model was first studied in the perfect recovery regime $\alpha \sim \kappa \log n$ in [1, 56], who showed that there exists a phase transition between a phase in which no algorithm can perfectly recover $\sigma$, and a phase in which efficient algorithms can. We will study the detectability regime of the planted spin glass in details in chapter 4. The perfect recovery regime of the general lSBM was studied in [163] who identified a phase transition in terms of a divergence function, generalizing the results of [3] for the SBM.

*Relation to the planted spin glass, or censored block model*

## 1.5 MEAN-FIELD APPROXIMATIONS

We now motivate and introduce a general class of approximate inference methods, and discuss the belief propagation algorithm and Bethe approximation which play a substantial role in the following.

### 1.5.1 *Intractability of exact inference*

We have shown on the example of the previous section that there is valuable information to be extracted from the marginals of a pairwise MRF. The gigantic caveat that we have swept under the rug is that the task of computing the marginals of a multivariate (discrete) distribution is in general exponentially hard[9] (meaning that this operation has a time complexity that is exponential in the number of variables). This follows directly from the definition, since the marginal of $\sigma_i$ for some $i \in [n]$ is given by

$$\mathbb{P}_i(\sigma_i) = \sum_{\sigma_1 \in \mathcal{X}} \dots \sum_{\sigma_{i-1} \in \mathcal{X}} \sum_{\sigma_{i+1} \in \mathcal{X}} \dots \sum_{\sigma_n \in \mathcal{X}} \mathbb{P}(\sigma_1, \dots, \sigma_n), \quad (1.48)$$

*Marginal*

---

9. Our applications will require computing single-variable marginals of the form $\mathbb{P}_i(\sigma_i)$, which is why we focus here on this problem. However, the following remarks apply to other inference problems, such as computing the marginal distribution of a set of variables (conditioned or not on another set of variables), or computing modes of a JPD.

so that computing this marginal requires summing over $|\mathcal{X}|^{n-1}$ terms. When the JPD has strong independence properties, as in the case of a pairwise MRF, it is possible to leverage this structure to *order* the sums in expression (1.48) in a clever way that reduces the overall complexity of the procedure. An extreme example of such a simplification is the case where the pairwise MRF factorizes according to a graph that is a tree. In this case, computing the marginals of the pairwise MRF can be done in time *linear* [10] in the number of variables using a dynamic programming procedure called the *sum-product algorithm*, or belief propagation (BP) [124, 125] [11].

*Tree graphical models*

Let us detail the procedure for the pairwise MRF of equation (1.11), defined on a given graph $G = ([n], E)$, with single-variable potentials $(\phi_i)_{i \in [n]}$ and pairwise potentials $(\psi_{ij})_{(ij) \in E}$. Here and throughout the dissertation, we denote by $\vec{E}$ the set of directed edges of the graph $G$. The BP algorithm iteratively updates *beliefs* $b_{i \to j}$ which are discrete distributions on $\mathcal{X}$, defined on the directed edges $(i \to j) \in \vec{E}$ of the graph $G$. These beliefs can be interpreted [103] as the marginal distribution of $\sigma_i$ in a "cavity" graph in which the potential $\psi_{ij}$ has been removed. When the graph $G$ is a tree, it is possible to show that that the beliefs must verify the following fixed point equation [12], for all $(i \to j) \in \vec{E}$

*Belief propagation*

*BP fixed point equations*

$$b_{i \to j}(\sigma_i) = \frac{1}{\mathcal{Z}_{i \to j}} \phi_i(\sigma_i) \prod_{k \in \partial i \setminus j} \sum_{\sigma_k \in \mathcal{X}} \psi_{ik}(\sigma_i, \sigma_k) b_{k \to i}(\sigma_k), \quad (1.49)$$

where $\partial i$ denotes the set of neighbors of vertex $i$ in the graph $G$, and the normalization $\mathcal{Z}_{i \to j}$ ensures that the belief $b_{i \to j}$ is normalized, i.e.

$$\mathcal{Z}_{i \to j} = \sum_{\sigma \in \mathcal{X}} b_{i \to j}(\sigma). \quad (1.50)$$

If we find beliefs that verify the fixed point equation (1.49), we can use them to recover the marginals of the pairwise MRF (1.11). More precisely, when the graph $G$ is a tree, we have $\mathbb{P}_i(\sigma_i) = b_i(\sigma_i)$ where

*BP estimates for the marginals*

$$b_i(\sigma_i) = \frac{1}{\mathcal{Z}_i} \phi_i(\sigma_i) \prod_{k \in \partial i} \sum_{\sigma_k \in \mathcal{X}} \psi_{ik}(\sigma_i, \sigma_k) b_{k \to i}(\sigma_k). \quad (1.51)$$

*BP is exact on trees*

In practice, the BP algorithm iterates equation (1.49) until a fixed point is reached, and outputs the marginals given by (1.51). As long as the

---

10. Each iteration of belief propagation (BP) is linear in the number of variables $n$ on a tree.

11. A variant of the sum-product algorithm allows to compute the modes of MRF.

12. Note that we write here BP in a "product-sum" rather than "sum-product" form. We recover the perhaps more usual form of [161], equation (14), by defining messages $m_{i \to j}(\sigma_j) = \sum_{\sigma_i \in \mathcal{X}} \psi_{ij}(\sigma_i, \sigma_j) b_{i \to j}(\sigma_i)$ for $(i \to j) \in \vec{E}$, so that the fixed point equation (1.49) becomes

$$m_{i \to j}(\sigma_j) = \frac{1}{\mathcal{Z}_{i \to j}} \sum_{\sigma_i \in \mathcal{X}} \phi_i(\sigma_i) \psi_{ij}(\sigma_i, \sigma_j) \prod_{k \in \partial i \setminus j} m_{k \to i}(\sigma_i).$$

graph G is a tree, this procedure is guaranteed to converge, and to give exact marginals [103].

When the graph G is not a tree, there exists a general procedure to associate the pairwise MRF with a distribution that factorizes according to a tree. This procedure, called the *junction tree algorithm*, allows to perform exact inference, *e. g.* by running BP on the resulting tree. A description of this procedure is beyond the scope of this dissertation, and we refer the interested reader to [151] for the details. We however remark that the complexity of the junction tree algorithm is exponential in a fundamental graph-theoretic quantity called the *treewidth* of the graph, which is one less than the size of the largest clique in the junction tree representation of the pairwise MRF. Note that when building a junction tree, we introduce additional edges in the graph G, so that the treewidth of G is *not* the size of the largest clique in G (it is larger). For many practical applications, including the computer vision problem of the previous section, defined on a regular lattice, and the problems studied in this dissertation, defined on random graphs, the treewidth is large [151], so that the junction tree algorithm is intractable.

*Junction tree algorithm*

*Treewidth of a graph*

In general, we therefore have to resort to approximate inference methods. For instance, we may close our eyes on the fact that our pairwise MRF does not factorize according to a tree, and run the BP algorithm, which in this context, is sometimes called *loopy* BP [13]. As before, we iterate equation (1.49) until we find a fixed point, and use it to output the *pseudo*-marginals $b_i$ for $i \in [n]$ given by equation (1.51). *If* the algorithm converges, although we do not expect that $\mathbb{P}_i(\sigma_i) = b_i(\sigma_i)$, we may hope that $\mathbb{P}_i(\sigma_i) \approx b_i(\sigma_i)$, especially if the graph G is not too different from a tree, i. e. if it has few and long loops. This will in particular be the case for the sparse random graphs we will consider in the following. Applying BP as a generic inference engine on general graphs has often been rewarded with impressive success, especially in coding theory [101, 129], but not only [32, 103]. It turns out that this approach is inscribed in a more general approximation framework originating in statistical physics, from which it inherited the name of *mean-field approximations*. It is the purpose of the present section to introduce this framework.

*"Loopy" BP*

### 1.5.2 *Variational mean-field*

We now show that the distribution $\mathbb{P}$ defined by the pairwise MRF is the unique solution of a variational problem. In this section we let $Q$ denote an arbitrary trial distribution over $\mathfrak{X}^n$.

---

13. In this dissertation, we will refrain from using the attribute "loopy" since it is after all exactly the same algorithm.

We define the KL divergence from $\mathbb{P}$ to $\mathbb{Q}$ as

$$D_{KL}(\mathbb{Q} \parallel \mathbb{P}) = \sum_{\sigma \in \mathcal{X}^n} \mathbb{Q}(\sigma) \log \frac{\mathbb{Q}(\sigma)}{\mathbb{P}(\sigma)} \,. \tag{1.52}$$

The KL divergence can be thought of as a sort of distance between probability distributions, with the caveat that it is not symmetric under the exchange of $\mathbb{P}$ and $\mathbb{Q}$. Its most useful property in the following is the fact that the KL divergence between any two distributions

$D_{KL}(\mathbb{Q} \parallel \mathbb{P}) \geqslant 0$    $\mathbb{P}$ and $\mathbb{Q}$ is always non-negative, and it vanishes if and only if $\mathbb{P} = \mathbb{Q}$. This result, known as *Gibb's inequality*, can be obtained as a consequence of Jensen's inequality. Inserting in the definition of the KL divergence the exponential form (1.13) of the pairwise MRF $\mathbb{P}$, with energies $(\epsilon_i)_{i \in [n]}, (\epsilon_{ij})_{(ij) \in E}$ given by (1.14), we find

$$D_{KL}(\mathbb{Q} \parallel \mathbb{P}) = \sum_{\sigma \in \mathcal{X}^n} \mathbb{Q}(\sigma) \log \mathbb{Q}(\sigma) + \sum_{\sigma \in \mathcal{X}^n} \mathbb{Q}(\sigma) \log \mathcal{Z}$$
$$+ \sum_{(ij) \in E} \sum_{\sigma \in \mathcal{X}^n} \mathbb{Q}(\sigma)\, \epsilon_{ij}(\sigma_i, \sigma_j) + \sum_{i=1}^{n} \sum_{\sigma \in \mathcal{X}^n} \mathbb{Q}(\sigma)\, \epsilon_i(\sigma_i) \,, \tag{1.53}$$

$$= \log \mathcal{Z} + E[\mathbb{Q}] - S[\mathbb{Q}] \,, \tag{1.54}$$

where $S[\mathbb{Q}]$ is the entropy (1.20) of the trial distribution $\mathbb{Q}$, and we have defined its *internal energy* $E[\mathbb{Q}]$ as the sum of the expectations (under $\mathbb{Q}$) of the energies $\epsilon_i, \epsilon_{ij}$, i.e.

$$E[\mathbb{Q}] = \sum_{(ij) \in E} \sum_{\sigma \in \mathcal{X}^n} \mathbb{Q}(\sigma)\, \epsilon_{ij}(\sigma_i, \sigma_j) + \sum_{i=1}^{n} \sum_{\sigma \in \mathcal{X}^n} \mathbb{Q}(\sigma)\, \epsilon_i(\sigma_i) \,. \tag{1.55}$$

We can finally define two quantities that play a major role in statistical physics. The first one is the *Helmholtz* free energy

$$F = -\log \mathcal{Z} \,, \tag{1.56}$$

and the second is the *Gibbs* free energy, which is a functional defined on the set of probability distributions over $\mathcal{X}^n$ and given by

$$G[\mathbb{Q}] = E[\mathbb{Q}] - S[\mathbb{Q}] \,. \tag{1.57}$$

We have therefore shown the following variational formulation. For any trial probability distribution $\mathbb{Q}$ over $\mathcal{X}^n$,

$$G[\mathbb{Q}] \geqslant F \,, \tag{1.58}$$

with equality if and only if $\mathbb{Q} = \mathbb{P}$. In particular, it holds that

$$\mathbb{P} = \operatorname{argmin}_{\mathbb{Q}} G[\mathbb{Q}] \,, \tag{1.59}$$

where the optimization is over the set of all probability distributions on $\mathcal{X}^n$. Once more, this optimization problem is intractable, since the mere evaluation of the Gibbs free energy of a distribution requires summing over $\mathfrak{X}^n$ terms. However, this variational formulation inspired a useful class of approximations termed *variational mean-field*.

The idea of variational mean-field is to make the optimization problem (1.59) tractable by restricting the set of possible choices for $Q$ to a smaller and manageable set. The simplest non-trivial choice is to consider *independent* (also called factorized [14]) distributions of the form

$$Q(\sigma) = \prod_{i=1}^{n} b_i(\sigma_i),\qquad(1.60)$$

where the $b_i$ for $i \in [n]$ verify $\sum_{\sigma_i \in \mathfrak{X}} b_i(\sigma_i) = 1$ and are therefore the marginals of the distribution $Q$. Solving the optimization problem (1.59) over this restricted class of distributions amounts to finding the independent distribution $Q^\star$ which is closest to the true pairwise MRF $\mathbb{P}$, in the sense of the KL divergence. By injecting the form (1.60) in the definition of the Gibbs free energy (1.57), it is straightforward to check that the solution of the restricted optimization problem is the distribution $Q^\star$ that maximizes the so-called naive mean-field (nMF) free energy

$$
\begin{aligned}
\mathcal{F}^{nMF}\big((b_i)_{i\in[n]}\big) = &\sum_{(ij)\in E}\sum_{\sigma_i,\sigma_j\in\mathfrak{X}} b_i(\sigma_i)\,b_j(\sigma_j)\,\epsilon_{ij}(\sigma_i,\sigma_j)\\
&+\sum_{i=1}^{n}\sum_{\sigma_i\in\mathfrak{X}} b_i(\sigma_i)\,\epsilon_i(\sigma_i) + \sum_{i=1}^{n}\sum_{\sigma_i\in\mathfrak{X}} b_i(\sigma_i)\,\log b_i(\sigma_i).
\end{aligned}
\qquad(1.61)
$$

The main advantage of this nMF approximation is that the evaluation of the Gibbs free energy is now tractable, and the optimization problem has become overall much easier to solve, since a distribution of the independent form (1.60) is specified by $n(\mathfrak{X}-1)$ real numbers instead of $\mathfrak{X}^n - 1$ for a generic distribution. Additionally, once we have found a solution to the restricted optimization problem, it is trivial to produce estimates for the marginals of $\mathbb{P}$, which we may approximate by the marginals $b_i^\star$ of $Q^\star$.

Unfortunately, in practice, employing the nMF approximation is often a bad idea [15], since we are effectively assuming that the variables $\sigma_i$ are independent. A better approach, which allows for correlation between the variables, is provided by the *Bethe approximation*, and is at the heart of the methods developed in this dissertation.

---

14. In the next chapter, we give a definition of factorized models that generalizes condition (1.60). This is why we use the non-standard adjective "independent" to refer to these trial distributions.

15. A notable exception is the Curie-Weiss, or fully connected ferromagnetic Ising model, for which the nMF approach is exact (see section 1.6.3).

### 1.5.3 *The Bethe approximation and belief propagation*

The Bethe approximation takes a slightly different route than the previous nMF approximation. More precisely, let us assume for now that the graph $G = ([n], E)$ is a tree. Then it can be shown [103] that a trial distribution $\mathbb{Q}$ that factorizes as a pairwise MRF according to the graph $G$ can be written in the form

*Factorization of a MRF on a tree*

$$\mathbb{Q}(\sigma) = \frac{\prod_{(ij) \in E} b_{ij}(\sigma_i, \sigma_j)}{\prod_{i=1}^n b_i(\sigma_i)^{|\partial_i|-1}}, \qquad (1.62)$$

where the $b_{ij}$ for $(ij) \in E$ are called the *beliefs* for $b_i$ for $i \in [n]$ reasons that will be made clear shortly, and are equal respectively to the pairwise and single-variable marginals of $\mathbb{Q}$. In particular, the beliefs should verify the normalization and consistency conditions

*Consistency constraints on the beliefs*

$$\sum_{\sigma_j} b_{ij}(\sigma_i, \sigma_j) = b_i(\sigma_i), \qquad \forall (ij) \in E, \sigma_i \in \mathcal{X},$$

$$\sum_{\sigma_i} b_i(\sigma_i) = 1, \qquad \forall i \in [n]. \qquad (1.63)$$

Note that these two sets of conditions ensure that the pairwise marginals are also correctly normalized. Injecting the form (1.62) in the definition (1.57) of the Gibbs free energy yields the so-called *Bethe free energy*

*Bethe free energy*

$$\mathcal{F}^{\text{Bethe}}\left((b_i)_{i \in [n]}, (b_{ij})_{(ij) \in E}\right) = \sum_{(ij) \in E} \sum_{\sigma_i, \sigma_j \in \mathcal{X}^n} b_{ij}(\sigma_i, \sigma_j) \, \epsilon_{ij}(\sigma_i, \sigma_j)$$

$$+ \sum_{i=1}^n \sum_{\sigma_i \in \mathcal{X}} b_i(\sigma_i) \, \epsilon_i(\sigma_i) + \sum_{(ij) \in E} \sum_{\sigma_i, \sigma_j \in \mathcal{X}^n} b_{ij}(\sigma_i, \sigma_j) \, \log b_{ij}(\sigma_i, \sigma_j)$$

$$+ \sum_{i=1}^n (|\partial_i| - 1) \sum_{\sigma_i \in \mathcal{X}} b_i(\sigma_i) \, \log b_i(\sigma_i). \qquad (1.64)$$

To sum up, as long as $G$ is a tree, we therefore have that $\mathbb{P} = \mathbb{Q}^\star$ where $\mathbb{Q}^\star$ is defined by the decomposition (1.62) where the beliefs are those that minimize the Bethe free energy (1.64), under the constraints (1.63). In the *Bethe approximation*, we assume that even when $G$ is not a tree, we can still approximate $\mathbb{P}$ by the "distribution" $\mathbb{Q}^\star$ thus defined. This is a strong assumption, since the form (1.62) does not even, in general, define a correctly normalized distribution when $G$ is not a tree [16].

*Bethe approximation*

*Relation between the Bethe approximation and BP*

However, there is reason to believe that the Bethe approximation is useful in many problems, because it is closely related to the very successful loopy BP strategy introduced previously. More precisely, it can be shown that there is a deep connection between the marginals estimated through BP and the stationary points of the Bethe free energy (1.64). This connection was first intimated by [73] in a special case, and later formalized by [161] in its full generality. Their result can be stated informally as follows.

---

16. In particular, the Bethe free energy is in general not a bound on the Helmholtz free energy.

**Proposition 1.5.1.** *There is a one-to-one correspondence between the fixed points of BP and the stationary points of the Bethe free energy.*

In order to make the correspondence explicit, one writes down a Lagrangian associated with the minimization of the Bethe free energy (1.64) under the constraints (1.63). By setting the derivatives of this Lagrangian to 0, one obtains a closed set of equations that allow to relate bijectively the fixed points of BP to the Lagrange multipliers [161]. Interestingly, when the graph G is a tree, or when it has a single loop, it is possible to show [62] that the Bethe free energy is convex over the set of beliefs verifying the constraints (1.63). In particular, in this case, BP has a unique fixed point, corresponding to the unique and global minimum of the Bethe free energy. In general however, the Bethe free energy is non-convex and has multiple local minima. Approximating these local minima will be our major concern starting from the next chapter. When the Bethe free energy is not convex, it is still possible to show that the *stable* fixed points of BP are local minima of the Bethe free energy [61]. The converse is in general not true, and we will see illustrations of this fact in chapter 2.

*(Non)-convexity of the Bethe free energy*

In the next section, we consider a special pairwise MRF, the Ising model, and discuss a different approach to mean-field approximations in this special case, before stating some general results about the phase diagram of the Ising model in its different flavors.

## 1.6 THE ISING MODEL: SOME PHYSICS

Pairwise models of the exponential form (1.13), also called Boltzmann distributions, have a century old history in statistical physics, where they have been studied extensively as idealized models of systems with interacting particles. In the special case where the alphabet $\mathfrak{X}$ consists of 2 possible values $\mathfrak{X} = \{\pm 1\}$, and the pairwise energies are symmetric, $\epsilon_{ij}(\sigma_i, \sigma_j) = \epsilon_{ij}(\sigma_j, \sigma_i)$, the distribution (1.13) can be written in the familiar form of the celebrated Ising model

$$\mathbb{P}(\sigma) = \frac{1}{\mathcal{Z}} \exp\left( \beta \sum_{(ij) \in E} J_{ij} \sigma_i \sigma_j + \sum_{i=1}^{n} h_i \sigma_i \right), \qquad (1.65)$$

*Ising model*

where the $(J_{ij})_{(ij) \in E}$ are called *couplings* and the $(h_i)_{i \in [n]}$ are called *fields*. The variables $(\sigma_i)_{i \in [n]}$ are usually called *spins*. The parameter $\beta$ controls the strength of the interactions between the spins, and is in-terpreted physically as the inverse of the temperature. The expected value of a spin [17]

$$m_i = \mathbb{E}[\sigma_i], \qquad \text{for } i \in [n] \tag{1.66}$$

is called its *magnetization*. Although the magnetizations alone do not fully determine the Ising model (1.65) (one also needs to specify the correlations), we will see that a great deal of insight lies in understanding the qualitative behavior of these quantities as we vary the inverse temperature $\beta$. The Ising model is a typical pairwise MRF, and captures most of the computational challenges of their study. Since this model will be central in the following, we list here some of its qualitative properties, and some of its variations, that will make it interesting for our applications. Before doing so, we introduce a different point of view on the mean-field approximation framework developed in the previous section. For simplicity (and also because it suits our upcoming applications), we consider throughout the section the case of vanishing fields $h_i = 0$ for $i \in [n]$.

### 1.6.1  *A different approach to mean-field approximations*

In the case of the Ising model, it is possible to derive another systematic approach to the class of mean-field approximations introduced in section 1.5, which we now present succinctly. This approach relies on an expansion of a free energy functional expressed in terms of the magnetizations $m = (m_i)_{i \in [n]}$. More precisely, from the variational formulation (1.58), it is straightforward [121] to see that

$$F = \inf_{m \in [-1,1]^n} \mathcal{F}(m), \tag{1.67}$$

where $F$ is the Helmholtz free energy of the Ising model, and we have defined a free energy functional $\mathcal{F}(m)$ by

$$\mathcal{F}(m) = \inf_{Q \,|\, \forall i \in [n], \, \mathbb{E}_Q[\sigma_i] = m_i} G[Q], \tag{1.68}$$

where $G[Q]$ is the Gibbs free energy of the trial distribution $Q$. This amounts to solving the variational problem (1.59) in two steps. We first minimize the Gibbs free energy over the trial distributions $Q$ with prescribed magnetizations, and we then minimize the resulting functional $\mathcal{F}(m)$ over the magnetizations. In particular, from (1.67), the magnetizations of the Ising model (1.65) are those that minimize the free energy functional $\mathcal{F}(m)$. Unfortunately, computing $\mathcal{F}(m)$ is just as intractable as minimizing the full Gibbs free energy.

However, the advantage of this two-step approach is that it is possible to perform a systematic small couplings expansions [51, 126]

---

17. There is a catch in this definition related to ergodicity breaking, which we explain in the following.

of the free energy functional $\mathcal{F}(\mathbf{m})$, which, when truncated at a correct order depending on the particular instance of the Ising model, provide tractable approximate free energies. One possible such expansion is the *Plefka expansion* [126], which first orders are

$$\mathcal{F}_0(\mathbf{m}) = \sum_{i=1}^{n} \frac{1+m_i}{2} \log \frac{1+m_i}{2} + \frac{1-m_i}{2} \log \frac{1-m_i}{2}, \qquad (1.69)$$

$$\mathcal{F}_1(\mathbf{m}) = -\sum_{(ij) \in E} J_{ij} m_i m_j, \qquad (1.70)$$

$$\mathcal{F}_2(\mathbf{m}) = -\frac{1}{2} \sum_{(ij) \in E} J_{ij}^2 \left(1 - m_i^2\right) \left(1 - m_j^2\right) \qquad (1.71)$$

*Plefka expansion up to order 2*

Note that the zeroth order is simply the entropy of $n$ independent random signs. Truncating the expansion to first order, we recover the nMF free energy of the Ising model

$$\begin{aligned}
\mathcal{F}^{\mathrm{nMF}}(\mathbf{m}) &= \mathcal{F}_0(\mathbf{m}) + \mathcal{F}_1(\mathbf{m}) \\
&= -\sum_{(ij) \in E} J_{ij} m_i m_j \\
&\quad + \sum_{i=1}^{n} \frac{1+m_i}{2} \log \frac{1+m_i}{2} + \frac{1-m_i}{2} \log \frac{1-m_i}{2}.
\end{aligned} \qquad (1.72)$$

*The first order in the Plefka expansion is the nMF approximation*

Indeed, it is straightforward to check that this last expression equals the nMF free energy $\mathcal{F}^{\mathrm{nMF}}((b_i)_{i \in [n]})$ of equation (1.61) after noting that for binary spins $\sigma_i \in \{\pm 1\}$, we can write

$$b_i(\sigma_i) = \frac{1 + m_i \sigma_i}{2}. \qquad (1.73)$$

Truncating the expansion at second order, we obtain a more accurate mean-field approximation called the Thouless-Anderson-Palmer (TAP) approximation [149]

$$\begin{aligned}
\mathcal{F}^{\mathrm{TAP}}(\mathbf{m}) &= \mathcal{F}_0(\mathbf{m}) + \mathcal{F}_1(\mathbf{m}) + \mathcal{F}_2(\mathbf{m}) \\
&= -\sum_{(ij) \in E} J_{ij} m_i m_j - \frac{1}{2} \sum_{(ij) \in E} J_{ij}^2 \left(1 - m_i^2\right)\left(1 - m_j^2\right) \\
&\quad + \sum_{i=1}^{n} \frac{1+m_i}{2} \log \frac{1+m_i}{2} + \frac{1-m_i}{2} \log \frac{1-m_i}{2}.
\end{aligned} \qquad (1.74)$$

*The second order in the Plefka expansion is the TAP approximation*

This free energy played an important role in the understanding of the Sherrington-Kirkpatrick model of spin glasses, which we introduce in the following. Interestingly, by summing a certain family of terms in the Plefka expansion, including the previous first order terms as well as other particular terms of arbitrary order in the couplings, one

*Recovering the Bethe approximation*

can also recover the Bethe free energy of equation (1.64) [126, 128] [18].
In section 2.3.3, we will compare spectral methods derived from the
nMF and the TAP approximations to those derived from the Bethe ap­
proximation, and discuss the superiority of the latter on the sparse
graphs we consider. In the rest of this section, we introduce vari­
ous types of Ising models and discuss the qualitative behavior of the
corresponding magnetizations.

### 1.6.2    The ferromagnetic Ising model

*Ferromagnetic Ising model*

When the couplings are positive ($J_{ij} > 0$ for $(ij) \in E$), the Ising
model (1.65) is called *ferromagnetic*. This case corresponds to the orig­
inal physical motivation of the Ising model, as a toy model for mag­
nets, in which each spin $\sigma_i$ for $i \in [n]$ represents the magnetic mo­
ment of an atom $i$ composing the medium. In this physical interpre­
tation, the graph $G$ is usually taken to be a regular lattice, mimicking
the positions of the atoms in an ordered medium, and the positive
couplings encourage neighboring atoms to have equal spins. In chap­
ter 5, we will however be interested in an Ising ferromagnet defined
on a sparse random graph. The magnetic quality of an Ising model
is controlled by its total average *magnetization*, defined as

*Total average magnetization*

$$M_n = \frac{1}{n} \sum_{i=1}^{n} m_i \,, \tag{1.75}$$

Note that the total average magnetization can also be written $M_n = \mathbb{E}[m[\sigma]]$ were the expectation is over the Boltzmann distribution (1.65),
and $m[\sigma]$ is the total magnetization of a single configuration

*Total magnetization of a configuration*

$$m[\sigma] = \frac{1}{n} \sum_{i=1}^{n} \sigma_i \,. \tag{1.76}$$

*Global order*

The average total magnetization (1.75) is therefore a measure of the
*global order* in the material: a large magnetization (in absolute value)
means that a large number of spins are typically in the same state. At
large temperature (small $\beta$), the spins are nearly independent, so that
we expect the magnetization to vanish for a typical configuration of
the material. As we decrease the temperature (increase $\beta$), the cou­
plings between neighboring spins become stronger, and large clusters
of aligned spins can appear. Physicists are usually interested in the
so-called *thermodynamic* limit where the number of atoms $n \to \infty$ [19].

*Thermodynamic limit and phase transitions*

---

18. Note that the Bethe free energy (1.64) is expressed, in the case of the Ising
model, in terms of both the magnetizations and the correlations between spins. The
expression we recover from Plefka's expansion corresponds to the Bethe free energy
evaluated at the correlations that minimize the Bethe free energy at fixed magnetiza­
tions [154, 155].

19. The thermodynamic limit is justified in practice by the value of the Avogadro
constant: one gram of matter contains approximately $10^{23}$ atoms.

In this limit, an interesting phenomenon known as a *phase transition*[20] can occur. More precisely, depending on the topology of the graph G and the values of the couplings, there may exist a critical (inverse) temperature $\beta_c$ such that,

$$\beta < \beta_c \implies \lim_{n \to \infty} M_n = 0 \text{ and } m_i = 0, \forall i \in [n] \qquad (1.77)$$

$$\beta > \beta_c \implies \lim_{n \to \infty} M_n \neq 0 \qquad (1.78)$$

The Ising model can therefore capture the dependence on temperature of the magnetic properties of a physical medium. When $\beta < \beta_c$, we say that the medium is in the *paramagnetic* phase, as the global magnetization vanishes. On the other hand, when $\beta > \beta_c$, the medium behaves like a magnet, and is said to be in the *ferromagnetic* phase. Interestingly, we will see in the following that exactly the same kind of phase transition phenomena explains why some inference problems are solvable while some others are not.

*Paramagnetic to ferromagnetic phase transition*

The alert reader may have noticed that, under the assumption of vanishing fields, the distribution (1.65) is invariant under a global flipping of the spins $\sigma_i \to -\sigma_i$, so that for any finite $n$ and any $\beta$, the magnetizations rigorously vanish. The existence of a ferromagnetic phase with finite magnetization therefore appears impossible at first sight. What makes it possible is a phenomenon known as *ergodicity breaking*. In the limit $n \to \infty$, when $\beta > \beta_c$, the support of the Boltzmann distribution (1.65) splits into two disconnected sets, called *pure states*, or ergodic components[21]. One of these sets consists of configurations $\sigma$ such that $m[\sigma] > 0$, while the other consists of configurations such that $m[\sigma] < 0$. The two pure states are separated by a set of configurations with vanishing probability. When multiple pure states exist, the Boltzmann distribution (1.65) must be restricted to one of them, and the magnetization must be computed within each of the pure states. Ergodicity breaking, and the fact that the Ising model can have non-vanishing magnetizations, will prove of paramount importance in the following. Apologizing to the reader familiar with statistical physics, we shall dwell a little more on this phenomenon by examining, in the next section, a simple special case of the Ising model for which we can compute analytically the magnetizations.

*The catch in the definition of the magnetizations*

*Ergodicity breaking*

*Pure states*

### 1.6.3    *The Curie-Weiss model*

To understand how ergodicity breaks and pure states emerge, it is instructive to look at one of the simplest realizations of the Ising model, called the *Curie-Weiss* model. The Curie-Weiss model corresponds to the Ising distribution (1.65) in the special case where the

*Definition of the Curie-Weiss model*

---

20. Phase transitions are signaled by a non-analyticity (as a function of $\beta$) of the Helmholtz free energy.

21. Ergodicity is assumed to hold within each of the pure states.

graph $G$ is complete (or fully-connected) and the couplings are uniform $J_{ij} = J/n$ for $1 < i < j < n$, for some $J > 0$. In particular, the model in invariant under any permutation of the spins, so that the magnetizations are uniform $m_i = M_n$, where $M_n$ is the total average magnetization defined in equation (1.75). The Boltzmann distribution of the Curie-Weiss model can be written as

*Boltzmann distribution*

$$
\begin{aligned}
\mathbb{P}(\sigma) &= \frac{1}{\mathcal{Z}} \exp\left( \frac{\beta J}{2\, n} \sum_{i,j=1}^{n} \sigma_i \sigma_j \right), \\
&= \frac{1}{\mathcal{Z}} \exp\left( \frac{n\, \beta J}{2}\, m[\sigma]^2 \right),
\end{aligned}
\tag{1.79}
$$

where $m[\sigma]$ is the total magnetization (1.76) of a single configuration $\sigma$. The probability of observing a configuration with total magnetization $m$ is therefore given by

*Distribution of the total magnetization*

$$
\mathbb{P}(m) = \frac{1}{\mathcal{Z}} \binom{n}{\frac{n(1+m)}{2}} \exp\left( \frac{n\, \beta J}{2}\, m^2 \right),
\tag{1.80}
$$

where the combinatorial factor is the number of configurations $\sigma$ with total magnetization $m[\sigma] = m$. Using Stirling's formula, we can write the distribution of the random variable $m$ for $n$ large enough as

*Large deviations of the total magnetization*

$$
\mathbb{P}(m) = \frac{1}{\mathcal{Z}} \exp\left( -n\, \beta\, f(m) \right),
\tag{1.81}
$$

where $f$ is the large deviation (rate) function of $m$, given for $m \in [-1, 1]$ by

$$
f(m) = -\frac{\beta J}{2}\, m^2 + \frac{1+m}{2} \log \frac{1+m}{2} + \frac{1-m}{2} \log \frac{1-m}{2}.
\tag{1.82}
$$

*Exactness of the nMF approximation*

Note that $n\, f(m) = \mathcal{F}^{\mathrm{nMF}}(m)$ where $\mathcal{F}^{\mathrm{nMF}}$ is the nMF free energy (1.72), so that the nMF approximation is exact for the Curie-Weiss model. Indeed, applying Laplace's method to the partition function $\mathcal{Z}$, it is straightforward to check that by (1.81), the distribution $\mathbb{P}(m)$ concentrates, in the limit $n \to \infty$ on the global minimum $m^\star$ of $f$. As a consequence, when sampling from the Curie-Weiss model, we observe w.h.p only configurations with total magnetization $m^\star$, so that, by symmetry, the magnetization of each spin equals $m^\star$.

*Shape of f and phase transition*

It turns out that depending on the value of the inverse temperature $\beta$, the free energy $f(m)$ of the Curie-Weiss model may take two qualitatively different shapes, shown on figure 1.8. For $\beta J < 1$, $f$ is a convex function of $m$, with a unique global minimum at $m^\star = 0$. The Curie-Weiss model is therefore in its paramagnetic phase. For $\beta J > 1$ on the other hand, $f$ is no longer convex and has two local minima $\pm m^\star \neq 0$. The Curie-Weiss model is then in its ferromagnetic phase, and has finite magnetizations. The transition from paramagnetic to ferromagnetic takes place at the critical inverse temperature $\beta_c = 1/J$.

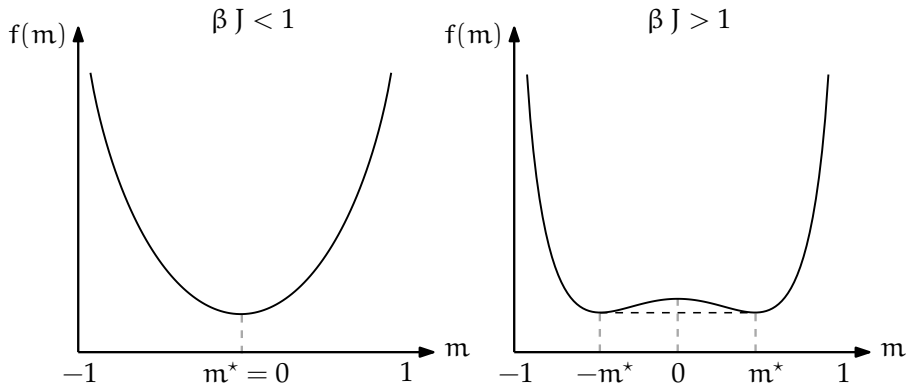

Figure 1.8 – Shape of the large deviation function $f$ of equation (1.82) for the Curie-Weiss model. The shape is qualitatively different depending on the value of the parameter $\beta J$. At high temperature (paramagnetic phase $\beta J < 1$), $f$ is convex and has a unique global minimum with vanishing magnetization $m^\star = 0$. At low temperature (ferromagnetic phase $\beta J > 1$), $f$ is no longer convex, and has two minima $\pm m^\star \neq 0$ with finite magnetization. These two local minima correspond to the two pure states in the ferromagnetic phase.

To understand how the existence of the ferromagnetic phase can be compatible with the fact that the magnetizations are, for any finite $n$, strictly vanishing, it is useful to imagine that we sample configurations from the Ising model using a Markov chain Monte Carlo algorithm such as the Metropolis algorithm [22]. When $\beta J > 1$, starting from a configuration $\sigma$ with arbitrary total magnetization $m[\sigma]$, the algorithm would soon start sampling configurations with total magnetization approximately equal to $\pm m^\star$, depending on which local minimum of $f$ is closest. After a number of iterations exponential in $n$, the Markov chain will make its way to the other local minimum, and remain in its vicinity for an exponential number of iterations. If we were to run the algorithm for a number of iterations exponential in $n$, we would therefore be able to explore the entire free energy landscape, and we would have as many configurations with positive total magnetization as configurations with negative total magnetization. Performing the empirical average of any given spin over this exponential number of realizations would yield a vanishing magnetization as expected. However, doing the same after running the algorithm for any reasonable amount of time would yield a finite magnetization, either $+m^\star$ or $-m^\star$. The two local minima of $f$ are therefore *metastable states* [165], escaped by a Markov chain only in exponentially large time. In the limit $n \to \infty$, the free energy barrier between the two

*A Markov chain point of view*

---

22. For the ferromagnetic Ising model, there exist so-called cluster Monte Carlo algorithms (see chapter 9) that, contrary to the Metropolis algorithm, can make non-local moves on the free energy landscape. Such algorithms do not remain stuck in the vicinity of a local minimum of $f$.

local minima, and therefore the escape time of the Markov chain, become infinite, causing the breaking of ergodicity, and the appearance of two pure states.

The analytical solution of the Curie-Weiss model leads us to identify the pure states with the local minima of a well-suited mean-field free energy [23]. The choice of the mean-field approximation that should be used depends on the instance of the Ising model we are considering. In our applications, we will be exclusively interested in models defined on sparse random graphs drawn from one of the ensembles introduced previously. Since such graphs are locally tree-like in the large $n$ limit, we will use use the Bethe free energy, and design spectral algorithms that approximate its local minima [24].

*The aim of this dissertation is to design spectral algorithms to approximate the local minima of the (Bethe) free energy.*

### 1.6.4   *The Ising spin glass*

When the couplings have arbitrary sign, the model (1.65) is usually called an *Ising spin glass*. Physically, it can be used to model disordered media, in which neighboring spins may be coupled positively or negatively depending on their distance. Such models have been studied in statistical physics mainly using the not (yet) rigorous replica and cavity methods [103], on cases where the couplings $J_{ij}$ are random variables with a prescribed distribution. Perhaps the best known spin glass model for which an analytical solution exists [122] is the Sherrington-Kirkpatrick model [146]. In this model, the graph G is the complete graph with $n$ vertices, and the couplings are drawn from a Gaussian distribution with mean $\mathbb{E}[J_{ij}] = J_0/n$ for some $J_0 \geqslant 0$, and variance $1/n$. Another model with qualitatively the same phenomenology is the Viana-Bray model [150], where the graph G is drawn from a sparse Erdős-Rényi model with finite average connectivity (see section 1.4.1), and the couplings $J_{ij} \in \{\pm 1\}$ are drawn from a Bernoulli distribution [25] with mean $\mathbb{E}[J_{ij}] \geqslant 0$. We say that an Ising spin glass has a ferromagnetic bias if $\mathbb{E}[J_{ij}] > 0$. Since we will be interested in pairwise models defined on sparse random graphs, this model will play an important role in the following.

*Sherrington-Kirkpatrick model*

*Viana-Bray model*

The phenomenology of the spin glass is more complex than the one of the ferromagnetic Ising model, and is in fact currently only understood in so-called infinite-dimensional models, such as the Sherrington-Kirkpatrick and Viana-Bray models. In the thermodynamic limit, in

---

23. This interpretation may break down in the spin glass phase introduced the following. However, in all our applications, we will always make sure to carefully tune the inverse temperature $\beta$ in such a way that we are never in the spin glass phase.

24. The actual free energy functional $\mathcal{F}(m)$ defined in equation (1.68) can be written as a Legendre transform and is in fact a convex (non-analytic) function, corresponding to the convex hull of the function $f$ of figure 1.8. One perhaps unexpected advantage of mean-field approximations is that they yield non-convex free energies whose local minima correspond to pure, or metastable states [165]

25. The Viana-Bray model is in fact more generic and allows for more general distributions of $J_{ij}$.

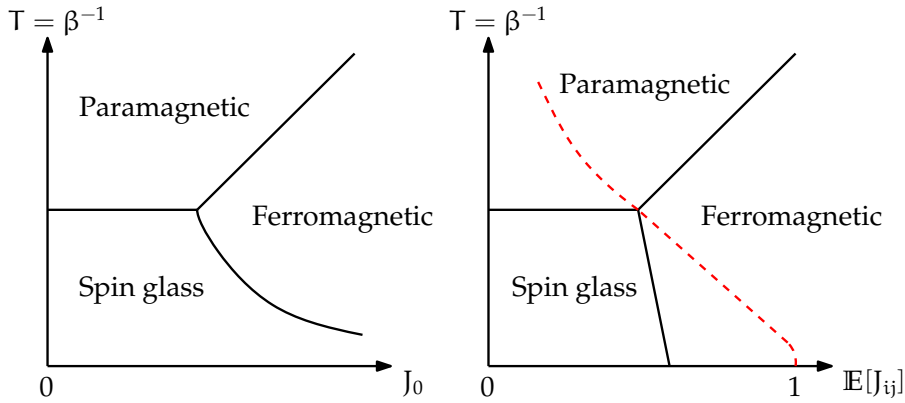

Figure 1.9 – Phase diagram of a spin glass in vanishing fields. On the left is the phase diagram of the Sherrington-Kirkpatrick model, defined on a fully connected graph, with i.i.d couplings drawn from a Gaussian distribution with mean $J_0/n$ and variance $1/n$. On the right is the phase diagram of the Viana-Bray model, defined on a sparse Erdős-Rényi graph with finite average connectivity $\alpha$, and $\pm 1$ couplings drawn from a Bernoulli distribution with mean $\mathbb{E}[J_{ij}]$. The equations of the paramagnetic to spin glass, and paramagnetic to ferromagnetic phase boundaries of the Viana-Bray model are given by equations (1.85) and (1.84). The dashed red line is the Nishimori line discussed in section 1.6.5. It can be noticed that it passes through the tricritical point without entering the spin glass phase. We note that the spin glass phase depicted on both phase diagrams in fact includes a mixed ferromagnetic/spin glass phase that we do not discuss here.

addition to the ferromagnetic and paramagnetic phases, there may exist, for large enough $\beta$, a third phase called the *spin glass phase*, in which the global magnetization (1.75) vanishes, but the following quantity, called the *Edwards-Anderson order parameter*, is finite

*Spin glass phase*

$$q_n = \frac{1}{n} \sum_{i=1}^{n} \mathbb{E}[\sigma_i]^2 \,. \tag{1.83}$$

*Edwards-Anderson order parameter*

In this definition, the expectation is taken first with respect to the Boltzmann distribution, and then with respect to the distribution of the graph and couplings, an operation usually termed *quenched disorder average*. Equation (1.83) implies that while there is no global ordering of the spins (the magnetization vanishes), each individual spin has a preferred orientation ($\mathbb{E}(\sigma_i) \neq 0$). We show on figure 1.9 a cartoon of the typical phase diagram of a spin glass. Note that generically, the existence of a ferromagnetic phase is conditioned on a strong enough ferromagnetic bias. This fact will prove important in the upcoming analyses.

The existence of a spin glass phase stems from the fact that, for couplings with arbitrary signs, the distribution (1.65) may have many modes. This feature differentiates the spin glass from the ferromagnet, for which the modes are trivially given by the two uniform configurations. In the spin glass, a highly non-trivial breaking of ergodicity may occur. Algorithmically, this leads to a dramatic slowing down of methods based on Markov chain Monte Carlo. It is in fact believed that the phenomenology of the spin glass phase could explain the computational hardness of some inference problems [103, 166]. The statistical properties of the spin glass phase are extremely complex, and cannot be fully stated without introducing additional concepts which will not be useful in the following. The reader is referred to [103, 105] for complete presentations. Our main point about the spin glass phase is that we should try to avoid it, which is exactly what all of the original inference methods presented in the following will do, by carefully choosing an adequate "temperature".

*We will avoid the spin glass phase*

In order to be able to avoid the spin glass phase, we need to be able to locate it. When the underlying graph G is drawn from the sparse Erdős-Rényi model with average connectivity $\alpha$, statistical physicists make a precise (though not yet fully rigorous [26]) prediction on the location of the paramagnetic to ferromagnetic, and paramagnetic to spin glass transitions. More precisely, when the couplings are i.i.d, they predict, based on the so-called replica-symmetric cavity method (see e. g. [168]), that the paramagnetic to ferromagnetic transition happens at an inverse temperature $\beta_F$ defined by

*Locating transitions*

*Para-Ferro transition*

$$\alpha \, \mathbb{E}_J \Big[ \tanh \big( \beta_F J \big) \Big] = 1 \,, \tag{1.84}$$

while the paramagnetic to spin glass transition occurs at the inverse temperature $\beta_{SG}$ defined by

*Para-SG transition*

$$\alpha \, \mathbb{E}_J \Big[ \tanh^2 \big( \beta_{SG} J \big) \Big] = 1 \,. \tag{1.85}$$

These equations are derived by analyzing the stability of the paramagnetic phase with respect to different types of perturbations. We detail and generalize these computations in chapter 3. We will regularly check throughout the dissertation that all of our results are in agreement with these predictions.

### 1.6.5   *The planted spin glass*

As stated previously, this dissertation in interested in designing inference methods to recover latent variables $(\sigma_i)_{i \in [n]}$ based on observations encoded in the form of a pairwise MRF. We give in this section a precise example of such a problem, called the *planted spin glass*, that will serve as our "drosophila", on which we will illustrate, in chapter 4, the effectiveness of the methods developed in chapters

*Our drosophila*

---

26. A fully rigorous proof for the paramagnetic to ferromagnetic transition of ferromagnetic Ising models on locally tree-like random graphs was given by [35].

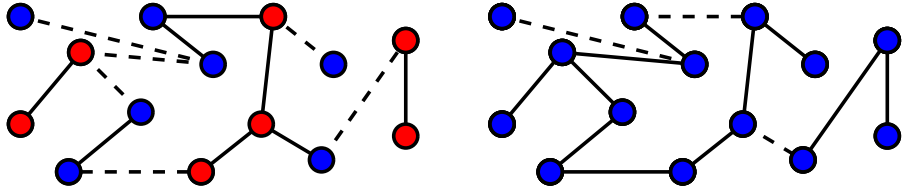

Figure 1.10 – The thought experiment of the planted spin glass, with $\epsilon \approx$ 23% of lying pairs (left). Solid edges are assigned to the pairs that answered "yes" to the question "do you have the same color?", and dashed edges are assigned to the pairs that answered "no". After the gauge transformation (1.92) (right), the problem is equivalent to an Ising spin glass with a strong ferromagnetic bias in the couplings. The dashed lines represent the negative couplings, and correspond to the pairs that lied.

2 and 3. This example corresponds to a special case of the lSBM previously introduced, and has also been known under the name of censored block model [1]. One way of introducing the planted spin glass is through the following thought experiment, borrowed from [168], and represented graphically on figure 1.10. Assume a playing card is given to each of $n$ players by a dealer. Each card is, with equal probability $1/2$, either red or blue. The dealer then asks a certain number of pairs of players whether they have been dealt a card of the same color or not. Each pair either answers the truth with probability $1 - \epsilon$, or lies with probability $\epsilon$. Can the dealer recover who was given a red card, and who was given a blue card? [27]

*A thought experiment*

The connection with the Ising spin glass might not appear obvious at this point. To make it precise, let us denote by $\sigma^\star = (\sigma_i^\star)_{i \in [n]} \in \{\pm 1\}^n$ the colors (+1 for blue, −1 for red) of the cards. We call this configuration the *planted* configuration. We assume that this configuration is chosen uniformly at random between the $2^n$ possible configurations, and we let $G = ([n], E)$, where $E$ is the set of pairs of players to whom we ask whether their cards are of the same color. To each of these pairs $(ij) \in E$, we assign a coupling $J_{ij} = \pm 1$, where +1 represents the answer "yes", and −1 represents "no". The distribution of the couplings is therefore given by

*Connection to the Ising model*

$$\mathbb{P}\left(J_{ij} = J \mid \sigma_i^\star, \sigma_j^\star\right) = (1 - \epsilon)\,\mathbf{1}(J = \sigma_i^\star \sigma_j^\star) + \epsilon\,\mathbf{1}(J = -\sigma_i^\star \sigma_j^\star). \quad (1.86)$$

Defining an inverse temperature $\beta^\star$ by

$$\beta^\star = \frac{1}{2} \log \frac{1 - \epsilon}{\epsilon}, \quad (1.87)$$

we can rewrite the distribution of the couplings as

$$\mathbb{P}\left(J_{ij} = J \mid \sigma_i^\star, \sigma_j^\star\right) = \frac{\exp \beta^\star J \sigma_i^\star \sigma_j^\star}{2 \cosh \beta}. \quad (1.88)$$

_______________

27. Note that the problem is invariant under the exchange of the colors blue and red, so that we can only hope to recover these colors up to a global flip.

Since the couplings are drawn independently from ([1.88](#)), we have by Bayes' theorem that the posterior distribution of the hidden assignment $\sigma^\star$ is given by

*Posterior distribution of the planted configuration*

$$\mathbb{P}\left(\sigma^\star = \sigma \mid (J_{ij})_{(ij) \in E}\right) = \frac{\mathbb{P}(\sigma^\star = \sigma) \prod_{(ij) \in E} \mathbb{P}\left(J_{ij} \mid \sigma_i, \sigma_j\right)}{\mathbb{P}\left((J_{ij})_{(ij) \in E}\right)}, \quad (1.89)$$

$$= \frac{1}{\mathcal{Z}} \exp\left(\beta^\star \sum_{(ij) \in E} J_{ij} \sigma_i \sigma_j\right), \quad (1.90)$$

where we have used that the prior distribution of $\sigma^\star$ is uniform. Thus, the posterior probability of the planted configuration $\sigma^\star$ given the observations $(J_{ij})_{(ij) \in E}$ is nothing but the Boltzmann distribution of the Ising spin glass, with the particular spin distribution ([1.88](#)) of the couplings. We will derive in chapter [4](#) efficient and provably optimal algorithms to recover the planted configuration from the knowledge of the couplings. For the time being, we would like to show that statistical physics offers a precise prediction as to whether the problem is solvable or not. Intuitively, the problem is solvable if a typical configuration sampled from the distribution $\mathbb{P}$ of equation ([1.90](#)) is correlated with the true planted configuration. The expected correlation between a configuration $\sigma$ sampled from $\mathbb{P}$ and the planted one is given by the following quantity, called *overlap*

*Overlap with the planted configuration*

$$O_n = \frac{1}{n} \sum_{i=1}^{n} \sigma_i^\star \, \mathbb{E}[\sigma_i], \quad (1.91)$$

where the expectation is over the distribution $\mathbb{P}$. In the thermodynamic limit $n \to \infty$, $O_n$ vanishes e. g. if the probability distribution $\mathbb{P}$ is uniform on $\{\pm 1\}^n$, and approaches unity as $\mathbb{P}$ becomes more peaked around $\sigma^\star$.

We now make use of a general property of the Ising model, called *gauge invariance*. This property relies on the fact that the Boltzmann distribution ([1.90](#)) is invariant under the following transformation, for any $\tau = (\tau_i)_{i \in [n]} \in \{\pm 1\}^n$

*Gauge invariance of the Ising model*

*Gauge transformation*

$$\begin{aligned} \sigma_i &\to \tilde{\sigma}_i = \sigma_i \tau_i & \text{for } i \in [n], \\ J_{ij} &\to \tilde{J}_{ij} = J_{ij} \tau_i \tau_j & \text{for } (ij) \in E. \end{aligned} \quad (1.92)$$

Indeed, it is readily checked that the thus gauge transformed distribution $\widetilde{\mathbb{P}}$ verifies $\widetilde{\mathbb{P}}(\tilde{\sigma}) = \mathbb{P}(\sigma)$ for any choice of $\tau$. In particular, we have $\widetilde{\mathbb{E}}[\tilde{\sigma}_i] = \mathbb{E}[\sigma_i]$ for any $i \in [n]$, with $\widetilde{\mathbb{E}}$ denoting expectation with respect to $\widetilde{\mathbb{P}}$. With the special choice $\tau = \sigma^\star$, we have the additional property that the overlap becomes equal to the total average magnetization (1.75), in the gauge transformed model

$$O_n = \frac{1}{n} \sum_{i=1}^{n} \sigma_i^\star \, \mathbb{E}[\sigma_i] = \frac{1}{n} \sum_{i=1}^{n} (\sigma_i^\star)^2 \, \mathbb{E}[\sigma_i^\star \sigma_i], \qquad (1.93)$$

$$= \frac{1}{n} \sum_{i=1}^{n} \widetilde{\mathbb{E}}[\tilde{\sigma}_i] = \widetilde{M}_n. \qquad (1.94)$$

*The overlap becomes the magnetization in the gauge transformed model*

Therefore we expect the original problem to be solvable in the thermodynamic limit if and only if the gauge transformed model is in the ferromagnetic phase. As seen in the previous section, this will depend on two things, namely the distribution of the couplings, and the value of the (inverse) temperature $\beta^\star$. The gauge transformation does not affect the value of the temperature, but changes the distribution of the couplings. Since $J_{ij}\sigma_i^\star\sigma_j^\star = \bar{J}$, we get that the couplings in the gauge transformed model are now i.i.d, with distribution $\widetilde{\mathbb{P}}$ given by

$$\widetilde{\mathbb{P}}\left(\bar{J}_{ij} = \bar{J}\right) = \frac{\exp \beta_\star}{2\cosh\beta^\star} \mathbf{1}\left(\bar{J} = 1\right) + \frac{\exp -\beta_\star}{2\cosh\beta^\star} \mathbf{1}\left(\bar{J} = -1\right) \qquad (1.95)$$

*Distribution of the gauge transformed couplings*

To sum up, we have shown that the dealer of our thought experiment can correctly guess the color of the cards of a finite fraction of the players if and only if the Ising spin glass, at temperature $\beta^\star$ given by (1.87) and with i.i.d couplings distributed according to (1.95), is in the ferromagnetic phase. This setting corresponds to the Viana-Bray model of the previous section, on a particular line of the phase diagram of figure 1.9, since the ferromagnetic bias $\mathbb{E}\left[\bar{J}_{ij}\right]$ of the gauge transformed couplings with distribution (1.95) is linked to the temperature by the identity

$$\mathbb{E}\left[\bar{J}_{ij}\right] = \tanh\beta^\star. \qquad (1.96)$$

It turns out that this particular line has a long history in statistical physics, where it is known under the name *Nishimori line* [117]. The analysis of a model on the Nishimori line is greatly simplified, mainly by the fact that it does not cross the spin glass phase. This can be seen by showing that on the Nishimori line, the magnetization $M_n$ equals the Edwards-Anderson parameter $q_n$, so that we cannot have $q_n > 0$ without having $M_n > 0$ [118, 168]. On the other hand, in the phase diagram of the Viana-Bray model, the Nishimori line interpolates between the points $(\beta^\star = 0, \widetilde{\mathbb{E}}[J_{ij}] = 0)$, and $(\beta^\star = +\infty, \widetilde{\mathbb{E}}[J_{ij}] = 1)$. The first point corresponds to a Boltzmann distribution uniform over $\{\pm 1\}^n$, which is therefore in the paramagnetic phase. The second point corresponds to the limit of vanishing temperature of a purely ferromagnetic model, and is therefore in the ferromagnetic phase. Consequently, the Nishimori line must cross the paramagnetic to ferromagnetic phase boundary. It turns out that it does so precisely at the tricritical point of the phase diagram of the Viana-Bray model of

*Nishimori line*

*The Nishimori line crosses the paramagnetic to ferromagnetic phase boundary at the critical point, and does not enter the spin glass phase (see figure 1.9).*

figure 1.9. To see this, one only needs to check that when the couplings are distributed according to (1.95), we have the identity

$$\alpha \, \widetilde{\mathbb{E}}_{\bar{J}} \left[ \tanh \left( \beta^\star \bar{J} \right) \right] = \alpha \, \widetilde{\mathbb{E}}_{\bar{J}} \left[ \tanh^2 \left( \beta^\star \bar{J} \right) \right] . \qquad (1.97)$$

At the paramagnetic to ferromagnetic transition, both quantities therefore equal unity, and we have by formulas (1.84) and (1.85) that the corresponding point in the phase diagram lies at the intersection of the paramagnetic to ferromagnetic, and paramagnetic to spin glass boundaries, i. e. at the tricritical point of figure 1.9.

*Nishimori line and Bayes optimality*

There is an important connection between the Nishimori line and *Bayes optimality*. In order to write the posterior probability (1.90) of the planted configuration $\sigma^\star$, we have assumed knowledge of $\beta^\star$, i. e. of the parameter $\epsilon$ of the model that was used to generate the data — here, the couplings. This setting, in which we know exactly all the parameters of the generative model is called *Bayes optimal*. When we do not know $\epsilon$, we have to consider the Ising model at general inverse temperature $\beta$, i. e. with a so-called *parameter mismatch*. In this case, the posterior probability is no longer on the Nishimori line, and may be in the spin glass phase. The definition of the Nishimori line can be extended to more complex models [50], and it is believed to be a general fact that in the Bayes optimal setting, a given inference problem always lies on a generalized version of the Nishimori line, that avoids the spin glass phase. We refer the reader to the excellent review [168] for arguments supporting this claim, and some of its algorithmic implications. We will come back to the problem of parameter mismatch in the planted spin glass in chapter 4.

In the Bayes optimal setting, we can read the statistical physics prediction about the solvability of our problem from the phase diagram of figure 1.9. Specifically, we expect that the problem is solvable if and only if the gauge transformed model with couplings distributed as in (1.95) is in the ferromagnetic phase, i. e.

*Transition predicted by statistical physics*

$$1 < \alpha \, \widetilde{\mathbb{E}}_{\bar{J}} \left[ \tanh^2 \left( \beta^\star \bar{J} \right) \right] = \alpha \, \tanh^2 \left( \beta^\star \right) \Longleftrightarrow \alpha > \frac{1}{(1 - 2\epsilon)^2} , \quad (1.98)$$

where we have used the definition (1.87) of $\beta^\star$. In chapter 4, we will show rigorously that this prediction is correct, and give a provably optimal algorithm that can partially recover the configuration $\sigma^\star$ as long as the condition (1.98) holds. Remarkably, this algorithm will not require the knowledge of $\beta^\star$ (i. e. $\epsilon$).

### 1.6.6  *The Hopfield model*

We finally introduce a variant of the Ising model that will be relevant to our application in matrix completion (chapter 8). This variant corresponds to the Ising spin glass of section 1.6.4 with yet another particular prescription for the couplings $(J_{ij})_{(ij) \in E}$. Specifically, we assume that there exist $r$ *patterns* $(\xi^\mu)_{\mu \in [r]}$ where each pattern is a vector of size $n$ with components $\xi^\mu = (\xi_i^\mu)_{i \in [n]}$, and we consider the Ising model (1.65) in vanishing fields, with couplings given, for $(ij) \in E$ by

*Patterns*

$$J_{ij} = \sum_{\mu=1}^{r} \xi_i^\mu \xi_j^\mu \, . \tag{1.99}$$

*Hebb's rule*

This defines a particular instance of the *Hopfield model*, in which the couplings are specified by the so-called *Hebbian rule*. The entries $\xi_i^\mu$ of the patterns are usually taken to be i.i.d binary signs $\pm 1$ with equal probability, although we will consider in chapter 8 i.i.d patterns drawn from a Gaussian distribution with mean 0. We take in this section patterns drawn from a general probability distribution of mean 0. The main interest of the Hopfield model stems from the fact that, under certain conditions, it has pure states correlated with the patterns $(\xi^\mu)_{\mu \in [r]}$. We say that a pure state is correlated with a pattern $\xi^\mu$ if the following *overlap* $O_n^\mu$ is finite in the thermodynamic limit

$$O_n^\mu = \frac{1}{n} \sum_{i=1}^{n} \mathrm{sign}(\xi_i^\mu) \, \mathbb{E}[\sigma_i] \, , \tag{1.100}$$

*Overlap between a pure state and a pattern*

where the expectation is, as in the case of the ferromagnetic Ising model of section 1.6.2, understood to be with respect to the Boltzmann distribution restricted to the pure state. We say that the pure state is positively correlated with the pattern $\xi^\mu$ if $O^\mu > 0$, and negatively correlated if $O^\mu < 0$.

To develop an intuition as to why this may be the case, we consider first the case of a single pattern ($r = 1$). The probability of a configuration $\sigma = (\sigma_i)_{i \in [n]}$ in the Hopfield model can then be written as $\mathbb{P}(\sigma) \propto \exp(-\beta E(\sigma))$ where the energy $E(\sigma)$ is defined by

*Case of a single pattern*

$$E(\sigma) = -\sum_{(ij) \in E} J_{ij} \sigma_i \sigma_j = -\sum_{(ij) \in E} \xi_i^1 \xi_j^1 \sigma_i \sigma_j \, . \tag{1.101}$$

Therefore the *ground-states*, i.e. the most probable configurations, are $\pm \sigma^1$, with $\sigma_i^1 = \mathrm{sign}(\xi_i^1)$ for $i \in [n]$. We would like to find out when the Boltzmann distribution concentrates around this ground-state. To do so, we use the invariance of the Boltzmann measure under the gauge transformation (1.92), which we take here to be

$$\sigma_i \to \tilde{\sigma}_i = \sigma_i \sigma_i^1 \qquad\qquad \text{for } i \in [n] \, , \tag{1.102}$$

$$J_{ij} \to \tilde{J}_{ij} = J_{ij} \sigma_i^1 \sigma_j^1 = |J_{ij}| \qquad \text{for } (ij) \in E \, . \tag{1.103}$$

The gauge transformed model is therefore a ferromagnetic Ising model, akin to the one of section 1.6.2. The overlap (1.100) with the unique pattern becomes

$$O_n^1 = \frac{1}{n} \sum_{i=1}^{n} \widetilde{\mathbb{E}}[\tilde{\sigma}_i] = \widetilde{M}_n \, , \tag{1.104}$$

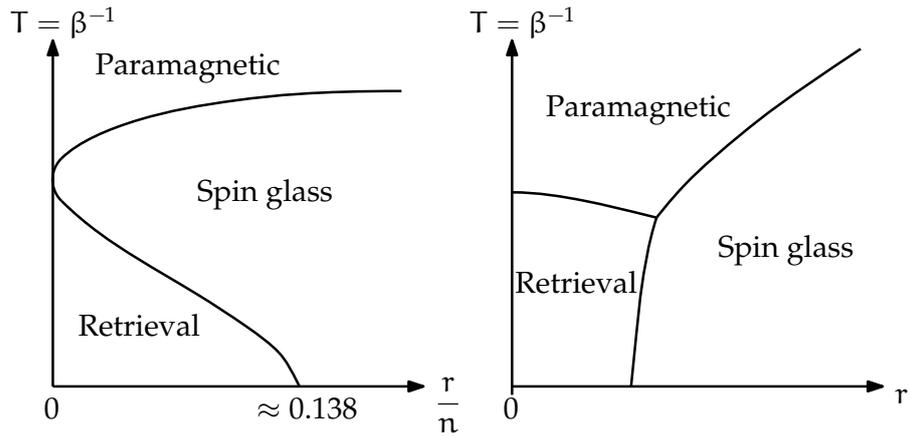

Figure 1.11 – Qualitative phase diagrams of the Hopfield model with binary patterns on the complete graph [11] (left) and on a sparse Erdős-Rényi graph with finite average connectivity [28] (right). Notice the different scaling of the number of patterns $r$ between the two cases.

where $\widetilde{M_n}$ is the total average magnetization, as defined in equation (1.75), of the gauge transformed model. Therefore, we expect from the results of section 1.6.2 that in the ferromagnetic phase (i. e. for $\beta$ large enough), the Boltzmann distribution will indeed split into two pure states, one of them positively correlated with the unique pattern, and the other one negatively.

*Case of $r$ patterns*    When $r$ patterns are used, the configurations $\pm\sigma^\mu$ with $\sigma_i^\mu = \text{sign}(\xi_i^\mu)$ for $\mu \in [r], i \in [n]$ are still approximately local minima of the energy $E(\sigma)$. Indeed, the energy $E(\pm\sigma^\mu)$ of any of these configurations decomposes into a term similar to (1.101), and a small *crosstalk* term, arising from random correlations between the different patterns. This second term is negligible for a small enough number $r$ of i.i.d patterns of mean $0$. In this case, and for a suitable range of values of $\beta$, the model is in a so-called *retrieval phase*, which generalizes the ferromagnetic phase of the Ising model, and in which the Boltzmann distribution splits into $r$ pairs of pure states, called in this context *retrieval states*, each pair correlated with one of the patterns[28]. When in the retrieval phase, the Hopfield model effectively acts like an associative memory, which is the function it was first introduced to perform [66]. In this context, we say that the $r$ patterns have been *memorized*, and we can *address* the associative memory to denoise or complete a noisy or partial version of one of the patterns. This is usually achieved using a Markov Chain Monte Carlo algorithm (often at vanishing

*Retrieval phase*

*Retrieval states*
*Associative memory*

---

28. This is an oversimplification, as upon lowering further the temperature, (many) spurious pure states appear, corresponding to superpositions of the patterns. However, these spurious states will not have an impact on the spectral method of chapter 8, since they correspond to linear combinations of the retrieval states, and hence do not affect the dimension of the vector space spanned by the retrieval states.

temperature), which can sample the retrieval state to which the observation belongs, without escaping to another, irrelevant retrieval state. In a tentative biological identification, the spins $(\sigma_i)_{i \in [n]}$ are identified with *neurons*, and their binary value reflects whether the neuron is spiking or not. The couplings $(J_{ij})_{(ij) \in E}$ are interpreted as *synaptic weights*.

As anticipated in the previous heuristic discussion, there is a limit, called *capacity*, to the number of patterns than can be memorized by a Hopfield model. This capacity depends on the distribution of the patterns, as well as the underlying graph. In the classical case of a complete graph with $n$ vertices, and random binary patterns, Amit, Gutfreund and Sompolinsky showed [11] using the replica method that the maximum number of patterns that can be stored in the thermodynamic limit is $C \, n$, where $C \approx 0.138$. The same authors also give the complete corresponding phase diagram, depicted qualitatively on figure 1.11. The situation is different for the Hopfield model on a sparse random graph with finite connectivity [29]. In this case, the capacity is finite in the limit of large number of neurons $n$. The phase diagram in the case of binary patterns has been computed using the replica method in [28], and is shown qualitatively on figure 1.11. In chapter 8, we compute numerically the capacity of a sparse Hopfield model with Gaussian patterns of mean 0.

*Capacity of the Hopfield model*

*Sparse Hopfield model*

## 1.7 GENERAL APPROACH OF THIS WORK

Before closing this review chapter and beginning the construction of our theory of spectral inference, let us give a high level description of the scope of the upcoming methods. As previously hinted at, we will be interested in particular problems whose solutions are given by the marginals of a pairwise MRF. Generically, our upcoming applications can be classified into two groups.

The first group consists of problems whose Bayes optimal solution is naturally written as a marginalization problem in a pairwise MRF. A typical example of such a situation is the LSBM of section 1.4.3. Indeed, the probability of generating a cluster assignment $\sigma = (\sigma_i) \in [q]^n$, a graph $G = ([n], E)$ and labels $\ell = (\ell_{ij})_{(ij) \in E}$ in this model is given by

*The "Bayes optimal" approach, exemplified in chapters 4,5 & 6*

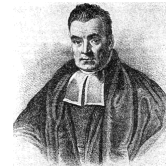

$$\mathbb{P}\left(G, \ell, \sigma \mid \theta\right) = \prod_{i=1}^{n} f_{\sigma_i} \prod_{(ij) \in E} p_{\sigma_i, \sigma_i}^{(e)} p_{\sigma_i, \sigma_i}^{(\ell)}(\ell_{ij}) \prod_{\substack{i<j \\ (ij) \notin E}} \left(1 - p_{\sigma_i, \sigma_j}^{(e)}\right),$$

---

29. The sparse Hopfield model has received comparatively less attention, probably both because it is irrelevant as a model of the brain, and because it has a much smaller capacity.

where $\theta = \left\{ (f_\sigma)_{\sigma \in [q]}, \left( p^{(e)}_{\sigma, \sigma'} \right)_{\sigma, \sigma' \in [q]}, \left( p^{(\ell)}_{\sigma, \sigma'}(\ell) \right)_{\sigma, \sigma' \in [q], \ell \in \mathcal{L}} \right\}$ is the set of parameters of the ISBM. The posterior probability of the cluster assignment $\sigma$ is therefore given by

$$\mathbb{P}\left( \sigma \mid G, \ell, \theta \right) = \frac{\mathbb{P}\left( G, \ell, \sigma \mid \theta \right)}{\mathbb{P}\left( G, \ell \mid \theta \right)},$$

$$= \frac{1}{\mathcal{Z}(G, \ell, \theta)} \prod_{i=1}^{n} f_{\sigma_i} \prod_{(ij) \in E} p^{(e)}_{\sigma_i, \sigma_i} p^{(\ell)}_{\sigma_i, \sigma_i}(\ell_{ij}) \prod_{\substack{i < j \\ (ij) \notin E}} \left( 1 - p^{(e)}_{\sigma_i, \sigma_j} \right). \tag{1.105}$$

From standard results in Bayesian inference, the optimal estimator of the cluster assignments $\sigma$, i.e. the estimator that minimizes the number of misclassified vertices, is given by

$$\hat{\sigma}_i = \mathrm{argmax}_{\sigma_i \in [q]} \, \mathbb{P}_i(\sigma_i), \qquad \forall i \in [n], \tag{1.106}$$

where $\mathbb{P}_i$ is the marginal distribution of $\sigma_i$. Therefore the Bayes optimal solution of the inference problem in the ISBM is naturally expressed in terms of the marginals of the pairwise MRF of equation (1.105) [30]. Chapters 4, 5 and 6 are devoted to the in-depth analysis of particular cases of this problem.

The second group of possible applications we consider consists of problems where we do not assume the data to be generated from a particular MRF [31], but we define an *ad hoc* pairwise MRF in which we expect the configuration $\sigma$ we are after to have high probability. An example of this approach is the computer vision problem of section 1.3.1, where we consider a pairwise MRF directly translating our belief that neighboring pixels should have similar colors. One way of achieving such a model is by assigning a cost (or energy) $E(\sigma)$ to each configuration such that the solution we are looking for is a (local) minimum of $E$. We may then consider the model

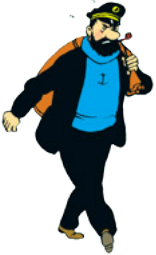

$$\mathbb{P}(\sigma) = \frac{1}{\mathcal{Z}} \exp\left( -\beta E(\sigma) \right), \tag{1.107}$$

where $\beta$ is a free parameter. As an example, let us consider the problem of similarity-based clustering, in which we try to cluster $n$ items into $q$ groups based on some measure of similarity between two given items. We may define a weighted graph $G = ([n], E)$ where each edge $(ij) \in E$ carries the similarity $s_{ij}$ between items $i$ and $j$. A standard approach (see e.g. [98]) is then to consider a cost function of the form

---

30. Note that, even when the graph $G$ is sparse, this pairwise MRF is, in general, defined on the fully connected graph with $n$ vertices, which causes numerical inefficiencies. We deal with this issue in section 3.1.1.

31. We may analyze the performance of our algorithms on data generated from a model, but we do not assume, when designing the algorithm, that the data comes from a model.

$$E(\sigma) = -\sum_{(ij) \in E} f(s_{ij}) \, \mathbf{1}(\sigma_i = \sigma_j) \,, \tag{1.108}$$

where f is an increasing function. This cost function encourages items with a large similarity to be in the same cluster. The probability distribution (1.107) is then a pairwise MRF, in fact a Potts model. This approach was considered in [17] who showed that, by computing the correlation of the Potts model thus defined using Monte Carlo, it is possible to cluster the items to very good accuracy. In chapter 7, we introduce a spectral algorithm that clusters the items by estimating the marginals of this model, eliminating the need to run costly Monte Carlo simulations. Note that instead of minimizing the cost function (1.108) (which would amount to taking the limit $\beta \to \infty$ in the distribution (1.107)), we follow a probabilistic approach based on the marginals of the model (1.107) at *finite* $\beta$. This allows us to overcome some severe overfitting limitations of methods based on the optimization of a cost function, which tend to find seemingly good clusters even when there are none [167]. Another example of this *ad hoc* approach in the context of matrix completion will be developed in chapter 8. Starting from the next chapter, we develop general spectral approximation schemes that will apply to both the "Bayes optimal" and the "ad hoc" types of problems.

*An important difference with optimization*

Part II

# SPECTRAL INFERENCE

In this second part, we outline a general theory of spectral inference. We identify a class of interesting pairwise MRFs exhibiting phase transitions, controlled by two linear operators, the non-backtracking operator and the Bethe Hessian. We study the properties of these operators with methods drawn from statistical physics, and use them to introduce new spectral algorithms. We finally illustrate the performance of these algorithms on the example of the planted spin glass.

# FACTORIZED AND SYMMETRIC MODELS

In this chapter, we introduce spectral relaxations that approximate the marginals of a graphical model using the eigenvectors of certain linear operators. Our starting point is the pairwise MRF introduced previously, and defined by the joint distribution, for $\sigma \in \mathcal{X}$

$$\mathbb{P}(\sigma) = \frac{1}{\mathcal{Z}} \prod_{i \in V} \phi_i(\sigma_i) \prod_{(ij) \in E} \psi_{ij}(\sigma_i, \sigma_j). \qquad (2.1)$$

We will derive in section 2.1 a set of conditions on this pairwise MRF such that the associated BP algorithm has a trivial and uninformative fixed point. In this setting, we will identify a necessary condition for BP to output non-trivial marginals. In section 2.2, we show that under additional conditions on the pairwise MRF (2.1), the stability of the trivial fixed point of BP is controlled by a matrix called the non-backtracking operator, acting on the directed edges of the graph. We introduce a spectral inference algorithm based on this operator and motivate it intuitively by relating it to the Ising model. Finally, in section 2.3, we change our perspective and show that the previous results can be related to the disappearance of a paramagnetic local minimum in the Bethe free energy. We show that this approach yields a smaller, symmetric operator called the Bethe Hessian, tightly related to the non-backtracking operator. We propose another spectral algorithm based on the Bethe Hessian, and discuss possible generalizations to other mean-field free energies. In particular, by considering other classical mean-field approximations, we make contact with other operators frequently used in graph clustering. To simplify some of the algebra, we assume throughout the chapter that

$$\sum_{\sigma_i \in \mathcal{X}} \phi_i(\sigma_i) = 1, \qquad \forall i \in [n]. \qquad (2.2)$$

This is achieved without loss of generality by rescaling the partition function $\mathcal{Z}$.

## 2.1 PHASE TRANSITION IN BELIEF PROPAGATION

We first identify a general condition on the potentials of a pairwise MRF such that the corresponding BP algorithm fails to provide useful marginals.

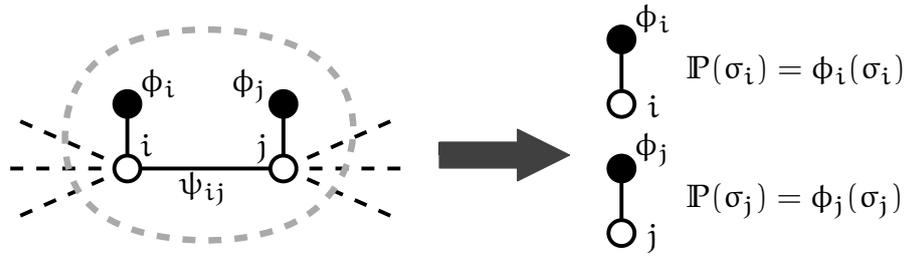

Figure 2.1 – Graphical representation of the factorized condition. A pairwise MRF is factorized if and only if, for any two nodes i and j such that (ij) ∈ E, the marginals of $\sigma_i$ and $\sigma_j$ in the subgraph enclosed by the dashed gray line are given by the single-variable potentials. Note that this does not imply that $\sigma_i$ and $\sigma_j$ are independent in the small graphical model with two variables.

### 2.1.1   *Factorized models*

*Factorized MRFs*    We introduce a subclass of pairwise MRFs, which we call *factorized* [1], for which BP has a phase transition. Given a model of the form (2.1), we define, for (ij) ∈ E, $\sigma_i \in \mathcal{X}$,

$$r_{i \to j}(\sigma_i) = \sum_{\sigma_j \in \mathcal{X}} \psi_{ij}(\sigma_i, \sigma_j)\, \phi_j(\sigma_j)\,. \tag{2.3}$$

We say that the model is factorized if for all (ij) ∈ E and for all $\sigma_i \in \mathcal{X}$,

*Factorized condition*

$$r_{i \to j}(\sigma_i) = r_{i \to j} \qquad \text{independent of } \sigma_i\,. \tag{2.4}$$

This property has a nice graphical representation depicted on figure 2.1. For any (ij) ∈ E and $\sigma_i \in \mathcal{X}$, $r_{i \to j}(\sigma_i)\phi_i(\sigma_i)$ can be interpreted as the (unnormalized) marginal probability distribution of $\sigma_i$ in a small graphical model containing only the nodes i and j. The factorized condition (2.4) therefore corresponds to assuming that in this small graphical model with two variables, each variable has its marginal determined by its single-point potential, i.e. the interaction with any other single variable alone does not influence its marginal.

At first sight, this condition seems very restrictive. Let us therefore give some examples of interesting problems that fall in the category of factorized models. A first non-trivial example is the Ising model in vanishing fields, which corresponds to $\mathcal{X} = \{\pm 1\}^n$, and

*The Ising model in vanishing fields is factorized*

$$\begin{aligned} \psi_{ij}(\sigma_i, \sigma_j) &= \exp J_{ij}\sigma_i\sigma_j\,, &\forall (ij) \in E, \sigma_i, \sigma_j \in \{\pm 1\},\\ \phi_i(\sigma_i) &= \frac{1}{2}\,, &\forall i \in [n], \sigma_i \in \{\pm 1\}. \end{aligned} \tag{2.5}$$

---

1. The word "factorized" does not imply here that the variables $\sigma_i$ are independent, as in the nMF approximation of section 1.5.2. Our wording is unfortunate, but motivated by the fact that our definition generalizes the factorized condition of [32].

In this case, we have for any $(ij) \in E, \sigma_i \in \{\pm 1\}$,

$$r_{i \to j}(\sigma_i) = \frac{1}{2} \sum_{\sigma_j \in \{\pm 1\}} \exp J_{ij} \sigma_i \sigma_j = \cosh \left( J_{ij} \sigma_i \right) = \cosh \left( J_{ij} \right),$$

which is indeed independent of $\sigma_i = \pm 1$.

A generalization of the previous example with an arbitrarily large alphabet $\mathcal{X} = [q]$ for some $q \in \mathbb{N}$ is given by the "symmetric" Potts model defined by

$$\psi_{ij}(\sigma_i, \sigma_j) = \begin{cases} \exp J_{ij}^{=} & \text{if } \sigma_i = \sigma_j \\ \exp J_{ij}^{\neq} & \text{if } \sigma_i \neq \sigma_j \end{cases}, \qquad \forall (ij) \in E, \sigma_i, \sigma_j \in [q],$$

$$\phi_i(\sigma_i) = \frac{1}{q}, \qquad \qquad \forall i \in [n], \sigma_i \in [q], \tag{2.6}$$

*The symmetric Potts model is factorized*

for arbitrary couplings $\left( J_{ij}^{=}, J_{ij}^{\neq} \right)_{(ij) \in E}$, so that for any $(ij) \in E, \sigma_i \in [q]$,

$$r_{i \to j}(\sigma_i) = \frac{1}{q} \sum_{\sigma_j \in [q]} \exp \left( J_{ij}^{=} \mathbf{1} \left( \sigma_i = \sigma_j \right) + J_{ij}^{+} \mathbf{1} \left( \sigma_i \neq \sigma_j \right) \right), \tag{2.7}$$

$$= \frac{1}{q} \left( \exp \left( J_{ij}^{=} \right) + (q-1) \exp \left( J_{ij}^{\neq} \right) \right), \tag{2.8}$$

again, independent of $\sigma_i \in [q]$. We will see that these two examples of factorized models alone provide significant insight into the solution (or absence of solution) of various distinct problems in inference and machine learning.

### 2.1.2 *Trivial fixed point of belief propagation*

Recall that a fixed point of the BP recursion for model (2.1) must verify

$$b_{i \to j}(\sigma_i) = \frac{1}{\mathcal{Z}_{i \to j}} \phi_i(\sigma_i) \prod_{k \in \partial i \setminus j} \sum_{\sigma_k \in \mathcal{X}} \psi_{ik}(\sigma_i, \sigma_k) b_{k \to i}(\sigma_k), \tag{2.9}$$

and the corresponding approximate marginals are given by

$$b_i(\sigma_i) = \frac{1}{\mathcal{Z}_i} \phi_i(\sigma_i) \prod_{k \in \partial i} \sum_{\sigma_k \in \mathcal{X}} \psi_{ik}(\sigma_i, \sigma_k) b_{k \to i}(\sigma_k). \tag{2.10}$$

The justification of the factorized condition (2.4) is that it implies the existence of a trivial solution of BP. Indeed, if we assume the model to be factorized, then the assignment

$$b_{i \to j}^{\star}(\sigma_i) = \phi_i(\sigma_i), \qquad \forall (i \to j) \in \vec{E}, \sigma_i \in \mathcal{X}, \tag{2.11}$$

*Trivial BP fixed point*

verifies the fixed point equation (2.9). Indeed, it holds that

$$\phi_i(\sigma_i) \prod_{k \in \partial i \setminus j} \sum_{\sigma_k \in \mathcal{X}} \psi_{ik}(\sigma_i, \sigma_k) b_{k \to i}^{\star}(\sigma_k) = \phi_i(\sigma_i) \prod_{k \in \partial i \setminus j} r_{i \to k},$$

$$= \mathcal{Z}_{i \to j}^{\star} b_{i \to j}^{\star}(\sigma_i),$$

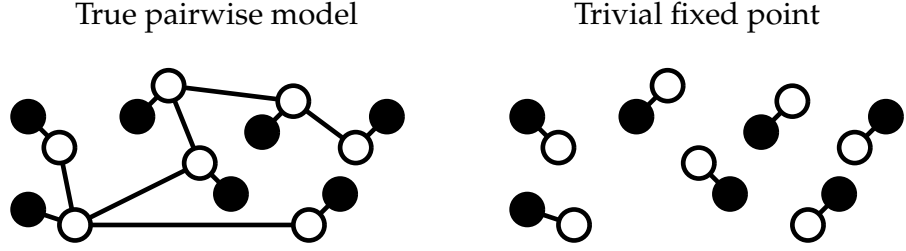

Figure 2.2 – The trivial fixed point (2.2) of BP estimates the marginals of the pairwise model on the left to be equal to the marginals of the independent model on the right.

with normalization constant given by

$$z^{\star}_{i \to j} = \prod_{k \in \partial i \setminus j} r_{i \to k}.$$  (2.12)

The corresponding beliefs output by BP are given by

$$b^{\star}_i(\sigma_i) = \phi_i(\sigma_i), \qquad \forall i \in [n], \sigma_i \in \mathcal{X}.$$  (2.13)

In other words, at the trivial fixed point, the marginals computed by BP on the pairwise model (2.1) are the same as the marginals of the independent model [2]

$$\mathbb{P}(\sigma) = \prod_{i=1}^{n} \phi_i(\sigma_i) = \prod_{i=1}^{n} \mathbb{P}_i(\sigma_i),$$  (2.14)

represented on figure 2.2 and which does not include any of the information encoded on the edges of the graphical model. In the various applications we consider, $\phi_i(\sigma_i)$ represents the prior probability of $\sigma_i$ in the absence of data in the form of pairwise potentials. The solution (2.2) is therefore *uninformative*. We will also call this solution *paramagnetic*, as it yields a vanishing magnetization (1.75) in the Ising model. Note that the paramagnetic solution exists for any choice of the couplings. In particular, in an Ising model with a temperature $\beta$ such that $J_{ij} = \beta J'_{ij}$, the fixed point exists for any value of $\beta$. However, as we will see in the next section, it is not always reachable by BP.

### 2.1.3   *Phase transition*

As explained in section 1.5, we run BP by iterating the fixed point equations (2.9) starting from an arbitrary initial condition. This iteration can be written in a compact form by introducing the mapping $F : \mathbb{R}^{|\vec{E}| \times |\mathcal{X}|} \to \mathbb{R}^{|\vec{E}| \times |\mathcal{X}|}$, such that, for any set of beliefs $b =$ $\left(b_{i\rightarrow j}(\sigma)\right)_{(i\rightarrow j)\in\vec{E},\sigma\in\mathcal{X}}\in\mathbb{R}^{|\vec{E}|\times|\mathcal{X}|}$, $F(b)=\left(f_{i\rightarrow j}^{(\sigma)}(b)\right)_{(i\rightarrow j)\in\vec{E},\sigma\in\mathcal{X}}$ with, for any $(i\rightarrow j)\in\vec{E}$, $\sigma\in\mathcal{X}$,

$$f_{i\rightarrow j}^{(\sigma)}(b)=\frac{1}{\mathcal{Z}_{i\rightarrow j}}\phi_i(\sigma)\prod_{k\in\partial i\backslash j}\sum_{\sigma_k\in\mathcal{X}}\psi_{ik}(\sigma,\sigma_k)\,b_{k\rightarrow i}(\sigma_k)\,. \quad (2.15)$$

---

2. Note that the correlations estimated through BP would by non-trivial, even at the trivial fixed point.

The BP recursion starting from an initial condition $b^0$ then simply writes, for $t\geqslant 0$

$$b^{t+1}=F(b^t)\,. \quad (2.16)$$

If we choose an initial condition $b^0=b^\star+\delta b^0\in\mathbb{R}^{|\vec{E}|\times\mathcal{X}}$ close to the uninformative fixed point $b^\star=F(b^\star)$, BP will converge to $b^\star$ if and only if it is a *stable* fixed point of BP. The reachability of the trivial fixed point is therefore controlled by the *Jacobian* $\mathcal{J}$ of $F$ at $b^\star$. Some algebra given in an appendix

$$\mathcal{J}_{i\rightarrow j,k\rightarrow l}^{(\sigma,\sigma')}=\phi_l(\sigma)\left(\frac{\psi_{kl}(\sigma',\sigma)}{r_{l\rightarrow k}}-\frac{r_{k\rightarrow l}}{r_{l\rightarrow k}}\right)\mathbf{1}\,(l=i)\,\mathbf{1}\,(k\neq j)\,, \quad (2.17)$$

and the perturbations evolve to linear order according to

$$\delta b^{t+1}=\mathcal{J}\,\delta b^t+o\left(\left\|\delta b^t\right\|\right)\,. \quad (2.18)$$

A perhaps friendlier, equivalent statement is the following. For $(i\rightarrow j)\in\vec{E}$, we define $\delta b_{i\rightarrow j}^t\in\mathbb{R}^{|\mathcal{X}|}$ to be the vector with elements $(\delta b_{i\rightarrow j}^t(\sigma_i))_{\sigma_i\in\mathcal{X}}$. The linearized system (2.18) can then be written equivalently, to linear order,

$$\delta b_{i\rightarrow j}^t=\sum_{k\in\partial i\backslash j}\mathcal{J}_{k\rightarrow i}\,\delta b_{k\rightarrow i}^{t-1}\,,\qquad\forall(i\rightarrow j)\in\vec{E}\,, \quad (2.19)$$

where for each $(k\rightarrow i)\in\vec{E}$, $\mathcal{J}_{k\rightarrow i}\in\mathbb{R}^{|\mathcal{X}|\times|\mathcal{X}|}$ is the matrix with entries

$$(\mathcal{J}_{k\rightarrow i})_{\sigma,\sigma'}=\phi_i(\sigma)\left(\frac{\psi_{ki}(\sigma',\sigma)}{r_{i\rightarrow k}}-\frac{r_{k\rightarrow i}}{r_{i\rightarrow k}}\right)\,. \quad (2.20)$$

We would like to find out if, upon iterating BP, the perturbations grow, or they decay. This question is readily answered in stability theory, when the initial perturbation can span the entire vector space $\mathbb{R}^{|\vec{E}|\times|\mathcal{X}|}$. In this case, the stability of the fixed point is controlled by the spectral radius $\rho(\mathcal{J})$ of the Jacobian at the fixed point. A small subtlety arises here from the fact that the beliefs must be normalized, so that the initial perturbation $\delta b^0$ must verify, for all $(i\rightarrow j)\in\vec{E}$,

$$\sum_{\sigma_i\in\mathcal{X}}\delta b_{i\rightarrow j}^0(\sigma_i)=0\,. \quad (2.21)$$

This condition defines a vector subspace $\mathcal{H}$ of dimension $|\vec{E}|\times(|\mathcal{X}|-1)$ of *admissible* perturbations. On the other hand, we observe that each column of $\mathcal{J}_{k\rightarrow j}$ sums to 0. Indeed,

$$\sum_{\sigma\in\mathcal{X}}(\mathcal{J}_{k\rightarrow i})_{\sigma,\sigma'}=\frac{\sum_{\sigma\in\mathcal{X}}\phi_i(\sigma)\psi_{ki}(\sigma',\sigma)}{r_{i\rightarrow k}}-\frac{r_{k\rightarrow i}}{r_{i\rightarrow k}}\sum_{\sigma_i\in\mathcal{X}}\phi_i(\sigma_i)\,, \quad (2.22)$$

$$=\frac{r_{k\rightarrow i}}{r_{i\rightarrow j}}-\frac{r_{k\rightarrow i}}{r_{i\rightarrow j}}=0\,. \quad (2.23)$$

This property can be used to show that $\mathrm{Im}\, \mathcal{J} \subset \mathcal{H}$, so that the restriction $\mathcal{J}|_{\mathcal{H}}$ of the Jacobian to the subspace of admissible perturbations defines a meaningful endomorphism of $\mathcal{H}$, whose spectral radius controls the stability of the trivial fixed point. Moreover, the fact that $\mathrm{Im}\, \mathcal{J} \subset \mathcal{H}$ implies that any eigenvalue of $\mathcal{J}$ is either an eigenvalue of $\mathcal{J}|_{\mathcal{H}}$ or is equal to zero. This implies that $\rho(\mathcal{J}|_{\mathcal{H}}) = \rho(\mathcal{J})$, so that the trivial fixed point of BP is unstable with respect to admissible perturbations if and only if

$$\rho(\mathcal{J}) > 1 \,, \tag{2.24}$$

where $\rho(\mathcal{J})$ denotes the spectral radius, i.e. the modulus of the largest eigenvalue of $\mathcal{J}$. There is therefore a phase transition in BP, and (2.24) is a necessary condition for BP to yield informative marginals.

## 2.2   SYMMETRIC MODELS

We now restrict further the class of factorized models by introducing a symmetry requirement on the potentials of a pairwise model. Specifically, we ask that the single-variable and pairwise potentials verify

$$\phi_i(\sigma_i) = \frac{1}{|\mathcal{X}|} \,, \qquad\qquad \forall i \in [n], \sigma_i \in \mathcal{X} \,,$$

$$\psi_{ij}(\sigma_i, \sigma_j) = \begin{cases} \psi_{ij}^{=} & \text{if } \sigma_i = \sigma_j \\ \psi_{ij}^{\neq} & \text{if } \sigma_i \neq \sigma_j \end{cases} \,, \quad \forall (ij) \in E, \sigma_i, \sigma_j \in \mathcal{X} \,, \tag{2.25}$$

for some constants $\psi_{ij}^{=}, \psi_{ij}^{\neq} > 0$. When this condition is fulfilled, we call the resulting model *symmetric*, and it is readily checked that symmetric models are factorized. Note that both the Ising and Potts examples of section 2.1.1 are of this particular form. Under the symmetry assumption (2.25), the trivial fixed point (2.11) of BP takes the simple form

$$b_{i \to j}(\sigma_i) = \frac{1}{|\mathcal{X}|} \,, \qquad \forall (i \to j) \in \vec{E}, \sigma_i \in [n] \,, \tag{2.26}$$

$$b_i(\sigma_i) = \frac{1}{|\mathcal{X}|} \,, \qquad \forall i \in [n], \sigma_i \in \mathcal{X} \,, \tag{2.27}$$

i.e. the beliefs output by BP are uniform, so that it is impossible to make a prediction about e.g. the most probable value of any particular variable $\sigma_i$. In this section, we show that this condition leads to simplifications in the study of the stability of the fixed point, and that this study leads to a new spectral algorithm for approximate inference.

### 2.2.1   *Tensor decomposition and the non-backtracking matrix*

Under the symmetry assumption (2.25), the parameters $r_{i \to j}$ of section 2.1.1 verify

$$r_{i \to j} = \sum_{\sigma_j \in \mathcal{X}} \psi_{ij}(\sigma_i, \sigma_j) \, \phi_j(\sigma_j) \, , \tag{2.28}$$

$$= \frac{1}{|\mathcal{X}|} \left( \psi_{ij}^{=} + (|\mathcal{X}| - 1) \, \psi_{ij}^{\neq} \right) \, , \tag{2.29}$$

$$= r_{j \to i} \, . \tag{2.30}$$

We therefore now denote $r_{ij} = r_{ji} = r_{i \to j} = r_{j \to i}$. The BP recursion linearized around the trivial fixed point, equation (2.19), reads

$$\delta b_{i \to j}^{t} = \sum_{k \in \partial i \setminus j} \mathcal{J}_{k \to i} \, \delta b_{k \to i}^{t-1} \, , \qquad \forall (i \to j) \in \vec{E}, \tag{2.31}$$

where the matrices $\mathcal{J}_{k \to i} \in \mathbb{R}^{|\mathcal{X}| \times |\mathcal{X}|}$ for $(k \to i) \in \vec{E}$ verify, under the symmetry assumption,

$$(\mathcal{J}_{k \to i})_{\sigma, \sigma'} = \phi_i(\sigma) \left( \frac{\psi_{ki}(\sigma', \sigma)}{r_{ik}} - 1 \right) \, ,$$

$$= \frac{\psi_{ik}^{=} - \psi_{ik}^{\neq}}{\psi_{ik}^{=} + (|\mathcal{X}| - 1) \, \psi_{ik}^{\neq}} \times \begin{cases} \dfrac{|\mathcal{X}| - 1}{|\mathcal{X}|} & \text{if } \sigma_i = \sigma_j \\[2mm] \dfrac{-1}{|\mathcal{X}|} & \text{if } \sigma_i \neq \sigma_j \end{cases} \, . \tag{2.32}$$

Defining *weights* $(w_{ij})_{(ij) \in E}$ by

$$w_{ij} = \frac{\psi_{ij}^{=} - \psi_{ij}^{\neq}}{\psi_{ij}^{=} + (|\mathcal{X}| - 1) \, \psi_{ij}^{\neq}} \, , \qquad \forall (ij) \in E, \tag{2.33}$$

we can rewrite (2.32) in full matrix form as

$$\mathcal{J}_{k \to i} = w_{ki} \left( I_{|\mathcal{X}|} - \frac{1}{|\mathcal{X}|} U_{|\mathcal{X}|} \right) \, , \tag{2.34}$$

where $I_{|\mathcal{X}|} \in \mathbb{R}^{|\mathcal{X}| \times |\mathcal{X}|}$ is the identity matrix and $U_{|\mathcal{X}|} \in \mathbb{R}^{|\mathcal{X}| \times |\mathcal{X}|}$ has all its entries equal to unity. The linearized BP recursion therefore reads

$$\delta b_{i \to j}^{t} = \left( I_{|\mathcal{X}|} - \frac{1}{|\mathcal{X}|} U_{|\mathcal{X}|} \right) \sum_{k \in \partial i \setminus j} w_{ki} \, \delta b_{k \to i}^{t-1} \, , \qquad \forall (i \to j) \in \vec{E}, \tag{2.35}$$

In other words, the full Jacobian of BP at the trivial fixed point decomposes as

$$\mathcal{J} = \left( I_{|\mathcal{X}|} - \frac{1}{|\mathcal{X}|} U_{|\mathcal{X}|} \right) \otimes B \, , \tag{2.36}$$

*Tensor decomposition of $\mathcal{J}$*

where $\otimes$ denotes the tensor (or Kronecker) product, and we have defined a new matrix $B \in \mathbb{R}^{|\vec{E}| \times |\vec{E}|}$, called the *non-backtracking matrix* (or *operator*), whose elements are given by

$$B_{i \to j, k \to l} = w_{kl} \, \mathbf{1}\,(l = i)\,\mathbf{1}\,(k \neq j)\,. \tag{2.37}$$

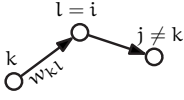

This operator mimics the way messages are passed in BP. In particular, it forbids backtracking immediately to the previous edge (see illustration in the margin). The non-backtracking matrix was first introduced in graph theory where it bears the name of Hashimoto matrix [59]. It was first used in the context of inference as the basis for a new spectral method for community detection in [89] — note that our definition corresponds to the transpose of the matrix they introduce. We will devote a considerable amount of time understanding the properties of the non-backtracking operator, and its applicability in various inference problems. For now, we simply remark that $B$ controls the stability of the trivial fixed point of BP. Indeed, we have by a standard result in linear algebra that

$$\rho(\mathcal{J}) = \rho\left(I_{|\mathcal{X}|} - \frac{1}{|\mathcal{X}|} U_{|\mathcal{X}|}\right)\rho(B)\,. \tag{2.38}$$

Since $\rho\left(I_{|\mathcal{X}|} - \frac{1}{|\mathcal{X}|} U_{|\mathcal{X}|}\right) = 1$, the instability of the trivial fixed point is equivalent to the condition

$$\rho(B) > 1\,. \tag{2.39}$$

### 2.2.2 *Reduction to an Ising model*

Let us compute explicitly the non-backtracking operator associated with the Ising model (2.5). In this case we have $\forall (ij) \in E$

$$\begin{aligned}
\psi_{ij}^{=} &= \exp J_{ij}\,,\\
\psi_{ij}^{\neq} &= \exp -J_{ij}\,,
\end{aligned} \tag{2.40}$$

so that

$$w_{ij} = \tanh(J_{ij})\,. \tag{2.41}$$

For a general symmetric pairwise model with alphabet $\mathcal{X}$ of arbitrary size, we observe that the weights introduced in equation (2.33) verify

$$-1 < w_{ij} < 1\,, \qquad \forall (ij) \in E\,, \tag{2.42}$$

as long as the potentials are finite, and therefore the associated pairwise model assigns a positive probability to each $\sigma \in \mathcal{X}^n$. Therefore we can write

$$w_{ij} = \tanh\left(J_{ij}^{\text{Ising}}\right)\,, \qquad \forall (ij) \in E\,, \tag{2.43}$$

for some couplings $\left(J_{ij}^{\text{Ising}}\right)_{(ij)\in E}$. In words, the non-backtracking operator of an arbitrary symmetric pairwise model is the same as the non-backtracking operator of an *associated Ising model* with properly chosen couplings. In particular, the onset of the instability of the paramagnetic solution in both models coincide. As an important example, the instability of the trivial fixed point in the Potts model (2.6) with $|\mathcal{X}| = q$ and couplings $\left(J_{ij}^{=}, J_{ij}^{\neq}\right)_{(ij)\in E}$ is equivalent to the instability of the trivial fixed point of the Ising model with couplings $\left(J_{ij}^{\text{Ising}}\right)_{(ij)\in E}$ defined by

$$\tanh\left(J_{ij}^{\text{Ising}}\right) = \frac{\exp\left(J_{ij}^{=}\right) - \exp\left(J_{ij}^{\neq}\right)}{\exp\left(J_{ij}^{=}\right) + (q-1)\exp\left(J_{ij}^{\neq}\right)}\,. \tag{2.44}$$

*Potts to Ising reduction*

This reduction to the simpler Ising model will guide our intuition when defining new approximate inference methods. It also provides a fruitful interpretation of the leading eigenvectors of the non-backtracking matrix, which we now explicit.

The BP algorithm takes a particularly simple form for the Ising model. Indeed, since $\sigma_i \in \{\pm 1\}$, each belief $b_i(\sigma_i)$ is completely specified by a single number,

$$m_i = \frac{b_i(+1) - b_i(+1)}{2}\,, \tag{2.45}$$

corresponding to the expectation of $\sigma_i$, and which we will therefore call (approximate) magnetization. We will denote $m \in \mathbb{R}^n$ the vector with entries $(m_i)_{i\in[n]}$. Similarly, the messages $b_{i\to j}(\sigma_i)$ are written in terms of their expectation $m_{i\to j}$, and the BP recursion reduces to

$$m_{i\to j}^t = \tanh\left(\sum_{k\in\partial i\setminus j} \operatorname{atanh}\left(\tanh\left(J_{ik}^{\text{Ising}}\right) m_{k\to i}^{t-1}\right)\right)\,. \tag{2.46}$$

*BP recursion for the Ising model*

To differentiate the messages from the final magnetizations (2.45), we will denote $m_\to \in \mathbb{R}^{|\vec{E}|}$ the vector whose entries $(m_{i\to j})_{(i\to j)\in\vec{E}}$ are the messages. The trivial fixed point with uniform beliefs corresponds to $m_\to = 0$. Linearizing around this fixed point, we find that for small $\|m_\to\|$, the BP recursion can be approximated by

$$m_{i\to j}^t \approx \sum_{k\in\partial i\setminus j} \tanh\left(J_{ik}^{\text{Ising}}\right) m_{k\to i}^{t-1} = \sum_{k\in\partial i\setminus j} w_{ik} m_{k\to i}^{t-1}\,, \tag{2.47}$$

or equivalently, in matrix form,

$$m_\to^t = B\, m_\to^{t-1} + o(\|m_\to^{t-1}\|)\,. \tag{2.48}$$

The magnetizations (2.45) are then given to linear order in the messages by

$$m_i \approx \sum_{k\in\partial i} \tanh\left(J_{ik}^{\text{Ising}}\right) m_{k\to i}^t = \sum_{k\in\partial i} w_{ik} m_{k\to i}^t\,, \tag{2.49}$$

i. e. in matrix form

$$m = P\, m_{\rightarrow}^{t} + o\left(\|m_{\rightarrow}^{t}\|\right) ,\qquad(2.50)$$

where we have defined a *pooling matrix* $P \in \mathbb{R}^{n \times |\vec{E}|}$, that maps, to linear order, the messages to the magnetizations. The elements of $P$ are given by

$$P_{i,k\rightarrow l} = w_{kl}\, \mathbf{1}\,(l = i) ,\qquad \forall i \in [n], (k \rightarrow l) \in \vec{E}.\qquad(2.51)$$

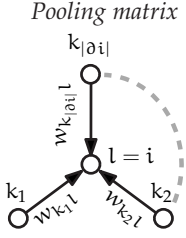

*Pooling matrix*

In the following, we refer to the operation of applying $P$ to an eigenvector of $B$ as *pooling*, and the result is a *pooled eigenvector*, the components of which are called *approximate magnetizations*. To linear order (admittedly a crude approximation), the BP recursion for the Ising model therefore reduces to a power iteration of the non-backtracking operator. When $t \rightarrow \infty$, one of two situations may arise. Either $\rho(B) < 1$, in which case all the perturbations converge to 0, or $\rho(B) > 1$, in which case $\|m_{\rightarrow}^{t}\| \rightarrow \infty$. When additionally $B$ has real eigenvalues larger than 1, the corresponding eigenvectors are *unstable directions* of BP. By unstable direction, we mean that a small perturbation $\|m_{\rightarrow}\| \ll 1$ of the trivial fixed point aligned with this direction will grow when iterating BP, and lead to a different, hopefully informative fixed point. In general, each of these unstable directions approximately "points" towards a distinct fixed point of BP. A cartoon of this argument is shown on figure 2.3.

*Unstable directions*

The leading eigenvectors of the non-backtracking operator therefore provide an approximation to the non-trivial fixed points of BP on the Ising model associated with a symmetric pairwise model. From an eigenvector $v \in \mathbb{R}^{|\vec{E}|}$ of $B$, we retrieve the corresponding approximate magnetizations by applying the pooling matrix of equation (2.51) to obtain a vector $m = P\,v \in \mathbb{R}^{n}$. This suggests the existence of a spectral algorithm based on $B$ that performs almost as well as BP. Before writing down the full algorithm, we need to understand more precisely what the non-trivial fixed points of BP might look like on an Ising model deriving from an inference problem.

*The leading eigenvectors of $B$ with real eigenvalues approximate the magnetizations of the associated Ising model*

### 2.2.3 *An example*

In this section, we consider a simple generalization of the planted spin glass (or censored block model) of section 1.6.5, and discuss how to solve it using the non-backtracking operator. Assume that $n$ vertices are partitioned into $q$ clusters by assigning each vertex $i \in [n]$ to a cluster $\sigma_i \in [q]$ chosen uniformly at random. We generate a sparse Erdős-Rényi random graph $G = ([n], E)$ with average connectivity $\alpha$, and we assign to each edge $(ij) \in E$ a label $\ell_{ij} = \pm 1$ drawn from the probability distribution

$$\mathbb{P}(\ell_{ij} = 1 \mid \sigma_i, \sigma_j) = \begin{cases} 1 - \epsilon & \text{if } \sigma_i = \sigma_j \\ \epsilon & \text{if } \sigma_i \neq \sigma_j \end{cases}.\qquad(2.52)$$

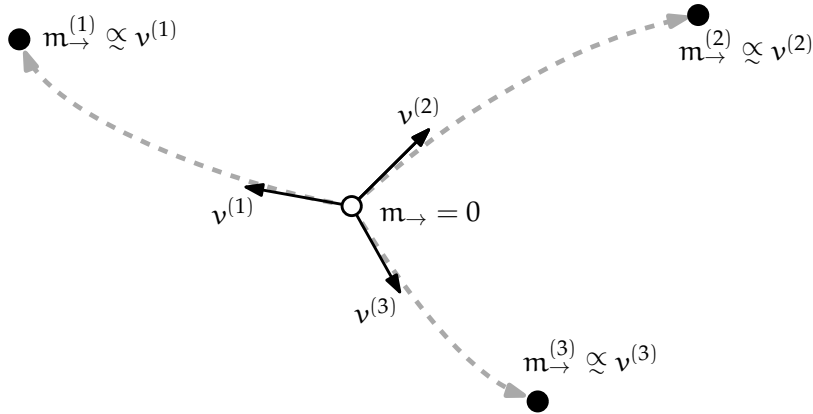

Figure 2.3 – Cartoon of the unstable directions of BP on the Ising model. Each of the black and white points represents a particular fixed point of BP, i.e. a vector in $\mathbb{R}^{|\vec{E}|}$. The white point at the center is the trivial fixed point $m_\rightarrow = 0$, i.e. $m_{i\rightarrow j} = 0$, $\forall (i \rightarrow j) \in \vec{E}$. The three black points are non-trivial fixed points of BP, represented as three vectors $m_\rightarrow^{(1)}, m_\rightarrow^{(2)}, m_\rightarrow^{(3)} \in \mathbb{R}^{|\vec{E}|}$. Here, the non-backtracking operator B has three real eigenvalues larger than 1, with corresponding eigenvectors $v^{(1)}, v^{(2)}, v^{(3)}$. Each of these eigenvectors corresponds to an unstable direction such that a small perturbation of the trivial fixed point aligned with this direction leads, upon iterating BP, to a different fixed point. The dashed gray arrows represent the true trajectory of the messages in $\mathbb{R}^{|\vec{E}|}$ as BP is iterated. In our approximate inference methods, we make the assumption that the these trajectories don't deviate too much from their linear approximation at the origin, so that the non-trivial fixed points are approximately proportional (denoted here $\propto$) to the eigenvectors of B.

Our aim is to recover the planted partition $\sigma$. Note that the case $q = 2$ is the planted spin glass. For general $q$, the distribution of the labels can be rewritten in the Boltzmann form

$$\mathbb{P}(\ell_{ij} \mid \sigma_i, \sigma_j) = \frac{\exp \beta \, \ell_{ij} \left( 2 \, \mathbf{1}(\sigma_i = \sigma_j) - 1 \right)}{2 \cosh(\beta)} \,, \qquad (2.53)$$

where the inverse temperature $\beta$ has the same definition as in the planted spin glass, and is given by (1.87), i. e.

$$\beta = \frac{1}{2} \log \left( \frac{1 - \epsilon}{\epsilon} \right) . \qquad (2.54)$$

Using Bayes' theorem, it is straightforward to show that the posterior probability distribution of $\sigma$ is

*Posterior distribution*

$$\mathbb{P}(\sigma) = \frac{1}{\mathcal{Z}} \prod_{(ij) \in E} \exp 2\beta \ell_{ij} \, \mathbf{1}(\sigma_i = \sigma_j) . \qquad (2.55)$$

Note that this model is a special case of the LSBM in which no information is encoded in the presence or absence of edges, and the labels are simple Bernoulli variables. This makes for a simpler first example, but we shall generalize the claims of the present section to the general setting in chapter 3. For the moment, let us remark that this model is of the symmetric form (2.25), so that BP has a trivial fixed point, whose stability is controlled by the spectral radius of the non-backtracking operator B (2.37), with weights (2.33) given here by

$$w_{ij} = \frac{\exp \left( 2\beta \ell_{ij} \right) - 1}{\exp \left( 2\beta \ell_{ij} \right) + q - 1} \,, \qquad \forall (ij) \in E . \qquad (2.56)$$

*Associated Ising model*

The associated Ising model has couplings $(J_{ij})_{(ij) \in E}$ defined through equation (2.43), i. e. $\tanh(J_{ij}) = w_{ij}$. In particular, the couplings assigned to edges carrying a label $\ell_{ij} = +1$ are positive, and they grow to infinity as $\beta \to \infty$. On the other hand, the couplings assigned to edges with a label $\ell_{ij} = -1$ are negative, and as $\beta \to \infty$, they decay to $-\infty$ if $q = 2$, and to a finite negative value if $q > 2$.

*Case $q = 2$*

Let us first consider the planted spin glass case $q = 2$. The Ising couplings are then given, as expected, by $J_{ij} = \beta \ell_{ij}$. As argued in section (1.6.5), when $\beta$ is large enough, this Ising model is in a partially ordered phase where typical configurations are correlated with the planted partition $\sigma$. BP should therefore have a stable ferromagnetic fixed point, at which the beliefs are correlated with the planted partition. In particular, the trivial, paramagnetic fixed point of BP should be unstable, and we expect BP to have one unstable direction $v \in \mathbb{R}^{\vec{E}}$, with corresponding approximate magnetizations $m = P v \in \mathbb{R}^n$ verifying equation (1.91), here

$$\lim_{n \to \infty} \frac{1}{n} \sum_{i=1}^{n} \left( 2 \, \mathbf{1}(\sigma_i = 1) - 1 \right) m_i \neq 0 . \qquad (2.57)$$

In words, we expect $m_i$ to take a different sign depending on whether $\sigma_i = 1$ or $\sigma_i = 2$. This unstable direction should be signaled a large, real eigenvalue of B. However, we do not expect this eigenvalue to be the only one with modulus larger than 1. Indeed, since the model in on the Nishimori line, statistical physics predicts (see discussion in section 1.6.5) that whenever the paramagnetic fixed point is unstable with respect to perturbations correlated with the ferromagnetic fixed point, it is also unstable with respect to perturbations leading to glassy states. Since there are potentially many glassy states, we expect many eigenvalues of B to become larger than 1 simultaneously. We can check on figure 2.4 that this is exactly what happens. For small $\beta$ (i. e. high noise $\epsilon$), the spectral radius of B is smaller than 1, and the trivial fixed point of BP is stable, so that it is impossible to detect the clusters using either BP or B. We will sometimes associate this fact with *unsolvability* of the problem in the following. On the other hand, for high enough $\beta$, there is one large and real eigenvalue, and many smaller (mostly complex) eigenvalues with modulus larger than 1. Additionally, in this case, the approximate magnetizations corresponding to the leading real eigenvector are strongly correlated with the planted partition $\sigma$, and can be used to recover them (see next section).

For a larger number of groups $q$, the situation is qualitatively the same. A given coupling $J_{ij}$ for $(ij) \in E$ is positive if $\ell_{ij}$ is, so that the ground states of the Ising model correspond to configurations where spins within the same cluster have typically the same value. For large enough $\beta$, we may therefore expect a *partially ordered phase* to appear, with magnetizations reflecting the true clustering of the graph. The numerical simulations of figure 2.4 suggest (on this particular model) that B has $q - 1$ large and real eigenvalues, well separated from the bulk of complex eigenvalues. It can also be checked that the corresponding approximate magnetizations take typically different values on the different clusters. We will come back in greater details on what the partially ordered phase looks like in chapter 3, and also show how to predict the appearance and the number of large and real eigenvalues of B on a general instance of the ISBM.

Notice the distinctive shape of the spectrum of B. There is a bulk of uninformative (mostly complex) eigenvalues that stays sharply confined in a disk, and well separated from the potential informative eigenvalues. This property, which we investigate in a special case in chapter 3, is the main reason why spectral algorithms based on B are superior to other, more traditional spectral methods, based e. g. on the adjacency matrix or the Laplacian, for *sparse* graphs. Indeed, in the case of the adjacency matrix for example, the bulk of the spectrum is constrained for large enough connectivity by Wigner's semi-circle law. However, for a constant connectivity $\alpha = O(1)$, this law is violated, and the spectrum exhibits tails that extend beyond the semi-circle, potentially drowning the informative eigenvalues [89]. By contrast, we will argue in chapter 3 that, on the sparse models we consider, the uninformative eigenvalues of B stay sharply confined, in the limit $n \to \infty$, in the circle of radius R given by

$$R = \sqrt{\alpha \, \mathbb{E}\left[w^2\right]}, \qquad (2.58)$$

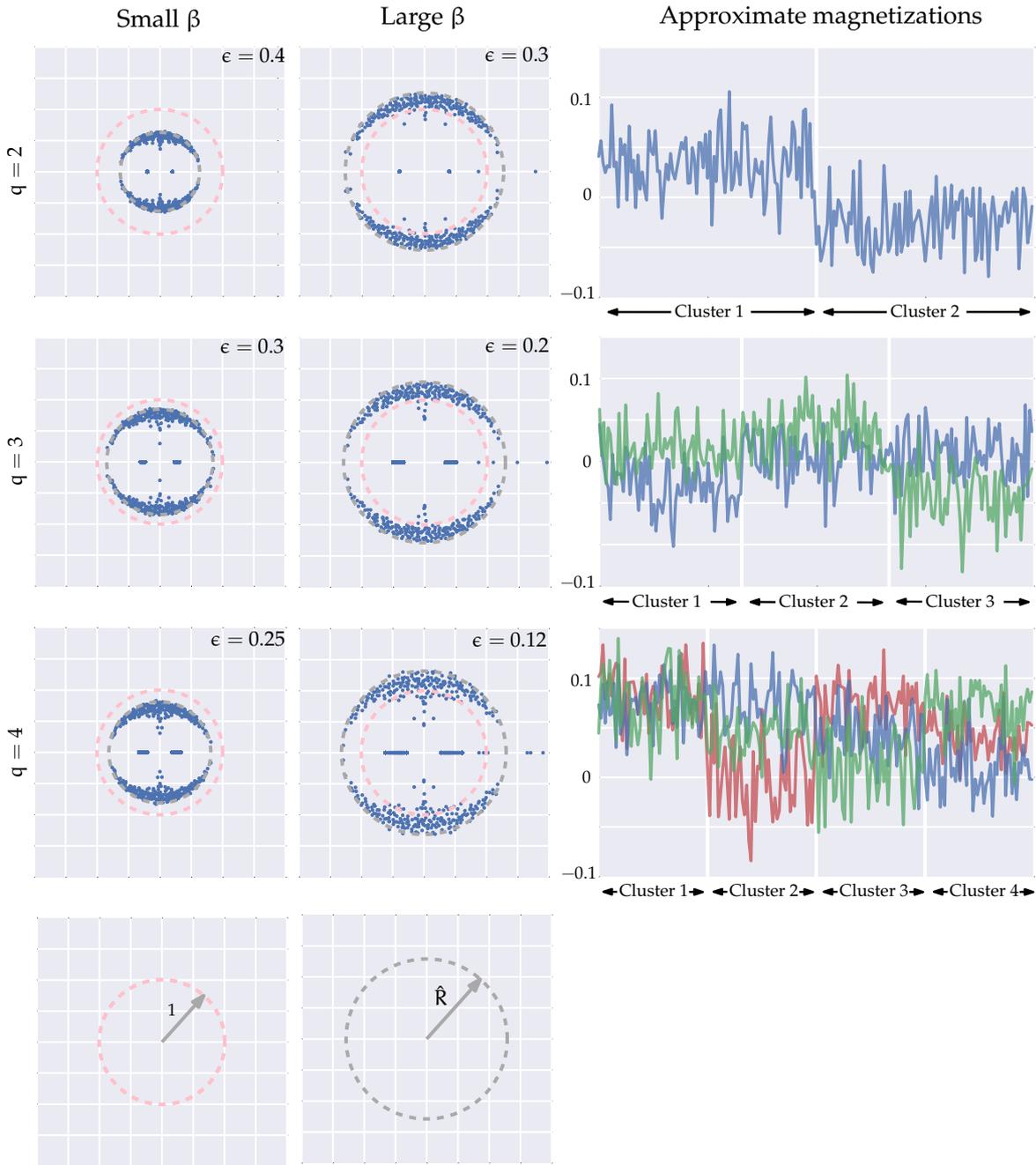

Figure 2.4 – Spectrum of the non-backtracking operator on the example of section 2.2.3, with $n = 200$ variables, on a random Erdős-Rényi graph of average connectivity $\alpha = 10$. The dashed pink circle has radius unity, and the dashed gray circle has radius $\hat{R}$ defined in equation (2.59). For $q = 2, 3, 4$, we represent the spectrum of B on a problem generated with a small $\beta$ (high noise $\epsilon$), and another generated with a large $\beta$ (low noise). In the rightmost column, we represent the approximate magnetizations obtained by pooling the top $q - 1$ eigenvectors of B obtained in the large $\beta$ problem. Generically, we observe that when the noise is too high, the spectrum of B is strictly included in the circle of radius unity, so that $\rho(B) < 1$, and the trivial fixed point of BP is stable. Upon decreasing the noise (increasing $\beta$), the situation changes, and for low enough noise, there are many eigenvalues of B that are larger than 1. However, there are only $q - 1$ eigenvalues that are larger than $\hat{R}$, they are real, and the corresponding (pooled) eigenvectors are correlated with the true cluster assignments of the vertices, in the sense that the approximate magnetizations take typically different values of the $q$ clusters. We characterize these approximate magnetizations in more details in chapter 3.

*Radius of the bulk*

where $w$ is a random variable that has the same distribution as the weights $w_{ij}$ of the non-backtracking operator. Note in particular that the associated Ising model has couplings $\tanh(J_{ij}) = w_{ij}$ so that the paramagnetic to spin glass transition of equation (1.85) is equivalent to $R > 1$. Conveniently, we can estimate this radius from the data

$$\hat{R} = \frac{1}{|E|} \sum_{(ij) \in E} w_{ij}^2 \underset{n \to \infty}{\longrightarrow} R. \qquad (2.59)$$

*Empirical estimator of the radius of the bulk*

We show on figure 2.4 the excellent agreement between this prediction and the the numerical simulations. We call the eigenvalues (and corresponding eigenvectors) of B that are outside of the circle of radius $\hat{R}$ *informative*.

### 2.2.4 *A linearized belief propagation algorithm*

In the previous sections, we have argued that the leading eigenvectors of B with real eigenvalues allow to approximate the marginals (or rather the mean of the marginals, i. e. the magnetizations) corresponding to the non-trivial fixed points of BP, in an Ising model associated with an arbitrary symmetric MRF. We show in this section how to use these observations to design a spectral algorithm for graph clustering.

As usual in spectral clustering methods (see *e. g.* [98]), we use the top eigenvectors of B to define a low dimensional embedding of the vertices of the graph. Let us detail the procedure on the example introduced in the previous section for the case of $q = 3$ clusters. We assume that $\beta$ is large enough so that there are 2 eigenvalues larger than $\hat{R}$, where $\hat{R}$ is defined in equation (2.59). We denote by $v^{(1)}$ and $v^{(2)}$ the top two eigenvectors of B. Using the pooling matrix P, we can define approximate magnetizations $m^{(1)} = P v^{(1)}$ and $m^{(2)} = P v^{(2)}$. We use these magnetizations as coordinates in $\mathbb{R}^2$. More precisely, to each vertex $i \in [n]$, we associate the point $\left(m_i^{(1)}, m_i^{(2)}\right) \in \mathbb{R}^2$. This defines an embedding of the vertices of the graph in $\mathbb{R}^2$ in which we expect points belonging to the same cluster to be close in the sense of the Euclidean distance of $\mathbb{R}^2$, since we have observed previously that the magnetizations reflect the true cluster memberships of the variables. The embedding is represented on the left of figure 2.5 using the eigenvectors shown on figure 2.4. To cluster the vertices of the graph, it only remains to cluster the $n$ points in $\mathbb{R}^2$, using *e. g.*

*Low dimensional embedding of the graph*

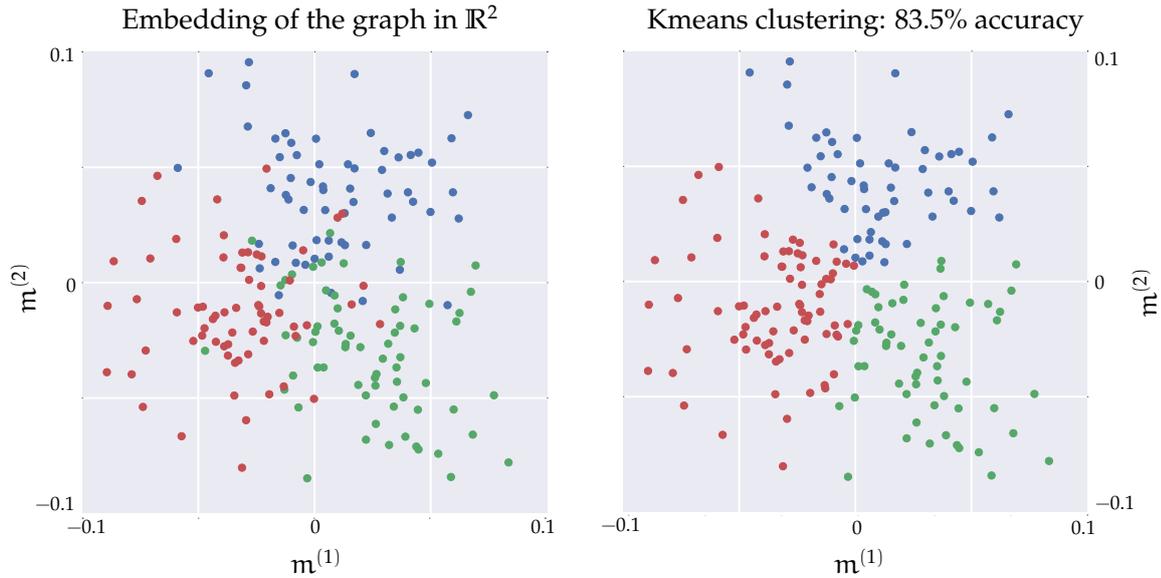

Figure 2.5 – Kmeans clustering of the eigenvectors of B on the example of section 2.2.3, for the case of $q = 3$ clusters. On the left is the two-dimensional embedding of the vertices obtained using the approximate magnetizations computed from the (pooled) top two eigenvectors of B represented on figure 2.4, with colors showing the ground true cluster assignment. On the right is the clustering obtained using kmeans on this two-dimensional embedding. The ground true cluster assignments are recovered with a 83.5% accuracy.

the *kmeans* algorithm. The result is shown on the right part of figure 2.5, and allows to recover the ground true cluster assignments with a 83.5% accuracy in this case. This procedure is readily adapted to the general case of $q$ clusters. As shown numerically in the previous section, there are in this case $q-1$ informative eigenvectors, with real eigenvalues lying outside of the circular bulk of complex eigenvalues. We may use these eigenvectors to define an embedding of the vertices in $\mathbb{R}^{q-1}$, and use kmeans to approximately recover the cluster assignments.

*A new spectral*
*inference algorithm*
Abstracting and generalizing from these examples, we consider a generic graph clustering problem defined on a graph $G = ([n], E)$, in which we are looking to assign each vertex $i \in [n]$ of G to a cluster $\sigma_i \in [q]$. We assume that the probability of a given assignment $\sigma = (\sigma_i)_{i \in [n]} \in [q]^n$ being correct takes the form of a pairwise MRF verifying the symmetry assumption (2.25). The optimal estimator of the cluster membership of vertex $i \in [n]$ is then given by $\hat{\sigma}_i = \mathrm{argmax}_{\sigma \in [q]} \, \mathbb{P}_i(\sigma)$ where $\mathbb{P}_i$ denotes the marginal probability distribution of $\sigma_i$. Note that since the symmetric MRF (2.25) is invariant under any permutation of the alphabet $\mathcal{X} = [q]$, we can only hope to recover the true cluster assignment up to a global permutation. To compute these estimators, we propose the procedure described in algorithm 1. The performance of this algorithm, which we illustrate on various examples in the following, stems from the spectral properties of the non-backtracking operator, which we study in chapter 3, in the case of random graphs. In particular, we will argue that for graphs generated from the ISBM, we can make this algorithm more precise by predicting the number of eigenvalues that should be computed (which is for instance $q-1$ in the previous example). This algorithm can be thought of as an alternative, linearized version of BP. Note that we have assumed here a *Bayes optimal* setting in which we know both the number of clusters $q$ and the true posterior probability distribution of the cluster assignment $\sigma$. We come back to these assumptions in the following and show that they can be relaxed in certain cases. Finally, we note that we expect this algorithm to perform well when BP accurately estimates the marginals of the MRF $\mathbb{P}$, therefore, particularly when the underlying graph $G$ is sparse so that it is locally tree-like. We briefly discuss other cases in the following.

---

**Algorithm 1** Linearized belief propagation for graph clustering

---

**Input:** Number of clusters $q$, probability $\mathbb{P}(\sigma)$ of each cluster assignment $\sigma \in [q]^n$ assumed to be a pairwise MRF of the symmetric form (2.25)

1: **Build** the non-backtracking operator B of (2.37) and the pooling matrix P of (2.51)
2: **Compute** the estimate $\hat{R}$ (2.59) of the radius of the bulk of the spectrum of B
3: **Compute** all the real eigenvalues of B which are larger (in absolute value) than $\hat{R}$. Let us call $r$ their number and $v^{(1)}, \ldots, v^{(r)}$ their corresponding eigenvectors. If $r = 0$, raise an error.
4: **Pool** the eigenvectors to compute the approximate magnetizations:
5:      **for** $\mu \in [r]$ **do** $m^\mu \leftarrow P v^{(\mu)}$
6: **Embed** the vertices of the graph in $\mathbb{R}^r$ by assigning to each vertex $i \in [n]$ the coordinates $(m_i^1, \ldots, m_i^r) \in \mathbb{R}^r$
7: **Cluster** the embedded vertices using (e. g.) *kmeans*

---

## 2.3 THE FREE ENERGY POINT OF VIEW

The non-backtracking operator is a large $\left(|\vec{E}| \times |\vec{E}|\right)$ and non-symmetric matrix. This causes numerical issues for large problems, as the fastest and most stable known numerical (sparse) eigensolvers, based *e. g.* on the Lanczos algorithm, work only for symmetric matrices. On the other hand, we are interested only in the eigenvectors $v$ with *real* eigenvalues of B, and even more precisely, in their pooled version $m = P v \in \mathbb{R}^n$. We may therefore hope that $m$ is in fact an eigenvector of a smaller $(n \times n)$, and symmetric operator H. This turns out to be true, in a certain sense which we explicit in the following.

Additionally, the smaller and symmetric operator H has a straightforward interpretation which yields insight into a more general class of spectral algorithms for clustering.

We assume in this section that we have reduced a symmetric MRF to its associated Ising model formulation, so that we write the non-backtracking operator as

$$B_{i \to j, k \to l} = \tanh(J_{ij}) \, \mathbb{1} \, (l = i) \, \mathbb{1} \, (k \neq j) \,, \tag{2.60}$$

where the couplings are defined by (2.43), so that the (pooled) principal eigenvectors of B approximate the magnetizations in the non-trivial fixed points of BP on the Ising model with couplings $(J_{ij})_{(ij) \in E}$, and vanishing fields.

### 2.3.1    *The Ihara-Bass formula*

Let $\lambda \geqslant 1$ be a real eigenvalue of B with associated eigenvector $v \in \mathbb{R}^{|\vec{E}|}$, with components $(v_{i \to j})_{(i \to j) \in \vec{E}}$. We are interested in finding an equation verified by $m = Pv \in \mathbb{R}^n$, where P is the pooling matrix defined in (2.51). The components $(m_i)_{i \in [n]}$ of m are therefore given by

$$m_i = \sum_{k \in \partial i} \tanh(J_{ki}) \, v_{k \to i} \,, \qquad \forall i \in [n] \,, \tag{2.61}$$

and the eigenvector $v$ verifies the set of equations

$$\lambda v_{i \to j} = \sum_{k \in \partial i \backslash j} \tanh(J_{ki}) \, v_{k \to i} \,, \qquad \forall (i \to j) \in \vec{E} \,. \tag{2.62}$$

This last equation implies the following closed set of equations relating, for each edge $(ij) \in E$, the magnetizations $m_i$ and $m_j$ on the one hand, and the messages $v_{i \to j}$ and $v_{j \to i}$ on the other

$$\begin{cases} \lambda v_{i \to j} = m_i - \tanh(J_{ij}) \, v_{j \to i} \\ \lambda v_{j \to i} = m_j - \tanh(J_{ij}) \, v_{i \to j} \end{cases} \,, \qquad \forall (ij) \in E \,. \tag{2.63}$$

For $\lambda \geqslant 1$ and finite couplings $(J_{ij})_{(ij) \in E}$, this system of equations is invertible. Inverting it allows to write the messages as a function of the magnetizations

$$v_{i \to j} = \frac{\lambda m_i - \tanh(J_{ij}) m_j}{\lambda^2 - \tanh^2(J_{ij})} \,, \qquad \forall (i \to j) \in \vec{E} \,. \tag{2.64}$$

Injecting this form of the messages in equation (2.61), we find that the magnetizations must verify

$$\left( 1 + \sum_{k \in \partial i} \frac{\tanh^2(J_{ij})}{\lambda^2 - \tanh^2(J_{ij})} \right) m_i - \sum_{k \in \partial i} \frac{\lambda \tanh(J_{ij})}{\lambda^2 - \tanh^2(J_{ij})} m_j = 0 \,. \tag{2.65}$$

To sum up, we have shown that $\left(\lambda \geqslant 1, \nu \in \mathbb{R}^{|\vec{E}|}\right)$ is an eigenpair of B if and only if, for $\mathfrak{m} = P\nu$, we have

$$H(\lambda)\mathfrak{m} = 0\,, \tag{2.66}$$

where, for any $x \in \mathbb{C} \setminus \left\{\pm \tanh(J_{ij})\right\}_{(ij)\in E}$, $H(x)$ is an $n \times n$ matrix with elements

$$H_{ij}(x) = \left(1 + \sum_{k\in\partial i} \frac{\tanh^2(J_{ik})}{x^2 - \tanh^2(J_{ik})}\right) \mathbf{1}(i = j) - \frac{x\tanh(J_{ij})}{x^2 - \tanh^2(J_{ij})} \mathbf{1}(j \in \partial i) \tag{2.67}$$

*The Bethe Hessian*

We will call this matrix *the Bethe Hessian*, for reasons that will be made clear in the next section. Note in particular that when $x \in \mathbb{R} \setminus \left\{\pm \tanh(J_{ij})\right\}_{(ij)\in E}$, $H(x)$ is a real and symmetric matrix.

The previous calculation is a (poor man's) version of the so-called Ihara-Bass formula[3], which, in its generalized form ([153]) relates the characteristic polynomial of B to the determinant of $H(x)$, for $x \in \mathbb{C} \setminus \left\{\pm \tanh(J_{ij})\right\}_{(ij)\in E}$ through

$$\det\left(xI_{|\vec{E}|} - B\right) = \det\left(H(x)\right) \prod_{(ij)\in E} \left(x^2 - \tanh^2(J_{ij})\right)\,. \tag{2.68}$$

*Generalized Ihara-Bass formula*

We provide a proof of this formula in chapter 3. The task of solving equation (2.66) simultaneously for $\lambda$ and $\mathfrak{m}$ is called a *nonlinear eigenproblem*. Although methods for solving such problems exist, they are typically slow because they require solving sequentially several (linear) eigenproblems. In the following, we will argue that it is actually not necessary to solve the nonlinear eigenproblem.

*Nonlinear eigenproblem*

The spectrum of $H(x)$ is tightly related to the spectrum of B. Let us restrict here to $x \in \mathbb{R}$, $x \geqslant 1$, so that the definition (2.67) of $H(x)$ has no poles. First, we observe that for large $x$, the diagonal entries of $H(x)$ stay larger than 1, while the non-diagonal entries become small. Using the Gershgorin circle theorem, we can therefore show that $H(x)$ has all its eigenvalues strictly positive for $x$ large enough. As we decrease $x$, the Ihara-Bass formula implies that $H(x)$ becomes singular when $x$ becomes equal to an eigenvalue of B. We find empirically (see figure 2.6) that as long as $x > R$, where R is the radius of the bulk in the spectrum of B, the smallest eigenvalues of $H(x)$ are a decreasing function of $x$. Therefore, $H(x)$ gains a new negative eigenvalue whenever $x$ becomes smaller than a real eigenvalue of B. As a consequence, at $x = R$, there is a one to one correspondence between the (potential) negative eigenvalues of $H(R)$ and the real eigenvalues of B that are outside of the bulk. We will therefore be particularly interested in the matrix $H\left(\hat{R}\right)$, where $\hat{R}$ is the estimator of $\mathbb{R}$ defined in equation (2.59). Note that this matrix can be safely defined whenever

*Spectrum of* $H(x)$

$\hat{R} > 1$. Indeed, if $\hat{R} < 1$, it is in general close (or maybe even equal) to a pole of the definition (2.67), so that the resulting matrix is at best very badly conditioned. On the other hand, when $\hat{R} < 1$, we have argued in 2.2.3 that the trivial fixed point of BP is stable so that both BP and B fail to output non-trivial solutions.

While there is no obvious relation between the eigenvectors of $H(\hat{R})$ and the (pooled) eigenvectors of B, the previous correspondence suggests that the eigenvectors of $H(\hat{R})$ with negative eigenvalues may also allow to approximate the magnetizations in the nontrivial fixed points of BP on the Ising model. In the following, we clarify in what sense this is the case, and propose an alternative to algorithm 1 based on $H(\hat{R})$.

### 2.3.2    *The Bethe Hessian*

We have obtained algorithm 1 as a linear relaxation of BP around a trivial fixed point. On the other hand, we have stated in section 1.5.3 that the fixed points of BP are stationary points of the Bethe free energy. The Bethe free energy of the Ising model in vanishing fields must therefore have a trivial stationary point, and we may wonder whether it is a local minimum of the Bethe free energy or not.

Recall that in the Bethe approximation, we approximate the magnetizations $\mathbb{E}[\sigma_i]$ for $i \in [n]$ and correlations between neighboring variables $\mathbb{E}[\sigma_i \sigma_j]$ for $(ij) \in E$ respectively by variables $m_i$ and $\chi_{ij}$ that minimize the Bethe free energy, i. e.

$$\left( (\mathbb{E}[\sigma_i])_{i \in [n]}, (\mathbb{E}[\sigma_i \sigma_j])_{(ij) \in E} \right) \approx \underset{m \in \mathbb{R}^n, \chi \in \mathbb{R}^{|E|}}{\mathrm{argmin}} \; \mathcal{F}^{\mathrm{Bethe}}(m, \chi), \quad (2.69)$$

where the Bethe free energy $\mathcal{F}^{\mathrm{Bethe}}$ can be written in the case of the Ising model ([153]) as

*Bethe free energy of the Ising model*

$$\mathcal{F}^{\mathrm{Bethe}}(m, \chi) = - \sum_{(ij) \in E} J_{ij} \chi_{ij} + \sum_{\substack{(ij) \in E \\ \sigma_i, \sigma_j \in \{\pm 1\}}} \eta \left( \frac{1 + m_i \sigma_i + m_j \sigma_j + \chi_{ij} \sigma_i \sigma_j}{4} \right)$$

$$+ \sum_{i \in [n]} (1 - |\partial_i|) \sum_{\sigma_i \in \{\pm 1\}} \eta \left( \frac{1 + m_i \sigma_i}{2} \right), \quad (2.70)$$

where for $x > 0$, $\eta(x) = x \log x$. The gradient of $\mathcal{F}^{\mathrm{Bethe}}$ is given ([107]), for all $i \in [n]$, $(ij) \in E$ by

*Gradient of the Bethe free energy*

$$\frac{\partial \mathcal{F}^{\mathrm{Bethe}}}{\partial m_i} = (1 - |\partial i|) \, \mathrm{atanh}(m_i)$$

$$+ \frac{1}{4} \sum_{j \in \partial i} \log \frac{\left(1 + m_i + m_j + \chi_{ij}\right)\left(1 + m_i - m_j - \chi_{ij}\right)}{\left(1 - m_i + m_j - \chi_{ij}\right)\left(1 - m_i - m_j + \chi_{ij}\right)},$$

$$\frac{\partial \mathcal{F}^{\mathrm{Bethe}}}{\partial \chi_{ij}} = -J_{ij} + \frac{1}{4} \log \frac{\left(1 + m_i + m_j + \chi_{ij}\right)\left(1 - m_i - m_j + \chi_{ij}\right)}{\left(1 + m_i - m_j - \chi_{ij}\right)\left(1 - m_i + m_j - \chi_{ij}\right)},$$

$$(2.71)$$

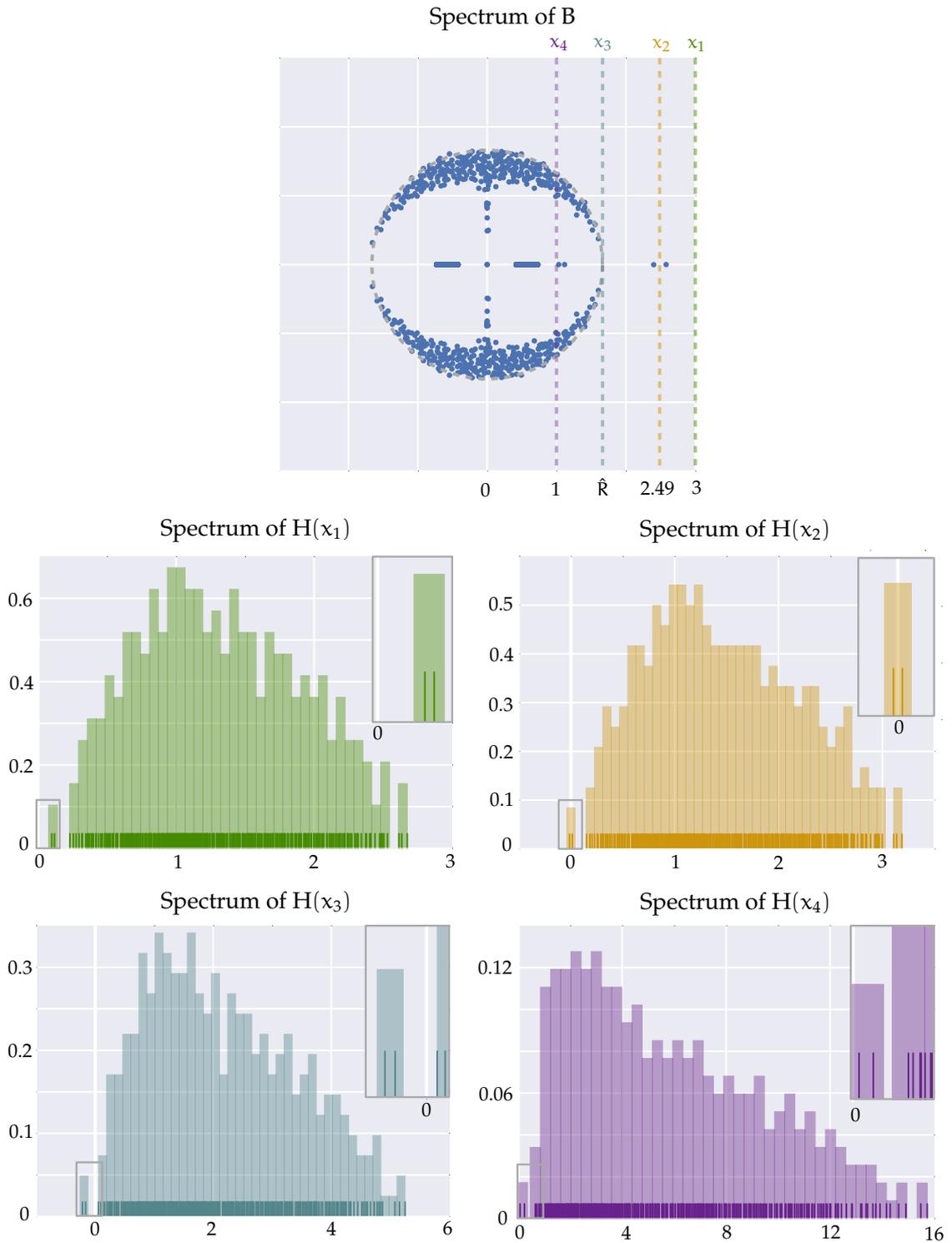

Figure 2.6 – Spectrum of the Bethe Hessian $H(x)$ on a single instance of the example of section 2.2.3, with $n = 300$ variables, average connectivity $\alpha = 9$, noise $\epsilon = 0.1$, and $q = 3$ clusters. On top is the spectrum of B, on which we have highlighted 4 real values $x_i \geqslant 1$ for $i \in [4]$. $x_1 > \rho(B)$, so that, as argued in the text, $H(x_1)$ is positive definite. As we decrease $x$, $H(x)$ gains a negative eigenvalue when $x$ becomes smaller than an eigenvalue $\lambda$ of B ($H(x_2), H(x_3)$). In particular, $H(x_3 = \hat{R})$ has two negative eigenvalues, corresponding to the informative eigenvalues of B. As we increase further $x$, these negative eigenvalues become positive again, as $x$ crosses real eigenvalues of B which are smaller than $\hat{R}$, and $H(x_4 = 1)$ is positive definite.

where we have used that for $|x| < 1$ $\operatorname{atanh}(x) = \frac{1}{2} \log \frac{1+x}{1-x}$. As anticipated, the following assignment $(m^\star, \chi^\star)$ is therefore always a trivial (or paramagnetic) stationary point of the Bethe free energy

*Trivial stationary point*

$$
\begin{aligned}
m_i^\star &= 0\,, & \forall i \in [n]\,, \\
\chi_{ij}^\star &= \tanh(J_{ij})\,, & \forall (ij) \in E\,.
\end{aligned}
\tag{2.72}
$$

Whether this stationary point is a local minimum of the Bethe free energy depends on the Hessian of $\mathcal{F}^{\text{Bethe}}$ evaluated at the stationary point. A straightforward computation yields

*Hessian of the Bethe free energy*

$$
\left.\frac{\partial^2 \mathcal{F}^{\text{Bethe}}}{\partial \chi_{ij} \partial \chi_{kl}}\right|_{m^\star, \chi^\star} = \frac{\mathbf{1}\big((ij) = (kl)\big)}{1 - \tanh^2(J_{ij})}\,, \qquad \forall (ij), (kl) \in E\,, \tag{2.73}
$$

$$
\left.\frac{\partial^2 \mathcal{F}^{\text{Bethe}}}{\partial \chi_{ij} \partial m_k}\right|_{m^\star, \chi^\star} = 0\,, \qquad \forall (ij) \in E, k \in [n]\,, \tag{2.74}
$$

$$
\left.\frac{\partial^2 \mathcal{F}^{\text{Bethe}}}{\partial m_i \partial m_j}\right|_{m^\star, \chi^\star} = H_{ij}(1)\,, \qquad \forall i, j \in [n]\,. \tag{2.75}
$$

The Hessian of the Bethe free energy therefore has a block diagonal structure, with a large "$\chi - \chi$" block given by equation (2.73) which is a diagonal matrix with strictly positive entries on the diagonal. The only non-trivial block in the Hessian is the "$m - m$" block of equation (2.75), which equals the matrix $H(1)$ defined in the previous section. In particular, the only possible negative eigenvalues $\lambda < 0$ of the full Hessian are negative eigenvalues of $H(1)$, and their corresponding eigenvectors are of the form $(v, \vec{0}) \in \mathbb{R}^{n+|E|}$, where $v$ is an eigenvector of $H(1)$ with eigenvalue $\lambda$, and $\vec{0} \in \mathbb{R}^{|E|}$ is the vector with all its entries equal to 0.

Reinterpreting the previous results, we have just shown that the matrix $H(x)$ is the (only non-trivial part of the) Hessian of the Bethe free energy, at the trivial, paramagnetic stationary point, of an *effective Ising model*, with vanishing fields, and couplings $\left(J'_{ij}\right)_{(ij) \in E}$ given by

*Effective Ising model*

$$
\tanh\left(J'_{ij}\right) = \frac{\tanh\left(J_{ij}\right)}{x}\,. \tag{2.76}
$$

*Pseudo-temperature*

The parameter $x$ therefore plays the role of a *pseudo-temperature*, controlling the strength of the interactions. In particular, the Hessian becomes singular at a phase transition, and we can therefore interpret the real eigenvalues of $B$ as critical pseudo-temperatures in this effective Ising model (assuming the limit $n \to \infty$). Indeed, from the fluctuation-dissipation theorem, $H(x)$ is nothing but the inverse of the susceptibility matrix of the effective Ising model.

Remarkably, the matrix $H\left(\hat{R}\right)$ corresponds to the Hessian of the Bethe free energy *at the paramagnetic to spin glass transition*. This fact justifies the following heuristic picture. As argued previously, BP and B give meaningful answers when the Ising model associated with the

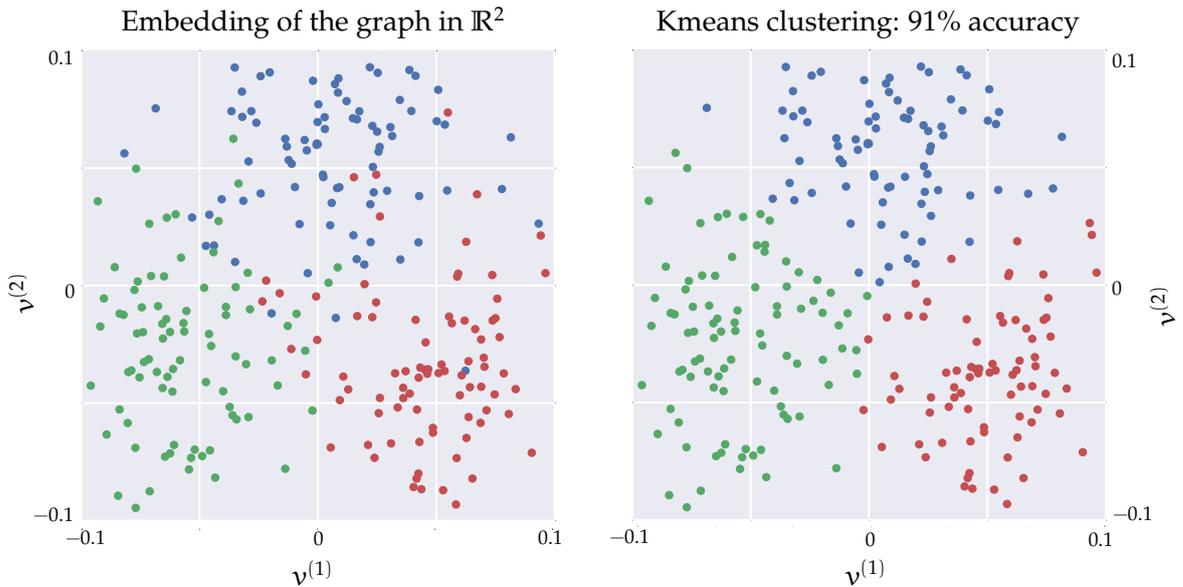

Figure 2.7 – Kmeans clustering of the eigenvectors of $H\left(\hat{R}\right)$ on the example of figure 2.6. We use the 2 eigenvectors $\nu^{(1)}$ and $\nu^{(2)}$ with negative eigenvalues of $H(x_3 = \hat{R})$ to embed the graph in $\mathbb{R}^2$. The embedding is shown on the left panel, where each vertex i is represented as a point $(m_i^{(1)}, m_i^{(2)}) \in \mathbb{R}^2$, and the colors represent the ground true cluster assignment. In the right panel, we show the result of the kmeans algorithm, which is able to correctly label 91% of the points.

pairwise MRF is in a partially ordered phase. By varying x in the effective Ising model, we explore its phase diagram. We find that for large x, H(x) is positive definite (as shown in the previous section), so that the paramagnetic stationary point is a local minimum of the Bethe free energy, and the system is in a paramagnetic phase. As we lower x from a large value to $\hat{R}$, the effective Ising model undergoes phase transitions if B has informative, real eigenvalues $\lambda > \hat{R}$. At $x = \hat{R}$, the model finally undergoes a paramagnetic to spin glass transition. From the typical phase diagrams of spin glasses presented in section 1.6.4, we expect that the ferromagnetic transitions, if any, must have happened before (i. e. for a larger value of x). Therefore, we expect that if the Bethe Hessian $H\left(\hat{R}\right)$ has negative eigenvalues, the associated eigenvectors can be used as proxies for the magnetizations of the effective Ising model in its partially ordered phases. One way of thinking about this approximation is by imagining that we minimize the Bethe free energy with a second order (e. g. Newton or quasi-Newton) algorithm, starting from the paramagnetic point. If $H\left(\hat{R}\right)$ has negative eigenvalues, then this paramagnetic point is a saddle point, and the first step of the optimization algorithm can decrease the Bethe free energy by moving along the direction of one of the eigenvectors of $H\left(\hat{R}\right)$ with negative eigenvalues. In a sense, we use $H\left(\hat{R}\right)$ to *probe* the Bethe free energy landscape, locally around the m = 0 point. Somewhat miraculously, in the examples we will

*Phase diagram of the effective Ising model*

*Probing the Bethe free energy landscape*

study, this probe is able to sense all the interesting local minima of the Bethe free energy, in the same way as the principal eigenvectors of B give away all the informative fixed points of BP on the associated Ising model.

By the same arguments as for the non-backtracking operator, we may use the eigenvectors of $H(\hat{R})$ with negative eigenvalues to embed the graph in an Euclidean space, and use kmeans to cluster the graph. The result for the example of figure 2.6 is shown on figure 2.7, and allows to recover the true clusters with a 91% accuracy. This hand-waving heuristic, which we shall detail more in the example of the planted spin glass in chapter 4, justifies algorithm 2. We will argue on the specific examples of the following sections that this algorithm is also optimal on some important models., in the sense that it can detect clusters as soon as it is possible to do so. However, compared to algorithm 1, this algorithm is based on a smaller ($n \times n$), real, and *symmetric* matrix, which is easy to build (while the non-backtracking matrix is not). This results in a sizable speed up of the whole procedure. Additionally, we will observe that algorithm 2 consistently provides a slightly better reconstruction accuracy, although a theoretical understanding of this fact is still missing. We will also discuss extensions of this algorithm in the non-Bayes optimal setting, where the parameters of the model are not known.

---

**Algorithm 2** Graph clustering with the Bethe Hessian

**Input:** Number of clusters q, probability $\mathbb{P}(\sigma)$ of each cluster assignment $\sigma \in [q]^n$ assumed to be a pairwise MRF of the symmetric form (2.25)

1: **Compute** the estimate $\hat{R}$ (2.59) of the radius of the bulk of the spectrum of B
2: **Build** the Bethe Hessian $H(\hat{R})$ where $H(x)$ is defined by equation (2.67)
3: **Compute** all the negative eigenvalues of $H(\hat{R})$. Let us call $r$ their number and $v^{(1)}, \ldots, v^{(r)}$ their corresponding eigenvectors. If $r = 0$, raise an error.
4: **Embed** the vertices of the graph in $\mathbb{R}^r$ by assigning to each vertex $i \in [n]$ the coordinates $(m_i^1, \ldots, m_i^r) \in \mathbb{R}^r$
5: **Cluster** the embedded vertices using (e. g.) *kmeans*

---

It is fair to wonder why we do not use instead the true Hessian of the Bethe free energy of the Ising model associated with the pairwise MRF, i. e. $H(1)$. The simple answer is that this method would not (always) work, in the sense that it may happen (as we will illustrate in the following) that $H(1)$ is positive definite, i. e. the trivial stationary point of the Bethe free energy is stable, but $\rho(B) > 1$, i. e. the trivial fixed point of BP is unstable. This is an illustration of the fact that, while stable fixed points of BP *are* local minima of the Bethe free en-ergy ([61]), the converse is not true in general. Note that this does not contradict the Ihara-Bass formula. We find numerically that, upon lowering $x$ from $+\infty$ to 1, $H(x)$ does indeed become singular whenever $x$ is equal to an eigenvalue $\lambda$ of $B$. However, what happens empirically is that for $\lambda > R$, the Hessian *gains* a new negative eigenvalue when $x$ becomes smaller than $\lambda$, while for $1 < \lambda < R$, the Hessian *loses* a negative eigenvalue. It can be checked on figure 2.4 that $B$ may indeed have real eigenvalues in the range $[1, R]$, so that $H(1)$ does not, in general, have as many negative eigenvalues as $B$ has informative eigenvalues. In particular, we find that the paramagnetic stationary point of the Bethe free energy may be a local minimum, while the trivial fixed point of BP is unstable. This is exemplified on figure 2.6, where $H(x_4 = 1)$ is positive definite while the spectral radius of $B$ is larger than 1.

### 2.3.3 *Free energy Hessians*

We have argued that, in order to cluster a graph into $q$ groups, it is interesting to consider an associated Ising model defined on this same graph, and to compute (approximately) its magnetizations in its ferromagnetic phases. On sparse graphs, a natural choice of approximation is BP or the Bethe free energy. However, on other types of graphs, it may be reasonable to resort to a different approximation. We show in this section that this point of view allows to make contact with more classical spectral clustering methods. To do so, we consider the Ising model

$$\mathbb{P}(\sigma) = \frac{1}{Z} \exp \sum_{(ij) \in E} \left( J_{ij} \sigma_i \sigma_j \right) , \qquad (2.77)$$

where each coupling $J_{ij}$ may be given by equation (2.44), or may simply be some number that reflects our confidence that $i$ and $j$ belong in the same cluster. We consider two mean-field approximations, different from the Bethe approximation, both introduced in section 1.6.1. The first is the simple nMF approximation, whose corresponding free energy reads

*The nMF approximation*

$$\mathcal{F}^{nMF}(m) = - \sum_{(ij) \in E} J_{ij} m_i m_j + \sum_{i=1}^{n} \sum_{\sigma_i = \pm 1} \eta \left( \frac{1 + \sigma_i m_i}{2} \right) , \quad (2.78)$$

where $\eta(x) = x \log(x)$ for $x > 0$. This free energy is defined for any set of magnetizations $m \in (0, 1)^n$. Computing the gradient of $\mathcal{F}^{nMF}$, we obtain

$$\begin{aligned} \frac{\partial \mathcal{F}^{nMF}}{\partial m_i} &= - \sum_{j \in \partial i} J_{ij} m_j + \frac{1}{2} \log \frac{1 + m_i}{1 - m_i} , \\ &= - \sum_{j \in \partial i} J_{ij} m_j + \operatorname{atanh}(m_i) . \end{aligned} \qquad (2.79)$$

Equating the gradient to 0, we find the familiar naive mean-field approximation to the magnetization

$$m_i = \tanh\left(\sum_{i \in \partial j} J_{ij} m_j\right), \tag{2.80}$$

which admits as a trivial solution the paramagnetic point $m = 0$. To understand whether this fixed point is a local minimum of the nMF free energy, we compute the Hessian (or inverse susceptibility) at this point

$$\left.\frac{\partial^2 \mathcal{F}^{\text{nMF}}}{\partial m_i \partial m_j}\right|_{m=0} = \mathbb{1}(i = j) - \mathbb{1}(j \in \partial i) J_{ij}. \tag{2.81}$$

Looking for the smallest eigenvalues of the nMF Hessian therefore amounts to looking for the largest eigenvalues of the matrix $A \in \mathbb{R}^{n \times n}$, with elements $A_{ij} = \mathbb{1}(j \in \partial i) J_{ij}$. This is nothing but the adjacency matrix of the graph $G$, potentially weighted. In the unweighted case $J_{ij} = 1$, $\forall(ij) \in E$, the adjacency matrix has been extensively used in community detection, and has been shown in [112] to correctly recover the planted partition in the SBM *provided* the graph is sufficiently dense, with average connectivity going to $\infty$ in the large $n$ limit. In particular, spectral methods based on the adjacency matrix are shown in [89] to fail significantly above the optimal threshold in the sparse case where the average connectivities $\alpha_{\text{in}}$, $\alpha_{\text{in}}$ are bounded, independent of $n$. Similarly, [94] showed that in the lSBM, a spectral method based on the adjacency matrix, with weights given by equation (2.44) (we write these weights explicitly for the lSBM in chapter 3), is optimal, again under the condition that the average connectivity of the graph tends to $\infty$ with $n$. In the sparse case, the same authors show that the adjacency matrix succeeds in detecting the planted partition only when above the threshold of the lSBM by a large constant. From a statistical physics point of view, all these results are not surprising. Indeed, as we have seen in section 1.6.3, the nMF approximation is exact on the Curie-Weiss model, i.e. on fully connected graphs with small ferromagnetic couplings. In general, the nMF approximation corresponds to assuming that each spin has a large number of neighbors with which it weakly interacts, so that the interaction of each spin with its neighbors can be replaced by an effective (mean) field, whose fluctuations are negligible. On sparse graphs, this assumption is strongly violated, and methods derived from the nMF approximation can therefore be expected to fail. These observations encourage us to look at more advanced mean-field approximations.

*The nMF approximation yields the adjacency matrix*

The next order approximation in the Plefka expansion of section 1.6.1 is the TAP free energy given by

$$\mathcal{F}^{\text{TAP}}(m) = -\sum_{(ij)\in E} \left( J_{ij} m_i m_j + \frac{1}{2} J_{ij}^2 (1-m_i^2)(1-m_j^2) \right)$$
$$+ \sum_{i=1}^{n} \sum_{\sigma_i = \pm 1} \eta \left( \frac{1+\sigma_i m_i}{2} \right). \tag{2.82}$$

As has become customary, we compute the gradient of $\mathcal{F}^{\text{TAP}}$

$$\frac{\partial \mathcal{F}^{\text{TAP}}}{\partial m_i} = -\sum_{j\in\partial i} \left( J_{ij} m_j - J_{ij}^2 m_i (1-m_j^2) \right) + \operatorname{atanh}(m_i), \tag{2.83}$$

so that a stationary point of the TAP must verify the following set of equations, known as the TAP equations

$$m_i = \tanh \left( \sum_{j\in\partial i} J_{ij} m_j - J_{ij}^2 m_i (1-m_j^2) \right). \tag{2.84}$$

Without surprise, we find that the paramagnetic point $m = 0$ is again a stationary point of the TAP free energy. The Hessian, or inverse susceptibility, is given by

$$\frac{\partial^2 \mathcal{F}^{\text{TAP}}}{\partial m_i \partial m_j} \bigg|_{m=0} = \left( 1 + \sum_{k\in\partial i} J_{ik}^2 \right) \mathbf{1}(i=j) - \mathbf{1}(j\in\partial i) J_{ij}. \tag{2.85}$$

The problem of finding the smallest eigenvalues of the TAP Hessian is equivalent to the problem of finding the smallest eigenvalues of the matrix $\widetilde{L} = \widetilde{D} - A$, where $A$ is the adjacency matrix defined previously, and $\widetilde{D}$ is a a diagonal matrix with entries

$$\widetilde{D}_{ii} = \sum_{k\in\partial i} J_{ik}^2, \qquad \forall i \in [n]. \tag{2.86}$$

The matrix $\widetilde{L}$ therefore resembles the classical *Laplacian* matrix, and is exactly equal to the Laplacian in the unweighted case $J_{ij} = 1$ for $(ij) \in E$, which arises naturally and is widely used in community detection [46, 114]. It is interesting to note that the usual motivation for the use of the Laplacian follows from a graph cut point of view [98]. Here, we have instead introduced it from a probabilistic perspective, as the Hessian of the TAP free energy of an Ising model defined on the graph to cluster. Note that the Laplacian matrix is always positive semi-definite, so that this matrix cannot have negative eigenvalues. The spectral algorithm instead consists in finding the smallest eigenvalues of the Laplacian (which are different from the trivial eigenvalue 0 [98]).

Like the adjacency matrix, the Laplacian is observed in [89] to fail to detect communities in the sparse SBM near the optimal threshold.

Once more, statistical physics gives an interpretation of this fact. Indeed, the TAP equations (2.84) can be obtained as a large connectivity limit of the BP equation for the Ising model. Therefore, they are expected to provide a good approximation of the magnetizations of the Ising model only in sufficiently dense graphs. For sparse graphs, the Bethe approximation should be used instead, i. e. the Bethe Hessian.

*Generalizing the Bethe approach*

Abstracting from these examples, it is in principle possible to generalize our approach to different types of graphs, *e. g.* those containing cliques. Indeed, the Bethe approximation, and therefore the non-backtracking operator and the Bethe Hessian, are expected to fail on graphs containing short loops. It was shown in [68] that adding cliques to the graph causes the appearance of negative eigenvalues of the Bethe Hessian with very localized eigenvectors, spoiling the ability of this Bethe Hessian to uncover global structure in a graph. This localization problem is a standard one, and affects all spectral methods [89]. On the other hand, the Bethe approximation can be generalized to take into account densely connected subgraphs. This class of mean-field approaches, which minimize a so-called Kikuchi free energy [85], has been used by [161] to propose a generalized belief propagation algorithm, tailored for graphs that contain regions with many (small) loops. An interesting direction for future work is the investigation of the existence of trivial fixed points of generalized BP, or trivial stationary points of the Kikuchi free energy. As we hope to have conveyed in this chapter, the local analysis of such trivial points may yield insight into the complexity of the problem at hand, as well as new spectral algorithms, obtained as linear relaxations of more involved algorithms such as (generalized) BP, or direct free energy minimization.

## 2.4   CONCLUSION

In this chapter, we have identified a general class of *factorized* pairwise MRF for which BP has a trivial fixed point. By studying the linear stability of this trivial fixed point, we have shown that BP undergoes a phase transition, expressed in terms of the spectral radius of its Jacobian. Restricting further the class of factorized models, we have defined a subclass of *symmetric* models, which contains in particular interesting special cases of the Potts and Ising models. On symmetric models, the Jacobian of BP factorizes and can be expressed in terms of a *non-backtracking* operator B, acting on the directed edges of the graph, and controlling the stability of the trivial fixed point of BP. We have then argued that the leading eigenvectors of B contain non-trivial information, and allow to design a spectral algorithm for approximate inference in symmetric pairwise MRFs. In particular, we have shown that these informative eigenvectors of B can be thought of as approximating the magnetizations of an associated Ising model.

Adopting a different point of view based on mean-field free energy approximations, we have related the non-backtracking operator B to a smaller, symmetric matrix $H(x)$ called the Bethe Hessian, which controls the stability of a paramagnetic stationary point in the associated Ising model, in the Bethe approximation. Generalizing our approach to other mean-field approximations, we have shown that we recover classical linear operators used in graph clustering, and we gave an interpretation of their expected performance based on the accuracy of the mean-field approximation from which they derive. In the next chapter, we study in details the spectral properties of the non-backtracking operator and the Bethe Hessian on sparse random graphs. These results will allow us, in the applications of the following chapters, to design and study the efficiency of spectral algorithms based on B and $H(x)$.

# SPECTRAL PROPERTIES OF THE NON-BACKTRACKING OPERATOR AND THE BETHE HESSIAN ON RANDOM GRAPHS

The performance of algorithms 1 and 2 relies crucially on their spectral properties on sparse graphs. In this chapter, we study heuristically these properties on graphs generated from the sparse ISBM, and show that both the spectra of B and H(x) are well-behaved, in the sense that we can control the location of both their informative and uninformative eigenvalues. In particular, their noisy eigenvalues are sharply separated from the informative ones in the limit where the size of the graph n grows to infinity, allowing algorithms 1 and 2 to perform well, even on very sparse graphs.

This is in sharp contrast with other matrices typically used for spectral clustering. As a simple example, let consider again the adjacency matrix A of a sparse Erdős-Rényi graph. As explained in the previous chapter, a typical spectral algorithm based on the adjacency matrix looks for its leading eigenvalues, and uses the corresponding eigenvectors to cluster the graph. On sparse Erdős-Rényi graphs, this strategy fails, due to the presence of large eigenvalues of A whose eigenvectors are localized on high degree vertices, and therefore do not allow to find any global structure in the graph. A simple argument to justify this claim is the following ([34]). Using classical results on the extreme value statistics of the Poisson distribution, it is straightforward to see that the largest degree $d_{max}$ of a vertex in the graph is of order $\log n / \log \log n$. Let us call $i$ the vertex with maximum degree, and $x \in \mathbb{R}^n$ the vector whose only non-zero component is $x_i = 1$. The quantity $x^\intercal A^2 x = (A^2)_{ii}$ is equal to the number of 2-steps walks on the graph G that start and end at $i$. Therefore, we have

$$d_{max} = x^\intercal A^2 x. \qquad (3.1)$$

Since $A^2$ is a real and symmetric matrix, $x$ can be orthogonally decomposed as a linear combination of eigenvectors of $A^2$, so that the previous argument shows that $A$ has an eigenvalue larger than $\sqrt{d_{max}}$, unbounded as $n \to \infty$. Therefore, due to the presence of heterogeneity in the degree of the vertices of a sparse random graph, $A$ has large eigenvalues, that are uninformative since they exist even when there is no cluster structure in the graph. Note that equation (3.1) holds because the adjacency matrix allows backtracking, and therefore coming back to the vertex $i$ right after leaving it.

In the following, we argue that the situation is different for the non-backtracking operator B and the Bethe Hessian H(x), in the sense

*Why classical spectral methods fail on sparse graphs*

that their uninformative eigenvalues are constrained in a well defined region, allowing to easily detect the potential useful eigenvalues. We first consider the non-backtracking operator in section 3.1, and turn to the Bethe Hessian in section 3.2.

### 3.1    THE NON-BACKTRACKING OPERATOR

We start by studying the spectral properties of the non-backtracking operator. We first derive the explicit expression of the matrix B arising from the stability analysis of the trivial BP fixed point on graphs G generated from the lSBM, and discuss some subtleties arising from the presence of interactions on so-called "non-edges" of G. We then use the cavity method of statistical physics to study analytically the stability of the trivial BP fixed point. This allows us to derive a precise picture of the spectrum of B on graphs generated from the lSBM. Finally, we show how to compute, using non-rigorous statistical physics methods, the spectral density of B on the unlabeled SBM.

#### 3.1.1    *Non-backtracking operator of the symmetric labeled stochastic block model*

In this section, we write down the non-backtracking operator of a graph G generated from the symmetric lSBM. Strictly speaking, the resulting non-backtracking matrix is defined on a fully connected graph, due to the presence of interactions on the "non-edges" of G (see following). However, we argue that in the limit $n \to \infty$, we can instead use a non-backtracking operator corresponding to the graph G, retaining only a subset of its eigenvectors.

*Sampling from the symmetric lSBM*

Recall that a graph $G = ([n], E)$ is generated, in the symmetric lSBM, as follows. First, we assign each vertex $i \in [n]$ to a cluster $\sigma_i \in [q]$ with uniform probability $1/q$. For each pair of vertices $i$ and $j$ with $1 \leqslant i < j \leqslant n$, we include the edge $(ij)$ in E with probability

$$\mathbb{P}\left((ij) \in E \mid \sigma_i, \sigma_j\right) = \frac{\alpha_{\text{in}}}{n}\, \mathbf{1}(\sigma_i = \sigma_j) + \frac{\alpha_{\text{out}}}{n}\, \mathbf{1}(\sigma_i \neq \sigma_j)\,. \tag{3.2}$$

The average connectivity of the graph is therefore

$$\alpha = \frac{\alpha_{\text{in}} + (q-1)\alpha_{\text{out}}}{q}\,. \tag{3.3}$$

Finally, we assign each edge $(ij) \in E$ a random label $\ell_{ij} \in \mathcal{L}$ drawn from the distribution

$$\mathbb{P}\left(\ell_{ij} = \ell \mid \sigma_i, \sigma_j\right) = p_{\text{in}}(\ell)\, \mathbf{1}(\sigma_i = \sigma_j) + p_{\text{out}}(\ell)\, \mathbf{1}(\sigma_i \neq \sigma_j)\,, \tag{3.4}$$

where $p_{\text{in}}$ and $p_{\text{out}}$ are two probability distributions over $\mathcal{L}$.

We have shown in chapter 1 that the posterior distribution (1.105) of the cluster assignment $\sigma$ given the labeled graph $G = ([n], E)$ is a pairwise MRF on the fully connected graph with $n$ vertices, with potentials given by

$$\forall (ij) \in E, \psi_{ij}(\sigma_i, \sigma_j) = \begin{cases} \dfrac{\alpha_{\text{in}}}{n} p_{\text{in}}(\ell_{ij}) & \text{if } \sigma_i = \sigma_j, \\[2mm] \dfrac{\alpha_{\text{out}}}{n} p_{\text{out}}(\ell_{ij}) & \text{if } \sigma_i \neq \sigma_j \end{cases},$$

$$\forall (ij) \notin E, \psi_{ij}(\sigma_i, \sigma_j) = \begin{cases} 1 - \dfrac{\alpha_{\text{in}}}{n} & \text{if } \sigma_i = \sigma_j \\[2mm] 1 - \dfrac{\alpha_{\text{out}}}{n} & \text{if } \sigma_i \neq \sigma_j \end{cases}, \qquad (3.5)$$

$$\forall i \in [n], \phi_i(\sigma_i) = \frac{1}{q}.$$

Note that, whenever $\alpha_{\text{in}} \neq \alpha_{\text{out}}$, the *absence* of an edge between two vertices also encodes information about the planted assignment $\sigma$. As a consequence, there are potentials $\psi_{ij}$ associated to *non-edges* $(ij) \notin E$, so that the labeled ISBM of equation (3.5) is a pairwise MRF with respect to the *complete* graph with $n$ vertices, but not with respect to the graph G generated from the ISBM. However, it is still of the symmetric form (2.25), so that the analysis of the previous section shows that BP has a trivial fixed point, and that its stability is controlled by the (huge) non-backtracking operator $B^c$ of the complete graph, with elements given by (2.33), for $i \neq j, k \neq l \in [n]$,

$$B^c_{i \to j, k \to l} = w_{kl}\, \mathbf{1}(l = i)\mathbf{1}(k \neq j), \quad \text{with weights} \qquad (3.6)$$

$$w_{kl} = \begin{cases} \dfrac{\alpha_{\text{in}} p_{\text{in}}(\ell_{kl}) - \alpha_{\text{out}} p_{\text{out}}(\ell_{kl})}{\alpha_{\text{in}} p_{\text{in}}(\ell_{kl}) + (q-1)\alpha_{\text{out}} p_{\text{out}}(\ell_{kl})} & \text{if } (kl) \in E, \\[3mm] -\dfrac{\alpha_{\text{in}} - \alpha_{\text{out}}}{nq - \alpha_{\text{in}} - (q-1)\alpha_{\text{out}}} & \text{if } (kl) \notin E. \end{cases}$$

$$(3.7)$$

An eigenpair $(\lambda, V)$ with real eigenvalue $\lambda$ of this large matrix must verify, for all $(i \to j)$ such that $(ij) \notin E$

$$\lambda V_{i \to j} = \sum_{k \in \partial i} w_{ki} V_{k \to i} + \sum_{\substack{k \notin \partial i \\ k \neq i,j}} w_{ki} V_{k \to i}, \qquad (3.8)$$

where only the last term depends on the target node $j$, i.e.

$$\sum_{\substack{k \notin \partial i \\ k \neq i,j}} w_{ki} V_{k \to i} = -\frac{\alpha_{\text{in}} - \alpha_{\text{out}}}{nq - \alpha_{\text{in}} - (q-1)\alpha_{\text{out}}} \sum_{\substack{k \notin \partial i \\ k \neq i,j}} V_{k \to i} \qquad (3.9)$$

$$= -\frac{\alpha_{\text{in}} - \alpha_{\text{out}}}{nq} \sum_{\substack{k \notin \partial i \\ k \neq i}} V_{k \to i} + O\left(\frac{1}{n}\right) \qquad (3.10)$$

Notice that in the last line, we have included back the contribution $V_{j \to i}$, which is negligible since the sum contains of the order of $n$ terms. As a consequence, in the limit $n \to \infty$, the components $V_{i \to j}$ corresponding to non-edges do not depend on the target node $j$, and we can write equation (3.8) as

$$V_{i \to j} = M_i + O\left(\frac{1}{n}\right), \qquad \forall (ij) \notin E, \tag{3.11}$$

where we have defined $n$ parameters $M_i$. The system of equations (3.8) then reduce to $n$ equations on $M_i$ for $i \in [n]$

$$\lambda M_i = \sum_{k \in \partial i} w_{ki} V_{k \to i} - \frac{\alpha_{in} - \alpha_{out}}{nq} \sum_{\substack{k \notin \partial i \\ k \neq i}} M_k + O\left(\frac{1}{n}\right), \tag{3.12}$$

$$= \sum_{k \in \partial i} w_{ki} V_{k \to i} - \frac{\alpha_{in} - \alpha_{out}}{nq} \sum_{i=1}^{n} M_k + O\left(\frac{1}{n}\right). \tag{3.13}$$

The components $V_{i \to j}$ for $(ij) \in E$ write

$$\lambda V_{i \to j} = \sum_{k \in \partial i \backslash j} w_{ki} V_{k \to i} + \sum_{\substack{k \notin \partial i \\ k \neq i}} w_{ki} V_{k \to i}, \tag{3.14}$$

$$= \sum_{k \in \partial i \backslash j} w_{ki} V_{k \to i} - \frac{\alpha_{in} - \alpha_{out}}{nq} \sum_{\substack{k \notin \partial i \\ k \neq i}} M_k + O\left(\frac{1}{n}\right), \tag{3.15}$$

$$= \sum_{k \in \partial i \backslash j} w_{ki} V_{k \to i} - \frac{\alpha_{in} - \alpha_{out}}{nq} \sum_{i=1}^{n} M_k + O\left(\frac{1}{n}\right). \tag{3.16}$$

Therefore, when $n \to \infty$, $(\lambda, V)$ is an eigenpair of $B^c$ if and only if it verifies the system

$$\begin{aligned} &\forall (i \to j) \in \vec{E}, \ \lambda V_{i \to j} = \sum_{k \in \partial i \backslash j} w_{ki} V_{k \to i} - \frac{\alpha_{in} - \alpha_{out}}{nq} \sum_{k=1}^{n} M_k, \\ &\forall (i \to j) \notin \vec{E}, \ V_{i \to j} = M_i, \\ &\forall i \in [n], \ \lambda M_i = \sum_{k \in \partial i} w_{ki} V_{k \to i} - \frac{\alpha_{in} - \alpha_{out}}{nq} \sum_{k=1}^{n} M_k. \end{aligned} \tag{3.17}$$

Let us now define the more natural non-backtracking operator $B$ of the graph $G$ by its elements, for $(i \to j), (k \to l) \in \vec{E}$

$$B_{i \to j, k \to l} = w_{kl} \ \mathbb{1}(l = i) \mathbb{1}(k \neq j), \tag{3.18}$$

with weights $w_{kl}$ given by equation (3.7) for $(kl) \in E$ only. Remark that if $(\lambda, v)$ is an eigenpair of $B$, and if $n^{-1} \sum_{i=1}^{n} m_i = 0$, were $m = P v$, then $(\lambda, V)$ is an eigenpair of $B^c$, where

$$\begin{aligned} &\forall (i \to j) \in \vec{E}, \ V_{i \to j} = v_{i \to j}, \\ &\forall i \in [n], \ M_i = \frac{m_i}{\lambda}. \end{aligned} \tag{3.19}$$

Indeed, it is readily checked that the previous assignment satisfies the set of equations (3.17). In particular, in the large $n$ limit, the (potential) real eigenvectors of $B$ with large real eigenvalue are unstable directions of the BP algorithm if the condition $n^{-1} \sum_{i=1}^{n} m_i = 0$ holds. In the next section, we show that we indeed expect to find such eigenvectors of $B$, so that a spectral algorithm algorithm based on the non-backtracking operator $B$ of the graph $G$ succeeds in recovering the hidden clustering of the vertices. The arguments exposed in the present section suggests that the non-edge interactions simply suppress the *globally ordered* eigenvectors of $B$, i.e. those whose total magnetization is finite (see next sections).

To sum up, we will therefore, somewhat improperly, call non-backtracking operator of the ISBM the matrix $B \in \mathbb{R}^{|\vec{E}| \times |\vec{E}|}$ defined by equation (3.18), which we write as

$$B_{(i \to j),(k \to l)} = w_{kl} \, \mathbf{1}(l = i)\mathbf{1}(k \neq j), \, \forall (i \to j), (k \to l) \in \vec{E}, \quad (3.20)$$

where the weights $w_{kl}$ for $(kl) \in E$ are given by

$$w_{kl} = \frac{\alpha_{in} \, p_{in}(\ell_{kl}) - \alpha_{out} \, p_{out}(\ell_{kl})}{\alpha_{in} \, p_{in}(\ell_{kl}) + (q - 1)\alpha_{out} \, p_{out}(\ell_{kl})}, \qquad \forall (kl) \in E. \quad (3.21)$$

The corresponding pooling matrix is

$$P_{i,(k \to l)} = w_{kl} \, \mathbf{1}(l = i), \quad \forall i \in [n], (k \to l) \in \vec{E}. \quad (3.22)$$

### 3.1.2 *Cavity approach to the stability of the paramagnetic fixed point*

We study heuristically the statistical properties of the eigenvectors of $B$ on the ISBM, in the limit $n \to \infty$. We stress that our approach in non-rigorous, and based on the cavity method of statistical physics [103]. However, we believe these claims to be correct, and will give rigorous arguments supporting them in some special cases in the following. We consider in this section a general non-backtracking operator $B \in \mathbb{R}^{|\vec{E}| \times |\vec{E}|}$ defined by its elements, for $(i \to j), (k \to l) \in \vec{E}$

$$B_{i \to j, k \to l} = w_{kl} \, \mathbf{1}(l = i)\mathbf{1}(k \neq j), \quad (3.23)$$

where for $(kl) \in E$, $w_{kl} = w(\ell_{kl})$ is an arbitrary weight, defined through a weighting function $w$ of the observed label $\ell_{kl}$. We call *Bayes optimal* the choice of weighting function $w = w^{\star}$, where $w^{\star}$ is given by

$$w^{\star}(\ell) = \frac{\alpha_{in} \, p_{in}(\ell) - \alpha_{out} \, p_{out}(\ell)}{\alpha_{in} \, p_{in}(\ell) + (q - 1)\alpha_{out} \, p_{out}(\ell)}, \qquad \forall \ell \in \mathcal{L}. \quad (3.24)$$

With this choice of weighting function, the non-backtracking operator is the one obtained in the previous section by linearizing BP in the Bayes optimal setting. We expect (and will check) that the performance of a spectral algorithm based on the non-backtracking operator will be best for this particular choice of weights. However, it is interesting to understand how the performance is degraded if we do not assume this Bayes optimal setting, e.g. in the case where the distributions $p_{in}$ and $p_{out}$ are unknown. We therefore consider in the following a generic weighting function $w$.

Starting from an arbitrary initial condition $v^{(0)} \in \mathbb{R}^{|\vec{E}|}$, we consider the power iteration, for $0 \leqslant l \leqslant t - 1$

*Power iteration*

$$v^{(l+1)} = \frac{1}{x} B v^{(l)},  \tag{3.25}$$

where $x > 0$ is a real constant. As shown in chapter 2, this power iteration controls the stability of the paramagnetic fixed point of BP in the Ising model with couplings $J_{ij}$ related to the weights $w_{ij}$ of the non-backtracking operator through

*Effective Ising model*

$$\tanh\left(J_{ij}\right) = \frac{w_{ij}}{x},  \qquad \forall (ij) \in E.  \tag{3.26}$$

As argued in section 2.3.2, when we vary the pseudo-temperature $x$, this Ising model undergoes phase transitions when $x$ equals an informative eigenvalue of B. As we will see, these phase transitions are signaled by an instability of the paramagnetic fixed point with respect to ordered, or partially ordered perturbations. We will compute explicitly the values of $x$ for which these instabilities happen, and this will give us a precise picture of the spectrum of B.

From the vector $v^{(t)}$ constructed by iterating equation (3.25), we obtain approximate magnetizations $m^{(t)} \in \mathbb{R}^n$ by using the pooling matrix of equation (3.22)

*Approximate magnetizations*

$$m^{(t)} = \frac{1}{x} P v^{(t)}.  \tag{3.27}$$

Our aim is to say something about the statistics of the components of $m^{(t)}$, as a function of the choice of the initial condition $v^{(0)}$. In particular, we are interested in finding out whether, starting from an initial condition "correlated" with the true cluster assignment of the vertices, this correlation grows or decays as we iterate B. We shall fix a vertex $I \in [n]$ chosen uniformly at random among the $n$ vertices, and investigate the distribution of its magnetization $m_I^{(t)}$ *conditioned* on the cluster assignment $\sigma_I \in [q]$ of vertex I.

As noted in section 1.4, the graph G is locally tree-like, so that it is possible to show ([103, 129]) that for any *finite* number of iterations t, $m_I^{(t)}$ converges in probability to a random variable

$$\mathbb{P}\left(m_I^{(t)} \mid \sigma_I = \sigma\right) \xrightarrow[n \to \infty]{} \mathbb{P}\left(m_\sigma^{(t)}\right),  \tag{3.28}$$

where the distribution of the random variables $m_\sigma^{(t)}$ for $\sigma \in [q]$ is given in terms of the distribution of other random variables $v_\sigma^{(t)}$ by

$$m_\sigma^{(t)} \stackrel{\mathcal{D}}{=} \frac{1}{x} \sum_{i=1}^{d_{in}} w_{i,in} v_{i,\sigma}^{(t)} + \frac{1}{x} \sum_{\tau \in [q] \setminus \sigma} \sum_{i=1}^{d_{\tau,out}} w_{i,\tau,out} v_{i,\tau}^{(t)}, \qquad (3.29)$$

where $\stackrel{\mathcal{D}}{=}$ means equality in distribution. In this expression, $d_{in}$ is a Poisson random variable with mean $\alpha_{in}/q$, and the $d_{\tau,out}$ for $\tau \in [q] \setminus \sigma$ are i.i.d Poisson random variables with mean $\alpha_{out}/q$. The $w_{i,in}$ (resp. $w_{i,\tau,out}$) are i.i.d copies of a random variable $w_{in}$ (resp. $w_{out}$) with distribution

$$\begin{aligned} \mathbb{P}(w_{in} = w) &= \mathbb{P}\left(w_{ij} = w \mid \sigma_i = \sigma_j\right), \\ \mathbb{P}(w_{out} = w) &= \mathbb{P}\left(w_{ij} = w \mid \sigma_i \neq \sigma_j\right), \end{aligned} \qquad (3.30)$$

*$w_{in}$ and $w_{out}$*

where the $w_{ij}$ are the weights of the non-backtracking operator of equation (3.23). Finally, the variables $v_\sigma^{(t)}$ have the same distribution as the components $v_{i \to j}^{(t)}$ for a randomly chosen directed edge $(i \to j)$, conditioned on $\sigma_i = \sigma$. Their distribution is given recursively, for $0 \leqslant l \leqslant t-1$ by

$$v_\sigma^{(l+1)} \stackrel{\mathcal{D}}{=} \frac{1}{x} \sum_{i=1}^{d_{in}} w_{i,in} v_{i,\sigma}^{(l)} + \frac{1}{x} \sum_{\tau \in [q] \setminus \sigma} \sum_{i=1}^{d_{\tau,out}} w_{i,\tau,out} v_{i,\tau}^{(l)}, \qquad (3.31)$$

*Distributional recursion*

Note in particular that $m_\sigma^{(t)}$ has the same distribution as $v_\sigma^{(t+1)}$. This follows from the fact that the excess degree (i. e. the quantity $|\partial i \setminus j|$ for $(ij) \in E$) in a Poissonian graph has the same distribution as the degree. Equation (3.31) is rigorously valid only for a finite number of iterations $l \leqslant t$. However, within the replica symmetric cavity method, we will assume that we can take the limit $l \to \infty$. Heuristically, since the smallest loops in a graph generated from the sBM are of size of the order of $\log n$, this is justified when the correlations between the components of the vector $v^{(l)}$ decay fast enough as a function of the distance between the edges. We will be interested in the evolution of the total magnetization $M^{(l)}$, and Edwards-Anderson parameter $Q^{(l)}$ defined here by

*Replica symmetric cavity method*

$$\begin{aligned} M^{(l)} &= \frac{1}{q} \sum_{\sigma \in [q]} \mathbb{E}\left[v_\sigma^{(l)}\right], \\ Q^{(l)} &= \frac{1}{q} \sum_{\sigma \in [q]} \mathbb{E}\left[\left(v_\sigma^{(l)}\right)^2\right]. \end{aligned} \qquad (3.32)$$

*Total magnetization $M^{(l)}$ and Edwards-Anderson parameter $Q^{(l)}$*

These quantities are analogous to those defined in chapter 1, and they control part of the phase diagram of the associated Ising model, as we will see. We also introduce another parameter which we call $P^{(l)}$, given by

$$P^{(l)} = \lim_{n \to \infty} = \frac{1}{q} \sum_{\sigma \in [q]} \mathbb{E}\left[v_\sigma^{(l)}\right]^2. \qquad (3.33)$$

*Partial order parameter $P^{(l)}$*

When $P^{(l)} \neq 0$ with $M^{(l)} = 0$, the approximate magnetizations are partially ordered, in the sense that their average on some of the clusters is finite, while their total average vanishes. In particular, the expected value of the magnetizations on different clusters must be different. This case corresponds to the informative unstable directions of chapter 2. Indeed, when the magnetizations take typically different values on the different clusters, they can be used as (noisy) indicator functions of the clusters. In the following, we investigate the evolution of the quantities $M^{(l)}, Q^{(l)}$ and $P^{(l)}$ upon iterating B, as a function of the distribution of the initial conditions $\nu_\sigma^{(0)}$, for $\sigma \in [q]$.

Equations of the form (3.31), sometimes called distributional recursions, or density evolution, allow in principle to compute all the moments of the random variables $\nu_\sigma^{(l)}$. In coding theory, they are the main ingredient allowing to study the performance of BP-based decoders on sparse graphical codes [103, 129]. The situation is here greatly simplified by the fact that our equations are linear, which stems from the fact that we have first linearized belief propagation. In particular, the first moments evolve according to the linear system, for $\sigma \in [q]$

*First moments*

$$\mathbb{E}\left[\nu_\sigma^{(l+1)}\right] = \frac{\alpha_{\text{in}}}{qx} \mathbb{E}[w_{\text{in}}] \, \mathbb{E}\left[\nu_\sigma^{(l)}\right] + \frac{\alpha_{\text{out}}}{qx} \mathbb{E}[w_{\text{out}}] \sum_{\tau \in [q] \setminus \sigma} \mathbb{E}\left[\nu_\tau^{(l)}\right] \, ,$$
(3.34)

This linear system can be written in terms of a $q \times q$ matrix whose eigenvalues are

*Eigenvalues of the first moments system*

$$\begin{aligned}
\lambda_1 &= \frac{1}{xq} \left(\alpha_{\text{in}} \, \mathbb{E}\left[w_{\text{in}}\right] + (q-1)\alpha_{\text{out}} \, \mathbb{E}\left[w_{\text{out}}\right]\right) &= \frac{\alpha \, \mathbb{E}[w]}{x} \, , \\
\lambda_2 &= \frac{1}{qx} \left(\alpha_{\text{in}} \, \mathbb{E}\left[w_{\text{in}}\right] - \alpha_{\text{out}} \, \mathbb{E}\left[w_{\text{out}}\right]\right) &= \frac{\alpha \, \Delta(w)}{x} \, ,
\end{aligned}$$
(3.35)

where $\alpha$ is the average connectivity of the graph, given by (3.3), and, with an abuse of notation, $\mathbb{E}[w]$ is the expected value of the weights in the non-backtracking operator, i. e. the expected value of a random variable $w$ with distribution

*Random variable $w$*

$$\mathbb{P}(w) = \mathbb{P}\left(w_{ij} = w\right) \, .$$
(3.36)

The parameter $\Delta(w)$ defined in equation (3.35) is given by

*Parameter $\Delta(w)$*

$$\Delta(w) = \frac{1}{\alpha q} \left(\alpha_{\text{in}} \, \mathbb{E}[w_{\text{in}}] - \alpha_{\text{out}} \, \mathbb{E}[w_{\text{out}}]\right) \, .$$
(3.37)

The eigenvector associated with $\lambda_1$ is $\nu_1 = (1, 1, \ldots, 1)^\intercal$, while $\lambda_2$ has an eigenspace of dimension $q - 1$ equal to the orthogonal of $\nu_1$. The second moments of the variables $\nu_\sigma^{(l)}$ for $\sigma \in [q]$ evolve according to

*Second moments*

$$\begin{aligned}
\mathbb{E}\left[\left(\nu_\sigma^{(l+1)}\right)^2\right] = {}& \frac{\alpha_{\text{in}}}{qx} \mathbb{E}\left[w_{\text{in}}^2\right] \, \mathbb{E}\left[\left(\nu_\sigma^{(l)}\right)^2\right] \\
& + \frac{\alpha_{\text{out}}}{qx} \mathbb{E}\left[w_{\text{out}}^2\right] \sum_{\tau \in [q] \setminus \sigma} \mathbb{E}\left[\left(\nu_\tau^{(l)}\right)^2\right] + \mathbb{E}\left[\nu_\sigma^{(l+1)}\right]^2 \, .
\end{aligned}$$
(3.38)

We examine three interesting choices of initial conditions, and their consequences on the spectrum of B. First, we study the stability of the paramagnetic point with respect to a *noisy* perturbation , of the form, for $\sigma \in [q]$

$$\begin{aligned} \mathbb{E}\left[v_\sigma^{(0)}\right] &= 0 \\ \mathbb{E}\left[\left(v_\sigma^{(0)}\right)^2\right] &= Q^{(0)} > 0 \text{, independently of } \sigma \end{aligned} \tag{3.39}$$

This perturbation is neither globally, nor partially ordered, and we have $M^{(0)} = P^{(0)} = 0$. From equation (3.34), it follows that for any $\sigma \in [q], l \geqslant 0$, $\mathbb{E}\left[v_\sigma^{(1)}\right] = 0$, so that $M^{(1)} = P^{(1)} = 0$. The evolution of the Edwards-Anderson parameter is given by equation (3.38), which implies

$$Q^{(l+1)} = \left(\frac{R(w)}{x}\right)^2 Q^{(l)}, \tag{3.40}$$

where $R(w)$ is defined by

$$R(w)^2 = \frac{1}{q}\left(\alpha_{in}\,\mathbb{E}\left[w_{in}^2\right] + (q-1)\alpha_{out}\,\mathbb{E}\left[w_{out}^2\right]\right) = \alpha\,\mathbb{E}\left[w^2\right]. \tag{3.41}$$

Here, $w$ is the random variable introduced in equation (3.36), so that $R(w)$ is the radius of the bulk of eigenvalues of B conjectured in equation (2.58). We find that that as long as $x > R(w)$, $Q^{(l)} \to 0$ when the number of iterations $l \to \infty$. By Chebyshev's inequality, this implies that the random variables $v_\sigma^{(l)}$ converge to 0 in probability. In statistical physics terms, the paramagnetic fixed point of the Ising model with couplings (3.26) is stable with respect to noisy perturbations as long as $x > R(w)$, and we recover that $x = R(w)$ is the critical pseudo-temperature associated with the paramagnetic to spin-glass transition. We expect that any real initial condition that can be written as a linear combination of the (complex) uninformative eigenvectors of B will verify the condition (3.39), in the limit $n \to \infty$. Therefore, the previous argument supports our conjecture that all the uninformative eigenvectors of B have their associated eigenvalues constrained to the disk of radius $R(w)$. In the following, we restrict to $x > R(w)$ and study other types of instability of the paramagnetic fixed point.

As a second choice of initial condition for the power iteration (3.25), we consider a *ferromagnetic*, or globally ordered perturbation of the paramagnetic fixed point, defined for $\sigma \in [q]$ by

$$\begin{aligned} \mathbb{E}\left[v_\sigma^{(0)}\right] &= M^{(0)} > 0 \text{, independently of } \sigma, \\ \mathbb{E}\left[\left(v_\sigma^{(0)}\right)^2\right] &= Q^{(0)} > 0 \text{, independently of } \sigma \end{aligned} \tag{3.42}$$

From the evolution of the expectations in equation (3.34), it is straightforward to check that for all $l \geqslant 0, \sigma \in [q]$, we have $\mathbb{E}\left[v_\sigma^{(l)}\right] = M^{(l)}$, where the magnetization $M^{(l)}$ is given by

$$M^{(l)} = \left(\frac{\alpha\,\mathbb{E}[w]}{x}\right)^l M^{(0)}. \qquad (3.43)$$

The evolution of the second moments in equation (3.38) implies that for all $l \geqslant 0, \sigma \in [q]$, $\mathbb{E}\left[\left(v_\sigma^{(l)}\right)^2\right] = Q^{(l)}$ where the Edwards-Anderson parameter $Q^{(l)}$ is given recursively for $l \geqslant 0$

$$Q^{(l+1)} = \left(\frac{R(w)}{x}\right)^2 Q^{(l)} + \left(\frac{\alpha\,\mathbb{E}[w]}{x}\right)^2 \left(M^{(l)}\right)^2. \qquad (3.44)$$

In the limit where the number of iterations $l$ grows to infinity, one of two things may happen. First, if $R(w) > |\alpha\,\mathbb{E}[w]|$, we have

$$\frac{\left(M^{(l)}\right)^2}{Q^{(l)}} \xrightarrow[n \to \infty]{} 0. \qquad (3.45)$$

In this case, the expected value of $v_\sigma^{(l)}$ is negligible with respect to its standard deviation, and the initial global order is washed away as the number of iterations increases. Additionally, for any $x > R(w)$, both $M^{(l)}$ and $Q^{(l)}$ tend to 0, so that the random variables $v_\sigma^{(l)}$ converge to 0 in probability by Chebyshev's inequality. We therefore say that for $x > R(w)$, the paramagnetic fixed point is stable with respect to a ferromagnetic perturbation. On the other hand, if $R(w) < |\alpha\,\mathbb{E}[w]|$, then

$$\frac{\left(M^{(l)}\right)^2}{Q^{(l)}} \xrightarrow[n \to \infty]{} \frac{(\alpha\,\mathbb{E}[w])^2 - R(w)^2}{(\alpha\,\mathbb{E}[w])^2} > 0. \qquad (3.46)$$

In this case, the standard deviation of $v_\sigma^{(l)}$ is of the same order as its expected value. Additionally, it holds that

$$Q^{(l+1)} \underset{l \to \infty}{\sim} \left(\frac{\alpha\,\mathbb{E}[w]}{x}\right)^2 Q^{(l)}. \qquad (3.47)$$

In particular, when $x > |\alpha\,\mathbb{E}[w]|$, both $M^{(l)}$ and $Q^{(l)}$ converge to 0, and the paramagnetic point is stable with respect to a ferromagnetic perturbation. On the other hand, when $R(w) < x < |\alpha\,\mathbb{E}[w]|$, both $\left|M^{(l)}\right|$ and $Q^{(l)}$ tend to $+\infty$, so that the paramagnetic point becomes unstable. Therefore, whenever $|\alpha\,\mathbb{E}[w]| > R(w)$, B should have a real eigenvalue $\alpha\,\mathbb{E}[w]$ outside of the bulk of radius R. Note that this eigenvalue can be positive or negative, depending on whether the associated Ising model has a ferromagnetic, or anti-ferromagnetic bias. The approximate magnetizations computed by pooling the corresponding eigenvectors are expected to exhibit global order, in the sense that their average should not vanish in the limit $n \to \infty$.

Finally, we consider a third class of initial conditions corresponding to a *partially ordered* perturbation of the trivial fixed point, defined by the conditions

$$\exists \sigma \in [q], \mathbb{E}\left[\nu_\sigma^{(0)}\right] \neq 0$$

$$\sum_{\sigma \in [q]} \mathbb{E}\left[\nu_\sigma^{(0)}\right] = 0 \tag{3.48}$$

$$\forall \sigma \in [q], \mathbb{E}\left[\left(\nu_\sigma^{(0)}\right)^2\right] = Q^{(0)} > 0.$$

The first condition ensures that this choice of initial condition is different from the noisy perturbation first considered. Together, the first two conditions imply that $M^{(0)} = 0$, but $P^{(0)} \neq 0$, so that the initial condition is partially ordered, but not globally ordered. Using equation (3.34), it is then straightforward to show that for all $l \geqslant 0$, $M^{(l)} = 0$, and

$$P^{(l)} = \left(\frac{\alpha \Delta(w)}{x}\right)^{2l} P^{(0)} \tag{3.49}$$

The evolution of the Edwards-Anderson parameter follows

$$Q^{(l+1)} = \left(\frac{R(w)}{x}\right)^2 Q^{(l)} + \left(\frac{\alpha \Delta(w)}{x}\right)^2 P^{(l)}. \tag{3.50}$$

Following the same reasoning as for the ferromagnetic perturbation, we can show that if $|\alpha \Delta(w)| < R(w)$, then $P^{(l)}$ becomes negligible compared to $Q^{(l)}$ as $l \to \infty$, so that the partial order disappears. Additionally, in this case, for any $x > R(w)$, the random variables $\nu_\sigma^{(l)}$ converge in probability to $0$. On the other hand, when $|\alpha \Delta(w)| > R$, then $Q^{(l)}$ and $P^{(l)}$ stay of the same order as we iterate, with

$$\frac{P^{(l)}}{Q^{(l)}} \underset{n \to \infty}{\longrightarrow} \frac{(\alpha \Delta(w))^2 - R(w)^2}{(\alpha \Delta(w))^2} > 0. \tag{3.51}$$

and the Edwards-Anderson parameter $Q^{(l)}$ verifies

$$Q^{(l+1)} \underset{l \to \infty}{\sim} \left(\frac{\alpha \Delta(w)}{x}\right)^2 Q^{(l)}, \tag{3.52}$$

so that when $x < |\alpha \Delta(w)|$, the paramagnetic fixed point is unstable with respect to partially ordered perturbations. In this case, we expect B to have an eigenvalue equal to $\alpha \Delta(w)$ outside of the bulk of radius $R(w)$. Additionally, we expect this eigenvalue to have multiplicity $q - 1$, since it is the dimension of the eigenspace associated with the eigenvalue $\alpha \Delta(w)$ in the linear system (3.34). Note that $\alpha \Delta(w)$ may be positive or negative. In contrast with the case of the ferromagnetic perturbation, the approximate magnetizations obtained from pooling the corresponding eigenvectors should not exhibit global order, but rather partial order, in the sense that their average on each cluster should not vanish in the limit $n \to \infty$, while their total average vanishes.

### 3.1.3 *Non-backtracking spectrum of the labeled stochastic block model*

Let us summarize the conclusions of the cavity reasoning of the previous section in the form of a conjecture about the spectrum of the non-backtracking operator. In the limit $n \to \infty$, we conjecture that the non-backtracking operator defined in equation (3.23) with a generic weighting function $w$ has the following rich spectral properties, depending on the parameters of the symmetric LSBM and the choice of the weighting function $w$.

— If $|\alpha \mathbb{E}[w]| < R(w)$ and $|\alpha \Delta(w)| < R(w)$, the whole spectrum of B lies inside of the disk of radius $R(w)$, and there is no informative eigenvector of B.

— If $|\alpha \mathbb{E}[w]| > R(w)$ and $|\alpha \Delta(w)| < R(w)$, there is one real eigenvalue of B outside of the bulk of radius $R(w)$. This eigenvalue, which we call *ferromagnetic* is equal to $\alpha \mathbb{E}[w]$. The corresponding approximate magnetizations $m \in \mathbb{R}^n$ verify

*Ferromagnetic eigenvalue with globally ordered eigenvector*

$$\lim_{n \to \infty} \frac{1}{n} \sum_{i=1}^{n} m_i \neq 0 \,. \qquad (3.53)$$

We call this eigenvector *globally ordered*, or *ferromagnetic*, and it does not allow to recover the clusters.

— If $|\alpha \mathbb{E}[w]| < R(w)$ and $|\alpha \Delta(w)| > R(w)$, there is one real eigenvalue outside of the bulk of radius $R(w)$. This eigenvalue equals $\alpha \Delta(w)$ and has multiplicity $q - 1$. In practice, for finite $n$, we observe $q - 1$ distinct real eigenvalues outside of the bulk, and we call them *informative*. For each of these eigenvalues, the corresponding approximate magnetizations $m \in \mathbb{R}^n$ are not globally ordered, but verify

*Informative eigenvalue with partially ordered eigenvector*

$$\lim_{n \to \infty} \sum_{\sigma \in [q]} \left( \frac{q}{n} \sum_{i, \sigma_i = \sigma} m_i^t \right)^2 \neq 0 \,. \qquad (3.54)$$

We call these eigenvectors *partially ordered*, and they allow to detect the clusters using the embedding described in algorithm 1.

— If $|\alpha \mathbb{E}[w]| > R(w)$ and $|\alpha \Delta(w)| > R(w)$, both the ferromagnetic and the informative eigenvalues are outside of the bulk of radius R, so that there are q eigenvalues outside of the bulk, counted with their multiplicity.

To be more explicit, let us write down the expression of the different parameters controlling the spectrum in terms of the distribution $p_{in}$ (resp. $p_{out}$) of the labels within a cluster (resp. between different clusters). The bulk of uninformative eigenvalues of B is constrained to a disk of radius $R(w)$ where

$$\begin{aligned}
R(w)^2 &= \alpha \mathbb{E}\left[w^2\right] \\
&= \frac{1}{q} \int_{\mathcal{L}} d\ell \left( \alpha_{in} \, p_{in}(\ell) + (q-1)\alpha_{out} \, p_{out}(\ell) \right) w(\ell)^2 \,.
\end{aligned} \qquad (3.55)$$

When $|\alpha \mathbb{E}[w]| > R(w)$, there is a ferromagnetic eigenvalue $\alpha \mathbb{E}[w]$ outside of the bulk of radius $R(w)$, and it equals

$$\alpha \mathbb{E}[w] = \frac{1}{q} \int_{\mathcal{L}} d\ell \left( \alpha_{\text{in}} \, p_{\text{in}}(\ell) + (q-1)\alpha_{\text{out}} \, p_{\text{out}}(\ell) \right) w(\ell) \,. \qquad (3.56)$$

Finally, when $|\alpha \Delta(w)| > R(w)$, there is an informative eigenvalue $\alpha \Delta(w)$ with multiplicity $q - 1$ outside of the bulk of radius $R(w)$, and it equals

$$\alpha \Delta(w) = \frac{1}{q} \int_{\mathcal{L}} d\ell \left( \alpha_{\text{in}} \, p_{\text{in}}(\ell) - \alpha_{\text{out}} \, p_{\text{out}}(\ell) \right) w(\ell) \,. \qquad (3.57)$$

As a corollary of the previous conjecture, we find that a spectral method based on the non-backtracking operator with weighting function $w$ succeeds in detecting the clusters if and only if it has informative eigenvalues outside of the bulk of radius $R(w)$, i. e. if and only if

$$\tau(\alpha, w) = \left( \frac{\alpha \Delta(w)}{R(w)} \right)^2 > 1 \,. \qquad (3.58)$$

Consequently, an optimal weighting function $w$ should maximize the quantity $\tau(\alpha, w)$. Unsurprisingly, it is straightforward to check that there exists a unique weighting function $w$ that maximizes $\tau(\alpha, w)$, and that it is equal to the Bayes optimal choice $w^\star$ of equation (3.24). With the Bayes optimal choice of weighting function, the radius of the bulk and the informative eigenvalue become equal to

$$
\begin{aligned}
R(w^\star)^2 &= \frac{1}{q} \int_{\mathcal{L}} d\ell \, \frac{\left( \alpha_{\text{in}} \, p_{\text{in}}(\ell) - \alpha_{\text{out}} \, p_{\text{out}}(\ell) \right)^2}{\alpha_{\text{in}} \, p_{\text{in}}(\ell) + (q-1)\alpha_{\text{out}} \, p_{\text{out}}(\ell)} \,, \\
\alpha \Delta(w^\star) &= \frac{1}{q} \int_{\mathcal{L}} d\ell \, \frac{\left( \alpha_{\text{in}} \, p_{\text{in}}(\ell) - \alpha_{\text{out}} \, p_{\text{out}}(\ell) \right)^2}{\alpha_{\text{in}} \, p_{\text{in}}(\ell) + (q-1)\alpha_{\text{out}} \, p_{\text{out}}(\ell)} \,.
\end{aligned}
\qquad (3.59)
$$

Remarkably, with the Bayes optimal weighting function, we have

$$R(w^\star)^2 = \alpha \Delta(w^\star) \,, \qquad (3.60)$$

so that the non-backtracking operator with Bayes optimal weights detects the clusters as soon as

$$\tau(\alpha, w^\star) = R(w^\star)^2 > 1 \,, \qquad (3.61)$$

This transition corresponds to the paramagnetic to spin glass transition of equation (1.85) in the associated Ising model. The fact that the paramagnetic to spin glass transition coincides with the detectability transition, like in the case of the planted spin glass of section 1.6.5, is specific to the Bayes optimal setting. Indeed, it stems from the fact that the paramagnetic to ferromagnetic transition on the one hand, and the paramagnetic to spin glass transition on the other, coincide, for a gauge transformed model on the Nishimori line. The onset of sensitivity to noise of BP, which corresponds to the spin glass transition, has been used *e. g.* in [60] to conjecture the detectability transition (1.47) in the ISBM, in the case of $q = 2$ clusters. In particular, for $q = 2$, detectability is information-theoretically impossible when $R(w^\star) < 1$, so that we conjecture the non-backtracking operator to be optimal, in that it achieves the information-theoretic threshold. For a larger number of clusters, the situation can more complicated, as we now discuss.

*Easy, hard and impossible inference*

When the non-backtracking operator with Bayes optimal weights fails to detect the clusters (i. e. when $R(w^\star) < 1$), it also has a spectral radius smaller than 1. In this case, as shown in chapter 2, the trivial, paramagnetic fixed point of BP is stable, so that BP also fails to detect the clusters. When this happens, detecting the clusters is either information-theoretically impossible, as in the case of $q = 2$ clusters, or (believed to be) possible but hard [32], depending on the number of clusters $q$. This distinction has a simple statistical physics explanation, based on the order of the paramagnetic to ferromagnetic phase transition in the planted Potts model. Here, the ferromagnetic phase refers to the detectable phase, in which the marginals of the Potts model are correlated with the planted clustering. For an "assortative" Potts model with $\Delta(w^\star) > 0$, this transition is second order for $q \leqslant 4$, and first order for $q \geqslant 5$. When the detectability transition is second order, the instability of the paramagnetic fixed point of BP coincides with the appearance of a ferromagnetic, informative and stable fixed point, so that the two fixed points are never both stable. In this case, detecting the clusters is either impossible (when below the detectability transition), or easy (when above), since nothing prevents BP from converging to its informative fixed point. When the detectability transition is first order, on the other hand, there is a range of values of the parameters of the model for which there are two stable fixed points of BP, namely the paramagnetic one, and another, informative one. In this range, inference is possible, since the informative fixed point is stable, but it is believed to be hard, in the sense that the basin of attraction of the informative fixed point is much smaller than the one of the paramagnetic fixed point, so that one has to resort to an exhaustive search to find it. In this case, there exists a information-computation gap, as explained in chapter 1. A typical way to test for the existence of this hard phase is to compare the result of BP when initialized randomly on the one hand, or close to the planted solution on the other. When the former fails to detect the clusters while the latter succeeds, the problem is in its hard phase. The problem becomes easy again above a second transition, the *spinodal*, at which the paramagnetic fixed point becomes unstable. Detailed considerations about the impossible, easy and hard phase in various inference and optimization problems can be found in [166, 168]. The condition $R(w^\star) > 1$ computed above corresponds to this spinodal, or hard to easy transition. In particular, we conjecture that the spectral method based on the non-backtracking operator with Bayes optimal weights is optimal, in the sense that when it fails to detect the clusters in the symmetric SBM, then no other polynomial time algorithm can succeed.

Let us make the link with section 3.1.1, where we have argued that we could consider the non-backtracking operator B of the graph G, while the stability analysis in fact yielded the non-backtracking operator $B^c$ of the complete graph. We have seen that the informative eigenvectors of B are also eigenvectors of $B^c$ provided the corresponding approximate magnetizations verified $n^{-1} \sum_{i=1}^{n} m_i = 0$. From the analysis of the previous section, it follows that the $q-1$ eigenvectors corresponding to the partially ordered instability do verify this condition, while the ferromagnetic eigenvector does not. Therefore, we can interpret the main purpose of the non-edges introduced in the previous section as to remove the purely ferromagnetic eigenvector (if it exists). In this case algorithms 1 and 2 can be adapted to define an embedding in $\mathbb{R}^{q-1}$ using only the partially ordered eigenvectors. In practice, it is in general safe to include this eigenvector and use algorithms 1 and 2 as they are [89].

It is fair to wonder how the condition (3.58) relates to the results of the previous chapter, where we have shown that a necessary condition for BP to recover the clusters is that $\rho(B^c) > 1$. This condition corresponds to the instability of the trivial, paramagnetic fixed point of BP. According to the previous conjecture, we expect that this condition is not sufficient when outside of the Bayes optimal setting, i.e. with weights $w$ different from the optimal weights $w^\star$. Indeed, we may then have $R(w) > 1$ with $\tau(\alpha, w) < 1$, so that we expect the spectrum of B to be fully contained in the disk of radius $R(w)$, except perhaps for an uninformative ferromagnetic eigenvalue, absent from the spectrum of $B^c$. In this case, the trivial fixed point of BP is unstable, since $B^c$ has eigenvalues close in norm to $R(w) > 1$, but neither B nor $B^c$ have any informative eigenvalue. Thus we have $\rho(B^c) > 1$, but the large eigenvalues of B then correspond to the noisy perturbations identified in the previous section. The behavior of BP in this situation is unclear. On the one hand, since the trivial fixed point is unstable, BP will not converge to it. However, the fact that the leading eigenvectors of $B^c$ are uninformative suggests that BP will either converge, under sufficient damping, to an uninformative ("glassy") solution, or more probably fail to converge at all. Interestingly, this problem is avoided when using the Bayes optimal weights $w^\star$. Indeed, in this case, we have by equation (3.61) that $R(w^\star) > 1$ if, and only if $\tau(\alpha, w^\star) > 1$. In particular, we find that both B and $B^c$ have informative eigenvalues as soon as the trivial fixed point of BP is unstable, so that the condition $\rho(B^c) > 1$ is both necessary and sufficient to recover the clusters. This fact is in fact not surprising, and stems from the previously mentioned properties of the Nishimori line.

Finally, we note that the conjecture presented in this section is consistent with previous rigorous results in some particular cases. The rigorous foundations for the study of the spectrum of the non-backtracking operator on random graphs were laid by Bordenave, Lelarge and Massoulié in [19], who studied the non-backtracking operator associated with the unlabeled SBM. In this case, the label set $\mathcal{L}$ is reduced to a single element, so that varying the weighting function $w$ only amounts to multiplying the non-backtracking operator by a global factor, without affecting its spectral properties. With the choice $w_{kl} = 1$, $\forall (kl) \in E$, the non-backtracking operator is said to be *unweighted*, and its elements are given, for $(i \to j), (k \to l) \in \vec{E}$, by

*Rigorous results*

*Spectrum of the unweighted non-backtracking operator on the SBM*

$$B_{i \to j, k \to l} = \mathbf{1}(l = i)\mathbf{1}(k \neq j). \tag{3.62}$$

This corresponds to the non-backtracking operator as first introduced in [89] and considered in the rigorous analysis of [19] [1]. In this case, our conjecture predicts that the ferromagnetic eigenvalue, the informative eigenvalue, and the edge of the bulk are given respectively by

$$\alpha \, \mathbb{E}[w] = \alpha,$$
$$\alpha \, \Delta(w) = \frac{\alpha_{\text{in}} - \alpha_{\text{out}}}{q}, \tag{3.63}$$
$$R(w) = \sqrt{\alpha},$$

so that the detectability condition becomes

$$\tau(\alpha, w) = \left( \frac{\alpha_{\text{in}} - \alpha_{\text{out}}}{q \sqrt{\alpha}} \right)^2 > 1, \tag{3.64}$$

which is equivalent to the conjecture (1.44) of [32]. When above this transition, [19] proved that the leading eigenvalue of B is $\alpha$, that the second largest eigenvalue is $\frac{\alpha_{\text{in}} - \alpha_{\text{out}}}{q}$, with multiplicity $q - 1$, and that all other eigenvalues are smaller than $\sqrt{\alpha}$ in modulus. When below the transition, they proved, in agreement with the previous conjecture, that the leading eigenvalue of B is $\alpha$, and that all other eigenvalues are smaller in modulus than $\sqrt{\alpha}$. The proof technique of [19] relies heavily on a certain symmetry of the non-backtracking operator, which they call "PT-invariance", where "PT" stands for parity-time. More precisely, we denote by M the involution of $R^{|\vec{E}|}$ with matrix elements

*PT-invariance*

$$M_{i \to j, k \to l} = \mathbf{1}(l = i)\mathbf{1}(k = j). \tag{3.65}$$

This operator maps a directed edge $(k \to l)$ onto its inverse $(l \to k)$, so that it is clear that $M^{\mathsf{T}} = M$ and $M^2 = I_{|\vec{E}|}$. The PT-invariance states that, while B is not a symmetric matrix, the matrix $MB^l$ for any $l \geqslant 0$ is symmetric. The proof of this invariance is straightforward, and uses the fact that all the non-zero entries of the unweighted non-backtracking operator are equal to unity. In the rest of this dissertation, we extend some of the techniques of [19] to a particular case of the SBM. An important step in these extensions will be the generalization of the PT-invariance to a weighted non-backtracking operator.

---

1. [19, 89] consider the transpose of this matrix, which has the same eigenvalue spectrum.

In the next section, we show that the spectral density of the unweighted non-backtracking operator can be computed using statistical physics methods.

### 3.1.4 *Cavity approach to the spectral density of the non-backtracking operator*

In this section, we continue our investigation of the spectral properties of the *unweighted* non-backtracking operator defined in equation (3.62). The results presented here were published in [140]. We investigate the spectral density of B on locally tree-like random graphs using the non-rigorous cavity method, and show that it is non-vanishing only in a disk of radius $\sqrt{\rho(B)}$ where $\rho(B)$ is the spectral radius of B. In the case of the SBM, $\rho(B) = \alpha$, so that we recover the results of [19, 89] that the bulk of the spectrum of B is constrained to the disk of radius $\sqrt{\alpha}$.

*We consider in this section the unweighted non-backtracking operator.*

Analytical results for spectral densities on sparse random graphs are largely based on the replica or cavity methods, and were mostly developed and studied for symmetric random matrices [40, 91, 130, 132, 145]. The result most relevant to the present section is that the tails of the spectrum of the commonly studied matrices associated with random graphs (including the adjacency and Laplacian matrices) are extended, see *e. g.* [84, 91, 145]. The spectral density of the non-backtracking operator, on the other hand, has a finite support, and we would like to provide a statistical-physics based explanation to this fact.

We derive the spectral density of the non-backtracking operator for random locally tree-like graphs $G = ([n], E)$ in the limit of large size $n \to \infty$. We use the methods of [40, 91, 130, 132] based on expressing the spectral density as the internal energy of a disordered system with quenched disorder. In particular we use the method applied to non-symmetric matrices as developed in [113, 131]. The corresponding disordered system is then studied using the cavity method and the associated BP algorithm.

Our main result is the discovery of a phase transition in the disordered system associated with the spectrum of the non-backtracking operator which translates to the fact that the spectral density can be finite only inside a circle of radius equal to the square root of the leading eigenvalue (a fact that was indeed proven in [89]). This is fundamentally different from the spectral properties of the commonly considered operators associated with a sparse random graph where the tails of the spectrum are unbounded in the limit of large size. The presence of this phase transition provides a physics-based explanation of the superior performance of spectral methods based on the non-backtracking operator.

*Statistical physics formulation*

We map the computation of the spectral density to a problem of statistical physics of disordered systems. We use straightforwardly the method of [131], with the only difference that we use the Ihara-Bass formula, which is specific to the non-backtracking operator, to simplify slightly the subsequent analysis. Using the results of section 2.3.1, it is straightforward to show that in this unweighted case, all the eigenvalues $\lambda_i$ of the B that are different from $\pm 1$ are the roots of the polynomial, for $z \in \mathbb{C}$,

$$\det \left[ D - zA - (1 - z^2)\, I_n \right] = \prod_{i}^{2n} (z - \lambda_i)\,. \tag{3.66}$$

where $A$ is the adjacency matrix of the graph, and $D$ is the diagonal matrix with entry $D_{ii} = |\partial i|$ for $i \in [n]$. We define the spectral density of B at $z \in \mathbb{C}$ as

*Spectral density*
$$\nu(z) = \frac{1}{2n} \sum_{i=1}^{2n} \delta(z - \lambda_i)\,. \tag{3.67}$$

We use the complex representation of the Dirac delta

$$\delta(z - \mu) = \frac{1}{\pi} \partial_{\bar{z}} (z - \mu)^{-1}\,, \tag{3.68}$$

where, for $z = x + iy$, $\partial_{\bar{z}} = 1/2 \left( \partial_x + i \partial_y \right)$ is the Wirtinger derivative, so that

$$\nu(z) = \frac{1}{2\pi n} \partial_{\bar{z}} \partial_z \log \det \left[ (D - zA - (1 - z^2)I_n)^{\dagger} (D - zA - (1 - z^2)I_n) \right]\,, \tag{3.69}$$

whenever $z$ is not an eigenvalue of B. Here and in the rest of this section, we denote by $X^{\dagger}$ the conjugate-transpose of $X$. To make this formula valid for all $z \in \mathbb{C}$, we add an infinitesimal regularizer $\epsilon^2 I_n$ in the determinant, so that one can rewrite

$$\nu(z) = \lim_{\epsilon \to 0} \frac{1}{2\pi n} \partial_{\bar{z}} \partial_z \log \det \mathcal{M}_\epsilon\,, \tag{3.70}$$

where $\mathcal{M}_\epsilon$ is a $2n \times 2n$ matrix given by

$$\mathcal{M}_\epsilon(z, A) = \begin{pmatrix} \epsilon \mathbf{1} & i(D - zA - (1 - z^2)I_n) \\ i(D - zA - (1 - z^2)I_n)^{\dagger} & \epsilon \mathbf{1} \end{pmatrix}\,.$$

All the eigenvalues of this matrix have a strictly positive real part equal to $\epsilon$, so we can use the complex Gaussian representation of the determinant

$$(\det \mathcal{M}_\epsilon)^{-1} = \left(\frac{1}{\pi}\right)^{2n} \int \exp\left(-\sum_{j,k=1}^{2n} \overline{\psi}_j \mathcal{M}_{jk} \psi_k\right) \prod_{j=1}^{2n} d^2\psi_j, \quad (3.71)$$

where for $z = x + iy \in \mathbb{C}$, $d^2z = dx dy$. It will prove convenient ([131]) to group the complex variables $\psi_j$ for $j \in [2n]$ in pairs

$$\chi_j = \begin{pmatrix} \psi_j \\ \psi_{j+n} \end{pmatrix}, \qquad \forall j \in [n]. \quad (3.72)$$

Finally, the computation of the spectral density has been mapped to a statistical physics problem

$$\nu(z) = -\lim_{\epsilon \to 0} \frac{1}{2\pi n} \partial_{\bar{z}} \partial_z \log \mathcal{Z}_\epsilon, \quad (3.73)$$

where we have introduced a partition function

$$\mathcal{Z}_\epsilon = \int \mathcal{D}\chi \exp\left(-\mathcal{H}_\epsilon(\chi)\right), \quad (3.74)$$

where

$$\mathcal{D}\chi = \prod_{j=1}^{2n} d^2\psi_j \quad (3.75)$$

and the corresponding Hamiltonian (or energy) is given by

$$\mathcal{H}_\epsilon(\chi) = \sum_{j=1}^{n} \chi_j{}^\dagger \begin{pmatrix} \epsilon & id_j - i(1-z^2) \\ id_j - i(1-\bar{z}^2) & \epsilon \end{pmatrix} \chi_j$$
$$+ i \sum_{(jk) \in E} \chi_j{}^\dagger \begin{pmatrix} 0 & -z \\ -\bar{z} & 0 \end{pmatrix} \chi_k. \quad (3.76)$$

*Hamiltonian*

In this last expression, and in the following, we denote by $d_j = |\partial j|$ the degree of vertex $j$ in the graph $G$. Computing the derivative with respect to $z$, we can express, as in [131], the spectral density in terms of "expectations" with respect to the (complex) Boltzmann distribution $\exp\left(-\mathcal{H}_\epsilon(\chi)\right)/\mathcal{Z}_\epsilon$, as

$$\nu(z) = \lim_{\epsilon \to 0} \frac{i}{\pi n} \partial_{\bar{z}} \left(z \sum_{j=1}^{n} \mathbb{E}\left[\chi_j^\dagger \sigma_+ \chi_j\right] - \sum_{(jk) \in E} \mathbb{E}\left[\chi_j^\dagger \sigma_+ \chi_k\right]\right), \quad (3.77)$$

*Spectral density as a function of expectations*

where $\sigma_+$ is a so-called Pauli matrix

$$\sigma_+ = \begin{pmatrix} 0 & 1 \\ 0 & 0 \end{pmatrix}, \quad (3.78)$$

and the expectation of a function $f(\chi)$ is defined formally as

$$\mathbb{E}\big[f(\chi)\big] = \int \mathcal{D}\chi \, f(\chi) \, \frac{\exp\left(-\mathcal{H}_\epsilon(\chi)\right)}{\mathcal{Z}_\epsilon} \tag{3.79}$$

While the "Boltzmann distribution" happens to be complex here, the problem is formally reduced to computing the expectations of the one and two-points marginals $\mathbb{P}_j(\chi_j)$ and $\mathbb{P}_{jk}(\chi_j, \chi_k)$, for $j \in [n]$, $(kl) \in E$, of a pairwise MRF defined by the energy function (3.76). Since the graph is assumed to be sampled from a locally tree-like random ensemble, we will compute these marginals using the replica symmetric cavity method, i. e. BP.

*The cavity method*

We write the energy (3.76) as

$$\mathcal{H}_\epsilon(\chi) = \sum_{j=1}^{n} \mathcal{H}_j(\chi_j) + \sum_{(jk) \in E} \mathcal{H}_{jk}(\chi_j, \chi_k), \tag{3.80}$$

where we have defined the single variable and pairwise energies

*Single variable and pairwise energies*

$$\begin{aligned}
\mathcal{H}_j(\chi_j) &= \chi_j^\dagger \begin{pmatrix} \epsilon & \mathrm{id}_j - i(1 - z^2) \\ \mathrm{id}_j - i(1 - \bar{z}^2) & \epsilon \end{pmatrix} \chi_j \\
\mathcal{H}_{jk}(\chi_j, \chi_k) &= -i\,\chi_j^\dagger \begin{pmatrix} 0 & z \\ \bar{z} & 0 \end{pmatrix} \chi_k - i\,\chi_k^\dagger \begin{pmatrix} 0 & z \\ \bar{z} & 0 \end{pmatrix} \chi_j \,.
\end{aligned} \tag{3.81}$$

The BP fixed point equations then read, for $(j \to k) \in \vec{E}$

*BP equations*

$$b_{j \to k}(\chi_j) \propto e^{-\mathcal{H}_j(\chi_j)} \prod_{l \in \partial j \setminus k} \int e^{-\mathcal{H}_{jl}(\chi_j, \chi_l)} \, b_{l \to j}(\chi_l) \, \mathcal{D}\chi_l, \tag{3.82}$$

where $\mathcal{D}\chi_l = d^2\psi_l \, d^2\psi_{l+n}$. When a fixed point of BP has been found, we obtain an approximation to the single point marginals $\mathbb{P}_j(\chi_j) \approx b_j(\chi_j)$ (with equality if the graph is a tree [133]) where

$$b_j(\chi_j) \propto e^{-\mathcal{H}_j(\chi_j)} \prod_{l \in \partial j} \int e^{-\mathcal{H}_{jl}(\chi_j, \chi_l)} \, b_{l \to j}(\chi_l) \, \mathcal{D}\chi_l. \tag{3.83}$$

Since the Hamiltonian (3.76) is quadratic, the variables $\chi_i$ for $i \in [n]$ are bi-variate complex Gaussians with mean 0, so that we can parametrize

*Gaussian parametrization of the beliefs*

$$b_{j \to k}(\chi_j) = \frac{1}{\pi^2 \det \Delta_{j \to k}} e^{-\chi_j^\dagger (\Delta_{j \to k})^{-1} \chi_j}, \tag{3.84}$$

$$b_j(\chi_j) = \frac{1}{\pi^2 \det \Delta_j} e^{-\chi_j^\dagger (\Delta_j)^{-1} \chi_j}, \tag{3.85}$$

where the $2 \times 2$ matrices $\Delta$ are constrained by symmetry arguments (see [131]) to the form

$$\Delta = \begin{pmatrix} x & i\bar{y} \\ iy & x \end{pmatrix} , \tag{3.86}$$

where $x$ is real and non-negative, and $y \in \mathbb{C}$. Injecting this form in (3.82)-(3.83), we find that the BP fixed point equations translate into the following fixed point equations for the parameters $x$ and $y$

$$\frac{x_{j \to k}}{x_{j \to k}^2 + |y_{j \to k}|^2} = \epsilon + |z|^2 \sum_{l \in \partial j \setminus k} x_{l \to j} , \tag{3.87}$$

$$\frac{\overline{y_{j \to k}}}{x_{j \to k}^2 + |y_{j \to k}|^2} = (1 - d_j - z^2) - z^2 \sum_{l \in \partial j \setminus k} y_{l \to j} , \tag{3.88}$$

*BP fixed point equations as a function of $x$ and $y$*

$$\frac{x_j}{x_j^2 + |y_j|^2} = \epsilon + |z|^2 \sum_{l \in \partial j} x_{l \to j} , \tag{3.89}$$

$$\frac{\overline{y_j}}{x_j^2 + |y_j|^2} = (1 - d_j - z^2) - z^2 \sum_{l \in \partial j} y_{l \to j} . \tag{3.90}$$

In the following, we take $\epsilon = 0$. To obtain an expression for the spectral density (3.77), it only remains to express the expectations $\mathbb{E}\left[\chi_j^\dagger \sigma_+ \chi_j\right]$ and $\mathbb{E}\left[\chi_j^\dagger \sigma_+ \chi_k\right]$ in terms of the variables $x$ and $y$. From (3.85), we find that

$$\mathbb{E}\left[\chi_j^\dagger \sigma_+ \chi_j\right] = iy_j . \tag{3.91}$$

To express $\mathbb{E}\left[\chi_j^\dagger \sigma_+ \chi_k\right]$, we need the two-points marginal $\mathbb{P}_{jk}(\chi_j, \chi_k)$. Since $(jk) \in E$, this is simple to estimate from the BP beliefs

$$\mathbb{P}_{jk}(\chi_j, \chi_k) \propto b_{j \to k}(\chi_j) b_{k \to j}(\chi_k) e^{-\mathcal{H}_{jk}(\chi_j, \chi_k)} . \tag{3.92}$$

Some algebra then yields

$$\mathbb{E}\left[\chi_j^\dagger \sigma_+ \chi_k\right] = i\left(-z y_{k \to j} y_j + \bar{z} x_{k \to j} x_j\right) . \tag{3.93}$$

Replacing in (3.77) and using the BP recursions (3.87) - (3.90), the spectral density takes the form

$$\nu(z) = -\frac{1}{2\pi n z} \sum_{j=1}^{n} \left(1 - d_j + z^2\right) \partial_{\bar{z}} y_j . \tag{3.94}$$

By differentiating (3.87)-(3.90), we obtain a set of closed form equations for the derivatives of the variables $x$ and $y$, which we iterate to

*Fixed point equations for the derivatives of $x$ and $y$*

compute these derivatives without resorting to numerical differentiation

$$\partial_{\bar{z}} x_{j \to k} = - x_{j \to k} \left( x_{j \to k} A_{j \to k} - \overline{y_{j \to k}} B_{j \to k} \right) \tag{3.95}$$
$$+ y_{j \to k} \left( x_{j \to k} C_{j \to k} + \overline{y_{j \to k}} A_{j \to k} \right),$$

$$\partial_{\bar{z}} y_{j \to k} = - x_{j \to k} \left( y_{j \to k} A_{j \to k} + x_{j \to k} B_{j \to k} \right) \tag{3.96}$$
$$- y_{j \to k} \left( -y_{j \to k} C_{j \to k} + x_{j \to k} A_{j \to k} \right),$$

$$\partial_{\bar{z}} \overline{y_{j \to k}} = - \overline{y_{j \to k}} \left( x_{j \to k} A_{j \to k} - \overline{y_{j \to k}} B_{j \to k} \right) \tag{3.97}$$
$$- x_{j \to k} \left( x_{j \to k} C_{j \to k} + \overline{y_{j \to k}} A_{j \to k} \right),$$

where we have defined messages $A, B$ and $C$ which verify

$$A_{j \to k} = \sum_{l \in \partial j \backslash k} \left( z x_{l \to j} + |z|^2 \partial_{\bar{z}} x_{l \to j} \right), \tag{3.98}$$

$$B_{j \to k} = 2 \bar{z} + \sum_{l \in \partial j \backslash k} \left( 2 \bar{z} \, \overline{y_{l \to j}} + \bar{z}^2 \partial_{\bar{z}} \overline{y_{l \to j}} \right), \tag{3.99}$$

$$C_{j \to k} = z^2 \sum_{l \in \partial j \backslash k} \partial_{\bar{z}} y_{l \to j} . \tag{3.100}$$

The expressions for the derivatives of the variables $x_j$ and $y_j$ are similar, expect that the quantities $A_{j \to k}, B_{j \to k}$ and $C_{j \to k}$ are replaced by $A_j, B_j$ and $C_j$, whose definition is given by (3.98)-(3.100), with the only difference that the sums become over all the neighbors of $j$.

Equations (3.87)-(3.88) and (3.95)-(3.100) are self-consistent BP equations which, when iterated, converge to a set of solutions $x_{j \to k}, y_{j \to k}$, $\partial_{\bar{z}} x_{j \to k}, \partial_{\bar{z}} y_{j \to k}, \partial_{\bar{z}} \overline{y_{j \to k}}$. We then compute the variables $x_j, y_j, \partial_{\bar{z}} x_j$, $\partial_{\bar{z}} y_j, \partial_{\bar{z}} \overline{y_j}$ using equations (3.89)-(3.90) and the counterpart of equations (3.95)-(3.100) for the variables $A_j, B_j$ and $C_j$ as explain previously. This finally allows us to compute the spectral density using expression (3.94).

*The paramagnetic phase*

It is straightforward to check that the following assignment of the messages is a fixed point of the belief propagation equations

$$x_{j \to k} = 0 \qquad \forall (j \to k) \in \vec{E}, \tag{3.101}$$
$$y_{j \to k} = -\frac{1}{z^2} \qquad \forall (j \to k) \in \vec{E}.$$

We call this the *factorized* fixed point. The corresponding variables $x_j$ and $y_j$ verify

$$x_j = 0 \qquad \forall j \in [n], \tag{3.102}$$
$$y_j = \frac{1}{1 - z^2} \qquad \forall j \in [n].$$

With this solution, we have $\partial_{\bar{z}} y_j = 0$ for all $j \in [n]$ so that, from (3.94), the spectral density is $\nu(z) = 0$. As a consequence, for any $z \in \mathbb{C}$ such that the fixed point (3.101) is stable, the corresponding spectral density $\nu(z)$ vanishes. We will refer to this region of the complex plane as the *paramagnetic* phase. To study the stability of this solution, we linearize the belief propagation equations (3.87)-(3.88) around the trivial fixed point. Writing $x_{j \to k} = 0 + \delta x_{j \to k}^0$, $y_{j \to k} = -1/z^2 + \delta y_{j \to k}^0$ where $\delta x_{j \to k}^0 \in \mathbb{R}$, $\delta y_{j \to k}^0 \in \mathbb{C}$ are infinitesimal initial perturbations, the evolution of these perturbations when iterating the BP equations is given by the system

$$\delta x_{j \to k}^{t+1} = \frac{1}{|z|^2} \sum_{l \in \partial j \setminus k} \delta x_{l \to j}^t \,,$$
$$\delta y_{j \to k}^{t+1} = \frac{1}{z^2} \sum_{l \in \partial j \setminus k} \delta y_{l \to j}^t \,. \tag{3.103}$$

As usual, we can write the perturbations as vectors $\delta x^t, \delta y^t \in \mathbb{R}^{|\vec{E}|}$, and we obtain that the evolution of the perturbations is governed by the unweighted non-backtracking operator $B$ through the iteration

$$\delta x^{t+1} = \frac{1}{|z|^2} \, B \, \delta x^t \,,$$
$$\delta y^{t+1} = \frac{1}{z^2} \, B \, \delta y^t \,. \tag{3.104}$$

Therefore, we find that the paramagnetic phase corresponds to the region $|z|^2 > \rho(B)$, where $\rho(B)$ is the spectral radius of the non-backtracking operator. As a consequence, the support of the spectral density, i. e. the bulk of eigenvalues of $B$, is contained in the disk

$$|z| \leqslant \sqrt{\rho(B)} \,, \tag{3.105}$$

at the boundary of which there is a phase transition in the model defined by the Hamiltonian (3.76). We expect the above result to hold for the *unweighted* non-backtracking operator of any graph such that the cavity method provides asymptotically correct predictions. We expect this to encompass at least all locally tree-like ensembles. In particular, for an Erdős-Rényi random graph, we have $\rho(B) = \alpha$ where $\alpha$ is the average connectivity of $G$, so that our results our consistent with the rigorous results of [19].

The existence of a factorized fixed point, and hence of a paramagnetic phase in which the spectral density is exactly 0, seems to be a special feature of the non-backtracking matrix. For instance, one can compute the spectral density for the (symmetric) adjacency matrix $A$, see e.g. [132]. The Hamiltonian is then again quadratic, and couples $n$ Gaussian variables, whose marginals are completely determined by their (complex) variance $\Delta_j$ for $j \in [n]$. The spectral density can then be shown to be proportional to the average of the quantities $\operatorname{Im}(\Delta_j)$ over the graph. More precisely, to compute the spectral density at $\lambda \in \mathbb{R}$, we need to find a solution to the fixed point equations, for $z \in \mathbb{C}$ such that $\mathrm{Re}\,(z) = \lambda$,

$$\Delta_{j \to k} = \left( z - \sum_{l \in \partial j \setminus k} \Delta_{l \to j} \right)^{-1}. \tag{3.106}$$

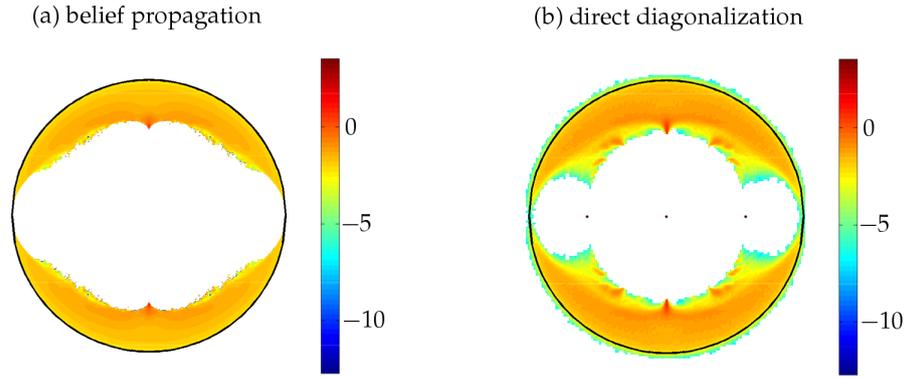

Figure 3.1 – Spectral density of the non-backtracking matrix in $\log z$-scale. Comparison between the result of belief propagation and direct diagonalization, on graphs of average degree $\alpha = 3$. Figure (a) is the result of applying BP to a single graph of size $n = 10000$, on a grid of $600 \times 600$ different points $z \in \mathbb{C}$. Figure (b) was obtained by diagonalizing 1000 matrices of size $3000 \times 3000$. The origins of the differences are discussed in the text. Figure taken from [140].

These equations in general do not admit a factorized (site-independent) solution, expect on regular graphs. The spectral density of the adjacency matrix instead exhibits Lifshitz tails [84] that spoil the gap between the bulk and the informative eigenvalues, associated with the cluster structure of the graph. Similar results hold for the other matrices commonly used for spectral clustering. In section (3.2.1), we show that the analog of equation (3.106) for the Bethe Hessian $H(x)$ of equation (2.67) does admit a "trivial solution", yielding a vanishing spectral density $\nu(\lambda) = 0$ for $\lambda$ in an open neighborhood of 0, for a certain range of values of $x$. Therefore, like the non-backtracking operator, and contrary to e. g. the adjacency matrix, we expect the Bethe Hessian to have a well-behaved bulk of uninformative eigenvalues, well separated from the potential informative eigenvalues.

*Numerical results*

We solve the belief propagation equations on a single graph. We discretize a chosen $z$-domain using a grid of points at which we compute the spectral density by iterating (3.87)-(3.88) and (3.95)-(3.97) until convergence, and finally outputting (3.94).

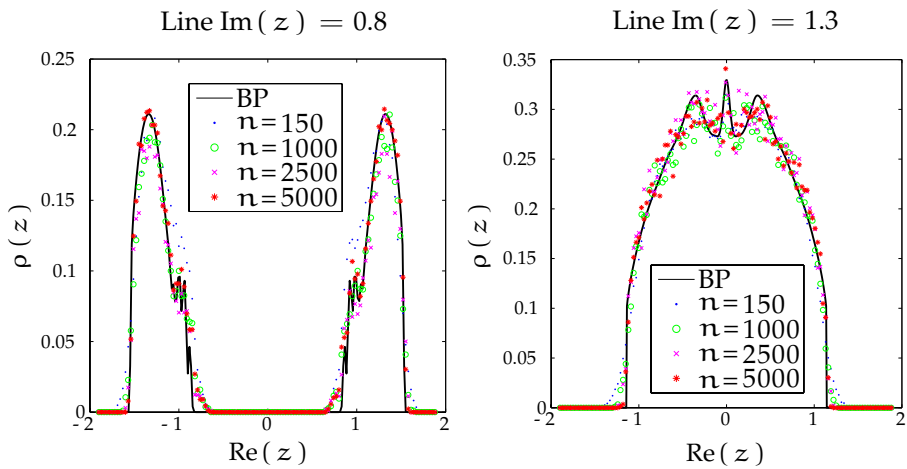

Figure 3.2 – Slices of the spectral density of B along two lines: comparison between BP (black line), and histograms of the eigenvalues of B on graphs of various sizes n, and average connectivity α = 3. For each n, we diagonalized a number of random matrices B such that the total number of eigenvalues obtained is equal to $10^6$. We then extracted the eigenvalues close to the line Im(z) = 0.8 for the left panel, and those close to the line Im(z) = 1.3 for the right panel. BP was run on a single graph of size n = $10^4$ for the figure of the right panel, and the spectral density was computed at 1000 points along the line Im(z) = 1.3. For the figure of the left panel, we used BP to compute the spectral density at 200 points along the line Im(z) = 0.8, and we averaged the results obtained for 500 different graphs of size n = $10^4$. For the left figure, we additionally smoothed the result of BP by setting to the spectral density to the average spectral density computed at 5 neighboring points, to reduce the strong finite size effects that we observed. Figure taken from [140].

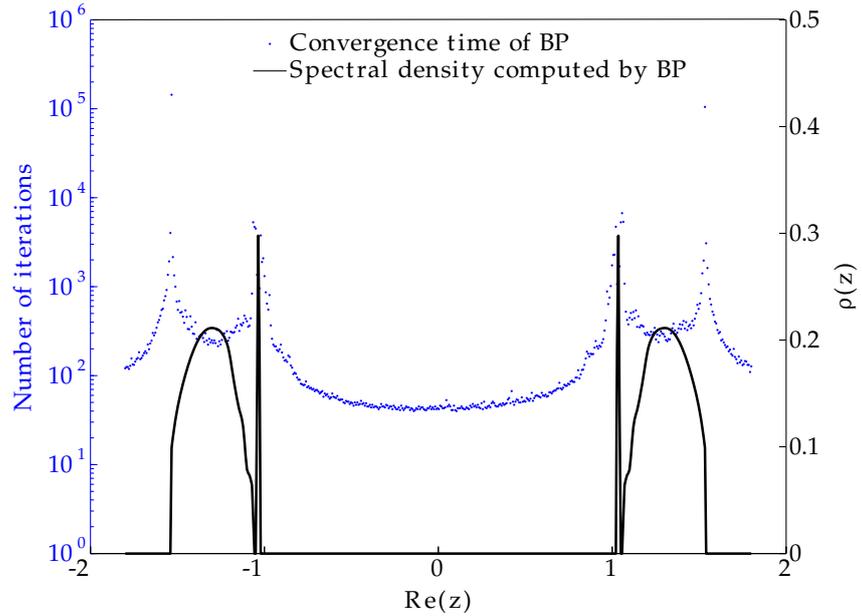

Figure 3.3 – Convergence time and spectral density computed by BP at 500 points along the line $\text{Im}(z) = 0.8$, on a single graph of size $n = 10000$ nodes, and with an average degree of $\alpha = 3$. Comparison with figure 3.2 shows that the location of the peaks inside the circle depends on the instance of the graph. The convergence time is defined as the number of iterations of BP that are needed for the variation of the messages of equations (3.87)-(3.88) and (3.95)-(3.100) to be smaller than $10^{-7}$, between two consecutive iterations. Figure taken from [140].

The left panel of figure 3.1 shows the results of BP for a typical random graph of size $n = 10000$, with average connectivity $\alpha = 3$. For comparison, we show on the right panel of figure 3.1 the spectral density estimated by histogramming the eigenvalues of the non-backtracking operator for many different realizations of the graph G. The discrepancies between the two figures are of two types. The first type consists of the tails that extend beyond the black circle in the direct diagonalization case. These represent sub-extensive contributions to the spectral density as can be seen from figure 3.2, that disappear in the thermodynamic limit, in agreement with the prediction of BP. The second type of discrepancy consist of the internal tails inside the circle in the right panel of figure 3.1, that are absent from the left panel of the same figure. As can be seen from 3.2, these tails do not seem to vanish in the large $n$ limit.

*Two types of discrepancies*

We investigated several possible explanations for this second kind of discrepancy. The first was the possibility of a replica symmetry breaking (RSB), for which we found no evidence. Figure 3.3 shows the convergence time of BP along the line $\text{Im}(z) = 0.8$, which stays

*No evidence of replica symmetry breaking*

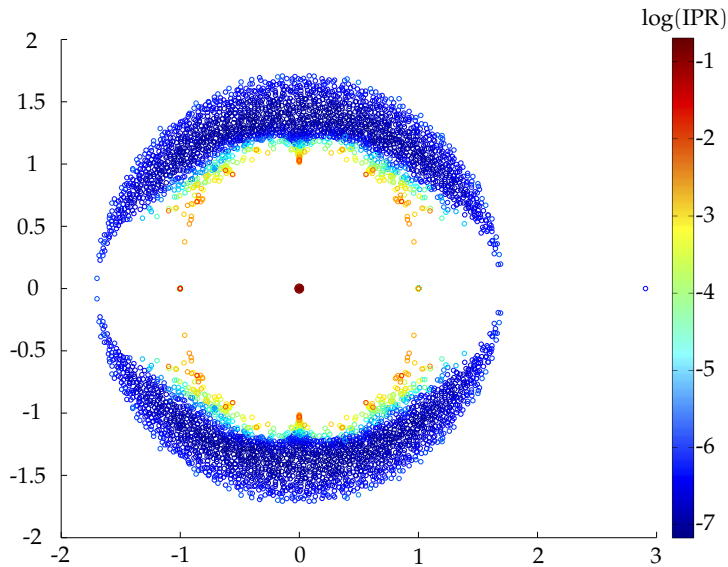

Figure 3.4 – Inverse participation ratio (IPR) of the eigenvectors of B, obtained by direct diagonalization of a graph of size $n = 3000$. The color scale represents the logarithm of the IPR of the (normalized) eigenvectors, defined as the sum of the fourth power of their components. The internal tails consist of eigenvalues whose corresponding eigenvectors are localized (IPR close to 1). Figure taken from [140].

finite on the internal tails. In our different numerical experiments, the algorithm always converged. Note that the peaks in the convergence time correspond to phase transitions in the statistical physics system, taking place at the boundary of the support of the spectral density. Outside of the disk of radius $\sqrt{\rho(B)}$, we find that BP does indeed converge to the trivial solution (3.101). However, there seems to be another region with vanishing spectral density *inside* the disk of radius $\sqrt{\rho(B)}$, where BP converges to a non-trivial fixed point such that the expression (3.94) also vanishes.

Our second line of investigation focused on very localized peaks that we see in rare cases in the BP solution for $z$ in the range of the internal tails, an example of which is shown on figure 3.3. If a peak in the spectral density is very localized then we might miss it simply because of the finite resolution of the grid used to discretize the $z$-domain. By averaging the result of BP over 500 graphs, as in figure 3.2, we recover a non-vanishing spectral density on the internal tails, although the finite size effects are strong. *Localized peaks*

Lastly, we investigated the localization of the eigenvectors corresponding to eigenvalues located in the inner tails. Figure 3.4 shows the inverse participation ratio of the eigenvectors of B. We found that the eigenvectors corresponding to the inner tails are more localized than the others. It is unclear to us how this influences the behavior of BP on a single graph and if it can explain the discrepancy *Localized eigenvectors of B inside the disk* we see between the single graph runs and their averages over many runs. On the other hand in population dynamics [103] (which is an alternate way of solving the cavity equations), which we also implemented, this is clearly a possible source of problems. Equation (3.94) expresses the spectral density as a sum of complex variables, the sum of which should have a vanishing imaginary part, since the spectral density must be a real and non-negative number. We found that the cancellation of the imaginary parts indeed takes place when we run BP on a single graph, but not in our implementation of population dynamics with populations of size up to $S = 10^6$ and equilibration time $10^4 \times S$ per point. It is possible that the events that correspond to localized eigenvectors are encountered only very rarely in the population dynamics and hence usual magnitudes of population sizes and equilibration time are not sufficient here.

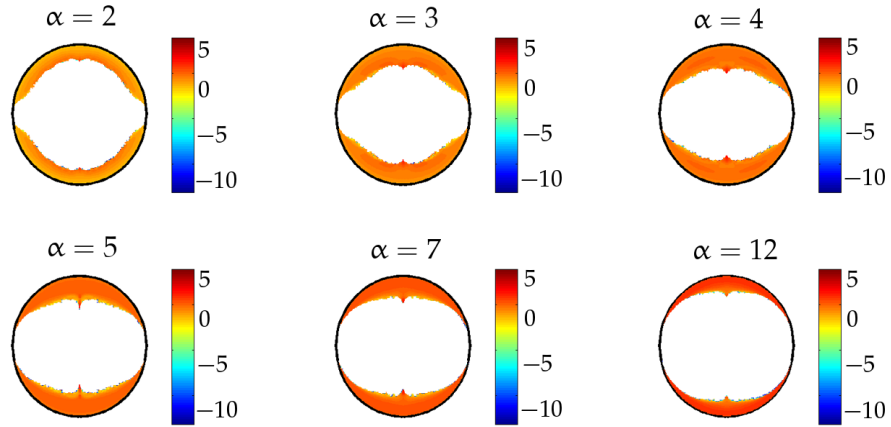

Figure 3.5 – Spectral density of B computed by BP for increasing values of the average connectivity $\alpha$. For each value of $\alpha$, BP was run on a single realization of an Erdős-Rényi graph with average connectivity $\alpha$ and $n = 10^4$ vertices. When $\alpha$ becomes larger, the support of the spectral density becomes closer to the circle of radius $\sqrt{\alpha}$, as expected by comparison with the case of random regular graphs. Figure taken from [140].

*Large connectivity limit*    In the case of random regular graphs of connectivity d, it is straightforward to show that the spectral density of B is non-zero only precisely on the circle of radius $\sqrt{d-1}$ ([89]). For an Erdős-Rényi graph, when the average connectivity $\alpha$ becomes larger, the fluctuations of the degrees of the nodes become negligible compared to the average connectivity, so that spectral density to become non-zero only in a region close to the circle of radius $\sqrt{\alpha}$ (recall that for a Poissonian graph, the excess degree has the same distribution as the degree). It can be checked on figure 3.5 that this is indeed what happens.

As a conclusion, the study of the spectral density of the non-backtracking operator in the thermodynamic limit by means of the cavity method allows to understand better its remarkable efficiency to perform spectral clustering. A phase transition-like behavior at the boundary of the circle of radius $\sqrt{\rho(B)}$ provides physical insight on why its spectral density vanishes sharply instead of exhibiting Lifshitz tails, like other popular choices of spectral methods. Additionally we identified puzzling properties of the interior tail of the spectrum. Their more complete understanding requires further work. In principle, the method used here in the case of the unweighted nonbacktracking operator can be extended to a general non-backtracking operator with arbitrary weights, although the analysis becomes more complex. In the following, we will instead study the spectral density of the Bethe Hessian with arbitrary weights, which will turn out to be much simpler, owing to the fact that it is a symmetric matrix.

## 3.2 THE BETHE HESSIAN

Recall the definition of the Bethe Hessian $H(x)$, with elements, for $x \in \mathbb{C} \setminus \left\{ \pm \tanh\left(J_{ij}\right) \right\}_{(ij) \in E}$

*Bethe Hessian and associated non-backtracking operator*

$$H_{ij}(x) = \left( 1 + \sum_{k \in \partial i} \frac{\tanh^2(J_{ik})}{x^2 - \tanh^2(J_{ik})} \right) \mathbf{1}(i = j) - \frac{x \tanh(J_{ij})}{x^2 - \tanh^2(J_{ij})} \mathbf{1}(j \in \partial i)$$
(3.107)

In the previous expression, the couplings $J_{kl}$ for $(kl) \in E$ are arbitrary real numbers. In chapter 2, we have shown that this matrix is tightly related to the non-backtracking operator with elements, for $(i \to j), (k \to l) \in \vec{E}$

$$B_{i \to j, k \to l} = w_{kl} \mathbf{1}(l = i) \mathbf{1}(k \neq j),$$
(3.108)

where the weights verify, for $(kl) \in E$,

$$w_{kl} = \tanh(J_{kl}).$$
(3.109)

Conversely, given an arbitrarily weighted non-backtracking operator of the form (3.108), we may, without affecting the spectral properties of B, rescale it by a global positive constant such that all the weights verify $|w_{kl}| < 1$, for $(kl) \in E$. We may then define couplings through equation (3.109), and the corresponding Bethe Hessian through equation (3.107). We will therefore say that the weighted non-backtracking operator (3.108) and the Bethe Hessian (3.107) are *associated*, provided the couplings of the Bethe Hessian are related to the weights of B by (3.109).

The point of this section is to explore and formalize the relationships between the two matrices, that we began to uncover in chapter 2. We first show that the spectral density of the Bethe Hessian can be computed via methods similar to those used for the nonbacktracking operator in section 3.1.4. This will allow us to relate the spectral properties of $H(x)$ to those of B, justifying in part the heuristic picture of section 2.3.1 about the evolution of the spectrum of $H(x)$ as we decrease $x$ from a large positive value. In the second part of this section, we show rigorous identities relating $H(x)$ and $B$, including the generalized Ihara-Bass formula used in section 2.3.1.

### 3.2.1 Cavity approach to the spectral density of the Bethe Hessian on the labeled stochastic block model

We consider here the same setting as in section 3.1.2. More precisely we assume that the graph $G$ is generated from the sparse lSBM, and take the non-backtracking operator of equation (3.108) with weights $w_{kl} = w(\ell_{kl})$ for $(kl) \in E$, where $w$ is an arbitrary weighting function of the observed labels $\ell_{kl}$. The associated Bethe Hessian is given by equation (3.107), with couplings defined through (3.109).

Using the same kind of techniques as those described in section 3.1.4, it is possible to argue [132, 133] that in the limit $n \to \infty$, the smooth part (in which potential delta peaks have been removed) of the spectral density $\nu_x(\lambda)$ of the Bethe Hessian $H(x)$ on sparse graphs verifies

$$
\begin{aligned}
\nu_x(\lambda) &= \lim_{n \to \infty} \frac{1}{n} \sum_{i=1}^{n} \delta(\lambda - \lambda_i), \\
&= \lim_{n \to \infty} \frac{1}{\pi n} \sum_{i=1}^{n} \mathrm{Im}(\Delta_i).
\end{aligned}
\tag{3.110}
$$

where the $\lambda_i$ are the eigenvalues of $H(x)$, and we have introduced complex variables $\Delta_i$, which depend on $x$ and $\lambda$ although we do not write the dependence explicitly to lighten the notations. The $\Delta_i \in \mathbb{C}$ for $i \in [n]$ verify

$$
\begin{aligned}
\Delta_i^{-1} = -\lambda + 1 &+ \sum_{k \in \partial i} \frac{\tanh^2(J_{ik})}{x^2 - \tanh^2(J_{ik})} \\
&- \sum_{k \in \partial i} \left( \frac{x \tanh(J_{ik})}{x^2 - \tanh^2(J_{ik})} \right)^2 \Delta_{k \to i},
\end{aligned}
\tag{3.111}
$$

and the $\Delta_{i \to j} \in \mathbb{C}$ for $(i \to j) \in \vec{E}$ are the linearly stable solution of the system

$$
\begin{aligned}
\Delta_{i \to j}^{-1} = -\lambda + 1 &+ \sum_{k \in \partial i} \frac{\tanh^2(J_{ik})}{x^2 - \tanh^2(J_{ik})} \\
&- \sum_{k \in \partial i \setminus j} \left( \frac{x \tanh(J_{ik})}{x^2 - \tanh^2(J_{ik})} \right)^2 \Delta_{k \to i}.
\end{aligned}
\tag{3.112}
$$

The steps to derive these equations are very similar (but simpler because $H(x)$ is symmetric) to the computations performed in section 3.1.4, and can be found in [132]. The main idea is once more to map the computation of the spectral density to a marginalization problem in a certain statistical physics problem with a complex, quadratic Hamiltonian, and then solve this marginalization problem using BP, which we believe to be exact on sparse random graphs, in the limit $n \to \infty$. In this case, running BP amounts to iterating the fixed point equation (3.112) until convergence, starting from a random initial condition. The solution found by BP must therefore be linearly stable. Numerical simulations showing the very high accuracy of this approach will be presented in chapters 5 and 8.

It is worth noting that, unlike the previous, non-symmetric case, this approach to the computation of the spectral density of a sparse symmetric matrix has been made rigorous in some cases. In particular, in the *unweighted* setting, corresponding to uniform couplings $J_{ij}$ for $(ij) \in E$, when the underlying graph is random and locally tree-like, it is possible to show that the variables $\Delta_i$ for $i \in [n]$ and $\Delta_{i \to j}$ for $(i \to j) \in \vec{E}$ converge in probability to random variables verifying a distributional fixed point equation. It was shown by [18] that this distributional fixed point equation has a unique solution, and that this solution allows to compute the spectral density in the limit $n \to \infty$[2]. The same authors show that these rigorous results generalize to *weighted* sparse and symmetric matrices when their diagonal elements vanish, which is not the case for the Bethe Hessian. However, we will see numerically in chapter 8 that, although not yet rigorous, the present approach based on BP allows to estimate very accurately the spectral density of $H(x)$.

*Rigorous results*

The aim of this section is to justify analytically part of the heuristic picture presented in section 2.3.1, where we have argued that, as we decrease $x$ from a large positive value, the bulk of uninformative eigenvalues of $H(x)$ is shifted to the left, until its left edge reaches the $\lambda = 0$ axis precisely at $x = R(w)$, where $R(w)$ the radius of the bulk of uninformative eigenvalues of the associated non-backtracking operator $B$. An idealized cartoon of this picture is presented in figure 3.6[3].

*Qualitative evolution of the spectral density of $H(x)$ with $x$*

From the results of section 3.1.3, the radius $R(w)$ is given by

$$
\begin{aligned}
R(w) &= \sqrt{\alpha \, \mathbb{E}\left[w^2\right]}, \\
&= \sqrt{\alpha \, \mathbb{E}\left[\tanh^2(J)\right]},
\end{aligned}
\tag{3.113}
$$

*Radius of the bulk of $B$*

where we recall that we abuse our notation by also calling $w$ a random variable with the same distribution as the weights $w_{kl}$ for $(kl) \in E$ of $B$, and similarly, $J$ is a random variable with the same distribution as the couplings $J_{kl}$ for $(kl) \in E$ of $H(x)$. In partial justification of this picture, we now show that whenever $x > R(w)$, there exists an open set around $\lambda = 0$ in which the spectral density vanishes.

---

2. We show in chapter 5 how to solve numerically this distributional equation.

3. We solve numerically equation (3.112) in two different settings in chapters 5 and 8, and show that the corresponding spectral density is indeed qualitatively close to the cartoon of figure 3.6.

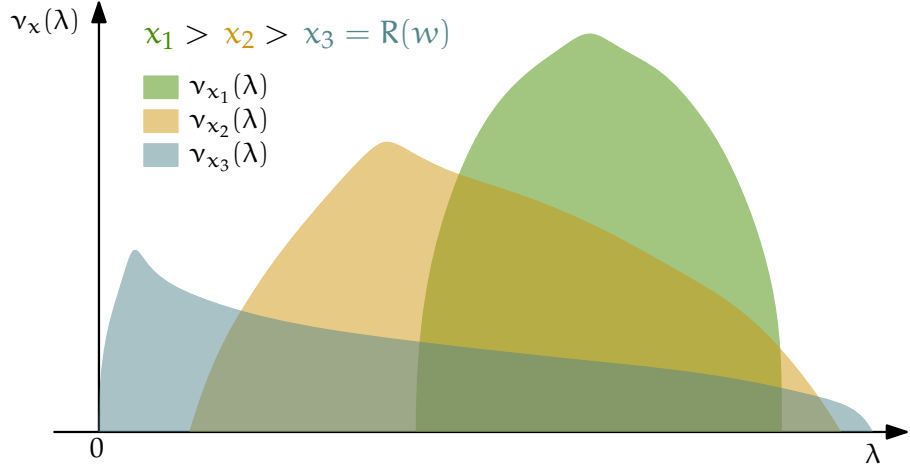

Figure 3.6 – Cartoon of the spectral density of the Bethe Hessian $H(x)$ for various values of $x$, in the limit $n \to \infty$. For large enough $x$, $H(x)$ is positive definite. As we decrease $x$, the bulk of uninformative eigenvalues if shifted to the left, and reaches the $\lambda = 0$ axis when $x = R(w)$, where $R(w)$ is the radius of the bulk of the spectrum of the associated non-backtracking operator $B$, with weights related to the couplings of the Bethe Hessian by (3.109).

From equation (3.110), a sufficient condition for the spectral density to vanish is for equation (3.112) to have a *real* and *stable* solution. We will show that such a solution exists for $\lambda$ close to $0$, whenever $x > R(w)$. To this end, we introduce first, for each $\lambda \in \mathbb{R}$, a function $G_\lambda : \Delta \in \mathbb{R}^{|\vec{E}|} \to G_\lambda(\Delta) \in \mathbb{R}^{|\vec{E}|}$, where for $\Delta \in \mathbb{R}^{|\vec{E}|}$ the components of $G_\lambda(\Delta)$ are given, for $(i \to j) \in \vec{E}$ by

$G_\lambda : \mathbb{R}^{|\vec{E}|} \to \mathbb{R}^{|\vec{E}|}$

$$(G_\lambda(\Delta))_{i \to j} = \left( -\lambda + 1 + \sum_{k \in \partial i} \frac{\tanh^2(J_{ik})}{x^2 - \tanh^2(J_{ik})} \right.$$
$$\left. - \sum_{k \in \partial i \setminus j} \left( \frac{x \tanh(J_{ik})}{x^2 - \tanh^2(J_{ik})} \right)^2 \Delta_{k \to i} \right)^{-1}, \quad (3.114)$$

so that $\Delta$ is a real solution of the fixed point equation (3.112) if, and only if

$$G_\lambda(\Delta) = \Delta. \quad (3.115)$$

We also require a function $F : (\lambda, \Delta) \in \mathbb{R}^{|\vec{E}|+1} \to F(\lambda, \Delta) \in \mathbb{R}^{|\vec{E}|}$ where, for $\lambda \in \mathbb{R}, \Delta \in \mathbb{R}^{|\vec{E}|}$

$F : \mathbb{R}^{|\vec{E}|+1} \to \mathbb{R}^{|\vec{E}|}$

$$F(\lambda, \Delta) = G_\lambda(\Delta) - \Delta. \quad (3.116)$$

The fixed point equation (3.115) is therefore equivalent to

$$F(\lambda, \Delta) = 0. \quad (3.117)$$

As has become customary in this dissertation, the first step of our argument is to remark that the previous equation has a trivial solution at $\lambda = 0$, namely $\Delta^\star$ with components

$$\Delta_{i \to j}^\star = \frac{x^2 - \tanh^2(J_{ij})}{x^2}, \qquad \forall (i \to j) \in \vec{E}. \qquad (3.118)$$

*Trivial fixed point $\Delta^\star$ at $\lambda = 0$*

Indeed, injecting this expression in (3.114), we find, for any directed edge $(i \to j) \in \vec{E}$

$$(G_0(\Delta^\star))_{i \to j} = \left( 1 + \sum_{k \in \partial i} \frac{\tanh^2(J_{ik})}{x^2 - \tanh^2(J_{ik})} \right.$$
$$\left. - \sum_{k \in \partial i \setminus j} \left( \frac{x \tanh(J_{ik})}{x^2 - \tanh^2(J_{ik})} \right)^2 \Delta_{k \to i}^\star \right)^{-1},$$
$$= \left( 1 + \sum_{k \in \partial i} \frac{\tanh^2(J_{ik})}{x^2 - \tanh^2(J_{ik})} - \sum_{k \in \partial i \setminus j} \frac{\tanh^2(J_{ik})}{x^2 - \tanh^2(J_{ik})} \right)^{-1},$$
$$= \left( 1 + \frac{\tanh^2(J_{ij})}{x^2 - \tanh^2(J_{ij})} \right)^{-1},$$
$$= \left( \frac{x^2}{x^2 - \tanh^2(J_{ij})} \right)^{-1} = \Delta_{i \to j}^\star,$$

so that $F(0, \Delta^\star) = 0$. We now show that this fact implies the existence of a real solution in the vicinity of $\lambda = 0$. To see this, we first note that the Jacobian $\mathcal{J}_F$ of $F$ at $(\lambda, \Delta) = (0, \Delta^\star)$ is an $|\vec{E}| \times (|\vec{E}| + 1)$ real matrix which can be written as

$$\mathcal{J}_F(0, \Delta^\star) = \left( \begin{array}{c|ccc} -1 & 0 & \cdots & 0 \\ \hline 0 & & & \\ \vdots & & \mathcal{J}_{G_0}(\Delta^\star) - I_{|\vec{E}|} & \\ 0 & & & \end{array} \right), \qquad (3.119)$$

*Jacobian of $F$*

where $\mathcal{J}_{G_0}(\Delta^\star)$ is the $|\vec{E}| \times |\vec{E}|$ Jacobian of $G_0$ at $\Delta = \Delta^\star$, given by

$$\mathcal{J}_{G_0}(\Delta^\star) = \mathcal{D}\mathcal{B}. \qquad (3.120)$$

In the last expression, we have introduced two matrices $\mathcal{D}, \mathcal{B} \in \mathbb{R}^{|\vec{E}| \times |\vec{E}|}$ whose elements, for $(i \to j), (k \to l) \in \vec{E}$, can be written as

$$\mathcal{D}_{i \to j, k \to l} = \left( \Delta_{i \to j}^\star \right)^2 \mathbf{1}(i = k)\mathbf{1}(j = l),$$
$$\mathcal{B}_{i \to j, k \to l} = \left( \frac{\tanh(J_{kl})}{x \Delta_{k \to l}^\star} \right)^2 \mathbf{1}(l = i)\mathbf{1}(k \neq j). \qquad (3.121)$$

Let us assume for now that the Jacobian $\mathcal{J}_F(0, \Delta^\star)$ is full rank. Then, since $F$ is continuously differentiable at $(0, \Delta^\star)$, by the implicit function theorem, there exists an open set $V \subset \mathbb{R}$ containing $0$ such that for any $\lambda \in V$, there exists a vector $\Delta(\lambda) \in \mathbb{R}^{|\vec{E}|}$ such that $F(\lambda, \Delta(\lambda)) = 0$. Additionally, the map $\lambda \to \Delta(\lambda)$ is continuous on $V$, and $\Delta(0) = \Delta^\star$. In other words, there exists a *real* fixed point $\Delta(\lambda)$ of $G_\lambda$ for $\lambda \in V$, where $0 \in V$. It is additionally a *stable* fixed point if and only if

$$\rho\big(\mathcal{J}_{G_\lambda}(\Delta(\lambda))\big) < 1 \,, \tag{3.122}$$

where $\rho$ denotes the spectral radius. By continuity of the spectral radius and of the map $\lambda \to \Delta(\lambda)$, it is straightforward to check that the map $\lambda \to \rho\big(\mathcal{J}_{G_\lambda}(\Delta(\lambda))\big)$ is continuous on an open set $U \subset V$ such that $0 \in U$. As a consequence, if we can show that

$$\rho\big(\mathcal{J}_{G_0}(\Delta^\star)\big) < 1 \,, \tag{3.123}$$

then there will finally exist an open set $W$ containing $0$ such that for all $\lambda \in W$, condition (3.122) holds. As a consequence, for any $\lambda \in W$, $\Delta(\lambda)$ is a *real* and *stable* fixed point of $G_\lambda$, so that the spectral density of $H(x)$ vanishes on $W$. Indeed, if (3.123) holds, then it also holds that the Jacobian $\mathcal{J}_F(0, \Delta^\star)$ of equation (3.119) is full rank, because then $\mathcal{J}_{G_0}(\Delta^\star) - I_{|\vec{E}|}$ is invertible, so that our previous application of the implicit function theorem was legitimate.

It turns out that we know how to compute $\rho\big(\mathcal{J}_{G_0}(\Delta^\star)\big)$. Indeed, we have that[4]

$$\begin{aligned}\rho\big(\mathcal{J}_{G_0}(\Delta^\star)\big) &= \rho(\mathcal{D}\mathcal{B}) \,, \\ &= \rho(\mathcal{B}\mathcal{D}) \,.\end{aligned} \tag{3.124}$$

The matrix $\mathcal{B}\mathcal{D} \in \mathbb{R}^{|\vec{E}| \times |\vec{E}|}$ has a familiar structure. Its elements, for $(i \to j), (k \to l) \in \vec{E}$, are given by

$$(\mathcal{B}\mathcal{D})_{i \to j, k \to l} = \frac{\tanh^2(J_{kl})}{x^2} \mathbf{1}(l = i)\mathbf{1}(k \neq j) \,. \tag{3.125}$$

This is a non-backtracking operator, with a particular choice of weights. Using the results of section 3.1.3, it is straightforward to check that its leading eigenvalue is unique, and given by

$$\alpha \, \mathbb{E}\left[\frac{\tanh^2(J)}{x^2}\right] = \frac{R(w)^2}{x^2} \,, \tag{3.126}$$

where $R(w)$ is the radius of the bulk of uninformative eigenvalues of the non-backtracking operator associated with the Bethe Hessian $H(x)$, recalled in equation (3.113). Note that this eigenvalue is a *ferromagnetic* one in the sense of section 3.1.3, i. e. with a corresponding globally ordered eigenvector. This is consistent with the Perron-Frobenius theorem which states that, since $\mathcal{B}\mathcal{D}$ is non-negative[5], it should have a unique eigenvalue equal to its spectral radius, with a corresponding eigenvector which whose are all positive.

---

4. Note that the matrix $\mathcal{D}\mathcal{B}$ does *not* have the same structure as the non-backtracking matrix, so that the inversion of the order of the matrix product is necessary here.

5. The Perron-Frobenius theorem applies to non-negative and *irreducible* matrices. For $\mathcal{B}\mathcal{D}$ to be irreducible, we need the graph G to be 2-connected, meaning that any vertex or pair of vertices are part of a cycle [19].

Let us wrap up our argument. From (3.126), it holds that whenever $x > R(w)$, we have $\rho(\mathcal{J}_{G_0}(\Delta^\star)) < 1$, so that there exists an open set around $\lambda = 0$ on which the spectral density of $H(x)$ vanishes. Coming back to the cartoon picture of figure 3.6, we have justified that as we lower $x$ from a large positive value, the moment where the bulk of uninformative eigenvalues of $H(x)$ reaches the $\lambda = 0$ axis is precisely $x = R(w)$. At this point, any potential informative eigenvalue of $H(x)$, i.e. any eigenvalue smaller than the uninformative eigenvalues, must have become negative. This gives an alternative justification of the graph clustering algorithm 2, based on the Bethe Hessian. We discuss what happens for $x < R(w)$ on the applications of the upcoming chapters 5 and 8.

### 3.2.2  *Ihara-Bass type formulas*

We prove here two identities relating the non-backtracking operator and the Bethe Hessian. The first one is a general matrix identity relating the resolvent of the non-backtracking operator to the inverse of the Bethe Hessian, i.e. the susceptibility (in the paramagnetic phase and in the Bethe approximation) of the Ising model. This formula will be useful in chapter 9, where we show in particular that this approximate susceptibility is an upper bound on the true susceptibility for the ferromagnetic Ising model, in a certain high temperature region. The second identity is the Ihara-Bass formula, which we show as a corollary of the first identity. This formula will also be useful in chapter 9, as the involved determinant will be shown to provide an upper bound on the partition function of a ferromagnetic Ising model in a high temperature region.

We stress that both the formulas stated in the two following propositions are algebraic identities, which are valid on any graph, deterministic or random, of any size, with any topology. In addition to the Bethe Hessian of equation (3.107) and the non-backtracking operator of equation (3.108), the following results are expressed in terms of the corresponding *pooling* matrix $P \in \mathbb{R}^{n \times |\vec{E}|}$ with elements

$$P_{i,(k\to l)} = \tanh(J_{kl})\, \mathbf{1}(l = i)\,, \ \forall i \in [n], (k \to l) \in \vec{E}\,, \qquad (3.127)$$

and another matrix $Q \in \mathbb{R}^{n \times |\vec{E}|}$ with elements, for $i \in [n], (k \to l) \in \vec{E}$

$$Q_{i,k\to l} = \mathbf{1}(i = k)\,. \qquad (3.128)$$

**Proposition 3.2.1.** *Let* $G = ([n], E)$ *be arbitrary couplings. Let* $B$ *be the non-backtracking operator of equation* (3.108), $P$ *be an arbitrary graph and* $(J_{ij})_{(ij)\in E}$ *the pooling operator of* (3.127), $H(x)$ *the Bethe Hessian of* (3.107), *and* $Q$ *the matrix defined in equation* (3.128). *The following equality* *holds for any* $x \in \mathbb{C} \setminus \{\pm \tanh(J_{ij})\}_{(ij) \in E}$ *such that* $x$ *is not an eigenvalue of* B

$$P \left( x I_{|\vec{E}|} - B \right)^{-1} Q^{\mathsf{T}} + I_n = H(x)^{-1} \qquad (3.129)$$

*Proof.* The fact that $x I_{|\vec{E}|} - B$ is invertible follows from the fact that $x$ is assumed *not* to be an eigenvalue of B. For an arbitrary $u \in \mathbb{C}^n$, we let

$$\nu = \left( x I_{|\vec{E}|} - B \right)^{-1} Q^{\mathsf{T}} u \in \mathbb{C}^{|\vec{E}|}. \qquad (3.130)$$

The fact that $H(x)$ is invertible, as well as the proposition, will follow if we show

$$H(x) P \nu + H(x) u = u. \qquad (3.131)$$

We denote $(P\nu)_i$ for $i \in [n]$ the $i$-th component of the vector $P\nu$, we have

$$(P\nu)_i = \sum_{k \in \partial i} \tanh(J_{ik}) \nu_{k \to i}. \qquad (3.132)$$

From (3.130), $\nu$ verifies $\left( x I_{|\vec{E}|} - B \right) \nu = Q^{\mathsf{T}} u$, so that for any $(i \to j) \in \vec{E}$,

$$x \nu_{i \to j} - \sum_{k \in \partial i \setminus j} \tanh(J_{ik}) \nu_{k \to i} = u_i \qquad (3.133)$$

For any $(ij) \in E$, we therefore have

$$x \nu_{i \to j} - (P\nu)_i + \tanh(J_{ij}) \nu_{j \to i} = u_i, \qquad (3.134)$$
$$x \nu_{j \to i} - (P\nu)_j + \tanh(J_{ij}) \nu_{i \to j} = u_j. \qquad (3.135)$$

Together, these two questions form a closed, linear system of equations which we solve for $\nu_{i \to j}$ and $\nu_{j \to i}$. The fact that this system is invertible follows from the assumption that $x \neq \pm \tanh(J_{ij})$ for any $(ij) \in E$. Carrying out the inversion, we obtain

$$\nu_{i \to j} = \frac{1}{x^2 - \tanh^2(J_{ij})} \Big( x \left( u_i + (P\nu)_i \right) - \tanh(J_{ij}) \left( u_j + (P\nu)_j \right) \Big). \qquad (3.136)$$

Inserting this expression in (3.132), we obtain

$$(P\nu)_i = \sum_{k \in \partial i} \tanh(J_{ik}) \nu_{k \to i},$$
$$= \sum_{k \in \partial i} \frac{\tanh(J_{ik})}{x^2 - \tanh^2(J_{ik})} \Big( x \left( u_k + (P\nu)_k \right) - \tanh(J_{ik}) \left( u_i + (P\nu)_i \right) \Big). \qquad (3.137)$$

Rearranging this last expression, we obtain

$$\left( 1 + \sum_{k \in \partial i} \frac{\tanh^2(J_{ik})}{x^2 - \tanh^2(J_{ik})} \right) (P\nu)_i - \sum_{k \in \partial i} \frac{x \tanh(J_{ik})}{x^2 - \tanh^2(J_{ik})} (P\nu)_k$$
$$= \sum_{k \in \partial i} \frac{x \tanh(J_{ik})}{x^2 - \tanh^2(J_{ik})} u_k - \left( \sum_{k \in \partial i} \frac{\tanh^2(J_{ik})}{x^2 - \tanh^2(J_{ik})} \right) u_i, \qquad (3.138)$$

or equivalently, in matrix form,

$$H(x) P \nu = -H(x) u + u \qquad (3.139)$$

which completes the proof. $\qquad\qquad\square$

We now turn to the proof of the generalized Ihara-Bass formula, first stated in [153], which we used in chapter 2. The proof uses the first identity, and is close to the one of [153].

**Proposition 3.2.2** (Generalized Ihara-Bass formula). *Let* $G = ([n], E)$ *be an arbitrary graph and* $(J_{ij})_{(ij) \in E}$ *be arbitrary couplings. Let* B *be the non-backtracking operator defined by equation* (3.108)*, and* $H(x)$ *the Bethe Hessian defined by equation* (3.107)*. Then the following equality holds for any* $x \in \mathbb{C} \setminus \left\{ \pm \tanh\left(J_{ij}\right) \right\}_{(ij) \in E}$

$$\det\left( x \, I_{|\vec{E}|} - B \right) = \det\left( H(x) \right) \prod_{(ij) \in E} \left( x^2 - \tanh^2(J_{ij}) \right) . \qquad (3.140)$$

*Proof.* We first note that the formula holds if $x$ is an eigenvalue of B, since in this case, $\det\left( x \, I_{|\vec{E}|} - B \right) = 0$, and the argument leading to equation (2.66) in section 2.3.1 is straightforwardly adapted to any eigenvalue $\lambda \in \mathbb{C}$ of B which is not in the set $\left\{ \pm \tanh\left(J_{ij}\right) \right\}_{(ij) \in E}$. We can therefore assume that $x$ is not an eigenvalue of B. From proposition 3.2.1, it holds that

$$\det\left( P \left( x \, I_{|\vec{E}|} - B \right)^{-1} Q^{\mathsf{T}} + I_n \right) = \det\left( H(x)^{-1} \right) . \qquad (3.141)$$

From Sylvester's determinant theorem, it holds that

$$\det\left( P \left( x \, I_{|\vec{E}|} - B \right)^{-1} Q^{\mathsf{T}} + I_n \right) = \det\left( \left( x \, I_{|\vec{E}|} - B \right)^{-1} Q^{\mathsf{T}} P + I_{|\vec{E}|} \right) ,$$

$$= \det\left( \left( x \, I_{|\vec{E}|} - B \right)^{-1} \right) \det\left( x \, I_{|\vec{E}|} - B + Q^{\mathsf{T}} P \right) . \qquad (3.142)$$

The crux of the proof is then to notice that

$$Q^{\mathsf{T}} P = B + M \qquad (3.143)$$

where M is an operator that maps a directed edge $(i \to j)$ onto the reciprocal edge $(j \to j)$. More precisely, the elements of M are given, for $(i \to j), (k \to l) \in \vec{E}$, by

$$M_{(i \to j),(k \to l)} = \tanh\left(J_{kl}\right) \mathbf{1}(i = l)\mathbf{1}(j = k) . \qquad (3.144)$$

We therefore have that

$$\det\left( H(x)^{-1} \right) = \det\left( \left( x \, I_{|\vec{E}|} - B \right)^{-1} \right) \det\left( x \, I_{|\vec{E}|} - B + Q^{\mathsf{T}} P \right) ,$$

$$= \det\left( \left( x \, I_{|\vec{E}|} - B \right)^{-1} \right) \det\left( x \, I_{|\vec{E}|} - B + B + M \right) ,$$

$$= \det\left( \left( x \, I_{|\vec{E}|} - B \right)^{-1} \right) \det\left( x \, I_{|\vec{E}|} + M \right) . \qquad (3.145)$$

*Generalized Ihara-Bass formula*

Since M is *almost* an involution, the last determinant is easy to compute. More precisely, we may order the canonical basis of $\mathbb{R}^{|\vec{E}|}$ by putting the vector corresponding to the directed edge $(i \to j)$ next to the one corresponding to the directed edge $(j \to i)$. In this basis, the matrix $x I_{|\vec{E}|} + M$ is block-diagonal, with diagonal blocks of size $2 \times 2$. The block corresponding to the subspace spanned by the vectors representing the two directed edges $(i \to j)$ and $(j \to i)$, for $(ij) \in E$ given by

$$\begin{pmatrix} x & \tanh\left(J_{ij}\right) \\ \tanh\left(J_{ij}\right) & x \end{pmatrix},$$
(3.146)

so that

$$\det\left(x I_{|\vec{E}|} + M\right) = \prod_{(ij)\in E}\left(x^2 - \tanh^2\left(J_{ij}\right)\right).$$
(3.147)

This completes the proof. □

## 3.3 CONCLUSION

In this chapter, we have analyzed non-rigorously the spectral properties of the non-backtracking operator and the Bethe Hessian on sparse weighted random graphs. We have argued, that contrary to more classical operators such as the adjacency matrix or the Laplacian, the spectra of these matrices are well-behaved in the limit where the size of the graph tends to infinity, even on very sparse graphs.

For the non-backtracking operator, we argued that its noisy, uninformative eigenvalues are constrained, in the complex plane, in a disk of a certain radius, depending on the statistical properties of the weighted random graph. We gave an analytical expression for the radius of this disk. We also discussed the possible presence of informative eigenvalues outside of this disk, and predicted a precise threshold for their appearance, again in terms of the statistical properties of the weighted graph. We computed, using the cavity method, the spectral density of the non-backtracking operator on Erdős-Rényi random graphs.

In a second part, we characterized the spectral density of the Bethe Hessian $H(x)$. In particular, we argued that its bulk of uninformative eigenvalues does not contain any eigenvalue close to 0 as long as $x < R$, where $R$ is the radius of the bulk of uninformative eigenvalues of the associated non-backtracking operator. We also gave rigorous results relating the non-backtracking operator and the Bethe Hessian.

Thanks to our newly gained understanding of the spectral properties of B and $H(x)$, we are now ready to look at particular applications. We start by going back to the simple planted spin glass introduced in section 1.6.5, and show how the methods introduced thus far allow to solve it.

# CASE STUDY: THE PLANTED SPIN GLASS, OR CENSORED BLOCK MODEL

In this chapter, we come back to the planted Ising spin glass problem introduced in section 1.6.5, and show how the ideas exposed thus far allow to design a provably optimal algorithm to recover the planted configuration. This chapter is based on the publication [137], although we adopt here a different perspective.

Recall that the planted spin glass (or *censored block model*) is a special case of the symmetric ISBM of equation (1.46) with $q = 2$ groups, average connectivity $\alpha = \alpha_{in} = \alpha_{out}$, edge labels set $\mathcal{L} = \{\pm 1\}$, and label distributions

$$p_{in}(\ell) = (1 - \epsilon) \mathbf{1}(\ell = 1) + \epsilon \mathbf{1}(\ell = -1),$$
$$p_{out}(\ell) = (1 - \epsilon) \mathbf{1}(\ell = -1) + \epsilon \mathbf{1}(\ell = 1). \tag{4.1}$$

*Distribution of the binary labels*

The posterior probability distribution of the hidden assignment $\sigma$ takes the form of an Ising spin glass

$$\mathbb{P}(\sigma) = \frac{1}{\mathcal{Z}} \exp\left( \beta \sum_{(ij) \in E} \ell_{ij} \sigma_i \sigma_j \right), \tag{4.2}$$

*Posterior distribution of the planted configuration*

where the inverse temperature $\beta = \beta^\star$ is related to the noise $\epsilon$ through

$$\beta^\star = \frac{1}{2} \log \frac{1 - \epsilon}{\epsilon}. \tag{4.3}$$

*Bayes optimal temperature*

As shown in section (1.6.5), with this choice of inverse temperature, corresponding to the Bayes optimal setting, the (gauge transformed) Ising spin glass is on the Nishimori line, so that it cannot be in the spin glass phase. However, defining $\beta^\star$ requires knowing the noise parameter $\epsilon$, which we will not assume in this chapter. Instead, we will consider the planted spin glass (4.2) at a *general* inverse temperature $\beta$, and show that we can still design an optimal algorithm.

Finally, recall that the statistical physics prediction for the *detectability* transition of this model (which also corresponds to the conjecture of [60]) is given by (1.98), i.e.

$$\alpha > \frac{1}{(1 - 2\epsilon)^2} = \alpha_{detect}. \tag{4.4}$$

*Conjectured transition*

It was shown in [94] that this condition is necessary, in the sense that when $\alpha < \alpha_{detect}$, it is impossible to distinguish a graph generated from the planted spin glass from a completely random graph with random $\pm 1$ labels, so that detectability is information-theoretically impossible. In this chapter, we show rigorously that this condition is also sufficient. More precisely, we introduce two algorithms, based respectively on the non-backtracking operator and the Bethe Hessian, that output an estimate $\hat{\sigma} \in \{\pm 1\}^n$ that has a finite overlap with the planted assignment $\sigma$ whenever (4.4) holds, where the overlap is defined as

*Overlap*

$$\lim_{n \to \infty} \frac{1}{n} \sum_{i=1}^{n} \sigma_i \hat{\sigma}_i . \qquad (4.5)$$

Note that this quantity is a rescaled version of the accuracy, that vanishes when $\hat{\sigma}$ is a random guess, and equals unity if the reconstruction is perfect.

There are various interpretations and models that connect to this problem. One example is a particular community detection problem [2], in which we try to recover the community membership of the nodes of a graph based on noisy (or censored) observations about their relationship. Another example is correlation clustering [12], in which we try to cluster the graph G by minimizing the number of "disagreeing edges" (here $\ell_{ij} = -1$) in each cluster. These examples, and others such as synchronization, are discussed in details in [1].

In the rest of this chapter, we introduce the particular forms of the non-backtracking and Bethe Hessian operators on the planted spin glass, and make some of the intuitive arguments of chapters 2 and 3 more precise. We end this chapter by sketching a rigorous proof of the optimality of the algorithm based on the non-backtracking operator, and discuss the implications of this result for the Bethe Hessian.

## 4.1  THE NON-BACKTRACKING OPERATOR

The non-backtracking operator of the symmetric ISBM was computed in section 3.1.3, and reads in general, for $(i \to j), (k \to l) \in \vec{E}$,

$$B_{(i \to j),(k \to l)} = \frac{\alpha_{in} p_{in}(\ell_{kl}) - \alpha_{out} p_{out}(\ell_{kl})}{\alpha_{in} p_{in}(\ell_{kl}) + (q-1)\alpha_{out} p_{out}(\ell_{kl})} \mathbf{1}(l = i)\mathbf{1}(k \neq j) .$$

On the particular case of the planted spin glass, this expression considerably simplifies to

$$B_{(i \to j),(k \to l)} = (1 - 2\epsilon) \, \ell_{kl} \, \mathbf{1}(l = i)\mathbf{1}(k \neq j) .$$

Interestingly, the noise parameter $\epsilon$ appears only as a global constant. We will therefore refer to the following matrix as the non-backtracking operator of the planted spin glass

*Non-backtracking operator of the planted spin glass*

$$B_{(i \to j),(k \to l)} = \ell_{kl} \, \mathbf{1}(l = i)\mathbf{1}(k \neq j) . \qquad (4.6)$$

This matrix has the same eigenvectors as the "true" non-backtracking operator (4.6), and their respective eigenvalues are related by a trivial scaling factor $(1 - 2\epsilon)$. The great advantage of this last expression is that it depends only on the data we are given, and not on the parameter $\epsilon$ that was used to generate it.

In chapter 3, we made precise predictions on the spectrum of B in a general instance of the symmetric lSBM. Their implications for the matrix B of equation (4.6) on a graph generated from the planted spin glass are as follows.

*Spectrum of the non-backtracking operator on the planted spin glass*

— The uninformative eigenvalues of B are constrained in a disk of radius R given here by

$$R = \sqrt{\alpha \, \mathbb{E}[\ell^2]} \, , \tag{4.7}$$

where $\ell$ denotes a random variable with the same distribution as the labels $\ell_{ij}$ for $(ij) \in E$. Since these labels are here equal to $\pm 1$, we have

$$R = \sqrt{\alpha} \, , \tag{4.8}$$

independently of the distribution of the labels.

— There is an informative eigenvalue $\alpha \, \Delta(\epsilon)$ outside of the bulk of radius R if and only if $\alpha \, \Delta(\epsilon) > R$, where

$$\Delta(\epsilon) = \frac{1}{2} \left( \mathbb{E}[\ell_{in}] - \mathbb{E}[\ell_{out}] \right) \, , \tag{4.9}$$

where $\ell_{in}$ (resp. $\ell_{out}$) is a random variable that has the same distribution as the labels between points in the same group (resp. different groups). From rules in equation (4.1), we have $\mathbb{E}[\ell_{in}] = -\mathbb{E}[\ell_{out}] = 1 - 2\epsilon$, so that the potential informative eigenvalue is equal to

$$\alpha \, \Delta(\epsilon) = \alpha \, (1 - 2\epsilon) \, . \tag{4.10}$$

— For any choice of $\epsilon$, there is no ferromagnetic eigenvalue, since $\alpha \, \mathbb{E}[\ell] = 0 < R$ for any $\epsilon$.

We therefore expect that a spectral method based on B can detect the planted partition if and only if

$$\tau(\alpha, \epsilon) = \left( \frac{\alpha \, \Delta(\epsilon)}{R} \right)^2 > 1 \, , \tag{4.11}$$

*Detectability threshold of B*

or equivalently, $\alpha \, (1 - 2\epsilon)^2 > 1$, i. e. from (4.4), as soon as it is information-theoretically possible. More precisely, the previous predictions imply that when below the previous transition, the spectrum of B is entirely constrained in the disk of radius $\sqrt{\alpha}$, while when above the transition, there is a single informative eigenvalue $\alpha \, (1 - 2\epsilon)$ outside of the bulk. An illustration of both situations in provided in figure 4.1. We will see in the following that we can, in this case, prove this picture rigorously.

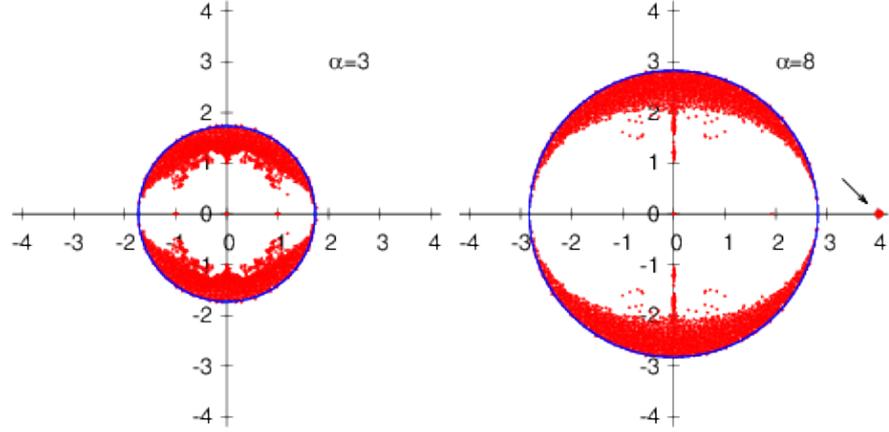

Figure 4.1 – Spectrum of B for a problem generated with $\epsilon = 0.25$, $n = 2000$. We used $\alpha = 3$ (left side) and $\alpha = 8$ (right side), to be compared with $\alpha_{\text{detect}} = 4$ for this value of $\epsilon$. In both cases, the bulk of the spectrum is confined in a circle of radius $\sqrt{\alpha}$. However, when $\alpha > \alpha_{\text{detect}}$, a single isolated eigenvalue appears out of the bulk at $\alpha(1 - 2\epsilon)$ (see the arrow on the right plot) and the corresponding eigenvector is correlated with the planted assignment. Figure taken from [137].

## 4.2 THE BETHE HESSIAN

The Bethe Hessian (3.107) associated with the non-backtracking operator defined in equation (4.6) reads here

$$
\begin{aligned}
H_{ij}(x) &= \left(1 + \sum_{k \in \partial i} \frac{\ell_{ik}^2}{x^2 - \ell_{ik}^2}\right) \mathbf{1}(i = j) - \frac{x\,\ell_{ij}}{x^2 - \ell_{ij}^2}\, \mathbf{1}(j \in \partial i)\,, \\
&= \frac{1}{x^2 - 1}\Big(\left(x^2 - 1 + |\partial_i|\right)\mathbf{1}(i = j) - x\,\ell_{ij}\,\mathbf{1}(j \in \partial i)\Big)\,.
\end{aligned}
\tag{4.12}
$$

Disregarding the trivial, positive factor $(x^2 - 1)^{-1}$, we will in fact call Bethe Hessian of the planted spin glass the matrix

*Bethe Hessian of the planted spin glass*

$$
H(x) = (x^2 - 1)\,I_n - x\,A + D\,,
\tag{4.13}
$$

where $A$ is the (weighted) adjacency matrix of the graph, with non-zero entries $A_{ij} = \ell_{ij}$ for $(ij) \in E$, and $D$ is the diagonal degree matrix, with entries $D_{ii} = |\partial_i|$ for $i \in [n]$.

We will be particularly interested in the matrix $H(R) = H\left(\sqrt{\alpha}\right)$. Note that as usual (see equation (2.59)), the quantity $R$ can be easily estimated from the data, without knowing the parameters of the model. Here, one only needs to take the empirical average of the degrees of the vertices in the graph. From the arguments of section 2.3.2, we expect that $H\left(\sqrt{\alpha}\right)$ has one negative eigenvalue whenever B has an informative eigenvalue (outside of the bulk). The corresponding eigen-vector is an unstable direction of the Bethe free energy at the paramagnetic point, pointing toward an informative, partially ordered phase (in fact the ferromagnetic phase of the gauge transformed model, as explained in section 1.6.5). More precisely, from the arguments of section 3.2, we expect that the bulk of uninformative eigenvalues of $H\left(\sqrt{\alpha}\right)$ is supported on a subset of $\mathbb{R}_+$, with its left edge precisely at $0$. Consequently, any negative eigenvalue of $H(x)$ is potentially informative. Once more, the advantage of using a spectral algorithm based on $H\left(\sqrt{\alpha}\right)$ rather than B is that $H\left(\sqrt{\alpha}\right)$ is a smaller [1] and symmetric matrix. Additionally, we will see numerically in the following that the overlap provided by the Bethe Hessian is slightly larger than the one given by B, although we lack a theoretical understanding of this fact.

*Spectrum of $H(R)$*

As explained in section 2.3.2, the presence of negative eigenvalues of $H\left(\sqrt{\alpha}\right)$ has an intuitive statistical physics interpretation, related to the instability of the paramagnetic phase in the Bethe approximation. On a model as simple as the planted spin glass, it is possible to give a transparent graphical representation of this interpretation which we sketch now. We have shown in section 2.3.2 that $H(x)$ is the Hessian of the Bethe free energy at the paramagnetic point of an associated Ising model with couplings $(J_{ij})_{(ij) \in E}$ given by equation (2.76), i. e. here

*Statistical physics interpretation of the Bethe Hessian*

$$\tanh(J_{ij}) = \frac{\ell_{ij}}{x}.  \qquad (4.14)$$

On the other hand, the true posterior (4.2) takes the form of an Ising model with true couplings $\beta^\star \ell_{ij}$, which verify

$$\tanh(\beta^\star \ell_{ij}) = (1 - 2\epsilon)\, \ell_{ij}.  \qquad (4.15)$$

Therefore, had we known $\epsilon$, we could have evaluated the Bethe Hessian $H(x)$ at $x = (1 - 2\epsilon)^{-1}$. Indeed, in the Bethe approximation, the matrix $H\left((1 - 2\epsilon)^{-1}\right)$ controls the stability of the paramagnetic stationary point of the *true* posterior distribution (4.2) , and the presence of a negative eigenvalue of $H\left((1 - 2\epsilon)^{-1}\right)$ indicates that the paramagnetic point is *not* a local minimum of the Bethe free energy, so that the problem is solvable. However, we dot not assume here that we know $\epsilon$. Comparing equations (4.14) and (4.15), we see that by considering the matrix $H(x)$, we are in fact investigating the stability of the paramagnetic stationary point of the planted spin glass (4.2) at a temperature $\beta$ *different* from $\beta^\star$, and defined by the equation

$$\tanh(\beta \ell_{ij}) = \tanh\left(J_{ij}\right) ,$$
$$\iff \tanh(\beta) = x^{-1} ,  \qquad (4.16)$$

where we have used that the labels are equal to $\pm 1$. Recall the phase diagram of the planted spin glass represented on figure 4.2, obtained from the phase diagram of the Viana-Bray model after gauge transforming, as discussed in section 1.6.5. The true (Bayes optimal) posterior of equation (4.2) lies somewhere on the Nishimori line, represented in red, say at a point $(\epsilon, \beta^{\star})$, where $\beta^{\star}$ is related to epsilon by equation (4.3). In section 1.6.5, we have argued that an instance of the planted spin glass is solvable if $(\epsilon, \beta^{\star}) \in \mathcal{F}$, and unsolvable if $(\epsilon, \beta^{\star}) \in \mathcal{P}$, where $\mathcal{F}$ and $\mathcal{P}$ denote respectively the ferromagnetic and the paramagnetic regions depicted on figure 4.2. By changing the temperature, we enter a parameter mismatch regime, and we venture outside the Nishimori line. More precisely, when using the matrix $H(x)$, we are effectively considering a model that is somewhere on the vertical line passing through the point $(\epsilon, \beta^{\star})$, specifically at the point $\left(\epsilon, \operatorname{atanh}\left(x^{-1}\right)\right)$ of this phase diagram. To design an optimal non-parametric approach based on $H(x)$, we need to find a magical value of $x$, independent of $\epsilon$ (since we dot not know it), and such that

$$\begin{aligned} (\epsilon, \beta^{\star}) \in \mathcal{P} &\iff \left(\epsilon, \operatorname{atanh}\left(x^{-1}\right)\right) \in \mathcal{P}, \\ (\epsilon, \beta^{\star}) \in \mathcal{F} &\iff \left(\epsilon, \operatorname{atanh}\left(x^{-1}\right)\right) \in \mathcal{F}. \end{aligned} \tag{4.17}$$

---

1. When the weights of B are equal to $\pm 1$, as is the case here, it is possible to reduce B to a smaller $(2n \times 2n)$ matrix B′, as explained in [89, 137]. This matrix is however still non-symmetric, and still larger than $H\left(\sqrt{\alpha}\right)$.

It turns out that there is one, and only one, value of $x$ that does the job, as can be seen from figure 4.2, and that this value is given by

$$x = \sqrt{\alpha}. \tag{4.18}$$

This corresponds to setting the Ising model (4.2) on the *edge* of the paramagnetic to spin glass transition. With this value, the conditions (4.17) hold. Consequently, the presence of a negative eigenvalue of $H\left(\sqrt{\alpha}\right)$ is equivalent to the presence of negative eigenvalue of $H\left((1-2\epsilon)^{-1}\right)$. In particular, a spectral algorithm based on $H\left(\sqrt{\alpha}\right)$ should be optimal in that it detects the hidden assignment as soon as the true posterior is in the ferromagnetic phase, i.e. as soon as (4.4) holds. However, remarkably, such an algorithm does *not* require the knowledge of $\epsilon$.

In the next section, we write down explicitly the spectral procedures (based on B and $H\left(\sqrt{\alpha}\right)$) suggested by the remarks of the previous sections, and show numerical simulations comparing the overlap achieved by these methods with the overlap achieved by BP.

## 4.3 ALGORITHMS

For completeness, we spell out explicitly the procedures described in algorithms 1 and 2 for the special case of the planted spin glass. Algorithm 3 presents our procedure for the non-backtracking operator. Recall that to form an estimate for each vertex from the eigenvectors of B, we use the pooling matrix $P \in \mathbb{R}^{n \times |\vec{E}|}$ whose definition is here

$$P_{i,k \to l} = \ell_{kl}\, \mathbf{1}(l=i). \tag{4.19}$$

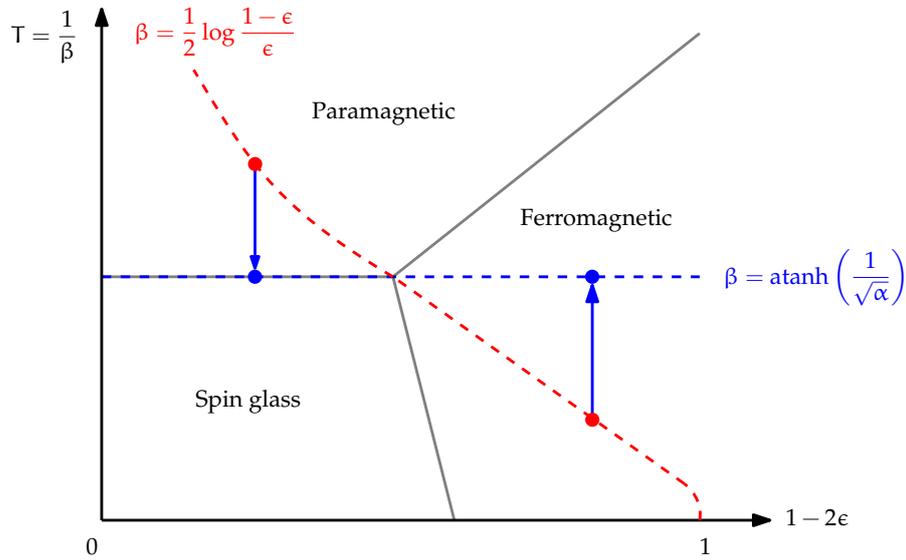

Figure 4.2 – Cartoon of the phase diagram of the planted spin glass, obtained from the phase diagram of the Viana-Bray model after performing a gauge transformation, as explained in section 1.6.5. The true posterior of equation (4.2), with Bayes optimal temperature $\beta^\star$ given by (4.3), lies somewhere on the red Nishimori line. The two red dots represent two instances of the true posterior, one corresponding to a solvable problem (in the ferromagnetic phase) and another corresponding to an unsolvable problem (in the paramagnetic phase). The Bethe Hessian $H(\sqrt{\alpha})$ controls the stability of the paramagnetic stationary point of a different model (blue dot), obtained by projecting the true posterior on the spin glass line $\beta = \text{atanh}(1/\sqrt{\alpha})$. This choice preserves the solvability (or unsolvability) of the model, as the projected model is always in the same region (paramagnetic or ferromagnetic) of the phase diagram as the true posterior.

Note that since we are dealing only with 2 clusters, we have replaced the use of kmeans in the last step of algorithm 1 by a partitioning based on the sign of the components of the pooled informative eigenvector of B.

---

**Algorithm 3** Non-backtracking spectral algorithm for the planted spin glass

---

**Input:** Graph $G = ([n], E)$, labels $\ell_{ij} = \pm 1$ for $(ij) \in E$

1: **Build** the non-backtracking operator B of (4.6) and the pooling matrix P of (4.19)
2: **Compute** the leading eigenvalue $\lambda_1$ of B and its corresponding eigenvector $v$. If $\lambda_1 \in \mathbb{C}\backslash\mathbb{R}$ or $\lambda_1 < \sqrt{\alpha}$, raise an error.
3: **Pool** the eigenvector $v$ to compute the approximate magnetizations $m = P v$
4: **Output** the assignments $\hat{\sigma}_i = \text{sign}(m_i)$ for $i \in [n]$

---

The spectral method based on the Bethe Hessian is presented in algorithm 4. Note that both algorithms 3 and 4 use the value of the average connectivity $\alpha$ which in practice is easily estimated from the graph.

---

**Algorithm 4** Bethe Hessian spectral algorithm for the planted spin glass

---

**Input:** Graph $G = ([n], E)$, labels $\ell_{ij} = \pm 1$ for $(ij) \in E$

1: **Build** the Bethe Hessian $H\left(\sqrt{\alpha}\right)$ where $H(x)$ is defined by equation (4.13)
2: **Compute** the (algebraically) smallest eigenvalue $\lambda_1$ of $H\left(\sqrt{\alpha}\right)$ and its corresponding eigenvector $v$. If $\lambda_1 > 0$, raise an error.
3: **Output** the assignments $\hat{\sigma}_i = \text{sign}(v_i)$ for $i \in [n]$

---

*Numerical simulations*    Before turning to proofs, we show on figure 4.3 the numerical performance of our two algorithms, and compare them with the asymptotic performance of BP, estimated with a population dynamics algorithm [103]. Here, BP is run on the true posterior distribution (4.2) with $\beta = \beta^\star$ so that its marginal (are believed to) provide an asymptotically correct estimate of the Bayes optimal assignment in the limit $n \to \infty$. We find that both algorithms 3 and 4 are able to achieve partial recovery as soon as $\alpha > \alpha_{\text{detect}}$, and that their overlap is similar to that of BP, though of course strictly smaller. Note again that BP requires here the knowledge of $\epsilon$ while the two spectral algorithms 3 and 4 do not. Additionally, they are trivial to implement, run faster, and avoid the potential non-convergence problem of belief propagation while remaining asymptotically optimal in detecting the hidden assignment. We also observe, empirically, that the overlap given by the Bethe Hessian seems to be always superior to the one provided by the non-backtracking operator.

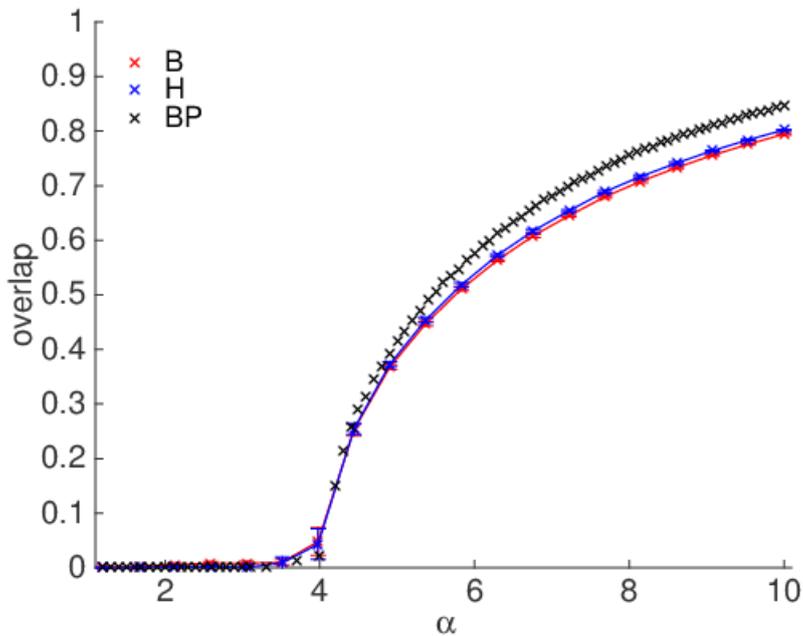

Figure 4.3 – Overlap as a function of $\alpha$: comparison between algorithm 3 (based on the non-backtracking operator B), algorithm 4 (based on the Bethe Hessian H $(\sqrt{\alpha})$ called here H), and BP. The noise parameter $\epsilon$ is fixed to 0.25 (corresponding to $\alpha_{\text{detect}} = 4$), and we vary $\alpha$. The overlap for B and H $(\sqrt{\alpha})$ is averaged over 20 graphs of size $n = 10^5$. The overlap for BP is estimated asymptotically using the standard method of population dynamics (see for instance [103]), with a population of size $10^4$. All three methods output a positively correlated assignment as soon as $\alpha > \alpha_{\text{detect}}$. The spectral algorithms 3 and 4 have an overlap similar to that of BP, with the same phase transition, while being simpler and not requiring the knowledge of the parameter $\epsilon$. Figure taken from [137].

## 4.4    SPECTRAL PROPERTIES OF THE NON-BACKTRACKING OPERATOR ON THE PLANTED SPIN GLASS

In this section, we state results concerning the spectrum of B and show that algorithm 3 outputs an assignment $\hat{\sigma}_i$ that has a finite overlap with the planted one, in the limit $n \to \infty$, whenever (4.4) holds. In fact, the following result makes rigorous the predictions of section 3.1.3 in the special case of the planted spin glass, extending some of the proof techniques of [19] to the case where the non-backtracking operator is weighted, with weights in $\{\pm 1\}$.

**Theorem 4.4.1.** *Consider an Erdős-Rényi random graph* $G = ([n], E)$ *with average degree* $\alpha$, *variables assigned to vertices* $\sigma_i = \pm 1$ *uniformly at random independently from the graph and where the edges carry labels* $(\ell_{ij})_{(ij) \in E}$ *sampled from (4.1). We denote by* B *the non-backtracking operator defined by (4.6), and by* $|\lambda_1| \geqslant |\lambda_2| \geqslant \cdots \geqslant |\lambda_{|\vec{E}|}|$ *the eigenvalues of* B *in order of decreasing magnitude. Then, with probability tending to* 1 *as* $n \to \infty$, *we have:*

*(i) if* $\alpha < \alpha_{detect}$ *then* $|\lambda_1| \leqslant \sqrt{\alpha} + o(1)$.

*(ii) if* $\alpha > \alpha_{detect}$, *then* $\lambda_1 \in \mathbb{R}$, $\lambda_1 = \alpha(1 - 2\epsilon) + o(1) > \sqrt{\alpha}$, *and* $|\lambda_2| \leqslant \sqrt{\alpha} + o(1)$. *Additionally, denoting* $v$ *the eigenvector associated with* $\lambda_1$, *the following assignment has a finite overlap with the planted variables* $\sigma_i$:

$$\hat{\sigma}_i = sign \left( \sum_{j \in \partial i} v_{j \to i} \right). \quad (4.20)$$

This theorem is illustrated on figure 4.1. It is then straightforward to show the following guarantee for algorithm 3.

**Corollary 4.4.2.** *The assignment output by algorithm 3 has a finite overlap with the planted assignment in the limit* $n \to \infty$ *if and only if*

$$\alpha > \alpha_{detect}. \quad (4.21)$$

We now give a brief sketch of proof for theorem 4.4.1. The proof relies heavily on the techniques developed in [19], and we try to use notations consistent with [19]. For an oriented edge $e = (u \to v) \in \vec{E}$, we set $e_1 = u$, $e_2 = v$ and $e^{-1} = (v \to u)$. We start with a simple observation, which translates the gauge invariance property exposed in section 1.6.5. If $t$ is the vector in $\mathbb{R}^{|\vec{E}|}$ defined by $t_e = \sigma_{e_2}$ and $\odot$ is the Hadamard product, i.e. $(t \odot x)_e = \sigma_{e_2} x_e$, we have

$$B x = \lambda x \iff \tilde{B}(t \odot x) = \lambda(t \odot x), \quad (4.22)$$

with $\tilde{B}$ defined by $\tilde{B}_{ef} = B_{ef}\sigma_{f_1}\sigma_{f_2}$. In particular, B an $\tilde{B}$ have the same spectrum and there is a trivial relation between their eigenvectors. It will be easier to work with $\tilde{B}$ so to lighten the notation, we will denote (throughout the proof)

$$B_{ef} = w_f \mathbf{1}(f_2 = e_1)\mathbf{1}(f_1 \neq e_2), \quad (4.23)$$

where $w_f = \sigma_{f_1} \ell_f \sigma_{f_2}$, and we write, with a abuse of notation, $\ell_f = \ell_{f^{-1}} = \ell_{f_1 f_2}$. Note that the random variables $w_f$ are now i.i.d with $\mathbb{P}(w_f = 1) = 1 - \mathbb{P}(w_f = -1) = 1 - \epsilon$. With this formulation, we have gauge transformed the problem which now lies on the Nishimori line [90, 117].

For the case $\alpha(1-2\epsilon)^2 < 1$, the proof is relatively easy. Indeed, from [94], we know that our setting is contiguous to the setting with $\epsilon = 1/2$. In this case, the random variables $w_f$ are centered and a version of the trace method will allow to upper bound the spectral radius of B. Note however, that one needs to condition on the graph to be k-tangle-free, i.e. such that every neighborhood of radius k contains at most one cycle in order to apply the first moment method.

We now consider the case $\alpha(1-2\epsilon)^2 > 1$ and denote by M the linear mapping on $\mathbb{R}^{|\vec{E}|}$ defined by $(Mx)_e = w_e x_{e^{-1}}$, i.e. the matrix associated with M has elements $M_{ef} = w_e \mathbf{1}\left(f = e^{-1}\right)$. Note that $M^{\intercal} = M$ and since $w_e^2 = 1$, M is an involution so that M is an orthogonal matrix. A simple computation shows that $B^k M = M \left(B^{\intercal}\right)^k$, hence $B^k M$ is a symmetric matrix. This symmetry generalizes the *PT- invariance* discussed in the unweighted case in equation (3.65), and will be crucial in adapting the analysis of [19].

We define $\tilde{\alpha} = \alpha(1-2\epsilon)$ and $\chi \in \mathbb{R}^{|\vec{E}|}$ with $\chi_e = 1$ for all $e \in \vec{E}$. The proof strategy is then similar to section 5 in [19]. Consider a sequence $k \sim \kappa \log_{\tilde{\alpha}} n$ for some small positive constant $\kappa$. Let

$$\varphi = \frac{B^k \chi}{\|B^k \chi\|}, \quad \theta = \|B^k M \varphi\|, \quad \zeta = \frac{B^k M \varphi}{\theta}. \tag{4.24}$$

If $R = B^k - \theta \zeta M \varphi^{\intercal}$ and we can prove that $\|R\|$ is small compared to $\theta$, then we can use a theorem on the perturbation of eigenvalues and eigenvectors adapted from the Bauer-Fike theorem (see section 4 in [19]) to show that $B^k$ has an eigenvalue close to $\theta$.

More precisely, for $y \in \mathbb{R}^{|\vec{E}|}$ with $\|y\| = 1$, write $y = s M \varphi + x$ with $x \in (M \varphi)^{\perp}$ and $s \in \mathbb{R}$. Then, we find

$$\|Ry\| = \|B^k x + s(B^k M \varphi - \theta \zeta)\| \leqslant \sup_{x : \langle x, M\varphi \rangle = 0, \|x\| = 1} \|B^k x\|. \tag{4.25}$$

This last quantity can be shown to be upper bounded by $(\log n)^c \alpha^{k/2}$ for some constant $c > 0$, similarly to proposition 12 in [19]. Moreover, we can also show that w.h.p, for some constants $c_1 > c_0 > 0$,

$$\zeta(M\varphi)^{\intercal} \geqslant c_0, \quad c_0 \tilde{\alpha}^k \leqslant \theta \leqslant c_1 \tilde{\alpha}^k. \tag{4.26}$$

Using the Bauer-Fike theorem, these bounds allow to show that B has an eigenvalue $|\lambda_1 - \tilde{\alpha}| = O(1/k)$ and that $|\lambda_2| \leqslant \sqrt{\alpha} + o(1)$. Note that

$$\theta = \frac{\|B^k (B^{\intercal})^k M \chi\|}{\|B^k \chi\|}, \tag{4.27}$$

so that in order to show the bounds (4.26), we need to compute quantities of the type $\|B^k\chi\|$. We now explain the main ideas to compute these quantities. First note that $(B^k\chi)_e$ depends only on the ball of radius $k$ around the edge $e$. For $k$ not too large, this neighborhood is w.h.p a tree, which can be coupled with a Galton-Watson branching process with offspring distribution Poisson($\alpha$). It is then natural to consider this Poisson Galton-Watson branching process with i.i.d weights $w_{uv} \in \{\pm 1\}$ with mean $1 - 2\epsilon$, on each edge $(uv)$. For $u$ in the tree, we denote by $|u|$ its generation and let $Y(u) = \prod_{s=1}^t w_{\gamma_s, \gamma_{s+1}}$ where $\gamma = (\gamma_1, \ldots, \gamma_t)$ is the unique path between the root $o = \gamma_1$ and $u = \gamma_t$. Then $(B^k\chi)_e$ is well approximated by

$$Z_k = \sum_{|u|=k} Y(u). \tag{4.28}$$

It is easy to check that $X_t = Z_t/\bar{\alpha}^t$ defines a martingale (with respect to the natural filtration) with unit mean. Moreover we have, by a computation similar to lemma 5.1 in [42], that

$$\mathbb{E}\left[Z_t^2\right] = \mathbb{E}\left[\sum_{u,v:|u|=|v|=t} Y(u)Y(v)\right] \tag{4.29}$$

$$= \sum_{i=0}^t \alpha^{t-i}(1-2\epsilon)^{2i}\alpha^{2i} = O\left(\bar{\alpha}^{2t}\right), \tag{4.30}$$

where the last equality is valid only if $\alpha(1-2\epsilon)^2 > 1$. So in this case, we have $\mathbb{E}\left[X_t^2\right] = O(1)$ and by Doob's martingale convergence theorem, the martingale $X_t$ converges almost surely and in $L^2$ to a limiting random variable $X(\infty)$ with unit mean and finite variance. Following the arguments of [19], this reasoning leads to (4.26).

We now consider the eigenvector associated with $\lambda_1$. It follows from the Bauer-Fike theorem (section 4 in [19]) that the eigenvector $x$ associated to $\lambda_1$ is asymptotically aligned with $\frac{B^k(B^\intercal)^k M\chi}{\|B^k(B^\intercal)^k M\chi\|}$. Thanks to the coupling with the branching process, we can prove that

$$\|B^k(B^\intercal)^k M\chi\| \approx \bar{\alpha}^{2k} \tag{4.31}$$

and moreover, we have for $e \in \vec{E}$,

$$\frac{(B^k(B^\intercal)^k M\chi)_e}{\bar{\alpha}^{2k}} \approx \frac{\bar{\alpha}}{\alpha(1-2\epsilon)^2 - 1} X(\infty), \tag{4.32}$$

where $X(\infty)$ is the limit of the martingale defined above and has mean one. We can now undo the gauge transformation to translate this result for the eigenvector of the original non-backtracking operator thanks to (4.22): $\nu_e = \sigma_{e_2} x_e$ where $x_e$ is approximated by (4.32). In particular, we see that $\sum_{e,e_2=v} \nu_e$ is correlated with $\sigma_v$. This completes our sketch of proof.

Let us finally briefly mention the implications of the previous theorem for the spectral method of algorithm 4 based on the Bethe Hessian. As we have already argued in chapter 2, for $x$ large enough, $H(x)$ is positive definite. Then as $x$ decreases, by the Ihara-Bass formula 2.68, the determinant of $H(x)$ vanishes if and only if $x$ becomes equal to an eigenvalue of B. Theorem 4.4.1 allows to control the sign of the characteristic polynomial of B, i. e., by the Ihara-Bass formula, the sign of the determinant of $H(x)$. Using continuity arguments, we can show the following corollary.

**Corollary 4.4.3.** *Under the same conditions as theorem 4.4.1, if $\alpha < \alpha_{detect}$, then for any $\eta > 0$, $H\left(\sqrt{\alpha} + \eta\right)$ is w.h.p definite positive. Conversely, if $\alpha > \alpha_{detect}$, then there exists $\eta_0 > 0$ such that for any $\eta_0 > \eta > 0$, $H\left(\sqrt{\alpha} + \eta\right)$ has w.h.p a unique strictly negative eigenvalue.*

Strictly speaking, if we denote by $\lambda_1$ the leading eigenvalue of B, theorem 4.4.1 combined the computation presented in section 2.3.1 only allows to show that the eigenvector with eigenvalue 0 of $H(\lambda_1)$ is correlated with the planted variables if $\alpha > \alpha_{detect}$. Indeed, in this case, the null space of $H(\lambda_1)$ is spanned by $P\nu$ where $P$ is the pooling matrix (4.19) and $\nu$ is the leading eigenvector of B. However, we observe numerically (as shown on figure 4.3) that the eigenvector with negative eigenvalue of $H\left(\sqrt{\alpha}\right)$ also has a positive, and in fact slightly larger overlap with the planted configuration. This fact will have to be clarified in future work.

## 4.5 CONCLUSION

In this chapter, we have illustrated the performance of spectral algorithms based on the non-backtracking operator and the Bethe Hessian on the planted spin glass problem. In particular, we have argued that these algorithms are optimal in that they can detect the hidden assignment as soon as it is information-theoretically possible to do so. For the case of the non-backtracking, we were able to rigorously and fully substantiate this claim, while the rigorous results for the Bethe Hessian are only partial. On the other hand, we were able to explicitly connect the algorithm based on the Bethe Hessian to the phase diagram of the planted spin glass, therefore giving a transparent graphical justification of the Bethe Hessian approach.

In the next, and final part of this dissertation, we examine various applications of our spectral methods, in the fields of community detection, similarity-based clustering, and matrix completion, before going back to statistical physics to prove rigorous bounds for the ferromagnetic Ising model on arbitrary graphs.

# Part III

## SOME APPLICATIONS

This last part is devoted to applications of the theory of spectral inference developed previously. We consider inference problems in community detection, unsupervised and semi-supervised similarity-based clustering and matrix completion before returning to statistical physics. In a final outlook chapter, we outline possible future applications in unsupervised learning.

# COMMUNITY DETECTION AND THE ISING FERROMAGNET

We consider community detection under the SBM, as introduced in section 1.4.2. This problem experienced a surge of interest after the work of [32] who conjectured the phase transition (1.44) for the symmetric SBM, and proposed an optimal inference algorithm based on BP. This algorithm, however, requires knowing the parameters of the model used to generate the problem, and, although these can be learned using an expectation-maximization approach [32, 33], this incurs a non-negligible cost in complexity [1] and makes the whole procedure non-trivial to implement. Additionally, since the SBM is, as argued in section 1.4.2, a bad model for real world network, fitting an instance of the SBM on a real world problem may not always make a lot of sense.

*Limitations of BP for community detection*

Among the popular non-parametric methods for community detection are spectral approaches, based *e. g.* on the computation of a few eigenpairs of the adjacency or modularity matrices [94, 112, 115], or the Laplacian [46, 98, 114, 116]. Such methods are popular because of their simplicity and scalability. Indeed, highly optimized sparse linear algebra libraries are widely available, and randomized or distributed matrix-free methods, akin to those developed for the PageRank algorithm, exist for (very) large scale problems. However, as explained in 2.3.3, these traditional spectral methods are severely suboptimal on the *sparse* SBM, and fail to detect the communities even hen significantly above the transition (1.44). This gap was closed in [89], who first proposed using a spectral method based on the non-backtracking operator, and conjectured that it achieves the optimal threshold (1.44). This conjecture was later proved rigorously in [19]. We review this approach in the first section of this chapter.

*Spectral methods, and their limitations*

These results are, however, not entirely satisfactory. First, the use a of a high-dimensional matrix (of dimension $|\vec{E}|$ rather than n, as for more traditional spectral methods) can be expensive, both in terms of computational time and memory. Secondly, numerical eigensolvers are typically faster and more efficient for symmetric matrices than non-symmetric ones. In the second section of this chapter, we show that there in fact exists a symmetric $n \times n$ operator that performs optimally well in the SBM, namely the Bethe Hessian. Finally, in a

---

1. When running BP with the correct parameters, we are on the Nishimori line, and BP converges [32]. On the other hand, when using an expectation-maximization approach, we explore a potentially wide region of the phase diagram and may end up (depending on the initial choice of parameters) in the spin glass phase where BP fails to converge.

third and last section, we show numerical simulations comparing the performance of various spectral algorithms as well as BP, on both synthetic and real world networks. The second and third part of this chapter are based on the publication [139], but follow a different route inspired by the theory developed previously in this dissertation.

## 5.1 THE NON-BACKTRACKING OPERATOR

*Non-backtracking operator of the SBM*

We quickly recall the results of section 3.1.3 on the spectrum of B in the symmetric SBM. Disregarding a global multiplicative factor, we define B by its elements

$$B_{(i \to j),(k \to l)} = \mathbf{1}(l = i)\mathbf{1}(k \neq j) \,. \tag{5.1}$$

The spectrum of B, in the limit $n \to \infty$ is composed of

— a bulk of uninformative eigenvalues constrained to the disk of radius $\sqrt{\alpha}$ where $\alpha = q^{-1}\left(\alpha_{in} + (q-1)\alpha_{out}\right)$ is the average connectivity of the graph.

*Spectrum of the unweighted non-backtracking matrix*

— a leading ferromagnetic eigenvalue $\alpha$, with a corresponding globally ordered eigenvector.

— a potential informative real eigenvalue $q^{-1}(\alpha_{in} - \alpha_{out})$ with multiplicity $q - 1$. This eigenvalue can be positive (assortative case) or negative (disassortative case), and exists if and only if

*Detectability transition*

$$|\alpha_{in} - \alpha_{out}| > q\sqrt{\alpha} \,, \tag{5.2}$$

which corresponds to the detectability transition of the symmetric SBM, equation (1.44).

Figure 5.1 shows the spectrum of B in the case of $q = 2$ communities, both in the detectable and undetectable regimes. This spectrum, consistent with our general derivation of chapter 3, was first conjectured by [89] and later proved by [19]. In the next section, we provide an additional independent derivation, in the case of $q = 2$ communities, based on a statistical physics analysis of the Ising model associated with B, introduced in section 2.2.2.

## 5.2 THE BETHE HESSIAN

*Bethe Hessian of the SBM*

The Bethe Hessian (3.107) associated with the non-backtracking operator (5.1) is given by

$$\begin{aligned} H_{ij}(x) &= \left(1 + \sum_{k \in \partial i} \frac{1}{x^2 - 1}\right) \mathbf{1}(i = j) - \frac{x}{x^2 - 1}\mathbf{1}(j \in \partial i) \,, \\ &= \frac{1}{x^2 - 1}\Big(\left(x^2 - 1 + |\partial_i|\right)\mathbf{1}(i = j) - x\,\mathbf{1}(j \in \partial i)\Big) \,. \end{aligned} \tag{5.3}$$

Just like for the planted spin glass, we will disregard the trivial, positive factor $(x^2 - 1)^{-1}$, and call Bethe Hessian of the SBM the matrix

$$H(x) = (x^2 - 1)I_n - xA + D \,, \tag{5.4}$$

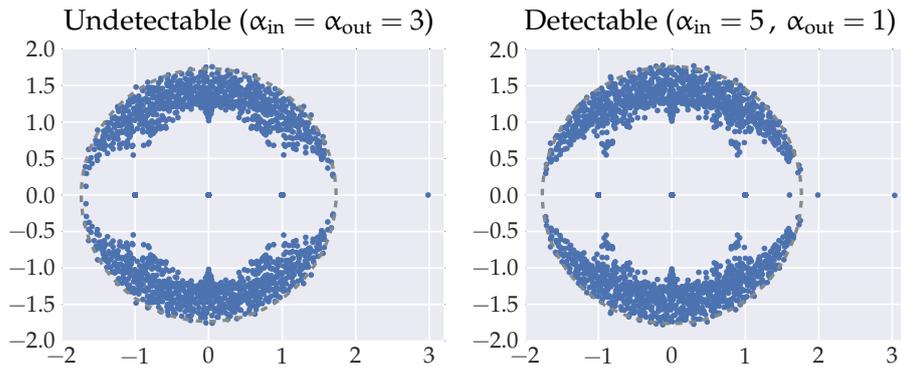

Figure 5.1 – Spectrum of B on two instances of the SBM with q = 2 communi-
ties, n = 1000 vertices and average connectivity α = 3. The left
figure corresponds to an undetectable problem, while the right
one is detectable (above the transition (5.2)). In both cases, the
spectrum of B has its bulk of uninformative eigenvalues con-
strained to the disk of radius $\sqrt{\alpha} \approx 1.732$, with a ferromagnetic
eigenvalue at α = 3. In the detectable case, there is in addition
an informative eigenvalue at $q^{-1}(\alpha_{in} - \alpha_{out}) = 2$.

where A is the adjacency matrix of the graph, with non-zero entries
$A_{ij} = 1$ for $(ij) \in E$, and D is the diagonal degree matrix, with en-
tries $D_{ii} = |\partial_i|$ for $i \in [n]$. This matrix is a deformed version of
the Laplacian, equal to the standard Laplacian when x = 1. We will
however be more interested in the matrix $H(\sqrt{\alpha})$, which, as we will
see, has much more desirable spectral properties than the Laplacian.
Indeed, according to the study of the spectral density of H(x) which
we performed in section 3.2.1, we expect that, as we lower x from a
large positive value, the left edge of the bulk of eigenvalues of H(x)
is shifted to the left, and reaches 0 exactly at $x = \sqrt{\alpha}$. As noted in
section 3.2.1, in the case of the unlabeled SBM, the fixed point equa-
tion (3.112) can be turned into a distributional fixed point equation
specifying the distribution of $\Delta_{i \to j}$ for a randomly chosen $(i \to j) \in \vec{E}$.
Moreover, this approach has been shown rigorously in [18] to give the
right spectral density in the limit $n \to \infty$ for locally tree-like graphs
such as those generated by the SBM. In particular, in the case of the
SBM, all the steps of the argument of section 3.2.1 can be made rigor-
ous, so that the spectral density $v_x(\lambda)$ of H(x) vanishes in an open set
around $\lambda = 0$ for any $x > \sqrt{\alpha}$.

*Spectral density of the Bethe Hessian*

To compute the spectral density of H(x), we solve the distributional
fixed point equation numerically using the standard population dy-
namics algorithm [103]. The results, presented in figure 5.2 are in
striking agreement with a direct estimation of the spectral density ob-
tained from a histogram of the eigenvalues of a realization of the ma-
trix H(x). In particular, this figure illustrates that, by taking $x = \sqrt{\alpha}$,
we maximize our chances of finding informative negative eigenvalues
(see the caption of figure 5.2).

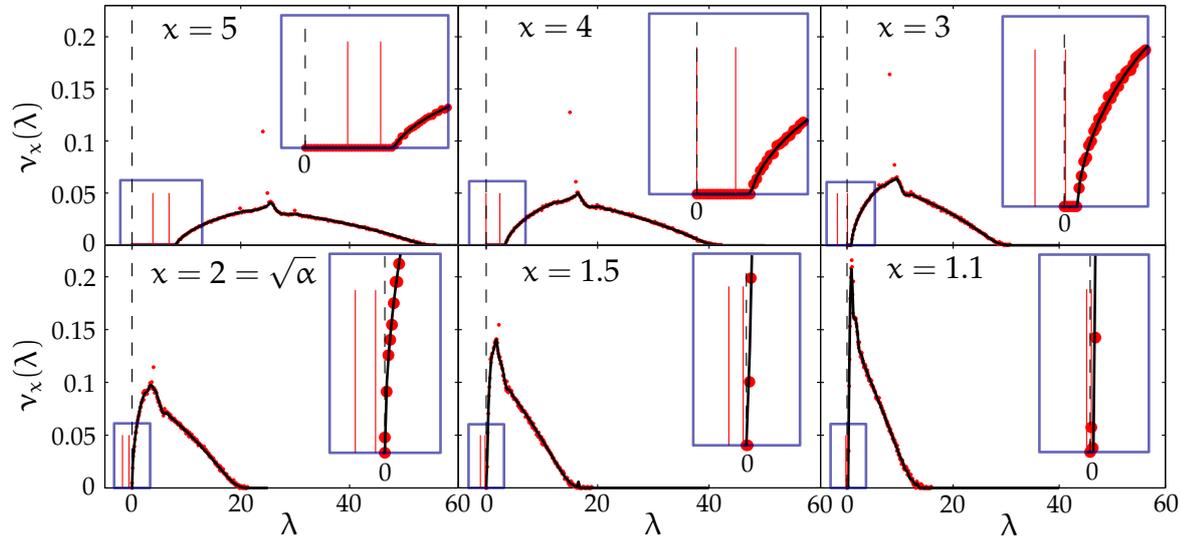

Figure 5.2 – Spectrum of the Bethe Hessian $H(x)$ of equation (5.4) for various values of $x > 1$, on graphs generated from the SBM. The red dots are the result of the direct diagonalization of the Bethe Hessian for a graph with $n = 10^4$ vertices and $q = 2$ communities, with $\alpha = 4, \alpha_{in} = 7, \alpha_{out} = 1$. The black curves are the solution of the distributional fixed point equation (see text), obtained from population dynamics (with a population of size $10^5$). We isolated the two smallest eigenvalues, represented as small red bars for convenience. The dashed black line marks the $\lambda = 0$ axis, and the inset is a zoom around this axis. At large value of $x$ ($x = 5$ in the top left panel), $H(x)$ is positive definite and all its eigenvalues are positive. As $x$ decays, the spectrum moves towards the $\lambda = 0$ axis. The smallest eigenvalue (corresponding to the uninformative ferromagnetic eigenvalue of B) reaches zero for $x = \alpha = 4$ (middle top), followed, as $x$ decays further, by the second (informative) eigenvalue at $x = (\alpha_{in} - \alpha_{out})/q = 3$, which is the value of the second largest eigenvalue of B in this case. Finally, as shown by the computation of section 3.2.1, the bulk reaches the $\lambda = 0$ axis at $x = \sqrt{\alpha} = 2$ (bottom left). At this point, the information is in the negative part, while the uninformative bulk is in the positive part. Interestingly, if $x$ decays further (bottom middle and right) the bulk of the spectrum remains positive, but the informative eigenvalues blend back into the bulk. The best choice is thus to consider the matrix $H\left(\sqrt{\alpha}\right)$. Figure adapted from [139].

The fact that $H(x)$ is the Hessian of the Bethe free energy at the paramagnetic stationary point of an Ising model in fact allows us to give a simple derivation of the transition (5.2) based on the gauge invariance property of the Ising model. This argument, which we now make explicit, also sheds a slightly different light on the informative eigenvectors of B than the original motivation of [89]. To make the argument simple, we consider the case of $q = 2$ communities. Recall (as shown in section 2.3.2) that the matrix $H(x)$ controls the stability of the paramagnetic phase (in the Bethe approximation) of an Ising model with couplings given by (2.76), here

*Statistical physics interpretation of the informative eigenvector of B for $q = 2$*

$$\tanh(J_{ij}) = \frac{1}{x}, \; \forall (ij) \in E, \tag{5.5}$$

where $x$ plays the role of a temperature, and we assume that $x > 1$. This is a ferromagnetic Ising model, with positive couplings that tend to infinity as $x \to 1$. From the classical results summed up in section 1.6.2, we expect that this model is for large enough $x$ (high temperature) in its paramagnetic phase, so that $H(x)$ is positive definite. As already mentioned, this is indeed the case, as can be shown using the Gershgorin circle theorem. As we lower $x$ (lower the temperature), we expect generically that the model will undergo a paramagnetic to ferromagnetic phase transition, corresponding to the onset of a global ordering of the spins. When this happens, the susceptibility diverges, i. e. $H(x)$ becomes singular and the paramagnetic stationary point of the Bethe free energy becomes a saddle point. The statistical physics prediction (1.84) (which is in fact rigorous for the sparse ferromagnetic Ising model [35]), is that this transition happens at a value of $x$ specified by the condition

*Associated (ferromagnetic) Ising model*

*Paramagnetic to ferromagnetic phase transition*

$$1 = \alpha \, \mathbb{E}\left[\tanh(J)\right] = \frac{\alpha}{x}, \tag{5.6}$$

where $J$ is a random variable with the same distribution as the couplings of equation (5.5), i. e. here a deterministic one. We find that the onset of the ferromagnetic instability is $x = \alpha$. On the other hand, by the Ihara-Bass formula, $\det H(x)$ vanishes if and only if $x$ is an eigenvalue of B so that we recover the ferromagnetic eigenvalue of B, equal to $\alpha$. From the previous arguments, we can interpret the corresponding eigenvector of B as an instability of the paramagnetic fixed point of BP, revealing the existence of a different, ferromagnetic fixed point.

Generically, the ferromagnetic Ising model has only two phases (the paramagnetic and the ferromagnetic ones). However, this does not mean that the paramagnetic phase does not have other types of instabilities. More precisely, let us suppose for the sake of intuition that the graph G is drawn from the SBM with $\alpha_{in} \gg \alpha_{out}$. In this case, although the ground state (i. e. the mode) of the Ising model is still a perfectly ordered configuration with all spins equal, there

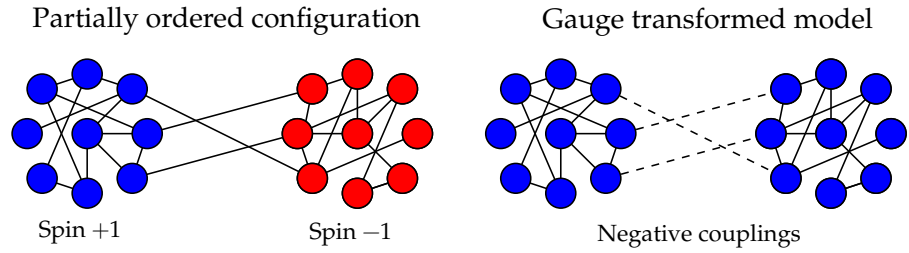

Figure 5.3 – In additional to the purely ferromagnetic ground state, an Ising model defined on a graph generated from a strongly assortative SBM has a low energy (high probability) partially ordered configuration, represented on the left. After gauge transforming, the resulting model is an Ising spin glass with a strong ferromagnetic bias (right).

*Partially ordered configuration*

is another, partially ordered configuration, that also has low energy (or high probability). This configuration, represented on figure 5.3, corresponds to setting the spins in one community to 1, and those in the other community to $-1$. Then, the only frustrated edges are those connecting the two communities, and there are comparatively few of them. For low enough pseudo-temperature $x$, we may therefore suspect the existence of a local minimum of the (Bethe) free energy, corresponding to partially ordered configurations [2]. These local minima are informative, since they correspond to configurations that have a finite overlap with the planted partition.

*Gauge transformation*

We can in fact even compute for which value of $x$ this local minimum starts having a lower free energy than the paramagnetic point. To do so, we use the gauge transformation introduced in section 1.6 to map the partially ordered configuration of figure 5.3 to the globally ordered, ferromagnetic one. In doing so, we create negative couplings on the edges connecting the two communities (see figure 5.3). We therefore have an Ising spin glass, but with a strong ferromagnetic bias. More precisely, the gauge transformed couplings $\tilde{J}_{ij}$ for $(ij) \in E$ now verify $\tanh\left(\tilde{J}_{ij}\right) = \pm 1/x$, and the fraction of negative couplings is $\alpha_{out}/(2\alpha)$, which is the fraction of edges that connect vertices from different communities. From equation (1.84), the paramagnetic to ferromagnetic transition in the gauge transformed model is given by

$$\begin{aligned} 1 = \alpha\,\mathbb{E}\left[\tanh\left(\tilde{J}\right)\right] &= \frac{\alpha}{x}\left(1 - \frac{\alpha_{out}}{2\alpha} - \frac{\alpha_{out}}{2\alpha}\right), \\ &= \frac{\alpha - \alpha_{out}}{x}, \\ &= \frac{\alpha_{in} - \alpha_{out}}{2x}. \end{aligned} \tag{5.7}$$

2. The global minimum of the free energy at low temperature is still the ferromagnetic one. In fact, it is not entirely clear that the partially ordered local minimum is really a local minimum, and not e.g. a saddle. The important point is that the paramagnetic stationary point is unstable with respect to a partially ordered perturbation.

The previous argument suggests that when $x < (\alpha_{in} - \alpha_{out})/2$, there is a second unstable direction of the paramagnetic stationary point of the Bethe free energy, pointing towards a partially ordered local minimum. By the Ihara-Bass formula, $(\alpha_{in} - \alpha_{out})/2$ is therefore an eigenvalue of B, and its corresponding eigenvector signals the existence of a partially ordered (therefore informative) fixed point of BP.

Interestingly, the choice $x = \sqrt{\alpha}$ corresponds to the onset of the paramagnetic to spin glass instability in the gauge transformed model. The detectability threshold (5.2) implies that if the partial ordered instability has not happened before the spin glass instability, the problem is not solvable, as in the case in the planted spin glass of the previous chapter.

## 5.3 ALGORITHM AND NUMERICAL SIMULATIONS

Based on the previous considerations, we propose the spectral algorithm 5 to perform community detection in the SBM. This algorithm includes two subtleties. First, in the case where the SBM is disassortative ($\alpha_{out} > \alpha_{in}$), we have shown in section 3.1.3 that the informative eigenvalues of B are negative, and smaller than $-\sqrt{\alpha}$. From the Ihara-Bass formula, to find these eigenvalues, we need to consider the Bethe Hessian at a negative pseudo-temperature $x = -\sqrt{\alpha}$ (see details in [139]). In the statistical physics interpretation of the previous section, this corresponds to an anti-ferromagnetic Ising model. In order to cluster a general graph generated from the SBM, we look at the negative eigenvalues of both $H\left(\sqrt{\alpha}\right)$ (which reveal the assortative communities) and $H\left(-\sqrt{\alpha}\right)$ (which reveal the disassortative ones). The other subtlety included in algorithm 5 is the estimation of the number of communities. By the heuristic analysis of section 3.1.3, and by the rigorous results of [19], when the problem is solvable, the non-backtracking operator has q eigenvalues outside of the disk of radius $\sqrt{\alpha}$. Their number therefore gives away the number of communities in the graph. We expect the number of negative eigenvalues of $H\left(\sqrt{\alpha}\right)$ and $H\left(-\sqrt{\alpha}\right)$ to be the same as the number of informative eigenvalues of B, allowing us to infer the number of communities. Note that when $q = 2$, the last step of the algorithm involving kmeans can be replaced by a classification based on the signs of the entries of the eigenvector corresponding to the second smallest eigenvector, similarly to algorithm 4.

*Disassortative communities*

*Estimating the number of communities*

We illustrate the accuracy of algorithm 5 on graphs generated by the SBM on figure 5.4, which also shows the performance of standard spectral clustering methods, as well as that of the BP approach of [32], believed to be asymptotically optimal in large tree-like graph.

*Experiments on synthetic networks*

---

**Algorithm 5** Bethe Hessian spectral algorithm for the SBM

---

**Input:** Graph $G = ([n], E)$

1: **Build** the Bethe Hessians $H\left(\sqrt{\alpha}\right)$ and $H\left(-\sqrt{\alpha}\right)$ where $H(x)$ is defined by equation (5.4)

2: **Compute** the negative eigenvalues of both $H\left(\sqrt{\alpha}\right)$ and $H\left(-\sqrt{\alpha}\right)$. If there are none, raise an error, otherwise call $\hat{q}$ their number, and $v_1, v_2, \ldots, v_{\hat{q}}$ their eigenvectors.

3: **Cluster** the eigenvectors $v_1, v_2, \ldots, v_{\hat{q}}$ into $\hat{q}$ clusters using kmeans.

---

The performance is measured in terms of the overlap with the true labeling, defined here as

$$\left(\frac{1}{n}\sum_{i=1}^{n}\mathbf{1}(\hat{\sigma}_i = \sigma_i) - \frac{1}{q}\right) \Big/ \left(1 - \frac{1}{q}\right) , \qquad (5.8)$$

where $\hat{\sigma}_i \in [q]$ is the estimate of the community membership of vertex $i$ output by algorithm 5, and $\sigma_i$ is the ground true community membership. This overlap is scaled so that a random guess yields a zero overlap, and a perfect clustering has an overlap of 1. Note that since we can only hope to cluster the vertices up to a global permutation of the cluster labels, we maximize this overlap over all $q!$ permutations of these labels.

Figure 5.4 shows that the Bethe Hessian systematically outperforms the non-backtracking operator B and does almost as well as BP, a more complicated algorithm, which we have run here assuming the knowledge of "oracle parameters": the number of communities, their sizes, and the parameters $\alpha_{in}$ and $\alpha_{out}$, which are needed to write down the posterior distribution (1.105). The Bethe Hessian, on the other hand is non-parametric and infers the number of communities in the graph by counting the number of negative eigenvalues (like the non-backtracking operator).

*Real-world experiments*     We finally turn to real world graphs to illustrate the performance of our approach, and to show that even though the SBM is not a good model for real networks, the Bethe Hessian remains a useful tool. Our results on common benchmarks are presented in table 1, where we also give the overlap achieved by the non-backtracking operator for comparison. We limited our experiments to this list of networks because they have known, "ground true" communities. For each case we observed a large correlation with the ground truth, and at least equal (and sometimes better) performance compared with the non-backtracking operator. The overlap was computed assuming knowledge of the number of ground true communities. We note however that the number of communities was correctly given by the number of negative eigenvalues of the Bethe Hessian in all the presented cases except for the political blogs network (10 predicted communities) and the football network (10 predicted communities). These differences either question the statistical significance of some of the human-decided labeling, or suggest the existence of additional relevant communities. It is also interesting to note that our approach works not only in the assortative case but also in the disassortative ones, for instance for the word adjacency network, whose two communities consist of adjectives on the one hand, and nouns on the other, so that neighbors in the graph are often in different communities.

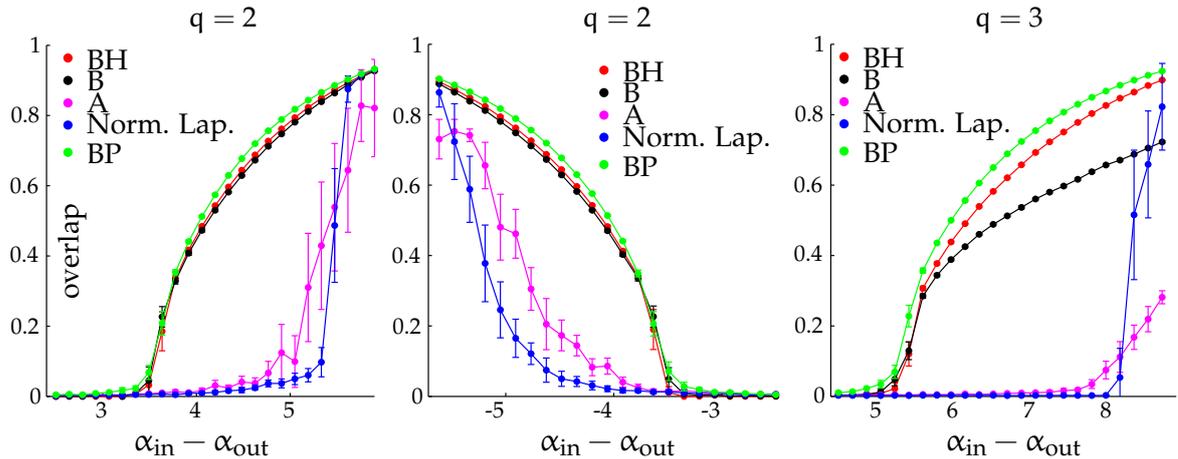

Figure 5.4 – Performance of various clustering methods on graphs of size $n = 10^5$ generated from the SBM. Each point is averaged over 20 such graphs. On the left is the assortative case with $q = 2$ communities (theoretical transition at $\alpha_{in} - \alpha_{out} = 3.46$). In the middle is the disassortative case with $q = 2$ (theoretical transition at $\alpha_{in} - \alpha_{out} = -3.46$). Finally, on the right is assortative case with $q = 3$ communities (theoretical transition at $\alpha_{in} - \alpha_{out} = 5.20$). While both the standard adjacency (A) and symmetrically normalized Laplacian $(D^{-1/2}(D - A)D^{-1/2})$ approaches fail to identify the communities in a large relevant region, both the non-backtracking operator (B) and the Bethe Hessian (BH) identify the communities almost as well as using the more complicated belief propagation algorithm (BP) with oracle parameters. Note, however, that the Bethe Hessian systematically outperforms the non-backtracking operator, and has a lower computational cost. We note that while clustering with the adjacency matrix and the normalized Laplacian requires first extracting the giant component of the graph $G$, the Bethe Hessian does not require any kind of preprocessing of the graph. While our theory explains why clustering with the Bethe Hessian gives a positive overlap whenever clustering with B does, we currently do not have an explanation as to why the Bethe Hessian overlap is actually larger. Figure taken from [137].

Table 1 – Overlap with human-made group assignments for some commonly used benchmarks for community detection. The non-backtracking operator detects communities in all these networks, with an overlap comparable to the performance of other spectral methods [89]. The Bethe Hessian systematically either matches or outperforms the results obtained by the non-backtracking operator. Table taken from [137].

| Network | Non-backtracking [89] | Bethe Hessian |
|---|---|---|
| Polbooks (q = 3) [8] | 0.742857 | 0.757143 |
| Polblogs (q = 2) [97] | 0.864157 | 0.865794 |
| Karate (q = 2) [164] | 1 | 1 |
| Football (q = 12) [52] | 0.924111 | 0.924111 |
| Dolphins (q = 2) [115] | 0.741935 | 0.806452 |
| Adjnoun (q = 2) [88] | 0.625000 | 0.660714 |

## 5.4 CONCLUSION

In this chapter, we have examined the problem of detecting communities in the SBM, for which an optimal spectral algorithm based on the non-backtracking operator was already introduced by [89]. We introduced a related algorithm based on the Bethe Hessian, which takes a particularly simple form for this problem. In particular, the computation of the spectral density can be made rigorous, thanks to techniques developed by [18]. While this fact does not guarantee the optimality of the algorithm presented here, it strongly supports our approach, which is also validated by statistical physics arguments, as well as numerical simulations on both synthetic and real world networks. In practice, we found that the Bethe Hessian systematically provides slightly better performance than the non-backtracking operator, and has lower computational cost. Additionally, we showed in the case of 2 communities how the phase transition can be understood in terms of the stability of the paramagnetic phase of an associated

Ising model, further motivating our algorithm. A rigorous guarantee of the optimality of the Bethe Hessian on the SBM is an interesting direction for future work. In the following, we turn to other types of applications, in which the available data is encoded as labels on the edges of a graph, rather than in its topology.

# RANDOMIZED CLUSTERING FROM SPARSE MEASUREMENTS

In this chapter, we consider the general framework of similarity-based clustering, in which the aim is to group items in a dataset based on the pairwise similarities between the items. After defining and motivating this problem, we introduce three algorithms for this task, based respectively on BP, the non-backtracking operator and the Bethe Hessian. We illustrate numerically the optimality of all three algorithms on a model, and substantiate this claim with rigorous arguments in the case of the non-backtracking algorithm. This chapter is based on results published in [143].

## 6.1 MOTIVATION AND DEFINITION OF THE PROBLEM

Similarity-based clustering is a standard approach to label items in a dataset based on some measure of their resemblance. In general, given a dataset $\{x_i\}_{i \in [n]} \in \mathcal{X}^n$, and a symmetric measurement function $s : \mathcal{X}^2 \to \mathbb{R}$ quantifying the similarity between two items, the aim is to cluster the dataset from the knowledge of the pairwise measurements $s_{ij} = s(x_i, x_j)$, for $1 \leqslant i < j \leqslant n$. This information is conveniently encoded in a similarity graph, whose vertices represent items in the dataset, and the weighted edges carry the pairwise similarities. Typical choices for this similarity graph are the complete graph and the nearest neighbor graph (see e.g. [98] for a discussion in the context of spectral clustering). In both cases, building the similarity graph requires computing (and storing if we use the fully connected graph) of the order of $n^2$ similarities.

*Similarity-based clustering*

Here, on the contrary, we will not assume knowledge of the measurements for all pairs of items in the dataset, but only for $O(n)$ of them chosen uniformly at random. In particular, instead of comparing each items in the dataset to all of the other items, we only compare each item to a constant (independent of $n$) number of other items, considerably reducing the cost of building the similarity graph. In fact, we will not even assume the measurement function $s$ to quantify the *similarity* between items, but more generally ask that the measurements be *typically different* depending on the cluster memberships of the items, in a way that will be made quantitative in the following. For instance, $s$ could be a distance in an Euclidean space or could take values in a set of colors (i. e. $s$ does not need to be real-valued).

*Our setting*

The ability to cluster data from as few pairwise comparisons as possible is of broad practical interest [29]. First, there are situations where all the pairwise comparisons are simply not available. This is particularly the case if a comparison is the result of a human-based experiment. For instance, in crowdclustering [53, 162], people are asked to compare a subset of the items in a dataset, and the aim is to cluster the whole dataset based on these comparisons. Clearly, for a large dataset of size $n$, we can't expect to have all $O(n^2)$ measurements. Second, even if these comparisons can be automated, the typical cost of computing all pairwise measurements is $O(n^2 d)$ where $d$ is the dimension of the data. For large datasets with $n$ in the millions or billions, or large dimensional data, like high resolution images where $d$ can easily equal tens of millions, this cost is often prohibitive. Storing all $O(n^2)$ measurements is also problematic. Our upcoming results support the idea that if the measurements between different classes of items are sufficiently different, a random subsampling of $O(n)$ measurements might be enough to accurately cluster the data. Random subsampling methods are a well-known means of reducing the complexity of a problem, and they have been shown to yield substantial speed-ups in clustering [38] and low rank approximation [5, 6, 57] The main challenge is to choose the lowest possible sampling rate while still being able to detect the signal of interest. In the following, we compute explicitly this fundamental limit for a simple probabilistic model and present three algorithms allowing partial recovery of the signal above this limit.

*Random subsampling*

The link between this problem and the results exposed thus far in this dissertation is also provided by the model, which we expose now. We consider $n$ items $i \in [n]$, each of them in a cluster $\sigma_i \in [q]$, where all $q$ clusters are assumed to be of equal size $n/q$. We choose the observed pairwise measurements uniformly at random, by generating an Erdős-Rényi random graph $G = ([n], E) \in \mathcal{G}(n, \alpha/n)$. The average degree $\alpha$ corresponds to the sampling rate: pairwise measurements are observed only on the edges of $G$, and $\alpha$ therefore controls the difficulty of the problem. From the base graph $G$, we build a measurement graph by weighting each edge $(ij) \in E$ with the measurement $s_{ij}$, assumed to be a random variable drawn from a probability distribution $p_{\sigma_i, \sigma_j}$, dependent only on the cluster assignments $\sigma_i$ and $\sigma_j$ of items $i$ and $j$. The aim is to recover the cluster assignments $\sigma_i$ for $i \in [n]$ from the measurement graph thus constructed.

*Model*

We consider the sparse regime $\alpha = O(1)$, and the limit $n \to \infty$ with fixed number of clusters $q$. As explained in section 1.4.1, with high probability, the graph $G$ is disconnected, so that *exact recovery* of the clusters, as considered e.g. in [29, 69], is impossible. As usual, we address here instead the question of how many measurements are needed to *partially recover*, or *detect* the clusters, i.e. to infer cluster assignments $\hat{\sigma}_i$ such that the following *overlap* is strictly positive:

$$\max_{\sigma \in \mathfrak{S}_k} \frac{\frac{1}{n} \sum_i \mathbf{1}(\sigma(\hat{\sigma}_i) = \sigma_i) - \frac{1}{q}}{1 - \frac{1}{q}}, \tag{6.1}$$

where $\mathfrak{S}_q$ is the set of permutations of $[q]$. This quantity, identical to the one considered in the previous chapter for the SBM, is monotonously increasing with the number of correctly classified items. In the limit $n \to \infty$, it vanishes for a random guess, and equals unity if the recovery is perfect. Finally, we note an important special case for which analytical results can be derived, which is the case of symmetric clusters: $\forall \sigma, \sigma' \in [q]$

$$p_{\sigma, \sigma'}(s) = \mathbf{1}(\sigma = \sigma') \, p_{\text{in}}(s) + \mathbf{1}(\sigma \neq \sigma') \, p_{\text{out}}(s), \qquad (6.2)$$

*Symmetric model*

where $p_{\text{in}}(s)$ (resp. $p_{\text{out}}(s)$) is the probability density of observing a measurement $s$ between items of the same cluster (resp. different clusters). In this case, the model we consider is nothing but a special case of the symmetric lSBM introduced in section 1.4.3, in the special case where

$$\alpha_{\text{in}} = \alpha_{\text{out}}, \qquad (6.3)$$

so that there is no information in the topology of the graph (which is simply a random Erdős-Rényi random graph), but only in the labels.

From the statistical physics analysis of section 3.1, we therefore predict that the problem is (efficiently) solvable if and only $\alpha > \alpha_c$, where

$$\frac{1}{\alpha_c} = \frac{1}{q} \int_{\mathcal{K}} ds \, \frac{\left(p_{\text{in}}(s) - p_{\text{out}}(s)\right)^2}{p_{\text{in}}(s) + (q-1) \, p_{\text{out}}(s)}, \qquad (6.4)$$

*Conjectured detectability threshold*

where $\mathcal{K}$ is the support of the function $p_{\text{in}} + (q-1) \, p_{\text{out}}$. This equation translates the general threshold (3.58) to the present setting. More precisely, as explained in section 3.1.3, we expect that for a small number of clusters $q$, equation (6.4) marks the transition between a phase were detection is information-theoretically impossible, and a phase where it is easy. For larger $q$, we expect that this transition separates the possible but hard phase from the easy phase.

## 6.2 ALGORITHMS

In this section, we list the three methods discussed previously for solving an instance of the lSBM and give their particular form and properties for the problem defined in the previous section. The three methods will be tested numerically on the experiments of the next section. We assume for now that the measurement distributions $p_{\text{in}}$ and $p_{\text{out}}$ are known and discuss ways to estimate them in the following.

### 6.2.1 *Belief propagation*

The posterior distribution of the cluster assignment $\sigma$ is a special case of the general posterior of the lSBM given by equation (1.105). The Bayes optimal clustering of the graph is given by the mode of the marginals of this pairwise MRF, which we can estimate using BP. The BP fixed point equation (2.9) reads here

$$b_{i \to j}(\sigma_i) = \frac{1}{Z_{i \to j}} \prod_{l \in \partial i \setminus j} \sum_{\sigma_l = 1}^{q} p_{\sigma_i, \sigma_l}(s_{il}) \, b_{l \to i}(\sigma_l) \,. \qquad (6.5)$$

As usual, starting from a random initial assignment of the beliefs, we iterate (6.5) until convergence, and use (2.10) to estimate the marginals.

### 6.2.2 *The non-backtracking operator*

As shown in section 3.1.1, the model introduced in section 6.1 is factorized and symmetric, so that the previous BP recursion has a trivial fixed point, whose stability is controlled by the non-backtracking operator (3.20) which writes here, for $(i \to j), (k \to l) \in \vec{E}$,

$$B_{(i \to j),(k \to l)} = w(s_{kl}) \, \mathbb{1}(l = i) \mathbb{1}(k \neq j) \,, \qquad (6.6)$$

with a weighting function $w$ corresponding to the Bayes optimal[1] setting of equation (3.24), given here by

$$\forall s, w(s) = \frac{p_{\mathrm{in}}(s) - p_{\mathrm{out}}(s)}{p_{\mathrm{in}}(s) + (q-1)p_{\mathrm{out}}(s)} \,. \qquad (6.7)$$

The predictions of section 3.1.3 concerning the spectrum of this matrix in the limit $n \to \infty$ are as follows.
— B has a bulk of uninformative eigenvalues constrained to the disk of radius $R(w)$ where $R(w)$ is given by

$$\begin{aligned} R(w) &= \sqrt{\alpha \, \mathbb{E}[w^2]} \,, \\ &= \sqrt{\frac{\alpha}{\alpha_c}} \,. \end{aligned} \qquad (6.8)$$

where we have used the definition (6.4) of $\alpha_c$.
— When $R(w) > 1$, i. e. when $\alpha > \alpha_c$, there is an informative eigenvalue of B with multiplicity $q - 1$, outside of the bulk of radius $R(w)$. With the notations of section 3.1.3, this eigenvalue is given by

$$\begin{aligned} \alpha \, \Delta(w) &= R(w)^2 \,, \\ &= \frac{\alpha}{\alpha_c} \,. \end{aligned} \qquad (6.9)$$

Once more, note that when $\alpha < \alpha_c$, there are no informative eigenvalues of B, but we also have that $\rho(B) < 1$, so that the trivial fixed point of BP is stable, and BP also fails to detect the clusters. The approach we propose based on the non-backtracking operator is summarized in algorithm 6. It requires the pooling matrix, for $i \in [n], (k \to l) \in \vec{E}$

$$P_{i,k \to l} = w(s_{kl})\, \mathbf{1}(l = i)\,. \tag{6.10}$$

---

1. In the next chapter, we discuss in a semi-supervised setting the non-Bayes optimal case where $p_{\mathrm{in}}, p_{\mathrm{out}}$ are unknown, and we consider a general weighting function $w$.

In the following, we will give rigorous arguments to support the previous claim about the spectrum of B in the special case of $q = 2$ clusters, justifying the performance of this algorithm.

---

**Algorithm 6** Non-backtracking spectral algorithm for randomized clustering from sparse measurements

---

**Input:** Graph $G = ([n], E)$, measurements $s_{ij}$ for $(ij) \in E$
  1: **Build** the non-backtracking operator B of (6.6) and the pooling matrix P of (6.10)
  2: **Compute** the eigenvalues of B which are larger in modulus than $R(w)$ defined by equation (6.8). If there are none, raise an error. Otherwise, call $\hat{q}$ their number and $v_1, v_2 \dots, v_{\hat{q}}$ the corresponding eigenvectors.
  3: **Pool** the eigenvectors $v_i$ for $i \in [\hat{q}]$ to compute the approximate magnetizations $m_i = P v_i$
  4: **Output** the assignments $\hat{\sigma}_i = \text{sign}(m_i)$ for $i \in [n]$

---

### 6.2.3 *The Bethe Hessian*

The Bethe Hessian associated with the non-backtracking operator (6.6) reads here

$$H_{ij}(x) = \left(1 + \sum_{k \in \partial i} \frac{w(s_{ik})^2}{x^2 - w(s_{ik})^2}\right) \mathbf{1}(i = j) - \frac{x\, w(s_{ik})}{x^2 - w(s_{ik})^2}\, \mathbf{1}(j \in \partial i) \qquad \textit{Bethe Hessian}$$

$$\tag{6.11}$$

From the analysis of section 3.2.1, we expect the left edge of the bulk of uninformative eigenvalues of $H(x)$ to reach 0 when $x = R(w)$ where $R(w)$ is given by equation (6.8). We therefore consider the matrix $H(R(w))$ and look for its negative eigenvalues.

---

**Algorithm 7** Bethe Hessian spectral algorithm for randomized clustering from sparse measurements

---

**Input:** Graph $G = ([n], E)$, measurements $s_{ij}$ for $(ij) \in E$
  1: **Build** the Bethe Hessians $H(R(w))$ where $R(w)$ is defined in equation (6.8)
  2: **Compute** the negative eigenvalues of $H(R(w))$. If there are none, raise an error, otherwise call $\hat{q}$ their number, and $v_1, v_2, \dots, v_{\hat{q}}$ their eigenvectors.
  3: **Cluster** the eigenvectors $v_1, v_2, \dots, v_{\hat{q}}$ into $\hat{q}$ clusters using kmeans.

---

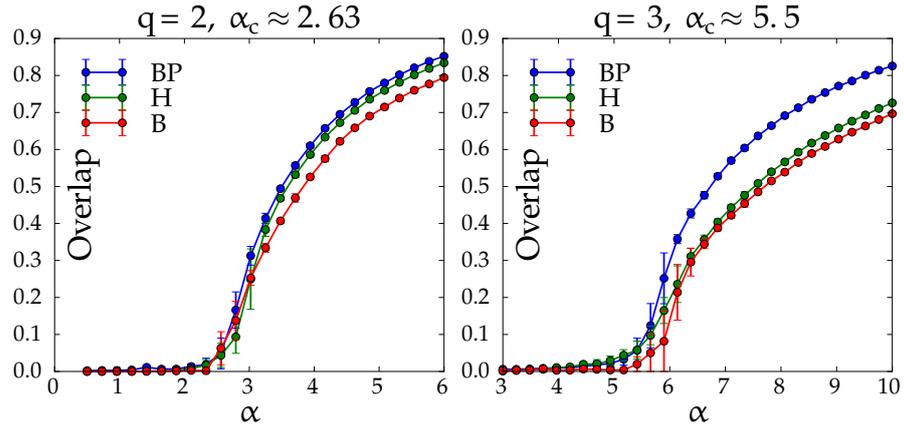

Figure 6.1 – Performance in clustering model-generated measurement graphs. The overlap is averaged over 20 realizations of graphs of size $n = 10^5$, with $q = 2, 3$ clusters, and Gaussian $p_{in}, p_{out}$ with mean respectively 1.5 and 0, and unit variance. The theoretical transition (6.4) is at $\alpha_c \approx 2.63$ for $q = 2$, and at $\alpha_c \approx 5.5$ for $q = 3$. All three methods achieve the theoretical transition, although the Bethe Hessian (H) and belief propagation (BP) achieve a higher overlap than the non-backtracking operator (B). Figure taken from [143].

The procedure based on the Bethe Hessian is summarized in algorithm 7. We investigate numerically the accuracy of this algorithm, as well as the two previous ones, in the next section.

## 6.3 NUMERICAL RESULTS

*Model-generated data*

Figure 6.1 shows the performance of all three algorithms (BP, algorithm 6 and algorithm 7) on model-generated problems. We consider the symmetric lSBM with $q = 2, 3$, fixed $p_{in}$ and $p_{out}$, chosen to be strongly overlapping Gaussians, and we vary the sampling parameters $\alpha_{in} = \alpha_{out} = \alpha$. All three algorithms achieve the theoretical threshold 6.4.

*Toy data not generated from the model*

While all the algorithms presented in this chapter assume the knowledge of the parameters of the model, namely the functions $p_{\sigma,\sigma'}$ for $\sigma, \sigma' \in [q]$, we argue that the BP algorithm is robust to large imprecisions on the estimation of these parameters. To support this claim, we show on figure 6.2 the result of the BP algorithm on standard toy datasets where the parameters were estimated on a small fraction of labeled data.

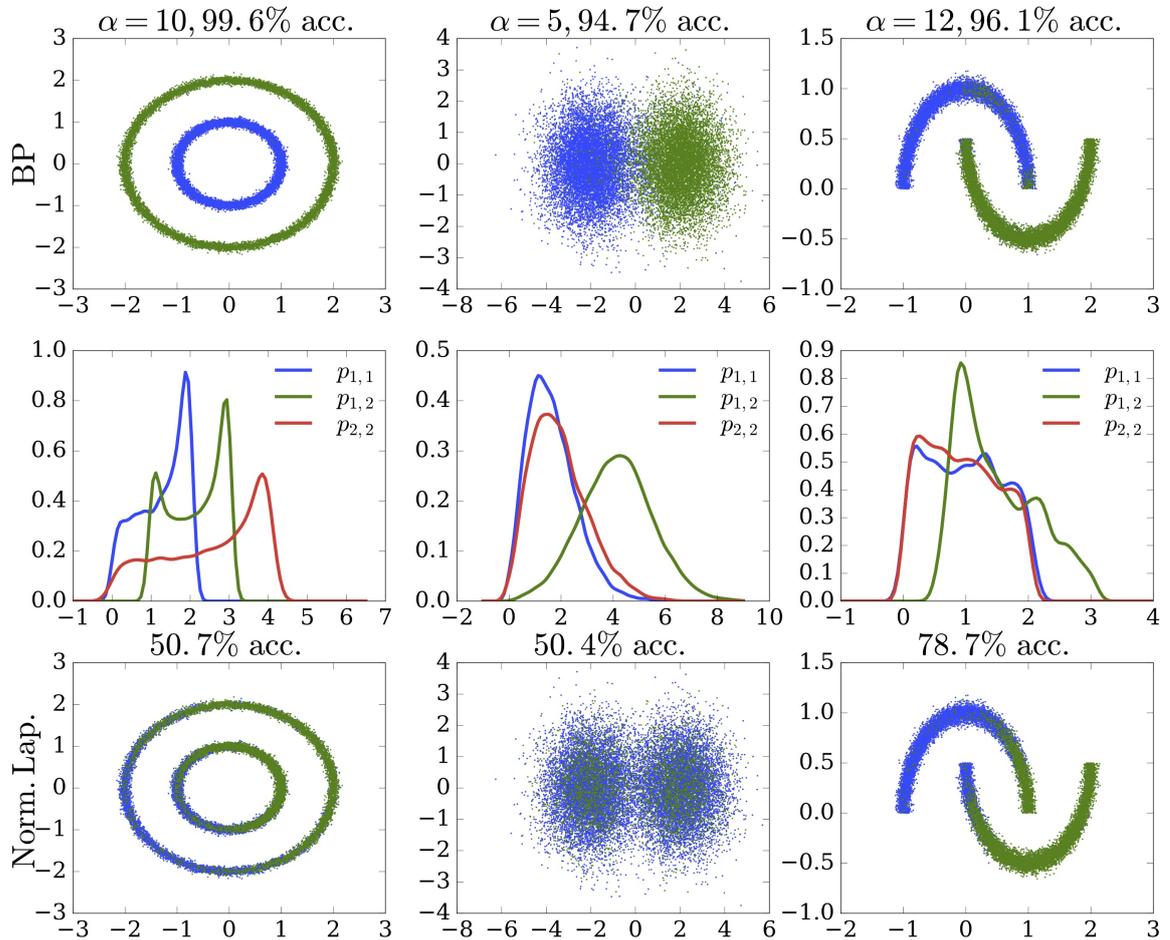

Figure 6.2 – Clustering of toy datasets with q = 2 clusters using BP. Each dataset is composed of n = 20000 points, 200 of which come labeled and constitute a training set used to learn the distribution of the measurements. We used the Euclidean distance as the measurement function s, and estimated the probability densities $p_{\sigma,\sigma'}$ for $\sigma, \sigma' \in [2]$ on the training set, using kernel density estimation (middle row). Note that we do not assume here the symmetric setting (6.2). Although these estimates are very noisy and overlapping, BP is able to achieve a very high accuracy using a random measurement graph G of small average degree α (top row). For comparison, we show in the third row the result of spectral clustering with the normalized Laplacian, using a 3-nearest neighbors similarity graph (see e.g. [98]) built from G, i.e. using only the available measurements. Figure taken from [143].

## 6.4 SPECTRAL PROPERTIES OF THE NON-BACKTRACKING OPERATOR

The accuracy of all three previous algorithms is related to the desirable spectral properties, conjectured in section 3.1.3, of the non-backtracking operator on graphs generated from the ISBM. In this section, we sketch some rigorous arguments supporting these claims on the graphs considered in this chapter, i. e. those generated from the symmetric ISBM with $\alpha_{in} = \alpha_{out}$. In the following, we restrict our analysis to the case of $q = 2$ clusters, and assume that $p_{in}$ and $p_{out}$ are distributions on a finite alphabet $\mathcal{S}$.

**Claim 1.** *Consider an Erdős-Rényi random graph on $n$ vertices with average degree $\alpha$, with variables assigned to vertices $\sigma_i \in [2]$ uniformly at random independently from the graph and measurements $s_{ij}$ between any two neighboring vertices drawn according to the probability density: $p_{\sigma_i, \sigma_j}(s) = \mathbf{1}(\sigma_i = \sigma_j) p_{in}(s) + \mathbf{1}(\sigma_i \neq \sigma_j) p_{out}(s)$ for two fixed (i. e. independent of $n$) discrete distributions $p_{in} \neq p_{out}$ on $\mathcal{S}$. Let $B$ be the matrix defined by (6.6) and denote by $|\lambda_1| \geqslant |\lambda_2| \geqslant \cdots \geqslant |\lambda_{|\vec{E}|}|$ the eigenvalues of $B$ in order of decreasing magnitude. Recall that $\alpha_c$ is defined by (6.4). Then, with probability tending to 1 as $n \to \infty$:*

*(i) If $\alpha < \alpha_c$, then $|\lambda_1| \leqslant \sqrt{\frac{\alpha}{\alpha_c}} + o(1)$.*

*(ii) If $\alpha > \alpha_c$, then $\lambda_1 = \frac{\alpha}{\alpha_c} + o(1)$ and $|\lambda_2| \leqslant \sqrt{\frac{\alpha}{\alpha_c}} + o(1)$. Additionally, denoting by $v$ the eigenvector associated with $\lambda_1$, $Pv$ is positively correlated with the planted variables $(\sigma_i)_{i \in [n]}$, where $P$ is the pooling matrix defined in equation (6.10).*

Note that this claim contains as a special case theorem 4.4.1 of chapter 4 for the planted spin glass, which it generalizes. The main idea which substantiates our claim is to introduce a new non-backtracking operator, with binary weights, and with spectral properties close to those of $B$, and then apply to it the techniques developed in [19] and extended in chapter 4. We use the same notations as in the proof of theorem 4.4.1: for an oriented edge $e = (u \to v)$, we set $e_1 = u$, $e_2 = v$ and $e^{-1} = (v \to u)$. For simplicity we also redefine the cluster labels $\sigma_i \in [2] \to 2\sigma_i - 3 \in \{\pm 1\}$, for $i \in [n]$.

We start by our usual gauge transformation: if $t$ is the vector in $\mathbb{R}^{|\vec{E}|}$ defined by $t_e = \sigma_{e_1}$ and $\odot$ is the Hadamard product, i. e. $(t \odot x)_e = \sigma_{e_1} x_e$, then we have

$$B^\intercal x = \lambda x \Leftrightarrow B^X(t \odot x) = \lambda(t \odot x),\tag{6.12}$$

with $B^X$ defined by $B^X_{ef} = B_{fe}\,\sigma_{e_1}\,\sigma_{e_2}$. In particular, $B^X$ and $B$ have the same spectrum and there is a trivial relation between their eigenvectors. Defining $X_e = \sigma_{e_1} w(s_e)\sigma_{e_2}$, we have:

$$B^X_{ef} = X_e\,\mathbf{1}(e_2 = f_1)\mathbf{1}(e_1 \neq f_2)\,.\tag{6.13}$$

Moreover, note that the random variables $(A_f = \sigma_{f_1}\sigma_{f_2})_{f \in E}$ are random signs with $\mathbb{P}(A_f = 1) = 1/2$ and the random variables $(X_f)_{f \in E}$ are such that

$$\mathbb{E}\left[X_f\right] = \mathbb{E}\left[X_f^2\right] = \frac{1}{2}\sum_s \frac{(p_{in}(s) - p_{out}(s))^2}{p_{in}(s) + p_{out}(s)} = \frac{1}{\alpha_c}. \qquad (6.14)$$

The fact that the random variables $X_f$ have equal first and second moments is a manifestation of the Nishimori conditions alluded to in section (1.6.5), see [168] for details.

The main difficulty in adapting the proof of theorem 4.4.1 to the present situation stems from the fact that we are now dealing with general weights (not only $\pm 1$ as before) in the non-backtracking operator. This breaks the PT-invariance, which plays a central role in the various bounds of [19]. To remedy the situation, we now define another non-backtracking operator $B^Y$. First, letting

$$\epsilon(s) = \frac{p_{out}(s)}{p_{in}(s) + p_{out}(s)} \in [0, 1], \qquad (6.15)$$

we define a sequence of independent random variables $\{\tilde{Y}_e\}_{e \in E}$ with $\mathbb{P}(\tilde{Y}_e = +1|s_e) = 1 - \mathbb{P}(\tilde{Y}_e = -1|s_e) = 1 - \epsilon(s_e)$, so that $\mathbb{E}\left[\tilde{Y}_e|s_e\right] = w(s_e)$. We define $Y_e = \tilde{Y}_e\sigma_{e_1}\sigma_{e_2}$, so that the expected value of the variables $Y_e$ equals the weights in the gauge transformed operator $B^X$. Finally, we define

$$B^Y_{ef} = Y_f \mathbf{1}(e_2 = f_1)\mathbf{1}(e_1 \neq f_2), \qquad (6.16)$$

so that $\mathbb{E}\left[B^Y|G, \{s_e\}_{e \in E}\right] = B^X$. We recover a non-backtracking operator with $\pm 1$ weights, whose conditional expectation is the matrix used in our algorithm. For the matrix $B^Y$, the PT-invariance holds, and its spectrum can be analyzed with the tools developed in [19], in a way similar to the proof of theorem 4.4.1. More precisely, we call $M$ the linear mapping on $\mathbb{R}^{|\vec{E}|}$ defined by $(Mx)_e = Y_e x_{e^{-1}}$, i.e. in matrix form $P_{ef} = Y_e \mathbf{1}(f = e^{-1})$. Since $M^{\mathsf{T}} = M$ and $(M_e)^2 = 1$, $M$ is an involution so that $M$ is an orthogonal matrix. As for the planted spin glass, a simple computation shows that for any integer $k \geqslant 0$, $(B^Y)^k M = M(B^{Y\mathsf{T}})^k$, hence $(B^Y)^k M$ is a symmetric matrix. If $(\tau_{j,k})$ for $1 \leqslant j \leqslant |\vec{E}|$ denote the eigenvalues of $(B^Y)^k M$ and $(x_{j,k})$ is a corresponding orthonormal basis of eigenvectors, we deduce as in [19] that

$$(B^Y)^k = \sum_{j=1}^{|\vec{E}|} \tau_{j,k}\, x_{j,k}\left(M x_{j,k}\right)^{\mathsf{T}}. \qquad (6.17)$$

Since $M$ is an orthogonal matrix $(Mx_{j,k})$ for $1 \leqslant j \leqslant |\vec{E}|$ is also an orthonormal basis of $\mathbb{R}^{|\vec{E}|}$. In particular, (6.17) gives the singular value decomposition of $\left(B^Y\right)^k$. Indeed, if we set $t_{j,k} = |\tau_{j,k}|$ and $y_{j,k} = \text{sign}(\tau_{j,k}) \, M \, x_{j,k}$, then we get

$$\left(B^Y\right)^k = \sum_{j=1}^{|\vec{E}|} t_{j,k} \, x_{j,k} \, y_{j,k}^\top, \tag{6.18}$$

which is precisely the singular value decomposition of $\left(B^Y\right)^k$. As shown in [19], for large $k$, the decomposition (6.17) carries structural information on the graph.

A crucial element in the proof of [19] is the result of Kesten and Stigum [82, 83] and we give now its extension required here which can be seen as a version of Kesten and Stigum's work in a random environment. We write $\mathbb{N}^\star = \{1, 2, \dots\}$ and $\mathbb{U} = \cup_{n \geqslant 0} (\mathbb{N}^\star)^n$ the set of finite sequences composed by $\mathbb{N}^\star$, where $(\mathbb{N}^\star)^0$ contains the null sequence $\emptyset$. For $u, v \in \mathbb{U}$, we denote $|u| = n$ the length of $u$ and we denote $uv$ the sequence obtained by juxtaposition of $u$ and $v$. Suppose that $\{(N_u, A_{u1}, A_{u2}, \dots)\}_{u \in \mathbb{U}}$ is a sequence of i.i.d random variables with value in $\mathbb{N} \times \{\pm 1\}^{\mathbb{N}^\star}$ such that $N_u$ is a Poisson random variable with mean $\alpha$ and the $A_{ui}$ are independent i.i.d random signs with $\mathbb{P}(A_{u1} = 1) = \mathbb{P}(A_{u1} = -1) = \frac{1}{2}$. We then define the following random variables: first $s_u$ such that $\mathbb{P}(s_u | A_u = 1) = p_{in}(s_u)$ and $\mathbb{P}(s_u | A_u = -1) = p_{out}(s_u)$, then $X_u = A_u \, w(s_u)$ and $Y_u = A_u \, \tilde{Y}_u$ where $\mathbb{P}(\tilde{Y}_u = 1 | s_u) = 1 - \mathbb{P}(Y_u = -1 | s_u) = 1 - \epsilon(s_u)$. We assume that for all $u \in \mathbb{U}$ and $i > N_u$, $A_{ui} = s_{ui} = X_{ui} = Y_{ui} = 0$. $N_u$ will be the number of children of node $u$ and the sequence $(s_{u1}, \dots, s_{uN_u})$ the measurements on edges between $u$ and its children. We set for $u = u_1 u_2 \dots u_n \in \mathbb{U}$,

$$\begin{aligned} P_\emptyset^X &= 1, & P_u^X &= X_{u_1} X_{u_1 u_2} \dots X_{u_1 \dots u_n}, \\ P_\emptyset^Y &= 1, & P_u^Y &= Y_{u_1} Y_{u_1 u_2} \dots Y_{u_1 \dots u_n}. \end{aligned} \tag{6.19}$$

We define (with the convention $\frac{0}{0} = 0$),

$$Q_0 = 1, \quad Q_t = \sum_{|u| = t} \frac{P_u^Y}{\alpha^t P_u^X}. \tag{6.20}$$

Then conditionally on the variables $(s_u)_{u \in \mathbb{U}}$, $Q_t$ is a martingale (generalizing the martingale $X_t$ of section 4.4) converging almost surely and in $L^2$ as soon as $\alpha > \alpha_c$. The fact that this martingale is bounded in $L^2$ follows from an argument given in the proof of theorem 3 in [60].

In order to apply the technique of [19], we need to deal with the $k$-th power of the non-backtracking operators. As in our proof of section 4.4 for the planted spin glass, for $k$ not too large (of the order of $\log n$), the local structure of the graph (up to depth $k$) can be coupled to a Poisson Galton-Watson branching process, so that the computations done for $Q_k$ above provide a good approximation of the k-th power of the non-backtracking operator and we can use the algebraic tools about perturbation of eigenvalues and eigenvectors, i. e. the Bauer-Fike theorem and its variations listed in Section 4 of [19].

## 6.5 CONCLUSION

In this chapter, we have considered the problem of clustering items based on a minimal number of comparisons between them, chosen uniformly at random. Our main result is that it is possible to partially cluster the $n$ items using only $O(n)$ comparisons, instead of computing all of the $O(n^2)$ comparisons. We introduced three related algorithms for this task, and gave a precise prediction about their performance on a model, supported by numerical experiments on synthetic networks. For one of the algorithms, based on the non-backtracking operator, we gave rigorous arguments supporting this prediction. Finally, we showed that our approach is useful outside of our model, by illustrating the ability of BP to cluster, to excellent accuracy, toy datasets *not* generated from the model, even in the presence of large imprecisions in its parameters. In the next chapter, we consider a semi-supervised variant of the same problem, and show that a highly efficient and *local* algorithm is rigorously guaranteed to achieve a low error from a small number of measurements.

# 7

As in the previous chapter we consider here a randomized similarity-based clustering problem where the aim is to cluster a dataset $\{x_i\}_{i \in [n]}$ of $n$ items from the knowledge of the pairwise similarities $s_{ij} = s(x_i, x_j)$, for $(ij) \in E$, where $s$ is a similarity function, and $E$ is a random subset of the $\binom{n}{2}$ possible pairs of items. In this chapter, we build upon the approach of chapter 6 by considering two variations motivated by real world applications.

The first question we address is how to incorporate the knowledge of the labels of a small fraction of the items to aid clustering of the whole dataset, resulting in a more efficient algorithm. This question, referred to as semi-supervised clustering, is of broad practical interest [15, 170]. For instance, in a social network, we may have pre-identified individuals of interest, and we might be looking for other individuals sharing similar characteristics. In biological networks, the function of some genes or proteins may have been determined by costly experiments, and we might seek other genes or proteins sharing the same function. More generally, efficient human-powered methods such as crowdsourcing can be used to accurately label part of the data [76, 77], and we might want to use this knowledge to cluster the rest of the dataset at no additional cost.

*Semi-supervised setting*

The second question we address is the number of randomly chosen pairwise similarities that are needed to achieve a given classification error. Previous work has mainly focused on two related, but different questions. One line of research has been interested in exact recovery, i.e. how many measurements are necessary to exactly cluster the data. Note that for exact recovery to be possible, as discussed in section 1.4.1, it is necessary to choose at least $O(n \log n)$ random measurements for the similarity graph to be connected with high probability. On simple models, [1, 56, 163] showed that this scaling is also sufficient for exact recovery. At the sparser end of the spectrum, [60, 94, 137, 143] have focused on the detectability threshold, i.e. how many measurements are needed to cluster the data better than chance. As shown in the previous chapter, on simple models, this threshold is typically achievable with $O(n)$ measurements only. While this scaling is certainly attractive for large problems, it is important for practical applications to understand how the expected classification error decays with the number of measurements.

*We are interested in the error decay rate in the detectability regime.*

To answer these two questions, we introduce a highly efficient, *local* algorithm based on a power iteration of the non-backtracking operator. For the case of two clusters, we show on a simple but reasonable model that the classification error decays exponentially with the number of measured pairwise similarities, thus allowing the algorithm to cluster data to arbitrary accuracy while being efficient both in terms of time and space complexities. We demonstrate the good performance of this algorithm on both synthetic and real world data, and compare it to the popular label propagation algorithm [170]. This chapter closely follows the presentation of the publication [138], with an added note on the link with the general approach outlined at the beginning of this dissertation.

## 7.1 ALGORITHM AND GUARANTEE

For simplicity, we start by introducing our procedure in the case of $q = 2$ clusters. On a model which generalizes the ISBM to the semi-supervised setting, we present strong guarantees on the performance of the algorithm. When then generalize the approach to a larger number $q > 2$ of clusters.

### 7.1.1 *Algorithm for 2 clusters*

*Similarity function s*

Consider $n$ items $\{x_i\}_{i \in [n]} \in \mathfrak{X}^n$ and a symmetric similarity function $s : \mathfrak{X}^2 \to \mathbb{R}$. The choice of the similarity function is problem dependent, and we will assume that one has been chosen. In this chapter, contrary to the previous one, we really think of $s$ as quantifying the resemblance between items, so that $s(x_i, x_j)$ is typically *larger* if $x_i$ and $x_j$ belong to the same cluster. For concreteness, $s$ can be thought of as a decreasing function of a distance if $\mathfrak{X}$ is an Euclidean space. We give explicit examples of such similarity functions in the upcoming numerical experiments. The analysis we perform, however, applies to a generic function $s$, and our bounds will depend explicitly on its statistical properties. We assume that the true labels

*Labeled set $\mathcal{L}$*

$(\sigma_i = \pm 1)_{i \in \mathcal{L}}$ of a subset $\mathcal{L} \subset [n]$ of items is known [1]. Our aim is to find an estimate $(\hat{\sigma}_i)_{i \in [n]}$ of the labels of all the items, using a

*Random subsampling*

small number of similarities. More precisely, let $E$ be a random subset of all the $\binom{n}{2}$ possible pairs of items, containing each given pair $(ij) \in [n]^2$ with probability $\alpha/n$, for some $\alpha > 0$. As in chapter 6, we compute only the $\alpha n/2$ similarities $(s_{ij} = s(x_i, x_j))_{(ij) \in E}$ of the pairs thus chosen.

We can now define a weighted similarity graph $G = ([n], E)$ where the vertices represent the items, and each edge $(ij) \in E$ carries a weight $w_{ij} = w(s_{ij})$, where $w$ is a weighting function. Once more,

*Weighting function w*

we will consider a generic function $w$ in our analysis, and discuss the performance of our algorithm as a function of the choice of $w$. In

---

1. In this chapter, $\mathcal{L}$ refers to the set of items $i \in [n]$ whose labels $\sigma_i$ are known, and not to a set of labels carried by the edges of the graph, as in the definition of the ISBM in section 1.4.3 particular, we show in section 7.1.2 that there is an optimal choice of $w$ when the data is generated from a model, corresponding to the Bayes optimal setting of the previous chapter. However, in practice, the main purpose of $w$ is to center the similarities, i.e. we will take in our numerical simulations $w(s) = s - \bar{s}$, where $\bar{s}$ is the empirical mean of the observed similarities. The advantage of centering the similarities is discussed in the following. Note that the graph G is a weighted version of an Erdős-Rényi random graph with average degree $\alpha$, which controls the sampling rate: larger $\alpha$ means more pairwise similarities are computed, at the expense of an increase in complexity. Algorithm 8 describes our clustering procedure for the case of 2 clusters.

*Centering the similarities*

---

**Algorithm 8** Non-backtracking local walk for 2 clusters

---

**Input:** $n, \mathcal{L}, (\sigma_i = \pm 1)_{i \in \mathcal{L}}, E, (w_{ij})_{(ij) \in E}, k_{max}$
**Output:** Cluster assignments $(\hat{\sigma}_i)_{i \in [n]}$

1: **Initialize** the messages $v^{(0)} = (v_{i \to j}^{(0)})_{(i \to j) \in \vec{E}}$
2:     **for all** $(i \to j) \in \vec{E}$ **do**
3:         **if** $i \in \mathcal{L}$ **then** $v_{i \to j}^{(0)} \leftarrow \sigma_i$
4:         **else** $v_{i \to j}^{(0)} \leftarrow \pm 1$ uniformly at random
5: **Iterate** for $k = 1, \ldots, k_{max}$
6:     **for** $(i \to j) \in \vec{E}$ **do** $v_{i \to j}^{(k)} \leftarrow \sum_{l \in \partial i \setminus j} w_{il} v_{l \to i}^{(k-1)}$
7: **Pool** the messages
8:     **for** $i \in [n]$ **do** $\hat{v}_i \leftarrow \sum_{l \in \partial i} w_{il} v_{l \to i}^{(k_{max})}$
9: **Output** the assignments
10:     **for** $i \in [n]$ **do** $\hat{\sigma}_i \leftarrow \mathrm{sign}(\hat{v}_i)$

---

This algorithm is in fact nothing but a power iteration of a non-backtracking operator B, with elements for $(i \to j), (k \to l) \in \vec{E}$

*Relation with the non-backtracking operator*

$$B_{(i \to j), (k \to l)} = w_{kl} \mathbf{1}(l = i) \mathbf{1}(k \neq j). \qquad (7.1)$$

If we let the number of iterations $k_{max}$ grow to infinity, then algorithm 8 computes the leading eigenvector of this non-backtracking operator. This procedure is therefore close in spirit to the unsupervised spectral methods introduced in chapters 4 and 6, which also rely on the computation of the principal eigenvectors of B. However, in contrast with these spectral approaches, algorithm 8 is local, in that the estimate $\hat{\sigma}_i$ for a given item $i \in [n]$ depends only on the messages on the edges that are at most $k_{max} + 1$ steps away from $i$ in the graph G. This fact will prove essential in the upcoming analysis. Indeed, we will show that in the semi-supervised setting, a finite number of iterations (independent of $n$) is enough to ensure a low classification error. On the other hand, in the unsupervised setting, we expect local algorithms not to be able to find large clusters in a graph, a limitation

*Local nature of algorithm 8*

that has already been highlighted on the related problems of finding large independent sets on graphs [48] and community detection [75].

It is possible to connect this algorithm with the general approach of this dissertation outlined in section 1.7 by noting that the leading eigenvector of the non-backtracking operator defined in equation (7.1) approximates the magnetizations of the Ising model defined by the Boltzmann distribution

$$\mathbb{P}(\sigma) = \frac{1}{\mathcal{Z}} \exp\left( \sum_{(ij)\in E} \operatorname{atanh}(w_{ij})\sigma_i\sigma_j \right), \qquad (7.2)$$

where we have assumed $|w_{ij}| < 1$ for $(ij) \in E$ (which can be assumed without loss of generality by rescaling B). Since atanh is an increasing function, this model promotes assigning the same label to pairs of vertices sharing a large weight (i. e. a large similarity if $w$ is an increasing function of the similarity $s$, e. g. $w(s) = s - \bar{s}$). The algorithm we describe here is therefore an instance of the "ad hoc" type of problems introduced in section 1.7. Indeed, instead of computing the marginals of the true posterior distribution of the labels $\sigma$, as we have done in the previous chapters, we are here trying to approximate the marginals of an ad-hoc Ising model which encodes the available data. Another example of such an approach will be provided in the next chapter, in the context of matrix completion.

### 7.1.2    *Model and guarantee*

To evaluate the performance of algorithm 8, we consider the following semi-supervised variant of the symmetric ISBM. Assign $n$ items to 2 predefined clusters of same average size $n/2$, by drawing for each item $i \in [n]$ a cluster label $\sigma_i \in \{\pm 1\}$ with uniform probability $1/2$. Choose uniformly at random $\eta n$ items to form a subset $\mathcal{L} \subset [n]$ of items whose label is revealed. $\eta$ is the fraction of labeled data. Next, choose which pairs of items are compared by generating an Erdős-Rényi random graph $G = ([n], E) \in \mathcal{G}(n, \alpha/n)$, for some constant $\alpha > 0$, independent of $n$. As usual, we will assume that the similarities are random variables, with symmetric distribution given by

$$P(s_{ij} = s|\sigma_i, \sigma_j) = \mathbf{1}(\sigma_i = \sigma_j)\, p_{in}(s) + \mathbf{1}(\sigma_i = -\sigma_j)\, p_{out}(s)\,, \qquad (7.3)$$

where $p_{in}$ (resp. $p_{out}$) is the distribution of the similarities between items within the same cluster (resp. different clusters). The properties of the weighting function $w$ will determine the performance of our algorithm through the two following quantities. Define $\Delta(w)$ and $\Sigma(w)$ as

$$\begin{aligned} 2\Delta(w) &= \mathbb{E}\left[w_{ij}|\sigma_i = \sigma_j\right] - \mathbb{E}\left[w_{ij}|\sigma_i \neq \sigma_j\right]\,, \\ \Sigma(w)^2 &= \mathbb{E}\left[w^2\right]\,. \end{aligned} \qquad (7.4)$$

$2\Delta(w)$ is the difference in expectation between the weights inside a cluster and between different clusters, and $\Sigma(w)^2$ is the second moment of the weights. We have come across these quantities before in the study of the spectrum of the non-backtracking operator B of equation (7.1). More precisely, according to the results of section 3.1.3, $R(w) = \sqrt{\alpha\,\Sigma^2(w)}$ is the radius of the bulk of uninformative eigenvalues of B. When $\alpha\,\Delta(w) > R(w)$, there is an informative eigenvalue $\alpha\,\Delta(w)$ of B outside of the disk of radius $R(w)$. In the unsupervised setting, when the number of iterations of algorithm 8 is allowed to grow to infinity, this algorithm detects the cluster if $\alpha\,\Delta(w) > R(w)$. Our first theorem extends this result to the semi-supervised setting, where algorithm 8 is run for a finite number of iterations.

*Relation with the spectrum of B*

**Theorem 7.1.1.** *Assume a similarity graph* G *with* $n$ *items and a labeled set* $\mathcal{L}$ *of size* $\eta n$ *to be generated from the symmetric model* (7.3) *with 2 clusters. Define* $\tau(\alpha, w) = \frac{\alpha\,\Delta(w)^2}{\Sigma(w)^2}$. *If* $\Delta(w) > 0$, *then there exists a constant* $C > 0$ *such that the estimates* $\hat{\sigma}_i$ *from* $k$ *iterations of algorithm 8 achieve*

*Guarantee in the small $\alpha$ regime*

$$\frac{1}{n}\sum_{i=1}^{n} P(\sigma_i \neq \hat{\sigma}_i) \leqslant 1 - r_{k+1} + C\,\frac{\alpha^{k+1}\log n}{\sqrt{n}}, \qquad (7.5)$$

*where* $r_0 = \eta^2$ *and for* $0 \leqslant l \leqslant k$,

$$r_{l+1} = \frac{\tau(\alpha, w) r_l}{1 + \tau(\alpha, w) r_l}. \qquad (7.6)$$

The proof relies on an analysis of the evolution of the distribution of the messages $v_{i \to j}^{(k)}$ and the Cantelli bound (see section 7.3). Note again that our bound depends on the same signal to noise ratio $\tau(\alpha, w)$ as identified in section 3.1.3. To understand the content of this bound, we consider the limit of a large number of items $n \to \infty$, so that the last term of (7.5) vanishes. Note first that if $\tau(\alpha, w) > 1$, then starting from any positive initial condition, $r_k$ converges to $(\tau(\alpha, w) - 1)/\tau(\alpha, w) > 0$ in the limit $k \to \infty$. A random guess on the unlabeled points yields an asymptotic error of

$$\lim_{n \to \infty} \frac{1}{n}\sum_{i=1}^{n} P(\sigma_i \neq \hat{\sigma}_i) = \frac{1 - \eta}{2}, \qquad (7.7)$$

so that a sufficient condition for algorithm 8 to improve the initial partial labeling, after a certain number of iterations $k(\tau(\alpha, w), \eta)$ independent of $n$, is

*Sufficient condition to improve the initial labeling*

$$\tau(\alpha, w) > \frac{2}{1 - \eta}. \qquad (7.8)$$

Let us compare this bound in the limit of small supervision $\eta \to 0$ with the transition of the previous chapter for the unsupervised case. We have already stated in section 3.1.3 that there is an optimal choice

*Comparison with optimal unsupervised threshold*

of weighting function $w$ which maximizes the signal to noise ratio $\tau(\alpha, w)$, namely the Bayes optimal choice

$$w^*(s) = \frac{p_{in}(s) - p_{out}(s)}{p_{in}(s) + p_{out}(s)} \quad \Longrightarrow \quad \tau(\alpha, w^*) = \frac{\alpha}{2} \int ds \, \frac{(p_{in}(s) - p_{out}(s))^2}{p_{in}(s) + p_{out}(s)},$$
(7.9)

whose definition however requires knowing $p_{in}$ and $p_{out}$. In the limit of vanishing supervision $\eta \to 0$, the bound (7.8) guarantees improving the initial labeling if $\tau(\alpha, w^*) > 2 + O(\eta)$. On the other hand, in the unsupervised setting, we have argued in the previous chapter that detectability is possible if and only if $\tau(\alpha, w^*) > 1$. In the limit of small supervision, the bound (7.8) is off by a constant factor of 2.

On the other hand, the bound (7.5) applies to any weighting function $w$. In particular, while the optimal choice (7.9) is not practical, theorem 7.1.1 guarantees that algorithm 8 retains the ability to improve the initial labeling from a small number of measurements, as long as $\Delta(w) > 0$. With the choice $w(s) = s - \bar{s}$ advocated for in section 7.1.1, we have $2\Delta(w) = \mathbb{E}\left[s_{ij}|\sigma_j = \sigma_j\right] - \mathbb{E}\left[s_{ij}|\sigma_i \neq \sigma_j\right]$. Therefore algorithm 8 improves over random guessing for $\alpha$ large enough if the similarity between items in the same cluster is larger in expectation than the similarity between items in different clusters, which is a reasonable requirement. Note that the hypotheses of theorem 7.1.1 do not require the weighting function $w$ to be centered. However, it is easy to check that if $\mathbb{E}[w] \neq 0$, defining a new weighting function by $w'(s) = w(s) - \mathbb{E}[w]$, we have $\tau(\alpha, w') > \tau(\alpha, w)$, so that the bound (7.5) is improved.

*Generality of the result for any reasonable $w$*

*Advantage of centering the weights*

While theorem 7.1.1 guarantees improving the initial clustering from a small sampling rate $\alpha$, it provides a rather loose bound on the expected error when $\alpha$ becomes larger. The next theorem addresses this regime. A proof is given in section 7.4.

*Exponentially decaying error*

**Theorem 7.1.2.** *Assume a similarity graph $G$ with $n$ items and a labeled set $\mathcal{L}$ of size $\eta n$ to be generated from the symmetric model (7.3) with 2 clusters. Assume further that the weighting function $w$ is bounded: $\forall s, |w(s)| \leqslant 1$. Define $\tau(\alpha, w) = \frac{\alpha \Delta(w)^2}{\Sigma(w)^2}$. If $\alpha \Delta(w) > 1$ and $\alpha \Sigma(w)^2 > 1$, then there exists a constant $C > 0$ such that the estimates $\hat{\sigma}_i$ from $k$ iterations of algorithm 8 achieve*

$$\frac{1}{n} \sum_{i=1}^n P(\sigma_i \neq \hat{\sigma}_i) \leqslant \exp\left[-\frac{q_{k+1}}{4} \min\left(1, \frac{\Sigma(w)^2}{\Delta(w)}\right)\right] + C \frac{\alpha^{k+1} \log n}{\sqrt{n}},$$
(7.10)

*where $q_0 = 2\eta^2$ and for $0 \leqslant l \leqslant k$,*

$$q_{l+1} = \frac{\tau(\alpha, w) q_l}{1 + 3/2 \max(1, q_l)}.$$
(7.11)

Note that by linearity of algorithm 8, the condition $\forall s, |w(s)| \leqslant 1$ can be relaxed to $w$ bounded. It is once more instructive to consider the limit of large number of items $n \to \infty$. Starting from any initial condition, if $\tau(\alpha, w) < 5/2$, then $q_k \underset{k \to \infty}{\longrightarrow} 0$ so that the bound (7.10) is uninformative. On the other hand, if $\tau(\alpha, w) > 5/2$, then starting from any positive initial condition, $q_k \underset{k \to \infty}{\longrightarrow} \frac{2}{3}(\tau(\alpha, w) - 1) > 0$. This bound therefore shows that on a model with a given distribution of similarities (7.3) and a given weighting function $w$, an error smaller than $\epsilon$ can be achieved from $\alpha n / 2 = O(n \log 1 / \epsilon)$ measurements, in the limit $\epsilon \to 0$, with a finite number of iterations $k(\tau(\alpha, w), \eta, \epsilon)$ independent of $n$. We note that this result is the analog, for a weighted graph, of the recent results of [24] who show that in the SBM, a local algorithm similar to algorithm 7.1.1 achieves an error decaying exponentially as a function of a relevant signal to noise ratio.

*Implications of the bound in the $n \to \infty$ limit*

### 7.1.3 *More than 2 clusters*

Algorithm 9 gives a natural extension of our algorithm to $q > 2$ clusters. In this case, as argued in section 3.1.3, we expect the non-backtracking operator B defined in equation (7.1) to have $q - 1$ large eigenvalues, with eigenvectors correlated with the types $\sigma_i$ of the items. We use a deflation-based power iteration method ([148]) to approximate these eigenvectors, starting from informative initial conditions incorporating the knowledge drawn from the partially labeled data. Numerical simulations illustrating the performance of this algorithm are presented in section 7.2. Note that each deflated matrix $B_c$ for $c \geqslant 2$ is a rank-$(c - 1)$ perturbation of a sparse matrix, so that the power iteration can be carried out efficiently using sparse linear algebra subroutines. In particular, both algorithms 8 and 9 have a time and space complexities linear in the number of items $n$.

*Deflation method to approximate $q - 1$ leading eigenvalues of B*

### 7.2 NUMERICAL SIMULATIONS

In addition to the theoretical guarantees presented in the previous section, we ran numerical simulations on two toy datasets consisting of 2 and 4 Gaussian blobs (figure 7.1), and two subsets of the MNIST dataset ([93]) consisting respectively of the digits in $\{0, 1\}$ and $\{0, 1, 2\}$ (figure 7.2). We also considered the 20 Newsgroups text dataset [92], consisting of text documents organized in 20 topics, of which we selected 2 for our experiments of figure 7.3. All of these datasets differ considerably from the model we have studied analytically. In particular, the random similarities are not identically distributed conditioned on the true labels of the items, but depend on latent variables, such as the distance to the center of the Gaussian, in the case of figure 7.1. Additionally, in the case of the MNIST dataset of figure 7.2, the clusters have different sizes (e.g. 6903 for the 0's and 7877 for the 1's). Nevertheless, we find that our algorithm performs well, and outper-

*Datasets used for the experiments*

---

**Algorithm 9** Non-backtracking local walk for q > 2 clusters

---

**Input:** $n, q, \mathcal{L}, (\sigma_i \in [q])_{i \in \mathcal{L}}, E, (w_{ij})_{(ij) \in E}, k_{max}$
**Output:** Cluster assignments $(\hat{\sigma}_i)_{i \in [n]}$

1: $B_1 \leftarrow B$ where $B$ is the non-backtracking matrix defined in equation (7.1)
2: **for** $c = 1, \cdots, q - 1$ **do**
3:     **Initialize** the messages $v^{(0)} = (v^{(0)}_{i \to j})_{(ij) \in E}$
4:         **for all** $(i \to j) \in \vec{E}$ **do**
5:             **if** $i \in \mathcal{L}$ and $\sigma_i = c$ **then** $v^{(0)}_{i \to j} \leftarrow 1$
6:             **else if** $i \in \mathcal{L}$ and $\sigma_i \neq c$ **then** $v^{(0)}_{i \to j} \leftarrow -1$
7:             **else** $v^{(0)}_{i \to j} \leftarrow \pm 1$ uniformly at random
8:     **Iterate** for $k = 1, \ldots, k_{max}$
9:         $v^{(k)} \leftarrow B_c v^{(k-1)}$
10:     **Pool** the messages in a vector $\hat{v}_c \in \mathbb{R}^n$ with entries $(\hat{v}_{i,c})_{i \in [n]}$
11:         **for** $i \in [n]$ **do** $\hat{v}_{i,c} \leftarrow \sum_{l \in \partial i} w_{il} v^{(k_{max})}_{l \to i}$
12:     **Deflate** $B_c$
13:         $B_{c+1} \leftarrow B_c - \dfrac{B_c v^{(k_{max})} v^{(k_{max})\intercal} B_c}{v^{(k_{max})\intercal} B_c v^{(k_{max})}}$
14: **Concatenate** $\hat{V} \leftarrow [\hat{v}_1 | \cdots | \hat{v}_{q-1}] \in \mathbb{R}^{n \times (q-1)}$
15: **Output** the assignments $(\hat{\sigma}_i)_{i \in [n]} \leftarrow \text{kmeans}(\hat{V})$

---

forms the popular label propagation algorithm ([170]) in a wide range of values of the sampling rate $\alpha$.

In all cases, we find that the accuracy achieved by algorithms 8 and 9 is an increasing function of $\alpha$, rapidly reaching a plateau at a limiting accuracy. Rather than the absolute value of this limiting accuracy, which depends on the choice of the similarity function, perhaps the most important observation is the rate of convergence of the accuracy to this limiting value, as a function of $\alpha$. Indeed, on these simple datasets, it is enough to compute, for each item, their similarity with a few randomly chosen other items to reach an accuracy within a few percents of the limiting accuracy allowed by the quality of the similarity function. As a consequence, similarity-based clustering can be significantly sped up. For example, we note that the semi-supervised clustering of the 0's and 1's of the MNIST dataset (representing $n = 14780$ points in dimension 784), from 1% of labeled data, and to an accuracy greater than 96% requires $\alpha \approx 6$ (see figure 7.2), and runs on a laptop in 2 seconds, including the computation of the randomly chosen similarities. Additionally, in contrast with our algorithms, we find that in the strongly undersampled regime (small $\alpha$), the performance of label propagation depends strongly on the fraction $\eta$ of available labeled data. We find in particular that algorithms

*A wall-clock timing example*

*Robustness in the small $\eta$ regime*

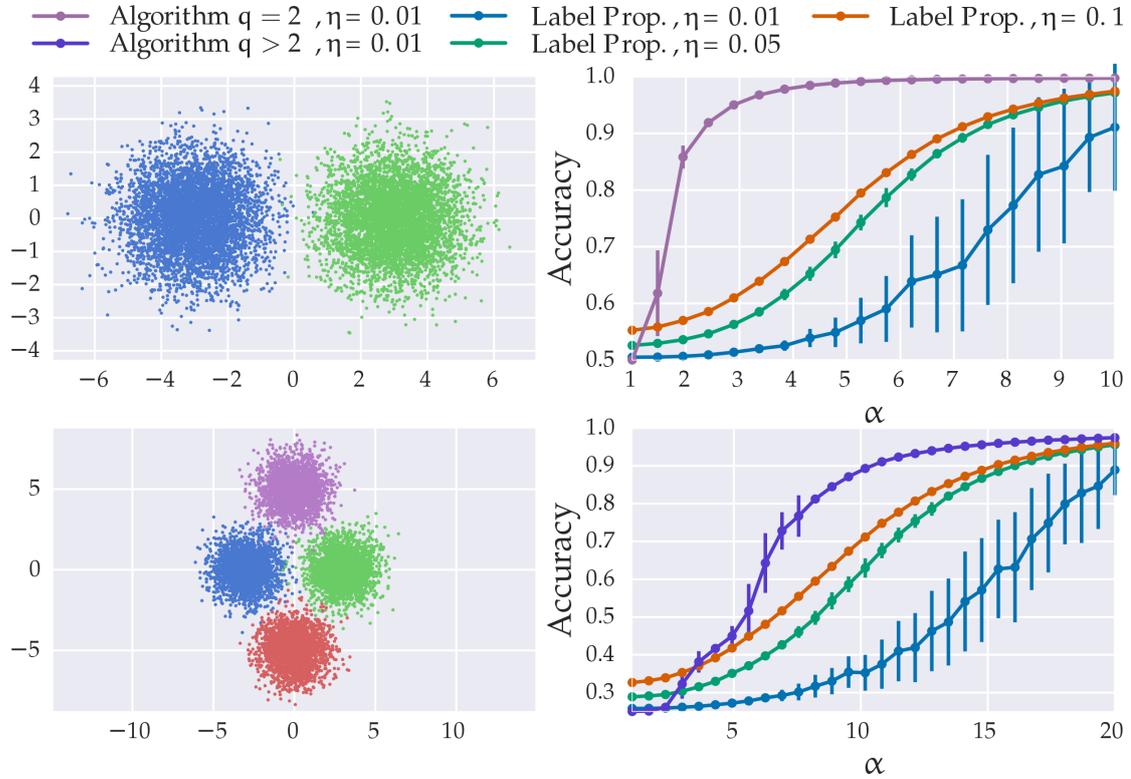

Figure 7.1 – Performance of algorithms 8 (denoted Algorithm $q = 2$) and 9 (denoted Algorithm $q > 2$) compared to label propagation on a toy dataset in two dimensions. The left panel shows the data, composed of $n = 10^4$ points, with their true labels. The right panel shows the clustering performance on random subsamples of the complete similarity graph. Each point is averaged over 100 realizations. The accuracy is defined as the fraction of correctly labeled points. We set the maximum number of iterations of our algorithms to $k_{max} = 30$. $\alpha$ is the average degree of the random similarity graph G, and $\eta$ is the fraction of labeled data. For all methods, we used the same similarity function $s_{ij} = \exp{-d_{ij}^2/\sigma^2}$ where $d_{ij}$ is the Euclidean distance between points $i$ and $j$ and $\sigma^2$ is a scaling factor which we set to the empirical mean of the observed squared distances $d_{ij}^2$. For algorithms 8 and 9, we used the weighting function $w(s) = s - \bar{s}$ (i.e. we simply center the similarities, see text). Label propagation is also run on the random similarity graph G. We note that we improved the performance of label propagation by using only, for each point, the similarities between this point and its three nearest neighbors in the random graph G. Figure taken from [138].

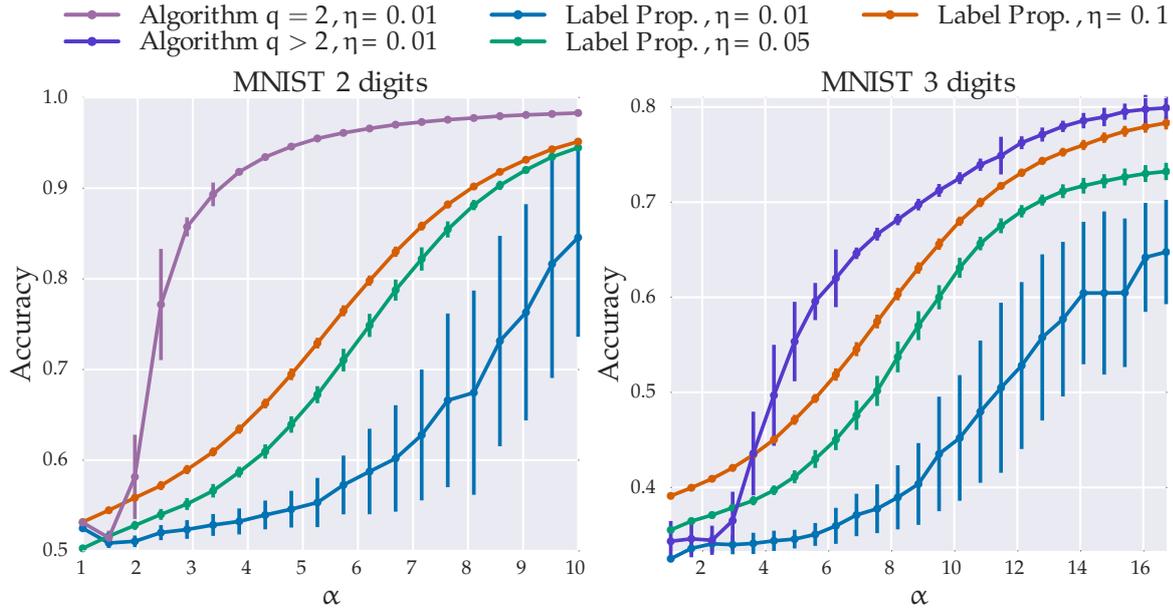

Figure 7.2 – Performance of algorithms 8 (denoted Algorithm q = 2) and 9 (denoted Algorithm q > 2) compared to label propagation on a subset of the MNIST dataset. The left panel corresponds to the set of 0's and 1's (n = 14780 samples) while the right panel corresponds to the 0's,1's and 2's (n = 21770). All the parameters are the same as in the caption of figure 7.1, except that we used the Cosine distance in place of the Euclidean one. Figure taken from [138].

8 and 9 outperform label propagation even starting from ten times fewer labeled data.

## 7.3   PROOF OF THEOREM 7.1.1

Consider the model introduced in section 7.1 for the case of two clusters. We will bound the probability of error on a randomly chosen node, and the different results will follow. Denote by $I$ an integer drawn uniformly at random from $[n]$, and by $\hat{\sigma}_I^{(k)} = \pm 1$ the decision variable after $k$ iterations of algorithm 8. We are interested in the probability of error at node $I$ conditioned on the true label of node $I$, i.e. $P(\hat{\sigma}_I^{(k)} \neq \sigma_I | \sigma_I)$. As noted previously, the algorithm is local in the sense that $\hat{\sigma}_I^{(k)}$ depends only on the messages in the neighborhood of $I$ consisting of all the nodes and edges of $G$ that are at most $k+1$ steps away from $I$. By bounding the total variation distance between the law of $G_{I,k+1}$ and a weighted Galton-Watson branching process, we show (see proposition 31 in [19])

$$P\left(\hat{\sigma}_I^{(k)} \neq \sigma_I | \sigma_I\right) \leqslant P\left(\sigma_I \hat{v}_{\sigma_I}^{(k)} \leqslant 0\right) + C\frac{\alpha^{k+1}\log n}{\sqrt{n}}, \qquad (7.12)$$

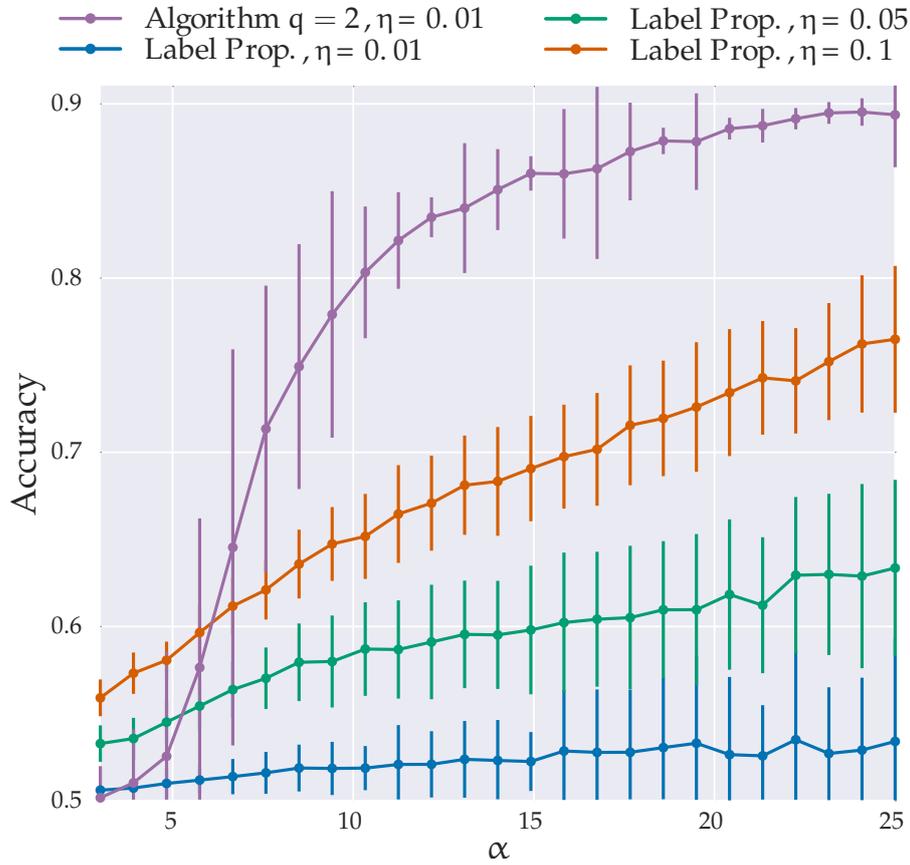

Figure 7.3 – Performance of algorithm 8 (denoted Algorithm q = 2) compared to label propagation on a subset of 20 Newsgroups dataset with q = 2 clusters. We consider the topics "misc.forsale" (975 documents) and " soc.religion.christian" (997 documents), which are relatively easy to distinguish, to illustrate the efficiency of our subsampling approach. The resulting dataset consists of n = 1972 text documents, to which we applied a standard "tf-idf" transformation (after stemming, and using word unigrams) to obtain a vector representation of the documents. We used the same similarity (based on the Cosine distance) and weighting functions as in figure 7.2.

where $C > 0$ is a constant, and the random variables $\hat{v}_\sigma^{(k)}$ for $\sigma = \pm 1$ are distributed according to

$$\hat{v}_\sigma^{(k)} \overset{\mathcal{D}}{=} \sum_{i=1}^{d_1} w_{i,\text{in}} v_{i,\sigma}^{(k)} + \sum_{i=1}^{d_2} w_{i,\text{out}} v_{i,-\sigma}^{(k)}, \qquad (7.13)$$

where $\overset{\mathcal{D}}{=}$ denotes equality in distribution. The random variables $v_{i,\sigma}^{(k)}$ for $\sigma = \pm 1$ have the same distribution as the message $v_{i \to j}^{(k)}$ after $k$ iterations of the algorithm, for a randomly chosen edge $(i \to j)$, conditioned on the type of node $i$ being $\sigma$. They are i.i.d. copies of a random variable $v_\sigma^{(k)}$ whose distribution is defined recursively, for $l \geqslant 0$ and $\sigma = \pm 1$, through

$$v_\sigma^{(l+1)} \overset{\mathcal{D}}{=} \sum_{i=1}^{d_1} w_{i,\text{in}} v_{i,\sigma}^{(l)} + \sum_{i=1}^{d_2} w_{i,\text{out}} v_{i,-\sigma}^{(l)}. \qquad (7.14)$$

In equations (7.13), $d_1$ and $d_2$ are two independent random variables with a Poisson distribution of mean $\alpha/2$, and $w_{i,\text{in}}$ (resp. $w_{i,\text{out}}$) are i.i.d copies of $w_{\text{in}}$ (resp. $w_{\text{out}}$) whose distribution is the same as the weights $w_{ij}$, conditioned on $\sigma_i = \sigma_j$ (resp. $\sigma_i \neq \sigma_j$). Note in particular that $\hat{v}_\sigma^{(k)}$ has the same distribution as $v_\sigma^{(k+1)}$.

Theorem 7.1.1 will follow by analyzing the evolution of the first and second moments of the distribution of $v_{+1}^{(k+1)}$ and $v_{-1}^{(k+1)}$. Equations (7.14) can be used to derive recursive formulas for the first and second moments. In particular, the expected values verify the following linear system

$$\begin{pmatrix} \mathbb{E}\left[v_{+1}^{(l+1)}\right] \\ \mathbb{E}\left[v_{-1}^{(l+1)}\right] \end{pmatrix} = \frac{\alpha}{2} \begin{pmatrix} \mathbb{E}[w_{\text{in}}] & \mathbb{E}[w_{\text{out}}] \\ \mathbb{E}[w_{\text{out}}] & \mathbb{E}[w_{\text{in}}] \end{pmatrix} \begin{pmatrix} \mathbb{E}\left[v_{+1}^{(l)}\right] \\ \mathbb{E}\left[v_{-1}^{(l)}\right] \end{pmatrix}. \qquad (7.15)$$

The eigenvalues of this matrix are $\mathbb{E}[w_{\text{in}}] + \mathbb{E}[w_{\text{out}}]$ with eigenvector $(1,1)^\mathsf{T}$, and $\mathbb{E}[w_{\text{in}}] - \mathbb{E}[w_{\text{out}}]$ with eigenvector $(1,-1)^\mathsf{T}$. With the assumption of our model, we have $\mathbb{E}\left[v_{+1}^{(0)}\right] = \eta = -\mathbb{E}\left[v_{-1}^{(0)}\right]$ which is proportional to the second eigenvector. Recalling the definition of $\Delta(w)$ from section 7.1, we therefore have, for any $l \geqslant 0$,

$$\mathbb{E}\left[v_{+1}^{(l+1)}\right] = \alpha \Delta(w) \mathbb{E}\left[v_{+1}^{(l)}\right], \qquad (7.16)$$

and $\mathbb{E}\left[v_{-1}^{(l)}\right] = -\mathbb{E}\left[v_{+1}^{(l)}\right]$. With the additional observation that

$$\mathbb{E}\left[\left(v_{+1}^{(0)}\right)^2\right] = \mathbb{E}\left[\left(v_{-1}^{(0)}\right)^2\right] = 1, \qquad (7.17)$$

a simple induction shows that for any $l \geqslant 0$,

$$\mathbb{E}\left[\left(v_{+1}^{(l)}\right)^2\right] = \mathbb{E}\left[\left(v_{-1}^{(l)}\right)^2\right], \qquad (7.18)$$

and that, recalling the definition of $\Sigma(w)^2$ from section 7.1, we have the recursion

$$\mathbb{E}\left[\left(\nu_{+1}^{(l+1)}\right)^2\right] = \alpha^2 \Delta(w)^2 \mathbb{E}\left[\nu_{+1}^{(l)}\right]^2 + \alpha \Sigma(w)^2 \mathbb{E}\left[\left(\nu_{+1}^{(l)}\right)^2\right]. \quad (7.19)$$

Noting that since we assume $\Delta(w) > 0$, we have $\sigma \mathbb{E}\left[\nu_\sigma^{(k+1)}\right] > 0$ for $\sigma = \pm 1$, the proof of theorem 7.1.1 is concluded by invoking Cantelli's inequality

$$P\left(\sigma \nu_\sigma^{(k+1)} \leqslant 0\right) \leqslant 1 - r_{k+1}, \quad (7.20)$$

with, for $l \geqslant 0$,

$$r_l = \mathbb{E}\left[\nu_\sigma^{(l)}\right]^2 \mathbb{E}\left[\left(\nu_\sigma^{(l)}\right)^2\right]^{-1}, \quad (7.21)$$

where $r_l$ is independent of $\sigma$, and is shown to verify the recursion (7.6) by combining (7.16) and (7.19).

## 7.4 PROOF OF THEOREM 7.1.2

The proof is adapted from a technique developed in [76]. We show that the random variables $\nu_\sigma^{(l)}$ are sub-exponential by induction on $l$. A random variable $X$ is said to be sub-exponential if there exist constants $K > 0, a, b$ such that for $|\lambda| < K$

$$\mathbb{E}[e^{\lambda X}] \leqslant e^{\lambda a + \lambda^2 b}. \quad (7.22)$$

Define $f_\sigma^{(l)}(\lambda) = \mathbb{E}\left[e^{\lambda \nu_\sigma^{(l)}}\right]$ for $l \geqslant 0$ and $\sigma = \pm 1$. Define two sequences $(a_l)_{l \geqslant 0}, (b_l)_{l \geqslant 0}$ by $a_0 = \eta, b_0 = 1/2$ and for $l \geqslant 0$

$$a_{l+1} = \alpha \Delta(w) a_l,$$
$$b_{l+1} = \alpha \Sigma(w)^2 \left(b_l + \frac{3}{2} \max(a_l^2, b_l)\right). \quad (7.23)$$

Note that since we assume $\alpha \Delta(w) > 1$ and $\alpha \Sigma(w)^2 > 1$, both sequences are positive and increasing. In the following, we show that

$$f_\sigma^{(k+1)}(\lambda) \leqslant e^{\sigma \lambda a_{k+1} + \lambda^2 b_{k+1}}, \quad (7.24)$$

for $|\lambda| \leqslant \left(2 \max\left(a_k, \sqrt{b_k}\right)\right)^{-1}$. Theorem 7.1.2 will follow from the Chernoff bound applied at

$$\lambda_\sigma^* = -\sigma \frac{a_{k+1}}{2 b_{k+1}} \min\left(1, \frac{\Sigma(w)^2}{\Delta(w)}\right). \quad (7.25)$$

The fact that $|\lambda_\sigma^*| \leqslant \left(2 \max\left(a_k, \sqrt{b_k}\right)\right)^{-1}$ follows from (7.23). Noting that $\sigma \lambda_\sigma^* < 0$ for $\sigma = \pm 1$, the Chernoff bound allows to show

$$P\left(\sigma \nu_\sigma^{(k+1)} \leqslant 0\right) \leqslant f_\sigma^{(k+1)}(\lambda_\sigma^*),$$
$$\leqslant \exp\left[-\frac{q_{k+1}}{4} \min\left(1, \frac{\Sigma(w)^2}{\Delta(w)}\right)\right], \quad (7.26)$$

where $q_{k+1} = a_{k+1}^2/b_{k+1}$ is shown using (7.23) to verify the recursion (7.11). We are left to show that $f_\sigma^{(k+1)}(\lambda)$ verifies (7.24). First, with the choice of initialization in algorithm 8, we have for any $\lambda \in \mathbb{R}$

$$f_{+1}^{(0)}(\lambda) = f_{-1}^{(0)}(-\lambda) = \frac{1+\eta}{2} e^\lambda + \frac{1-\eta}{2} e^{-\lambda} \leqslant e^{\eta\lambda + \lambda^2/2}. \qquad (7.27)$$

where the last inequality follows from the fact that for $x \in [0,1], \lambda \in \mathbb{R}$

$$x e^\lambda + (1-x) e^{-\lambda} \leqslant e^{(2x-1)\lambda + \lambda^2/2}. \qquad (7.28)$$

Therefore we have $f_\sigma^{(0)}(\lambda) \leqslant \exp\left(\sigma\lambda a_0 + \lambda^2 b_0\right)$. Next, assume that for some $l \geqslant 0$, and $\forall |\lambda| \leqslant \left(2\max\left(a_{l-1}, \sqrt{b_{l-1}}\right)\right)^{-1}$

$$f_\sigma^{(l)}(\lambda) \leqslant \exp\left(\sigma\lambda\, a_l + \lambda^2\, b_l\right), \qquad (7.29)$$

with the convention $a_{-1} = b_{-1} = 0$ so that the previous statement is true for any $\lambda \in \mathbb{R}$ if $l = 0$. The density evolution equations (7.14) imply the following recursion on the moment-generating functions, for any $\lambda \in \mathbb{R}, \sigma = \pm 1$,

$$f_\sigma^{(l+1)}(\lambda) = \exp\left(-\alpha + \frac{\alpha}{2}\left(\mathbb{E}_{w_{in}}\left[f_\sigma^{(l)}(\lambda w_{in})\right] + \mathbb{E}_{w_{out}}\left[f_{-\sigma}^{(l)}(\lambda w_{out})\right]\right)\right). \qquad (7.30)$$

We claim that for $|\lambda| \leqslant \left(2\max\left(a_l, \sqrt{b_l}\right)\right)^{-1}$ and $\sigma = \pm 1$,

$$\frac{1}{2}\left(\mathbb{E}_{w_{in}}\left[f_\sigma^{(l)}(\lambda w_{in})\right] + \mathbb{E}_{w_{out}}\left[f_{-\sigma}^{(l)}(\lambda w_{out})\right]\right) \leqslant \\ 1 + \sigma a_l\, \Delta(w)\,\lambda + \lambda^2 \Sigma(w)^2\left(b_l + \frac{3}{2}\max(a_l^2, b_l)\right). \qquad (7.31)$$

Injecting (7.31) in (7.30) yields $f_\sigma^{(l+1)}(\lambda) \leqslant \exp\left(\sigma\lambda a_{l+1} + \lambda^2 b_{l+1}\right)$, for $|\lambda| \leqslant \left(2\max\left(a_l, \sqrt{b_l}\right)\right)^{-1}$, with $a_{l+1}, b_{l+1}$ defined by (7.23). The proof is then concluded by induction on $0 \leqslant l \leqslant k$. To show (7.31), we start from the following inequality: for $|a| \leqslant 3/4$,

$$e^a \leqslant 1 + a + (2/3)a^2. \qquad (7.32)$$

With $|w| \leqslant 1$ as per the assumption of theorem 7.1.2, we have that for $|\lambda| \leqslant \left(2\max\left(a_l, \sqrt{b_l}\right)\right)^{-1}$ and $\sigma = \pm 1$, $|\sigma\lambda w a_l + \lambda^2 w^2 b_l| \leqslant 3/4$. Additionally, since $a_l$ and $b_l$ are non-decreasing in $l$, we also have $|\lambda| \leqslant \left(2\max\left(a_{l-1}, \sqrt{b_{l-1}}\right)\right)^{-1}$, so that by our induction hypothesis, for $\sigma = \pm 1$,

$$f_\sigma^{(l)}(\lambda w) \leqslant e^{\sigma\lambda w a_l + \lambda^2 w^2 b_l}, \qquad (7.33)$$

$$\leqslant 1 + \sigma\lambda w a_l + \lambda^2 w^2 b_l + \frac{2}{3}\left(\sigma\lambda w a_l + \lambda^2 w^2 b_l\right)^2, \qquad (7.34)$$

$$\leqslant 1 + \sigma\lambda w a_l + \lambda^2 w^2 b_l + \frac{2}{3}\lambda^2 w^2\left(a_l + |\lambda|b_l\right)^2, \qquad (7.35)$$

$$\leqslant 1 + \sigma\lambda w a_l + \lambda^2 w^2 b_l + \frac{3}{2}\lambda^2 w^2 \max(a_l^2, b_l), \qquad (7.36)$$

where we have used that $\left(a_l + |\lambda|b_l\right)^2 \leqslant 9/4\max(a_l^2, b_l)$. (7.31) follows by taking expectations.

7.5 CONCLUSION

In this chapter, we have considered a semi-supervised similarity-based clustering problem, and have argued that a small number of randomly chosen similarity measures are enough to cluster the data to high accuracy. In particular, we introduced a highly efficient local algorithm based on a power iteration of the non-backtracking operator, and gave rigorous guarantees regarding its performance on a simple model. We also illustrated the good performance of this algorithm on other types of data, including real world datasets, and showed that it compares favorably with the popular label propagation algorithm. In the next chapter, we come back to the Bethe Hessian and consider an application in matrix completion.

# MATRIX COMPLETION AND THE HOPFIELD MODEL

---

We consider here the (noiseless) matrix completion problem, where the aim is to infer the missing entries of a (low rank) matrix from a small number of revealed entries. More precisely, we are looking to reconstruct a rank-$r$ matrix $\mathcal{M}^{\text{true}} \in \mathbb{R}^{n \times m}$ written as

$$\mathcal{M}^{\text{true}} = XY^{\mathsf{T}} , \tag{8.1}$$

for some (unknown) tall matrices $X \in \mathbb{R}^{n \times r}$ and $Y \in \mathbb{R}^{m \times r}$. We assume that we observe only a small fraction of the elements of $\mathcal{M}^{\text{true}}$, chosen uniformly at random, and we wish to infer the missing entries using the low rank assumption. This problem has witnessed a recent burst of activity (see e.g. [26, 27, 80]) motivated by many applications such as collaborative filtering [26], quantum tomography [55] in physics, or the analysis of a covariance matrix [26]. Perhaps the most widely considered question in this setting is *perfect recovery*, i.e. how many entries need to be revealed in order for the matrix to be completed exactly in a computationally efficient way [26, 80]. The main questions we investigate in this chapter are different.

The first question we address is *detectability*, which here can be stated as: how many random entries do we need to reveal in order to be able to estimate the rank $r$ reliably? This is motivated by the more generic problem of detecting (in our case, low rank) structure hidden in partially observed data. As argued in section 1.4.1, it is reasonable to expect the existence of a region where exact completion is hard or even impossible yet the rank estimation is tractable. A second question we address is what is the minimum achievable root mean square error (RMSE) in estimating the unknown elements of the matrix. In practice, even if exact reconstruction is not possible, having a procedure that provides a very small RMSE might be quite sufficient.

Like in the previous chapter, our approach to this problem differs considerably from the strategy we adopted for the first clustering problems presented in this dissertation. For these, we were able to write down the posterior probability of the true cluster assignment on a probabilistic model, and to solve it either using BP or spectral methods tightly related to BP. For matrix completion, the situation is different. While we can still consider a generative model and write the posterior distribution of the missing entries, the marginalization problem turns out to be intractable, even using BP. This is because, unlike the previous clustering problems, the hidden assignment is here real valued, so that the BP messages are now full-fledged real distributions, instead of finite distributions, which could be represented

*Detectability*

*Reconstruction error*

*Unlike previously, we do not attempt here to approximate the marginals of a "true" posterior.*

as vectors. To solve the BP equations, one would have store a large number of populations, representing the marginals of the problem, and iteratively update these populations which is very expensive in practice [1]. Instead, we use here the available data to define a tractable pairwise model, whose marginals, we believe, may help us solve the problem. This approach is similar in spirit to the computer vision example of section 1.3.1, and corresponds to the "ad hoc" type of applications introduced in section 1.7 when describing the general approach of this dissertation. Anticipating on the following, the model we will consider is a *Hopfield model*, where the patterns will correspond to the factors X and Y of equation (8.1). When in the retrieval phase, as explained in section 1.6.6, the free energy of this model has interesting local minima, correlated with the patterns. The number of these local minima gives away the number of patterns, i. e. here the rank $r$ of the matrix. We use a spectral method based on the Bethe Hessian to detect all these interesting local minima at once.

The resulting algorithm, called MaCBetH (for Matrix Completion with the Bethe Hessian), allows to infer the rank and reconstruct the missing entries of $\mathcal{M}^{\text{true}}$, and is the main subject of this chapter. For a random model, we will compute explicitly the number of observations required for the algorithm to correctly infer the rank $r$. In particular, we will show that MaCBetH efficiently detects the rank $r$ of a large $n \times m$ matrix from $C(r)\, r\, \sqrt{nm}$ entries, where $C(r)$ is a constant close to 1, which we compute numerically. Additionally, we will show on numerical simulations that MaCBetH compares favorably to state-of-the-art matrix completion approaches in terms of reconstruction RMSE. Our presentation follows the publication [141], with some remarks and numerical simulations. We formally define the matrix completion problem and present generally our approach in the context of existing work in section 8.1. In section 8.2, we describe our algorithm and motivate its construction via a spectral relaxation of the Hopfield model. Next, in section 8.3.1 we show how to compute the phase transition in rank estimation, and also compute the spectral density of the Bethe Hessian on the sparse Hopfield model. Finally, in section 8.4 we present numerical simulations that demonstrate the accuracy of MaCBetH. Implementations of our algorithm in the Julia [16] and Matlab programming languages are available on the SPHINX webpage `http://www.lps.ens.fr/~krzakala/WASP.html`.

## 8.1   PROBLEM DEFINITION AND RELATION TO OTHER WORK

Let $\mathcal{M}^{\text{true}}$ be given by equation (8.1), where X and Y are called the *factors*. Recall that we observe only a small fraction of the elements of $\mathcal{M}^{\text{true}}$, chosen uniformly at random. We call $E$ the subset of observed entries, and $\mathcal{M}$ the (sparse) matrix supported on $E$ whose nonzero elements are the revealed entries of $\mathcal{M}^{\text{true}}$. The aim is to reconstruct the rank $r$ matrix $\mathcal{M}^{\text{true}} = XY^{\mathsf{T}}$ given $\mathcal{M}$. An important parameter which controls the difficulty of the problem is $\alpha = |E|/\sqrt{nm}$. In the case of a square matrix $\mathcal{M}$, this is the average number of revealed entries per line or column. In our numerical examples and theoretical justifications we shall generate the low rank matrix $\mathcal{M}^{\text{true}} = XY^{\mathsf{T}}$, using tall matrices $X$ and $Y$ with i.i.d Gaussian elements, and call this the random matrix setting. The MaCBetH algorithm is, however, non-parametric and does not use any prior knowledge about $X$ and $Y$. The analysis we perform applies in the limit $n \to \infty$ while $m/n$ is fixed and $r = O(1)$.

*Definition of $\alpha$*

*Random matrix setting*

The matrix completion problem was popularized by [26] who proposed nuclear norm minimization as a convex relaxation of the problem. The algorithmic complexity of the associated semidefinite programming approach is, however, $O\left(n^2 m^2\right)$. A low complexity procedure to solve the problem was later proposed by [23] and is based on a singular value decomposition (SVD) of the observed matrix $\mathcal{M}$. A considerable step towards the theoretical understanding of matrix completion from few entries was achieved by [80] who proved that with the use of *trimming* (see following), the performance of SVD-based matrix completion can be improved and a RMSE proportional to $\sqrt{nr/|E|}$ can be achieved. The algorithm of [80] is referred to as OptSpace, and empirically it achieves state-of-the-art RMSE in the regime of very few revealed entries.

OptSpace proceeds in three steps [80]. First, one trims the observed matrix $\mathcal{M}$ by setting to zero all rows (resp. columns) with more revealed entries than twice the average number of revealed entries per row (resp. per column). Second, a SVD of the matrix $\mathcal{M}$ is used to compute the $r$ pairs of singular vectors with largest singular value. When the rank $r$ is unknown, it is estimated as the index for which the ratio between two consecutive singular values has a minimum. Third, a local minimization of the discrepancy between the observed entries and the estimate is performed (see following). The initial condition for this minimization is given by the top $r$ left and right singular vectors from the second step.

*The OptSpace algorithm*

The advantage of trimming is that it allows to avoid the localization phenomenon discussed in the introduction of chapter 3 by removing the high degree vertices of the Erdős-Rényi graph $G = (n, E)$. However, this improvement comes at the cost of erasing part of the available data. On the other hand, we have been discussing at length two operators that do not suffer from the localization problem, even on very sparse graphs, namely the non-backtracking operator and the Bethe Hessian. In this work we improve upon OptSpace by replacing the first two steps by a different spectral procedure, based on

*On trimming*

the Bethe Hessian, that detects the rank and provides a better initial condition for the discrepancy minimization. We find that this new approach to matrix completion provides an improvement analogous to the one obtained in clustering by using the Bethe Hessian (or the non-backtracking) operator instead of more traditional matrices.

## 8.2    ALGORITHM AND MOTIVATION

We start by introducing our algorithm, and motivate it by making the connection with the Hopfield model explicit.

### 8.2.1    *The MaCBetH algorithm*

A standard approach to the completion problem (see e.g. [80]) is to minimize the cost function

*Cost function*
$$\min_{X,Y} \sum_{(ij) \in E} [\mathcal{M}_{ij} - (XY^{\mathsf{T}})_{ij}]^2 \,, \tag{8.2}$$

over factors $X \in \mathbb{R}^{n \times r}$ and $Y \in \mathbb{R}^{m \times r}$. This function is non-convex, and global optimization is hard [2]. One therefore resorts to a local optimization technique with a careful choice of the initial conditions $X_0, Y_0$. In our method, given the matrix $\mathcal{M}$, we consider a weighted bipartite undirected graph with adjacency matrix $A \in \mathbb{R}^{(n+m) \times (n+m)}$

*Adjacency matrix of the associated bipartite graph*
$$A = \begin{pmatrix} 0 & \mathcal{M} \\ \mathcal{M}^{\mathsf{T}} & 0 \end{pmatrix} \,. \tag{8.3}$$

We will refer to the graph thus defined as $G$. We now define the Bethe Hessian matrix $H(\beta) \in \mathbb{R}^{(n+m) \times (n+m)}$ to be the matrix with elements, for $i, j \in [n]$

*Bethe Hessian*
$$\begin{aligned} H_{ij}(\beta) &= \left(1 + \sum_{k \in \partial i} \frac{\tanh^2 \beta A_{ik}}{1 - \tanh^2 \beta A_{ik}}\right) \mathbf{1}(i = j) - \frac{\tanh \beta A_{ij}}{1 - \tanh^2 \beta A_{ij}} \,, \\ &= \left(1 + \sum_{k \in \partial i} \sinh^2 \beta A_{ik}\right) \delta_{ij} - \frac{1}{2} \sinh(2\beta A_{ij}) \,, \end{aligned} \tag{8.4}$$

where $\beta$ is a parameter that we will fix to a well-defined value $\hat{\beta}_{SG}$ depending on the data, and $\partial i$ stands for the neighbors of $i$ in the graph $G$. We justify the form of this matrix in the next section. The MaCBetH algorithm that is the main subject of this chapter is described in algorithm 10. In step 1 of this procedure, $\hat{\beta}_{SG}$ is an approximation

---

2. It was recently proved [49] that when $\alpha = \text{polylog}(n)$, this cost function has no spurious local minimum, so that a simple local optimization of the cost function (8.2) starting from any initial condition allows to perfectly recover the matrix $\mathcal{M}$. In practice, and for the very sparse regimes we consider in our numerical tests, we will see that the initial condition plays an important role.

---

**Algorithm 10** MaCBetH algorithm for matrix completion

---

**Input:** Partially observed matrix $\mathcal{M}$, which we assume to be *centered*

1: **Solve** for the value of $\hat{\beta}_{SG}$ such that $F\left(\hat{\beta}_{SG}\right) = 1$, where

$$F(\beta) = \frac{1}{\sqrt{nm}} \sum_{(i,j) \in E} \tanh^2(\beta \mathcal{M}_{ij}). \qquad (8.5)$$

2: **Build** the Bethe Hessian $H\left(\hat{\beta}_{SG}\right)$ of equation (8.4).

3: **Compute** all the negative eigenvalues $\lambda_1, \cdots, \lambda_{\hat{r}}$ of $H\left(\hat{\beta}_{SG}\right)$ and their corresponding eigenvectors $v_1, \cdots, v_{\hat{r}}$. $\hat{r}$ is our estimate for the rank $r$. Set $X_0$ (resp. $Y_0$) to be the first $n$ lines (resp. the last $m$ lines) of the matrix $[v_1 \ v_2 \ \cdots \ v_{\hat{r}}]$.

4: **Perform** local optimization of the cost function (8.2) with rank $\hat{r}$ and initial condition $X_0, Y_0$.

---

to the optimal value of $\beta$, for which $H(\beta)$ has a maximum number of negative eigenvalues (see section 8.3.1). Instead of this approximation, $\beta$ can be chosen in such a way as to maximize the number of negative eigenvalues. We however observed numerically that the algorithm is robust to some imprecision on the value of $\hat{\beta}_{SG}$. Note that in step 2, we could also use the associated non-backtracking matrix with weights $\tanh\left(\beta \mathcal{M}_{ij}\right)$ (see section 8.3.1). The Bethe Hessian is however smaller and symmetric, which reduces the complexity of the spectral part of our algorithm. In the next section, we will motivate and analyze this algorithm in the setting where $\mathcal{M}^{\text{true}}$ was generated from element-wise random factors $X$ and $Y$. We will argue that in this case MaCBetH is able to infer the rank whenever $\alpha > \alpha_c$, where $\alpha_c$ is a (small) constant, independent of $n$. Figure 8.1 illustrates the spectral properties of the Bethe Hessian that justify this algorithm. With the choice $\beta = \hat{\beta}_{SG}$, the informative eigenvalues of $H(\beta)$ are negative, and well separated from the bulk of uninformative eigenvalues, which stays positive.

### 8.2.2 *Motivation from a Hopfield model*

As shown in section 2.3.2, the matrix $H(\beta)$ is closely related to the pairwise MRF

$$\mathbb{P}(\sigma, \tau) = \frac{1}{\mathcal{Z}} \exp\left(\beta \sum_{(ij) \in E} \mathcal{M}_{ij} \sigma_i \tau_j\right), \qquad (8.6) \qquad \textit{Hopfield model}$$

where the $\sigma = (\sigma_i)_{i \in [n]}$ and $\tau = (\tau)_{i \in [n]}$ are two collections of binary variables, and $\beta$ is an inverse temperature controlling the strength of the interactions. This model is a (generalized) Hebbian Hopfield model (see section 1.6.6) on the bipartite sparse graph G, with r patterns given by

$$\xi^{\mu} = \begin{pmatrix} X_{\mu} \\ Y_{\mu} \end{pmatrix} \in \mathbb{R}^{n+m}, \qquad \forall \mu \in [r], \qquad (8.7)$$

where $X_{\mu}$ (resp. $Y_{\mu}$) denotes the $\mu$-th column of the factor $X$ (resp. $Y$). From the analysis of section 2.3.2, $H(\beta)$ is the Hessian (or susceptibility) of the Bethe free energy at the trivial, paramagnetic stationary point of this model.

*Phase diagram of the Hopfield model*

Let us recall here the salient features of the Hopfield model, summarized in the phase diagram 1.11, which justify algorithm 10. For $\beta$ small enough, model (8.6) is in its paramagnetic phase, meaning that the paramagnetic stationary point of the Bethe free energy is a (global) minimum. Therefore, we expect (and can check by the Gershgorin circle theorem) that $H(\beta)$ is positive definite. As we lower $\beta$,

*Paramagnetic phase*

*Retrieval phase*

model (8.6) may enter its *retrieval* phase, in which we expect the paramagnetic stationary point to become a saddle point. In this phase, the Bethe free energy has local minima correlated with the factors $X$ and $Y$. We therefore expect that in the corresponding range of (inverse) temperature $\beta$, $H(\beta)$ has negative eigenvalues, with corresponding eigenvectors approximating the magnetizations of model (8.6) in the various retrieval states. These retrieval states have a finite overlap (1.100) with the patterns (8.7). Therefore, the eigenvectors of $H(\beta)$ with negative eigenvalues are suitable initial conditions for the optimization of the cost function (8.2). Additionally, as argued in section 1.6.6, we expect the dimension of the subspace of $\mathbb{R}^{n+m}$ spanned by these retrieval states to be $r$, so that we expect $H(\beta)$ to have exactly $r$ negative eigenvalues, from which we can infer the rank. Finally, as we increase $\beta$ above a critical value $\beta_{SG}$ the Hopfield model

*Spin glass phase*

eventually enters its spin glass phase, marked by the appearance of many spurious minima. In particular, for low enough $\beta$, the paramagnetic stationary point may become a local minimum again. Figure 8.1 summarizes the consequences of the phase diagram of the Hopfield model for the spectrum of $H(\beta)$. The choice of $\beta = \hat{\beta}_{SG}$ used in algorithm 10 approximates the critical temperature corresponding to the paramagnetic to spin glass instability (see section 8.3.1). As in the previous clustering examples, we expect from the phase diagram of figure 1.11 that *if* the retrieval phase exists, then the paramagnetic to retrieval transition happens for $\beta < \beta_{SG}$. Another justification for this particular choice of $\beta$ is provided in section 8.3.1.

Note that a similar approach was used in [169] to detect the retrieval states of a Hopfield model using the weighted non-backtracking matrix, which linearizes the belief propagation equations rather than the Bethe free energy, resulting in a larger, non-symmetric matrix. The Bethe Hessian, while mathematically closely related, is also simpler to handle in practice.

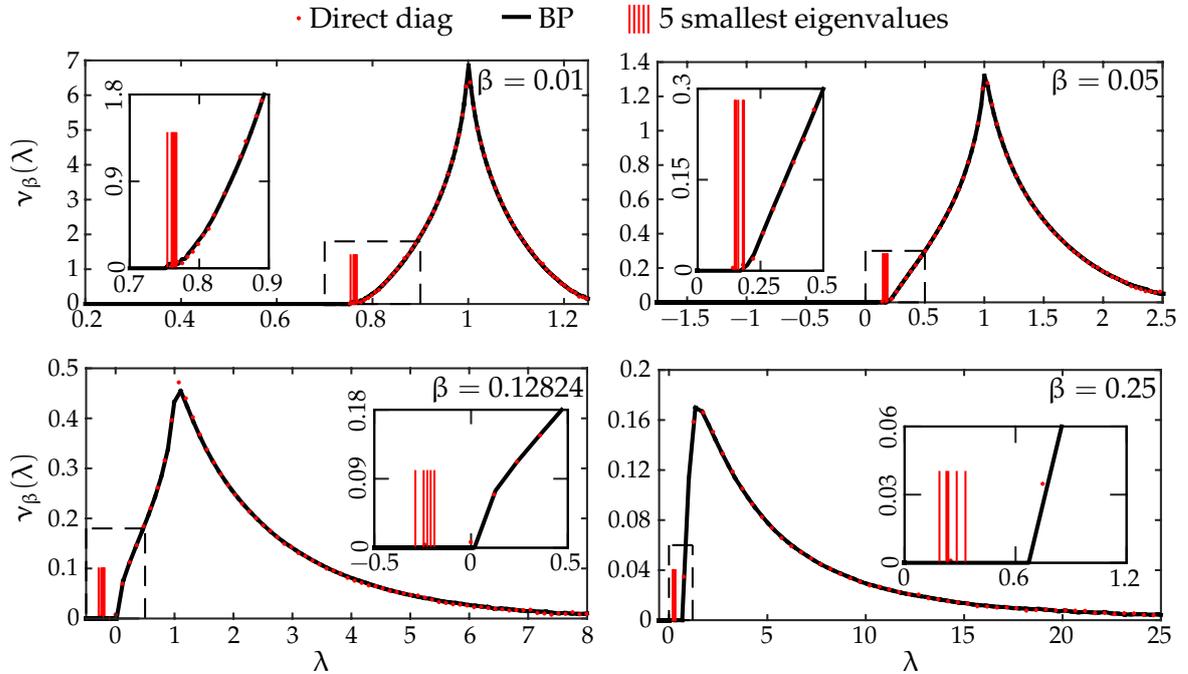

Figure 8.1 – Spectrum of the Bethe Hessian $H(\beta)$ for various values of the parameter $\beta$. The red dots are the result of the direct diagonalization of the Bethe Hessian for a rank $r = 5$ and $n = m = 10^4$ matrix, with $\alpha = 15$ revealed entries per row on average. The black curves are the solutions of (8.18) computed with belief propagation on a graph of size $10^5$. We isolated the 5 smallest eigenvalues, represented as small bars for convenience, and the inset is a zoom around these smallest eigenvalues. For $\beta$ small enough (top plots), the Bethe Hessian is positive definite, signaling that the paramagnetic stationary point is a local minimum of the Bethe free energy. As $\beta$ increases, the spectrum is shifted towards the negative region and has 5 negative eigenvalues at the approximate value of $\hat{\beta}_{SG} = 0.12824$ (to be compared to $\beta_R = 0.0832$ for this case) evaluated by our algorithm (lower left plot). These eigenvalues, corresponding to the retrieval states, eventually become positive again as $\beta$ is further increased (lower right plot), while the bulk of uninformative eigenvalues remains at all values of $\beta$ in the positive region. Figure taken from [141].

Interestingly, if we solve the Hopfield model (8.6) in the nMF approximation instead of the Bethe approximation, we have seen in section 2.3.3 that the Hessian of the free energy at the paramagnetic point is then straightforwardly related to the weighted adjacency matrix (8.3) of the bipartite graph G, i. e. (a symmetrized version of) the observed matrix M. Finding the smallest eigenvalues of the nMF Hessian is therefore equivalent to solving for the largest singular values of M, which sheds a new light on the matrix completion algorithms based on the SVD of M, such as OptSpace. Consistently with the discussion of section 2.3.3, when there are enough revealed entries so that the graph G is dense enough, the spectrum of the nMF Hessian is well behaved, and the informative eigenvalues are those lying outside of the bulk that can be derived form the Marchenko-Pastur law [20, 99]. In the sparse case, however, we will see that replacing the nMF approximation by the Bethe approximation leads, once more, to substantial improvements.

## 8.3    ANALYSIS OF PERFORMANCE IN DETECTION

We now show how to analyze the performance of MaCBetH using statistical physics tools to compute the rank detectability transition. We then give an independent justification of our choice of $\beta = \hat{\beta}_{SG}$ in algorithm 10 based on the analysis of the spectral density of $H(\beta)$.

### 8.3.1    *Analysis of the phase transition*

We investigate here the minimal value of $\alpha$ from which algorithm 10 starts inferring the rank r correctly. As noted previously, the Bethe Hessian of equation (8.4) is closely related to the non-backtracking operator defined for $(i \to j), (k \to l) \in \vec{E}$ by

$$B_{i \to j, k \to l} = \tanh(\beta M_{kl}) . \tag{8.8}$$

In particular, by the Ihara-Bass formula, the appearance of negative eigenvalues of $H(\beta)$ is related to the appearance of eigenvalues of B with modulus larger than 1, i. e. to the instability of the paramagnetic fixed point of BP for model (8.6). Assuming a random matrix setting in which the entries of the factors X and Y are i.i.d random variables with mean 0, the stability of this paramagnetic fixed point can be studied by means of the cavity method, using the same arguments as in section 3.1.2 [10, 28, 156, 169]. The only two differences here are, first, that we are considering a bipartite graph, and second, that since the patterns of the Hopfield model are real-valued, the weights of the non-backtracking operator (8.8) are correlated random variables. As a consequence, the definition of the various parameters of section 3.1.2 need to be adapted. In particular, we find that the bulk of uninformative eigenvalues of B is constrained in the disk of radius $R(\beta) = \sqrt{\alpha \, \Sigma(\beta)^2}$, where

$$\Sigma(\beta)^2 = \lim_{t \to \infty} \mathbb{E}\left[\prod_{l=1}^{t} \tanh^2\left(\beta \sum_{\mu=1}^{r} x_l^\mu y_l^\mu\right) \tanh^2\left(\beta \sum_{\mu=1}^{r} x_{l+1}^\mu y_l^\mu\right)\right]^{\frac{1}{2t}}.$$
(8.9)

In the last expression, the expectation is with respect to the distribution of the i.i.d random variables $x_l^\mu \overset{\mathcal{D}}{=} x$ and $y_l^\mu \overset{\mathcal{D}}{=} y$ for $l \in \mathbb{N}$, $\mu \in [r]$, where x (resp. y) has the same distribution as the entries of the factor X (resp. Y). The parameter $R(\beta) = \sqrt{\alpha \, \Sigma(\beta)^2}$ thus defined equals the growth rate of a noisy perturbation of the paramagnetic fixed point, uncorrelated with the patterns of the Hopfield model, as we iterate the non-backtracking operator. Additionally, there is an informative eigenvalue $\alpha \, \Delta(\beta)$ of B, with multiplicity r, outside of the disk of radius $R(\beta)$ if and only if $\alpha \, \Delta(\beta) > R(\beta)$, where

$$\Delta(\beta) = \lim_{t \to \infty} \mathbb{E}\left[\prod_{l=1}^{t} \tanh\left(\beta \left|x_l^1 y_l^1\right| + \beta \sum_{\mu=2}^{r} x_l^\mu y_l^\mu\right)\right.$$
$$\left. \times \tanh\left(\beta \left|x_{l+1}^1 y_l^1\right| + \beta \sum_{\mu=2}^{r} x_{l+1}^\mu y_l^\mu\right)\right]^{\frac{1}{2t}},$$
(8.10)

where the expectation is positive since we assume the variables x and y to be centered. With this definition, the parameter $\alpha \, \Delta(\beta)$ controls the growth rate of the overlap (1.100) with the first pattern $\xi^1$ defined in (8.7), as we iterate the non-backtracking operator, starting from an initial condition correlated with the patterns of the Hopfield model. We may now define the critical temperature $\beta_R$ (resp. $\beta_{SG}$) marking the onset of the paramagnetic to retrieval instability (resp. paramagnetic to spin glass). These temperature are defined implicitly as

$$\alpha \, \Sigma(\beta_{SG})^2 = 1$$
$$\alpha \, \Delta(\beta_R) = 1,$$
(8.11)

so that for $\beta > \beta_R$ (resp. $\beta > \beta_{SG}$), a perturbation initially correlated with the patterns (resp. a noisy perturbation) grows when iterating B. From the phase diagram of the sparse Hopfield model 1.11, the retrieval phase exists if the paramagnetic to retrieval instability happens at larger temperature (smaller $\beta$) than the paramagnetic to spin glass instability. This is equivalent to the condition

$$\beta_{SG} > \beta_R.$$
(8.12)

Note that this condition implies the existence of a range of values of $\beta$ for which B has an informative eigenvalue outside of the bulk of radius $R(\beta)$. More precisely, since $\Sigma$ and $\Delta$ are increasing functions of $\beta$, for $\beta = \beta_{SG}$, we expect under condition (8.12) that B has an informative eigenvalue $\alpha \Delta(\beta_R) > 1$ with multiplicity $q$ outside of the bulk of uninformative eigenvalues of radius $R(\beta) = \sqrt{\alpha \Sigma(\beta_{SG})^2} = 1$. As a consequence, we expect that under condition (8.12), $H(\beta_{SG})$ has an informative, negative eigenvalue with multiplicity $r$.

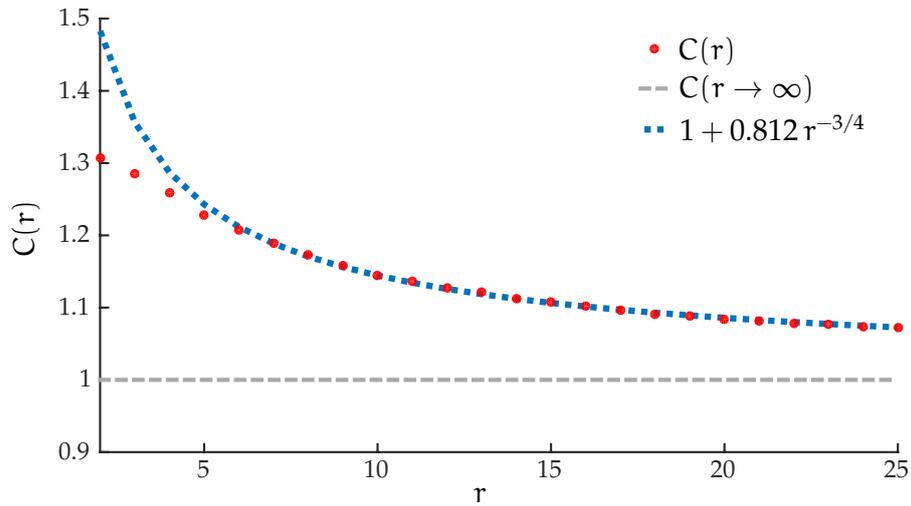

Figure 8.2 – Rank detectability threshold for factors $X, Y$ with Gaussian entries of mean $0$ and unit variance. MaCBetH is able to estimate the correct rank from $|E| > C(r)r\sqrt{nm}$ known entries, where the value of $C(r)$ is represented as red dots. The gray dashed line gives the asymptotic value $C(r) = 1$ in the limit $r \to \infty$ (see text). We used a population dynamics algorithm with a population of size $10^6$ to compute the functions $\Sigma$ and $\Delta$ from equations (8.9, 8.10). The dotted line is a fit suggesting that $C(r) - 1 = O(r^{-3/4})$. Figure taken from [141].

*Critical value $\alpha_c$*      We define the critical value $\alpha_c$ of $\alpha$ such that $\beta_{SG} > \beta_R$ if and only if $\alpha > \alpha_c$. In general, there is no closed-form formula for this critical value, which is defined implicitly in terms of the functions $\Sigma$ and $\Delta$. We can however compute $\alpha_c$ numerically using a population dynamics[3] algorithm [102] which allows to compute the parameter $C(r) = \alpha_c/r$. The results are presented on figure 8.2. Quite remarkably, with the definition $\alpha = |E|/\sqrt{nm}$, the critical value $\alpha_c$ does not depend on the ratio $m/n$, only on the rank $r$.

*Explicit value of $\alpha_c$ for large $r$*      In the limit of large $\alpha$ and $r$, it is possible to obtain a simple closed-form formula for $\alpha_c$. In this case the observed entries of the matrix become jointly Gaussian distributed, and uncorrelated, and therefore independent. Expression (8.9) then simplifies to

$$\Sigma(\beta)^2 \underset{r\to\infty}{\sim} \mathbb{E}\left[\tanh^2\left(\beta\sum_{\mu=1}^{r}x^\mu y^\mu\right)\right].\qquad(8.13)$$

---

3. We use a population dynamics algorithm to iterate the linear distributional recursion (3.31) for the non-backtracking operator (8.8). We monitor the evolution of the overlap with a pattern to compute $\Delta$, and the evolution of the Edwards-Anderson parameter to compute $\Sigma$.

Note that in step 1 of algorithm 10, we use an empirical estimator $F(\beta) \simeq \alpha\Sigma(\beta)^2$ of the quantity (8.13) to compute a simple approximation $\hat{\beta}_{SG}$ of $\beta_{SG}$ from the revealed entries. In the large $r, \alpha$ regime, both $\beta_{SG}, \beta_R$ decay to 0, so that we can further approximate

$$1 = \alpha\Sigma(\beta_{SG})^2 \underset{r\to\infty}{\sim} \alpha\,r\,\beta_{SG}^2\,\mathbb{E}[x^2]\,\mathbb{E}[y^2],\qquad(8.14)$$

$$1 = \alpha\Delta(\beta_R) \underset{r\to\infty}{\sim} \alpha\,\beta_R\,\sqrt{\mathbb{E}[x^2]\,\mathbb{E}[y^2]},\qquad(8.15)$$

so that we obtain the simple asymptotic expression $\alpha_c = r$, in the large $\alpha, r$ limit, or equivalently $C(r) = 1$. Interestingly, the same result was obtained for the *detectability* threshold in the completion of rank $r = O(n)$ matrices from $O(n^2)$ entries in the Bayes optimal setting in [72]. Notice, however, that *exact* completion in the setting of [72] is only possible for $\alpha > r(m+n)/\sqrt{nm}$. Clearly detection and exact completion are different phenomena, with different transitions. The previous analysis can in principle be extended beyond the random setting assumption, as long as the empirical distribution of the entries is well defined, and the lines of $X$ (resp. $Y$) are approximately orthogonal and centered. This condition is related to the standard incoherence property of matrix completion [26, 80].

### 8.3.2 *Computation of the spectral density*

The spectral density $\nu_\beta(\lambda)$ of $H(\beta)$ can be computed using the method presented in section 3.2.1. We find that $\nu_\beta(\lambda)$ verifies

$$\nu_\beta(\lambda) = \lim_{n,m\to\infty}\frac{1}{\pi(n+m)}\sum_{i=1}^{n+m}\mathrm{Im}\,\Delta_i(\lambda),\qquad(8.16)$$

where the $\Delta_i$ are complex variables living on the vertices of the graph $G$, given by

$$\Delta_i = \left(-\lambda + 1 + \sum_{k\in\partial i}\sinh^2\beta A_{ik} - \sum_{l\in\partial i}\frac{1}{4}\sinh^2(2\beta A_{il})\Delta_{l\to i}\right)^{-1}.\qquad(8.17)$$

The $\Delta_{i\to j}$ for $(i\to j)\in\vec{E}$ are the (linearly stable) solution of the following fixed point equation

$$\Delta_{i\to j} = \left(-\lambda + 1 + \sum_{k\in\partial i}\sinh^2\beta A_{ik} - \sum_{l\in\partial i\backslash j}\frac{1}{4}\sinh^2(2\beta A_{il})\,\Delta_{l\to i}\right)^{-1}.\qquad(8.18)$$

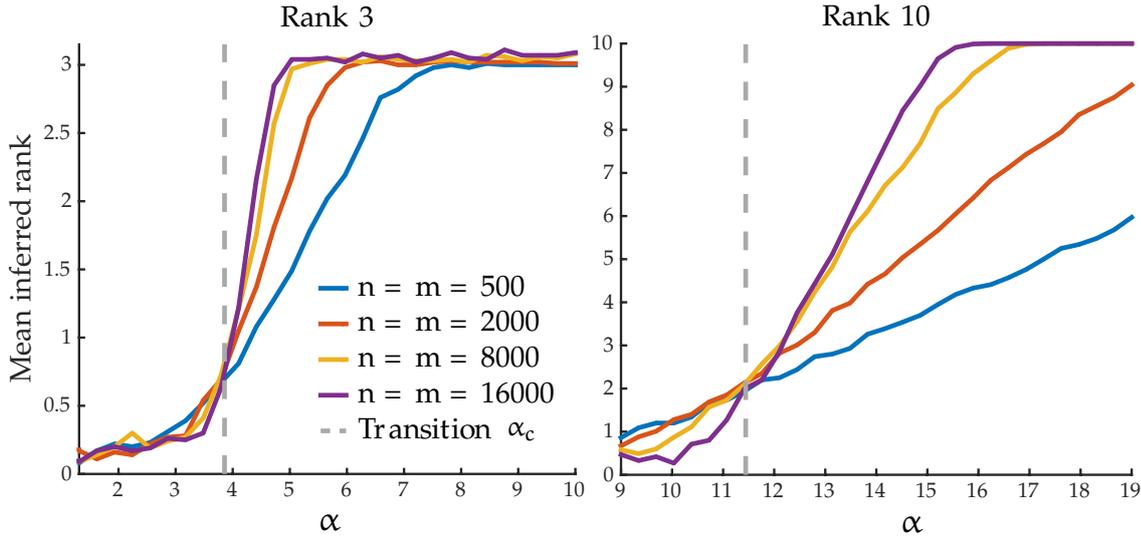

Figure 8.3 – Mean inferred rank as a function of $\alpha$, for different sizes, averaged over 100 samples of $n \times m$ matrices $XY^{\intercal}$. The entries of $X, Y$ are drawn from a Gaussian distribution of mean 0 and variance 1. The theoretical transition $\alpha_c$ is computed with a population dynamics algorithm (see section 8.3.1), and is in good agreement with the empirical performance of MaCBetH in rank estimation. Figure taken from [141].

The previous fixed point equation can be solved using BP, which allows to accurately compute the spectral density of $H(\beta)$, as shown on figure 8.1. Using the same arguments as in section 3.2.1, we can argue that the spectral density $\nu_{\beta}(\lambda)$ vanishes on an open set surrounding $\lambda = 0$ for any $\beta < \beta_{SG}$, where $\beta_{SG}$ has been defined in the previous section. To see this, we note that $\Delta_{i \to j} = \cosh^{-2}(\beta A_{ij})$ is a real solution of (8.18), yielding a vanishing spectral density. By following the same steps as in section 3.2, we can show that this implies the existence of a real and stable solution of (8.18) in a an open set surrounding $\lambda = 0$ whenever $\rho(\mathcal{B}) < 1$, where $\mathcal{B}$ is defined by

$$\mathcal{B}_{i \to j, k \to l} = \tanh^2(\beta \mathcal{M}_{kl}) . \tag{8.19}$$

This is a non-backtracking operator whose spectral radius can be computed using the methods explained in section 3.1.2. As noted in the previous section, we only have to deal with the additional subtlety that the graph $G$ is a bipartite graph, and that the weights carried by the non-backtracking operator (8.19) are correlated random variables. We find that

$$\rho(\mathcal{B}) = \sqrt{\alpha \, \Sigma(\beta)^2} , \tag{8.20}$$

where $\Sigma(\beta)^2$ is defined in equation (8.9). This formula generalizes equation (3.126) to the case of correlated weights. Since $\Sigma$ is an increasing function of $\beta$, we have $\rho(\mathcal{B}) < 1$ for any $\beta < \beta_{SG}$, implying the existence of an open set $U \ni 0$ such that for $\lambda \in U$, $\nu_{\beta}(\lambda) = 0$. As we increase $\beta$ from a small value, we therefore expect the bulk of uninformative eigenvalues of $H(\beta)$ to reach $\lambda = 0$ at $\beta = \beta_{SG}$, so that the informative eigenvalues of $H(\beta_{SG})$ are negative, as illustrated on figure 8.1. As explained in the previous section, the choice $\beta = \hat{\beta}_{SG}$ in algorithm 10 approximates the critical value. Interestingly, this figure suggests that for $\beta > \beta_{SG}$, the bulk of uninformative eigenvalues shifts back to the right on this problem.

## 8.4 NUMERICAL TESTS

Figure 8.3 illustrates the ability of the Bethe Hessian to infer the rank $r$ above the critical value $\alpha_c$ defined in section 8.3.1, in the limit of large size $n, m \to \infty$.

*Rank inference*

In figure 8.4, we demonstrate the suitability of the informative eigenvectors of the Bethe Hessian as starting point for the minimization of the cost function (8.2). We compare the final RMSE achieved on the reconstructed matrix $XY^\intercal$ with 4 other initializations of the optimization, including the top singular vectors of the trimmed matrix $\mathcal{M}$ as used in the OptSpace algorithm [80]. In all cases, the optimization of the cost function is carried out using an off-the-shelf quasi-Newton algorithm (see caption of figure 8.4). MaCBetH systematically outperforms all the other choices of initial conditions, providing a better initial condition for the optimization of (8.2). Remarkably, the performance achieved by MaCBetH with the inferred rank is essentially the same as the one achieved with an oracle rank. By contrast, estimating the correct rank from the (trimmed) SVD is more challenging. We note that for the choice of parameters we considered, trimming had in practice a negligible effect.

*Reconstruction accuracy*

We also investigated the effect of choosing a different optimization procedure for the minimization of the cost function (8.2). As shown in figure 8.5, when using the popular Alternating Least Squares (ALS) method [58, 67] as the optimization method, we also obtained a similar improvement in reconstruction by using the eigenvectors of the Bethe Hessian, instead of the singular vectors of $\mathcal{M}$ as initial condition. Along the same lines, OptSpace [80] uses yet another minimization procedure, close in spirit to ALS, and that had, in our tests, the same performance.

*Effect of the optimization routine*

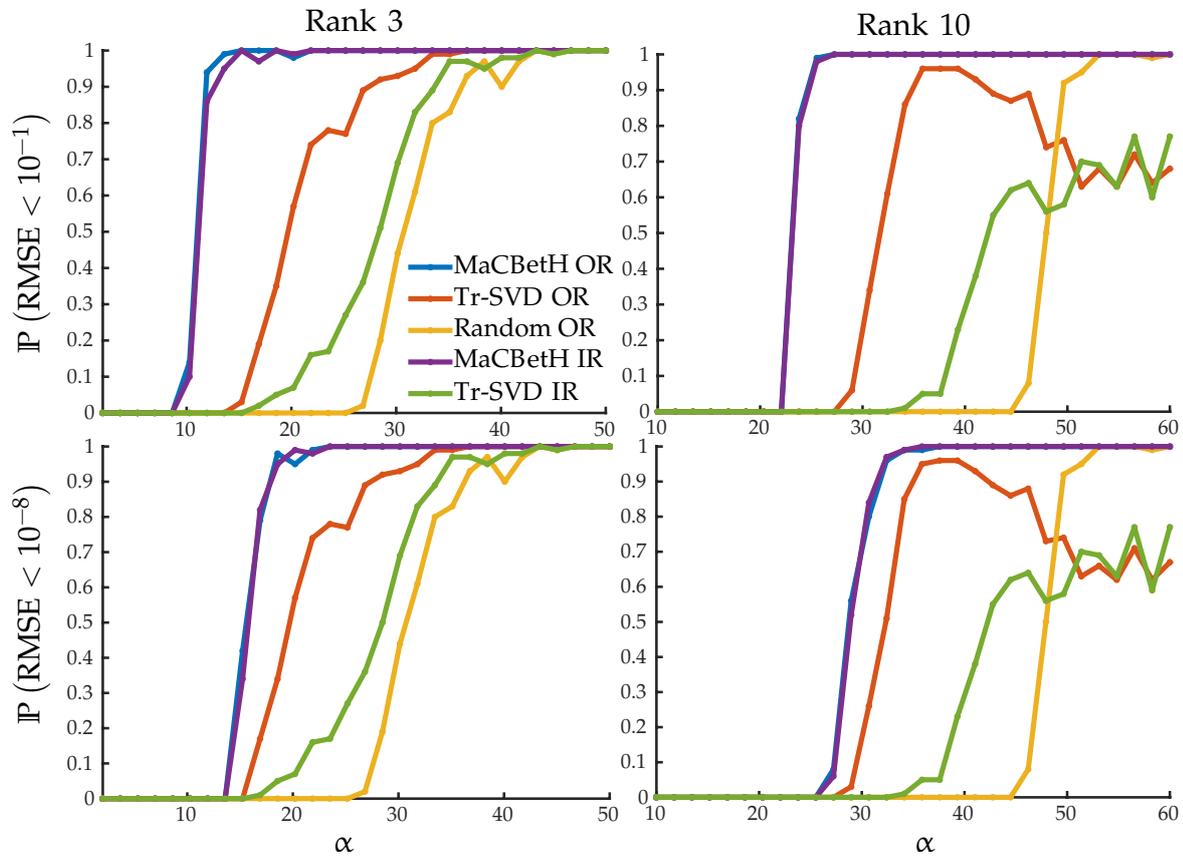

Figure 8.4 – RMSE as a function of the average number of revealed entries per row α: comparison between different initializations for the optimization of the cost function (8.2). The top row shows the probability that the achieved RMSE is smaller than $10^{-1}$, while the bottom row shows the probability that the final RMSE is smaller than $10^{-8}$. These probabilities were estimated as the frequency of success over 100 samples of matrices $XY^\intercal$ of size $10000 \times 10000$, with the entries of $X, Y$ drawn from a Gaussian distribution of mean 0 and variance 1. All methods optimize the cost function (8.2) using a low-storage BFGS algorithm [96] part of NLopt [71], starting from different initial conditions. The maximum number of iterations of the L-BFGS algorithm was set to 1000. The initial conditions compared are MaCBetH with oracle rank (MaCBetH OR) or inferred rank (MaCBetH IR), SVD of the observed matrix $\mathbb{M}$ after trimming, with oracle rank (Tr-SVD OR), or inferred rank (Tr-SVD IR, note that this is equivalent to OptSpace [80]), and random initial conditions with oracle rank (Random OR). For the Tr-SVD IR method, we inferred the rank from the SVD by looking for an index for which the ratio between two consecutive eigenvalues is minimized, as suggested in [79]. Figure taken from [141].

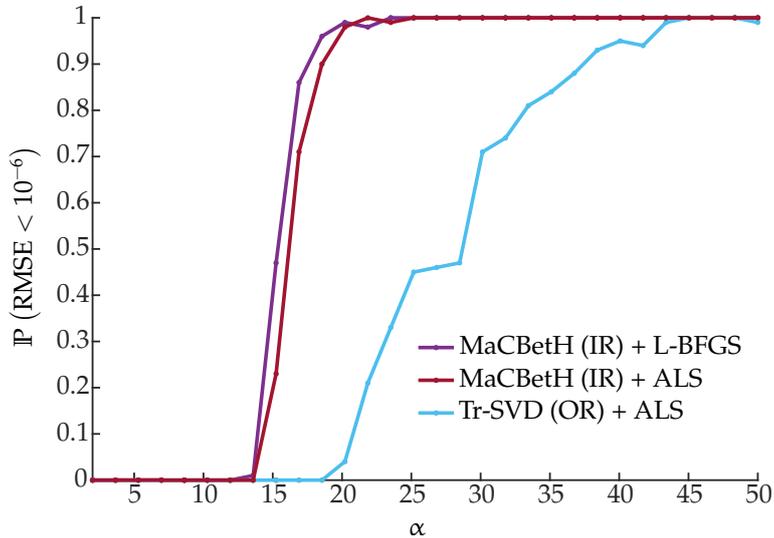

Figure 8.5 – RMSE as a function of the average number of revealed entries per row $\alpha$: comparison between different choices of optimization routines for the minimization of the cost function (8.2). Like in figure 8.4, the probability that the output RMSE is smaller than $10^{-6}$ is estimated as the frequency of success over 100 samples of matrices $XY^\intercal$ of size $10000 \times 10000$, with the entries of $X, Y$ drawn from a Gaussian distribution of mean 0 and variance 1. The rank is here fixed to $r = 3$. The blue curve (Tr-SVD (OR) + ALS) was obtained by using Alternating Least Squares (ALS) as the optimization routine, initialized from the singular value decomposition of the trimmed observed matrix $M$, using the oracle rank. By replacing this initialization by the eigenvectors of the Bethe Hessian, we obtain a significant improvement in reconstruction accuracy as shown by the red curve, even using the rank inferred by our algorithm instead of the oracle rank (MaCBetH (IR) + ALS). We show for comparison in purple the result of using MaCBetH with the off-the-shelf L-BFGS algorithm (MaCBetH (IR) + L-BFGS) used for figure 8.4.

8.5   CONCLUSION

In this chapter, we have introduced MaCBetH, a matrix completion algorithm based on the Bethe Hessian. We have argued that MaC-BetH is efficient for two distinct and complementary tasks. First, it has the ability to infer the rank of a strongly subsampled matrix from fewer entries than any existing approach. In particular, on synthetic problems, we were able to make a precise prediction about the exact number of entries that are needed, and showed numerical simulations in excellent agreement with our predictions. Second, we found empirically that MaCBetH provides a lower reconstruction RMSE than its competitors.

In the next chapter, we come back to statistical physics, and show how the non-backtracking operator and the Bethe Hessian can be used to prove rigorous bounds on several quantities of interest in the ferromagnetic Ising model.

# SPECTRAL BOUNDS ON THE ISING FERROMAGNET

We have used extensively the non-backtracking operator and the Bethe Hessian to approximate the magnetizations of an Ising model associated with a given inference problem. For both operators, this approach is based on the Bethe approximation, which, although it is believed to be accurate for sparse random graphs, does not in general provide bounds on the magnetizations (see section 1.5.3). In this chapter, we show that in the special case of the Ising ferromagnet, we can establish *algorithmic* upper bounds on the partition function, the magnetizations, and the correlations, which are expressed in terms of the non-backtracking operator and the Bethe Hessian. The general approach we follow is not new, and is based on a bound on the so-called high temperature expansion due to Fisher [44], which we review in section 9.3.1. However, up to now, and to the best of our knowledge, this approach has not yielded explicit bounds that could be computed efficiently, except in very particular regular graphs [44] with particular couplings. In the following, we *explicitly* compute the upper bound obtained using the approach of [44] on *arbitrary* graphs, yielding a simple algorithmic upper bound expressed in terms of the non-backtracking and Bethe Hessian operators, which can be computed efficiently.

## 9.1 INTRODUCTION

Our starting point is the general Ising model in non-vanishing fields defined by the pairwise MRF

$$\mathbb{P}(\sigma) = \frac{1}{\mathcal{Z}} \exp\left( \sum_{(ij) \in E(G)} J_{ij} \sigma_i \sigma_j + \sum_{i \in V} h_i \sigma_i \right), \qquad (9.1)$$

*Ising model*

where $G = ([n], E(G))$ is an arbitrary graph, with $n$ vertices and edge set $E(G)$, $\sigma \in \{\pm 1\}^n$ is a collection of binary spins, and the *partition function* $\mathcal{Z}$ is defined by

$$\mathcal{Z} = \sum_{\sigma \in \{\pm 1\}^n} \prod_{(ij) \in E(G)} \exp J_{ij} \sigma_i \sigma_j \prod_{i \in V} \exp h_i \sigma_i. \qquad (9.2)$$

*Partition function*

We say that the Ising model (9.1) *ferromagnetic* if both the couplings $(J_{ij})_{(ij) \in E(G)}$ and the fields $(h_i)_{i \in [n]}$ are all positive. Recall ([151]) that an Ising model of the form (9.1) is completely specified by its

*Magnetizations and susceptibility matrix*

*magnetizations* $(\mathfrak{m}_a)_{a \in [n]}$ and its *susceptibility matrix* $\chi \in \mathbb{R}^{n \times n}$, defined by

$$\mathfrak{m}_a = \mathbb{E}(\sigma_a) = \sum_{\sigma \in \{\pm 1\}^n} \sigma_a \, \mathbb{P}(\sigma) \qquad \text{for } a \in [n], \quad (9.3)$$

$$\chi_{ab} = \mathbb{E}(\sigma_a \sigma_b) = \sum_{\sigma \in \{\pm 1\}^n} \sigma_a \sigma_b \, \mathbb{P}(\sigma) \qquad \text{for } a, b \in [n]. \quad (9.4)$$

We are interested in upper bounding the quantities (9.2,9.3,9.4).

### 9.1.1 *Significance and prior work*

*A short reminder of the importance of the Ising model*

The task of computing the quantities (9.2, 9.3, 9.4) knowing the joint probability distribution (9.1) is a typical *inference* problem. Its applications, alluded to in chapter 1, include e. g. reconstructing partially observed binary images in computer vision [54], or similarity-based clustering into 2 groups [17]. Conversely, the task of computing the joint probability distribution (9.1) knowing the magnetizations (9.3) and correlations (9.4) is usually called *learning*, or inverse problem. The practical importance of this inverse task stems in part from the fact that model (9.1) is the maximum entropy model with constrained means and correlations, as shown in section 1.3.2. In particular, it has found numerous applications in biology [13, 109], from predicting the three-dimensional folding of proteins, to identifying neural activity patterns. In machine learning, a variant of this inverse problem is usually referred to as Boltzmann machine learning [7], and is a typical unsupervised learning problem. In practice, learning is done through a local optimization of the likelihood function, which involves the partition function (9.2), and whose gradients involve the magnetizations (9.3) and correlations (9.4) (see next chapter).

*The planar case*

Despite their considerable practical importance, the estimation of these three quantities in general is a notoriously intractable problem, except in particular cases, notably when the graph G is planar and the external fields $(h_i)_{i \in [n]}$ vanish. In the latter case, explicit expressions exist, originating in the work of Kac and Ward [43, 74, 78], and allowing to devise polynomial-time inference algorithms [144]. Recently, [70] proposed a greedy algorithm to approximate an arbitrary graph by a planar one, thus making inference and learning tractable in a more general setting. Their approach, however, does not provide bounds on their estimates.

*Ferromagnetic Ising model and graph cuts*

For the case of the ferromagnetic Ising model (9.1) with positive couplings $(J_{ij})_{(ij) \in E(G)}$ and fields $(h_i)_{i \in [n]}$, it is well known [87] that the ground-state, i.e. the configuration of the spins with minimum energy, can be found in polynomial time via graph cuts. This method has been used successfully in a number of computer vision problems, (see e.g. [21, 22, 54]), where a ferromagnetic Ising model is used to denoise a partially observed image. Under the same ferromagnetic assumption, sometimes called attractive or log-supermodular in the machine learning community, [135, 157] showed that the stationary points of the Bethe free energy allow to *lower* bound the partition function (9.2). The proof of [157] relies on a loop series expansion of the partition function first derived by [30]. In this paper, we show how to *upper* bound the marginals, correlations and partition function of the ferromagnetic Ising model. Interestingly, our results also have a strong connection with the Bethe approximation (see section 9.2.3).

The difficulty of inference and learning in the Ising model (9.1) has prompted the development of a wealth of numerical methods to approximate the quantities (9.2, 9.3, 9.4). In the ferromagnetic case, a particularly important breakthrough came from the so-called cluster Monte Carlo methods of [147, 158]. By updating a whole cluster of spins instead of a single one, these algorithms can make non-local moves in the space of spin configurations, while still verifying the detailed balance condition. Therefore, they provided a substantial speed-up over conventional Markov Chain Monte Carlo methods, especially when the model (9.1) is near criticality. Other numerical methods are typically based on a low or high temperature expansion of the quantities of interest, and an exhaustive enumeration of an increasing number of the terms contributing to this expansion [25]. On the learning side of the problem, [31] introduced a principled and accurate approximation scheme based on the identification of the clusters of spins that contribute the most to the entropy of the model. While potentially allowing to reach an arbitrary accuracy in the determination of the quantities (9.2,9.3,9.4), these numerical approaches do not, by nature, admit a closed-form solution, and do not in general provide bounds on their estimates.

In this chapter, we prove explicit upper bounds on the three quantities of interest (9.2,9.3,9.4) for *arbitrary* graphs. These bounds are valid under two assumptions. First, we consider the particular case where the couplings $(J_{ij})_{(ij) \in E(G)}$ and the fields $(h_i)_{i \in [n]}$ are positive. Second, we require the model (9.1) to be in a well-defined high temperature region, specified by a condition on the spectral radius of the non-backtracking operator of model (9.1). Our results use the same starting point as some the intensive numerical methods described previously, but are much simpler in nature, and provide a simple and efficient closed-form bound, valid on any finite graph, as long as the couplings and the fields are positive. We therefore expect these results to find natural applications, e.g. in the inference and learning examples listed above, where the computation of quantities such as the partition function (9.2) and the expectations (9.3, 9.4) play a central role.

Our approach is based on the high temperature expansion of the Ising model, reviewed in section 9.3.1. For ferromagnetic models, this expansion is composed of positive contributions corresponding to cer-

*Numerical methods*

tain paths on the graph G. Our upper bounds are obtained by noting that the set of such paths is included in a tractable set of more general walks, for which an analytical expression can be derived. In this sense, our work is closely related to [44] who derived similar upper bounds. However, unlike [44] who derived formulas only for some regular lattices, our approach generalizes to arbitrary topologies and arbitrary positive couplings.

### 9.1.2  *Assumptions and definitions*

It will prove convenient to restrict our analysis to the Ising model in vanishing external fields, defined by

*Reduction to an Ising model in vanishing fields*

$$P(\sigma) = \frac{1}{\mathcal{Z}} \exp\left( \sum_{(ij) \in E(G)} J_{ij} \sigma_i \sigma_j \right) . \tag{9.5}$$

While this may look like a severe restriction, the addition of a single spin makes model (9.5) as expressive as model (9.1) [44]. More precisely, we will use proposition 1 in [70], which we recall here for completeness.

**Proposition 9.1.1.** *(proposition 1 in [70]) Consider the Ising model (9.1) on the graph* $G = ([n], E(G))$, *with couplings* $(J_{ij})_{(ij) \in E(G)}$ *and fields* $(h_i)_{i \in [n]}$, *and corresponding partition function* $\mathcal{Z}$, *magnetizations* $(m_a)_{a \in [n]}$, *and correlations* $(\chi_{ab})_{(a,b) \in [n]^2}$. *Define another Ising model on the graph* $\hat{G} = ([n+1], E(G) \cup \{(i, n+1)\}_{i \in [n]})$ *with vanishing fields, and couplings*

$$\hat{J}_{ij} = \begin{cases} J_{ij} \ if \ j < n+1 \\ h_i \ if \ j = n+1 \end{cases} . \tag{9.6}$$

*Call* $\hat{\mathcal{Z}}$ *its partition function, and* $(\hat{\chi}_{ab})_{(a,b) \in [n+1]^2}$ *its correlations. Then* $\hat{\mathcal{Z}} = 2\mathcal{Z}$ *and*

$$\hat{\chi}_{ab} = \begin{cases} \chi_{ab} \ if \ a < n+1 \ and \ b < n+1 \\ m_a \ if \ a < n+1 \ and \ b = n+1 \end{cases} . \tag{9.7}$$

We will therefore consider model (9.5) in the following, and bound its partition function and correlations, which will yield a bound on the partition function, magnetizations and correlations of model (9.1). Note also that from proposition 9.1.1, the new couplings $(\hat{J}_{ij})_{(ij) \in E(\hat{G})}$ are positive if and only if both the original couplings $(J_{ij})_{(ij) \in E(G)}$ and original fields $(h_i)_{i \in [n]}$ are positive.

*Non-backtracking operator and Bethe Hessian*

The upcoming results are expressed in terms of the non-backtracking operator and Bethe Hessian of model (9.5), which definition we now recall. For a given graph G, we denote throughout the chapter by E(G) its set of edges, and by $\vec{E}(G)$ its set of directed edges. The non-backtracking operator $B \in \mathbb{R}^{|\vec{E}(G)| \times |\vec{E}(G)|}$ of model (9.5) is defined by its elements, for $(i \rightarrow j), (k \rightarrow l) \in \vec{E}(G)$

$$B_{(i \rightarrow j),(k \rightarrow l)} = \tanh(J_{kl}) \, \mathbf{1}(i = l)\mathbf{1}(j \neq k) \, . \qquad (9.8)$$

In this chapter, we will call Bethe Hessian the matrix $H \in \mathbb{R}^{n \times n}$ defined by its elements [1]

$$H_{ij} = \mathbf{1}(i = j) \left( 1 + \sum_{k \in \partial i} \frac{\tanh(J_{ik})^2}{1 - \tanh(J_{ik})^2} \right) - \mathbf{1}(j \in \partial i) \frac{\tanh(J_{ij})}{1 - \tanh(J_{ij})^2} \, . \qquad (9.9)$$

As shown in section 2.3.2, this matrix is (the only non-trivial block of) the Hessian of the Bethe free energy at the paramagnetic fixed point of model (9.5). In other words, it is also the inverse of the susceptibility matrix, in the Bethe approximation.

Our results hold for arbitrary finite graphs $G$, provided the two following conditions hold.

*Assumptions*

— First, we assume that all the couplings $(J_{ij})_{(ij) \in E(G)}$ as well as the external fields $(h_i)_{i \in [n]}$ are non-negative, so that the Ising model in vanishing field (9.5) given by proposition 9.1.1 has positive couplings.

— Second, we assume that the model (9.1) is in a high temperature region, specified by the condition $\rho(B) < 1$. This condition corresponds to assuming that the paramagnetic fixed point of BP is stable.

## 9.2 MAIN RESULTS

We now state our main results and some of their consequences, leaving the proofs to the next section. Recall that throughout the chapter, we denote by $E(G)$ the set of edges of a graph $G$, and by $\vec{E}(G)$ its set of directed edges.

### 9.2.1 *Bound on the partition function*

Our first result is an upper bound on the partition function (9.2).

*Bound on the partition function*

**Theorem 9.2.1.** *Consider model (9.5) with positive couplings* $J_{ij} > 0$ *for* $(ij) \in E(G)$. *Assume that* $\rho(B) < 1$, *where* $B$ *is the non-backtracking matrix defined in (9.8), and let* $H$ *be the Bethe Hessian defined in (9.9). Then*

$$\begin{aligned} \mathcal{Z} &\leqslant 2^n \det(I - B)^{-1/2} \prod_{(ij) \in E(G)} \cosh(J_{ij}) \, , \\ &\leqslant 2^n \det(H)^{-1/2} \prod_{(ij) \in E(G)} \cosh(J_{ij})^2 \, . \end{aligned} \qquad (9.10)$$

______

1. This matrix corresponds to the Bethe Hessian $H(x)$ of equation (3.107) evaluated at $x = 1$.

Note that for graphs that are not sparse, the non-backtracking operator is a large $\left(|\vec{E}(G)| \times |\vec{E}(G)|\right)$, non-symmetric matrix, non-trivial to build or manipulate. The equality in (9.10) allows to compute this bound without having to build the non-backtracking matrix, by using instead the smaller $(n \times n)$ and symmetric Bethe Hessian. This last equality is a simple consequence of the Ihara-Bass formula 3.2.2, which states here that

$$\det(I - B) = \det(H) \prod_{(ij) \in E(G)} \cosh(J_{ij})^{-2}, \qquad (9.11)$$

where I is the identity matrix in dimension $|\vec{E}(G)|$. In particular, this formula implies that on any *finite* graph, H is non-singular under the condition $\rho(B) < 1$. In the thermodynamic limit $n \to \infty$, some subtleties arise, as discussed in the next section.

As will be apparent from the proof, this bound is based on an over counting of the subgraphs of G contributing to the high temperature expansion of the partition function (see section 9.3.1). These subgraphs correspond to closed paths (i.e. closed self-avoiding walks), and are notoriously hard to count [45]. To derive an analytical upper bound on the partition function, the set of contributing subgraphs must be included in a larger set of walks on the graph G whose contribution can be computed analytically. A simple bound can be obtained by including the set of closed paths in the set of all closed walks on the graph G. This approach yields an analytical bound similar to theorem 9.2.1 expressed in terms of the adjacency matrix of the graph, namely

$$\mathcal{Z} \leqslant 2^n \det(I - A)^{-1} \prod_{(ij) \in E(G)} \cosh(J_{ij}), \qquad (9.12)$$

whenever $\rho(A) < 1$, where A is the weighted adjacency matrix of the graph G, with entries $A_{ij} = \tanh(J_{ij}) \, \mathbf{1}(i \in \partial j)$ (see section 9.3.2 for details). This bound is however too loose, because it counts many spurious contributions, in particular walks that are allowed to go back and forth on the same edge. In particular, by the argument explained in the introduction of chapter 3, on sparse Erdős-Rényi random graphs, the spectral radius of A in unbounded with the size $n$ of the graph, so that the bound (9.12) is not useful. We considerably improve this bound by forbidding that the walk immediately backtracks to the previous edge. This is achieved by replacing the adjacency matrix with the non-backtracking operator. From the proof of section 9.3.2, it is straightforward to see that the bound (9.10) is less tight if the graph G contains more loops. In practice, however, figure 9.1 shows that using the non-backtracking operator instead of the adjacency matrix leads to a dramatic improvement, even in the case of a 3D lattice Ising model, which has many loops.

### 9.2.2 *Bound on the susceptibility*

We now state our upper bound on the correlations, encoded in the susceptibility matrix $\chi$ of equation (9.4).

**Theorem 9.2.2.** *Consider model (9.5) with positive couplings $J_{ij} > 0$ for $(ij) \in E(G)$. Assume that $\rho(B) < 1$, where $B$ is the matrix defined in (9.8). Define the matrices $P, Q \in \mathbb{R}^{n \times |\vec{E}(G)|}$ by their elements, for $a \in [n], (ij) \in E(G)$*

$$P_{a,(i \to j)} = \tanh(J_{ij})\mathbf{1}(a = j), \qquad Q_{a,(i \to j)} = \mathbf{1}(a = i). \tag{9.13}$$

*Then*

$$\chi \leqslant P(I - B)^{-1}Q^{\mathsf{T}} + I_{n \times n}, \tag{9.14}$$

*where the inequality holds element-wise.*

Note that $P$ is the pooling matrix (3.127) associated with the non-backtracking operator. Once more, it is possible to express this last result in terms of the Bethe Hessian rather than the non-backtracking operator, yielding a surprisingly simple result.

**Corollary 9.2.3.** *Let $H$ be the matrix defined in (9.9). Under the same assumptions as theorem 9.2.2, it holds that $H$ is invertible, and*

$$\chi \leqslant H^{-1}, \tag{9.15}$$

*where the inequality holds element-wise.*

This corollary results from applying proposition 3.2.1 to the inequality (9.14). As for the partition function, these bounds rely on the inclusion of the set of paths between two spins in a larger, tractable set of walks on the graph $G$ (see section 9.3.3). More precisely, expression (9.14) follows from the inclusion of the set of paths between two fixed spins $a, b \in [n]$ in the set of non-backtracking walks starting at $a$ and ending at $b$. If we allow these walks to backtrack, i.e. if we include the set of paths in the set of *all* walks, we get the looser bound

$$\chi \leqslant (I_n - A)^{-1}, \tag{9.16}$$

whenever $\rho(A) < 1$ where $A$ is the weighted adjacency matrix with elements $A_{ij} = \tanh(J_{ij})\mathbf{1}(i \in \partial j)$. Once more, forbidding backtracking allows to dramatically improve the bound on the susceptibility, as shown in figure 9.1. In fact, we have seen in section 2.3.3 that the Hessian of the nMF free energy at the paramagnetic point is given by $I - A$, so that equation (9.16) bounds $\chi$ by the nMF susceptibility. By contrast, the corollary (9.15) bounds $\chi$ by the susceptibility in the *Bethe* approximation. We come back to this point in section 9.2.3.

While valid only on finite graphs, the previous results allow us, in certain cases, to derive a bound on the paramagnetic to ferromagnetic phase transition of the Ising model. More precisely, for a sequence of Ising models, the previous results allow to bound the scalar susceptibility, defined as

$$\bar{\chi} = \frac{1}{n} \sum_{x,y=1}^{n} \chi_{xy} \, . \tag{9.17}$$

**Corollary 9.2.4.** *Consider a sequence* $(I_p)_{p \in \mathbb{N}}$ *of Ising models of the form* *(9.5), each of them defined on a graph* $G_p$*, with positive couplings* $J_{ij}^{(p)} > 0$*, for* $(ij) \in E(G_p)$*, and scalar susceptibility (9.17) denoted* $\bar{\chi}_p$*. Denote* $B_p$ *the non-backtracking operator defined in (9.8) for the Ising model* $I_p$*, and assume that* $\forall p \in \mathbb{N}$*,* $\rho(B_p) < 1$*. Define* $H_p$ *to be the Bethe Hessian defined in (9.9) for the Ising model* $I_p$*, and let* $\lambda_{\min}(H_p)$ *denote its smallest eigenvalue. Assume that there exists* $\epsilon > 0$ *such that* $\forall p \in \mathbb{N}$*,* $\lambda_{\min}(H_p) > \epsilon$*. Then for any* $p \in \mathbb{N}$*,*

$$\bar{\chi}_p \leqslant \frac{1}{\epsilon} \tag{9.18}$$

*Proof:* for any fixed $p \in \mathbb{N}$, let $n_p$ be the number of vertices of the graph $G_p$. Defining $U_p \in \mathbb{R}^{n_p}$ to be the vector with all its entries equal to 1, and denoting by $\|.\|_p$ the euclidean norm, we have from corollary 9.2.3

$$\bar{\chi}_p \leqslant \frac{1}{n_p} \sum_{x,y=1}^{n_p} (H_p)_{xy}^{-1} = \frac{U_p^{\mathsf{T}} H_p^{-1} U_p}{\|U_p\|_p^2} \leqslant \rho\left(H_p^{-1}\right) = \frac{1}{\lambda_{\min}(H_p)} \leqslant \frac{1}{\epsilon} \, , \tag{9.19}$$

where we have used that the matrix $H_p$ is symmetric.

*Bound on the paramagnetic to ferromagnetic phase transition*

For sequences of graphs that admit a thermodynamic limit, corollary 9.2.4 implies a condition under which the infinite Ising model is in the paramagnetic phase, therefore yielding a bound on the paramagnetic to ferromagnetic transition. As an explicit example, let us consider the case of a sequence $(G_p)_{p \in \mathbb{N}}$ of d-regular graphs with uniform couplings $J_{ij} = \beta$. We assume that the number of vertices in $G_p$ goes to infinity as $p \to \infty$. It is straightforward to check that, for any $p \in \mathbb{N}$,

$$\rho(B_p) = (d-1) \tanh(\beta) \qquad \text{and} \qquad \lambda_{\min}(H_p) = 1 - \frac{d \tanh(\beta)}{1 + \tanh(\beta)} \, . \tag{9.20}$$

By application of corollary 9.2.4, it follows that if

$$(d-1) \tanh(\beta) < 1 \, , \tag{9.21}$$

the scalar susceptibility $\bar{\chi}_p$ remains bounded. Equivalently, if the sequence of Ising models admits a thermodynamic limit with a phase transition at $\beta_c$, then

$$\beta_c \geqslant \operatorname{atanh} \frac{1}{d-1} \, . \tag{9.22}$$

Interestingly, the right hand side of the last equation corresponds to the critical inverse temperature in the Bethe approximation, which is already known to be a lower bound on $\beta_c$ for certain ferromagnetic models on hypercubic lattices [44]. Our contribution generalizes such previous results, giving a simple algorithm to derive a lower bound on arbitrary graphs, with arbitrary (positive) couplings. It is worth noting that, perhaps unsurprisingly, this bound is tight on sparse random d-regular graphs with uniform couplings $\forall (ij) \in E(G), J_{ij} = \beta$. Indeed, on such locally tree-like graphs, the existence of the thermodynamic limit has been proved, and the critical temperature has been shown to verify $(d-1)\tanh(\beta_c) = 1$ [35, 37].

From the Ihara-Bass formula (9.11), it holds that on any *finite* graph, $\lambda_{min}(H) > 0$ as long as $\rho(B) < 1$. Therefore one may expect that the condition on the smallest eigenvalue of $H_p$ in corollary 9.2.4 could be relaxed, e.g. by assuming instead that $\rho(B) < 1 - \epsilon$ for some $\epsilon > 0$. This relaxation turns out to be not possible, because it may happen that the spectral radius of B is bounded away from 1, while the smallest eigenvalue of H tends to 0, the limit $p \to \infty$. As an example, take $G_p$ to be the star graph with $n_p = p + 1$ spins and edges $(i, p + 1)$ for $i \in [p]$, and uniform couplings $J_{ij} = \beta$. Since $G_p$ is a tree, and the non-backtracking operator is nilpotent on trees, it holds that $\rho(B_p) = 0$ for all $p$. On the other hand, the spectrum of $H_p$ can be computed explicitly, and its smallest eigenvalue shown to verify

$$\lambda_{min}(H_p) \underset{p \to \infty}{\sim} \frac{1}{p \tanh(\beta)^2} \underset{p \to \infty}{\longrightarrow} 0. \tag{9.23}$$

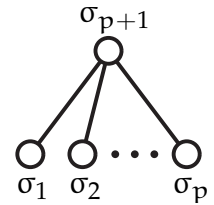

### 9.2.3 *Relation to belief propagation and susceptibility propagation*

Recall from section 2.2.2 that a standard approach to compute the magnetizations of the Ising model (9.5) is to use BP, which approximates the magnetizations $m_i = \mathbb{E}(\sigma_i)$ as

$$m_i \approx \tanh\left(\sum_{l \in \partial i} \text{atanh}(m_{l \to i} \tanh J_{il})\right), \tag{9.24}$$

where the messages $(m_{i \to j})_{(i \to j) \in \vec{E}(G)}$ verify the fixed point equation

$$m_{i \to j} = \tanh\left(\sum_{l \in \partial i \setminus j} \text{atanh}(m_{l \to i} \tanh J_{il})\right). \tag{9.25}$$

In practice, starting from a random initial condition, one iterates equation (9.25) until convergence, and outputs the result of (9.24). BP is known to be exact on trees, and widely believed to yield asymptotically accurate results for sparse, locally-tree like graphs with few, large loops, as well as models with small couplings [103].

This message-passing approach can be extended to allow the com-putation of the correlations $\chi_{ij}$ for any $i, j \in [n]$ using the fluctuation dissipation theorem. The resulting algorithm is called susceptibility propagation [104, 108], and approximates $\chi_{ij}$ as

$$\chi_{ij} \approx \left(1 - m_i^2\right) \left( \mathbf{1}(i = j) + \sum_{l \in \partial i} \frac{\chi_{l \to i, j} \tanh J_{il}}{1 - m_{l \to i}^2 \tanh^2 J_{il}} \right), \qquad (9.26)$$

where the $(m_{i \to j})_{(i \to j) \in \vec{E}(G)}$ are the fixed point of (9.25), the $(m_i)_{i \in [n]}$ are the BP estimates of the magnetization derived from (9.24) and the $(\chi_{i \to j, k})_{(i \to j) \in \vec{E}(G), k \in [n]}$ verify the fixed point equation

$$\chi_{i \to j, k} = \left(1 - m_{i \to j}^2\right) \left( \mathbf{1}(i = k) + \sum_{l \in \partial i \backslash j} \frac{\chi_{l \to i, k} \tanh J_{il}}{1 - m_{l \to i}^2 \tanh^2 J_{il}} \right), \qquad (9.27)$$

Similarly to BP, starting from a random initial condition, equation (9.27) is first iterated until convergence, and the correlations are then estimated from equation (9.26). Note that it is possible to invert the relation between the susceptibilities and the couplings, resulting in an inference algorithm for solving the inverse Ising model. This algorithm has been shown to yield better results than other mean field approaches on certain problems [108, 128].

While only exact on trees, both BP and susceptibility propagation have been observed to yield good approximate results on more general topologies, when they converge. However, these algorithms are based on the Bethe approximation [160], which, unlike the naive mean-field approach, does not provide bounds on the actual partition function [151]. The following result might therefore appear surprising.

*Behavior of susceptibility propagation in the paramagnetic phase of the ferromagnetic Ising model*

**Corollary 9.2.5.** *Consider model (9.5) with positive couplings $J_{ij} > 0$ for $(ij) \in E(G)$. Assume that $\rho(B) < 1$, where $B$ is the non-backtracking matrix defined in (9.8). Then $(m_{i \to j} = 0)_{(i \to j) \in \vec{E}(G)}$ is a stable fixed point of the BP recursion (9.25). Additionally, the corresponding susceptibility propagation algorithm converges, and yields an upper bound on the true correlations, regardless of the topology of the graph.*

*Proof:* the fact that $(m_{i \to j} = 0)_{(i \to j) \in \vec{E}(G)}$ is a fixed point of BP is readily checked on equation (9.25). To see that it is stable, we repeat the arguments developed at length in chapter 2. Starting from a small perturbation $\delta m_{i \to j}^0$, we have at first order at iteration $t \geqslant 1$

$$\delta m_{i \to j}^t = \sum_{l \in \partial i \backslash j} \tanh J_{il} \, \delta m_{l \to i}^{t-1}, \qquad (9.28)$$

which in matrix form can be written $\delta m^t = B \, \delta m^{t-1}$. The stability of this fixed point follows from the assumption $\rho(B) < 1$. The corresponding BP solution is $m_i = 0, \forall i \in [n]$. The corresponding susceptibility propagation recursion reads

$$\chi_{i \to j,k} = \mathbf{1}(i = k) + \sum_{l \in \partial i \setminus j} \chi_{l \to i,k} \tanh J_{il} . \qquad (9.29)$$

This defines a linear system which we now write in matrix form. For $k \in [n]$, define a vector $\chi_k \in \mathbb{R}^{|\vec{E}(G)|}$ with elements $\chi_{i \to j,k}$, for $(i \to j) \in \vec{E}(G)$, and call $Q_k$ the k-th line of the matrix $Q$ defined in theorem 9.2.2. Then equation (9.29) can be rewritten in matrix form as

$$\chi_k = Q_k^\intercal + B \chi_k . \qquad (9.30)$$

whose solution $\chi_k = (I - B)^{-1} Q_k^\intercal$ exists and is unique, since $\rho(B) < 1$. Iterating equation (9.29) starting from the initial condition $\chi_k^0$, we get at iteration $t \geqslant 1$

$$\chi_k^t - \chi_k = B \left( \chi_k^{t-1} - \chi_k \right) , \qquad (9.31)$$

so that $\chi_k^t \to \chi_k$ as $t \to \infty$, using again that $\rho(B) < 1$. Finally, using equation (9.26), it is straightforward to check that the correlations output by susceptibility propagation are given in matrix form by $P(I - B)^{-1} Q^\intercal + I_{n \times n}$, which is, by theorem 9.2.2, an upper bound on the true correlations [2].

## 9.3 PROOFS

We now detail the proof of our two main theorems, involving the non-backtracking operator. We start by reviewing the high temperature expansion of the Ising model. We then use it to prove theorems 9.2.1 and 9.2.2.

### 9.3.1 *High temperature expansion*

Central to our results is the high temperature expansion of the partition function, which we quickly rederive here. We use the following rewriting of the Boltzmann weight, relying on the fact that the spins are binary variables equal to $\pm 1$

$$\exp J_{ij} \sigma_i \sigma_j = a_{ij}(1 + b_{ij} \sigma_i \sigma_j) , \qquad (9.32)$$

---

2. The fact that, when susceptibility propagation converges, its estimates for the correlations can be expressed in terms of the inverse of the Bethe susceptibility, is still true outside of the paramagnetic phase, and was first noticed by [128].

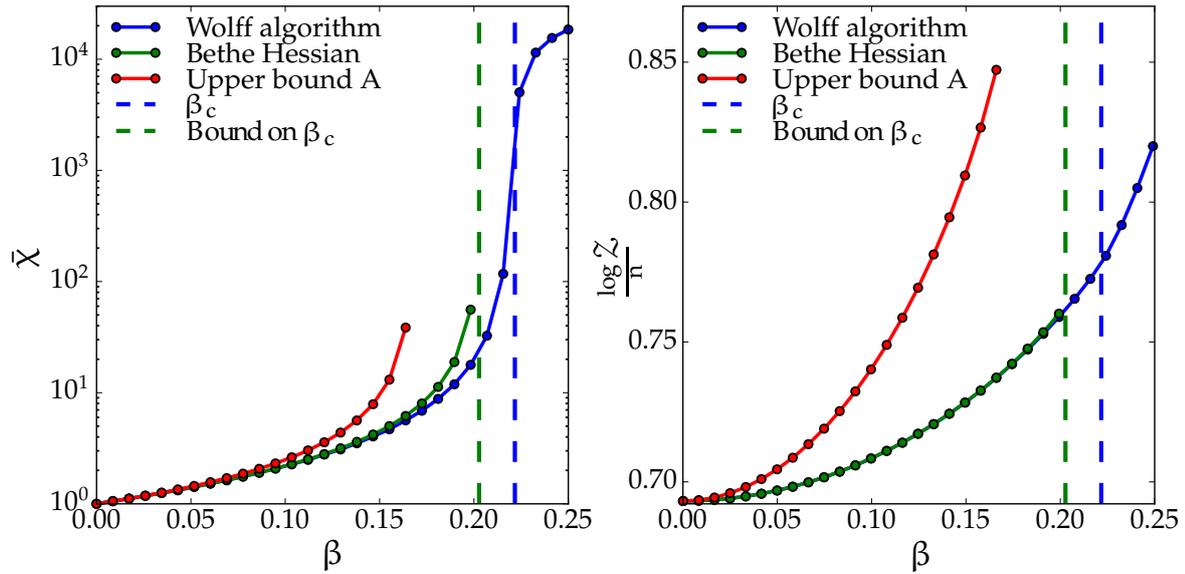

Figure 9.1 – Numerical simulation of the 3D lattice Ising model with $n = 32^3$ spins, with periodic boundary conditions, and uniform couplings $J_{ij} = \beta$. The left panel is the scalar susceptibility $\bar{\chi}$ of equation (9.17), and the right panel is the log of the partition function defined in equation (9.2). For both quantities we represented the exact value computed via a Monte Carlo simulation (using the Wolff algorithm [158]), and the upper bounds (9.10) and (9.15) expressed in terms of the Bethe Hessian operator. The dashed lines signal the critical temperature $\beta_c$ of the 3D Ising model as computed numerically in [127], and the lower bound on this critical temperature provided by the Bethe Hessian (9.22). Finally, we included for comparison the upper bound obtained by using the adjacency matrix $A$ instead of the non-backtracking matrix (equations (9.12) and (9.16)).

with $a_{ij} = \cosh(J_{ij})$, $b_{ij} = \tanh(J_{ij})$. The partition function of model (9.5) is given by

$$\mathcal{Z} = \sum_{\sigma \in \{\pm 1\}^n} \prod_{(ij) \in E(G)} \exp J_{ij} \sigma_i \sigma_j \tag{9.33}$$

$$= \left( \prod_{(ij) \in E(G)} a_{ij} \right) \sum_{\sigma \in \{\pm 1\}^n} \prod_{(ij) \in E(G)} (1 + b_{ij} \sigma_i \sigma_j) \tag{9.34}$$

$$= \left( \prod_{(ij) \in E(G)} a_{ij} \right) \sum_{\sigma \in \{\pm 1\}^n} \left[ 1 + \sum_{(ij) \in E(G)} b_{ij} \sigma_i \sigma_j \right.$$
$$\left. + \sum_{(ij),(kl) \in E(G)} b_{ij} b_{kl} \sigma_i \sigma_j \sigma_k \sigma_l + \cdots \right]. \tag{9.35}$$

After summing on the spin configurations $\sigma \in \{\pm 1\}^n$, the only terms in the expansion that do not vanish are the ones supported on a subgraph of G where all nodes have even degree. These are closed paths (not necessarily connected). Each of these closed paths contribute a factor $2^n$ to the partition function. We can therefore rewrite the partition function in the following form, called *high temperature expansion*

$$\mathcal{Z} = 2^n \left( \prod_{(ij) \in E(G)} a_{ij} \right) \left[ 1 + \sum_{g \in \mathcal{C}} \prod_{(ij) \in E(g)} b_{ij} \right], \tag{9.36}$$

where $\mathcal{C}$ is the set of closed paths, possibly disconnected.

### 9.3.2  *Proof of theorem 9.2.1*

We now introduce the set $\mathcal{C}_l^c$ of *connected* closed paths of length $l$. We denote by $\mathcal{W}_l^c$ the sum of all contributions to the partition function coming from connected closed paths of length $l \geqslant 1$, i.e.

$$\mathcal{W}_l^c = \sum_{g \in \mathcal{C}_l^c} \prod_{(ij) \in E(g)} b_{ij}. \tag{9.37}$$

We have the inequality

$$\mathcal{Z} \leqslant 2^n \left( \prod_{(ij) \in E(G)} a_{ij} \right)$$
$$\times \left[ 1 + \sum_{l \geqslant 1} \sum_{n_c \geqslant 1} \frac{1}{n_c!} \sum_{l_1 + l_2 + \cdots + l_{n_c} = l} \mathcal{W}_{l_1}^c \mathcal{W}_{l_2}^c \cdots \mathcal{W}_{l_{n_c}}^c \right],$$
$$\leqslant 2^n \left( \prod_{(ij) \in E(G)} a_{ij} \right) \left[ 1 + \sum_{n_c \geqslant 1} \frac{1}{n_c!} \left( \sum_{l \geqslant 1} \mathcal{W}_l^c \right)^{n_c} \right]. \tag{9.38}$$

Indeed, the product $\mathcal{W}_{l_1}^c \mathcal{W}_{l_2}^c \cdots \mathcal{W}_{l_{n_c}}^c$ contains all the contributions coming from disconnected closed paths whose connected components are of size $l_1, \cdots, l_{n_c}$. The factor $n_c!$ accounts for all the permutations of the factors in this product. This is only an inequality because we are counting graphs in which some edges appear more than once. However, the factors $\mathcal{W}_l^c$ are still hard to compute, and we will look for an upper bound. A simple upper bound can be derived by considering a weighted version of the adjacency matrix of the graph. Define $A \in \mathbb{R}^{n \times n}$ by its entries

$$A_{ij} = \begin{cases} b_{ij} & \text{if } (ij) \in E(G) \\ 0 & \text{otherwise} \end{cases}, \qquad (9.39)$$

then it holds that

$$\mathcal{W}_l^c \leqslant \frac{\mathrm{Tr} A^l}{l}. \qquad (9.40)$$

Indeed, the right hand side contains all contributions coming from closed walks that start and end at the same point. The factor $1/l$ accounts for the choice of the starting point. However, as hinted at in section 9.2.1, this upper bound is loose, because the adjacency matrix allows backtracking, and therefore going back and forth on the same edge, whereas such contributions do not appear in the partition sum. To improve this bound, we use a matrix that forbids backtracking: the operator $B$ of equation (9.8). Since connected closed paths are closed non-backtracking walks (although the converse is not true), it holds that

$$\mathcal{W}_l^c \leqslant \frac{\mathrm{Tr} B^l}{2l}. \qquad (9.41)$$

where the additional factor $1/2$ accounts for the degeneracy due to the orientation of the non-backtracking closed walk. Under the assumption that $\rho(B) < 1$, the following bound holds

$$\sum_{l \geqslant 1} \mathcal{W}_l^c \leqslant \sum_{l \geqslant 1} \frac{\mathrm{Tr} B^l}{2l} = -\frac{1}{2} \mathrm{Tr} \log(I - B) = -\frac{1}{2} \log \det(I - B), \qquad (9.42)$$

where $I \in \mathbb{R}^{|\vec{E}(G)| \times |\vec{E}(G)|}$ is the identity matrix. This, along with the Ihara-Bass formula (9.11) completes the proof of theorem 9.2.1.

### 9.3.3 *Proof of theorem 9.2.2*

The correlation functions are given, for any $x, y \in [n]$, by

$$\mathbb{E}(\sigma_x \sigma_y) = \frac{\sum_{\sigma \in \{\pm 1\}^n} \sigma_x \sigma_y \prod_{(ij) \in E(G)} (1 + b_{ij} \sigma_i \sigma_j)}{\sum_{\sigma \in \{\pm 1\}^n} \prod_{(ij) \in E(G)} (1 + b_{ij} \sigma_i \sigma_j)} = \frac{\sum_{g \in \mathcal{P}_{xy} + \mathcal{C}} \prod_{(ij) \in E(g)} b_{ij}}{1 + \sum_{g' \in \mathcal{C}} \prod_{(ij) \in E(g')} b_{ij}} \qquad (9.43)$$

where

— $\mathcal{P}_{xy}$ is the set of paths from $x$ to $y$.

— $\mathcal{C}$ is, as in the previous section, the set of closed paths.

— $\mathcal{P}_{xy} + \mathcal{C}$ is the set of diagrams made of a path from $x$ to $y$ and any number of disconnected closed paths such that each edge is selected at most once. The closed paths therefore do not intersect the path from $x$ to $y$.

We assume without loss of generality that $x \neq y$. From the previous definitions, we have that

$$\mathbb{E}(\sigma_x \sigma_y) \leqslant \frac{\left( \sum\limits_{g \in \mathcal{P}_{xy}} \prod\limits_{(ij) \in E(g)} b_{ij} \right) \left( 1 + \sum\limits_{g' \in \mathcal{C}} \prod\limits_{(ij) \in E(g')} b_{ij} \right)}{1 + \sum\limits_{g' \in \mathcal{C}} \prod\limits_{(ij) \in E(g')} b_{ij}}, \qquad (9.44)$$

$$\leqslant \sum\limits_{g \in \mathcal{P}_{xy}} \prod\limits_{(ij) \in E(g)} b_{ij}.$$

Indeed, developing the product at the numerator gives a sum of positive contributions including the ones in $\mathcal{P}_{xy} + \mathcal{C}$ and also (positive) spurious contributions coming from diagrams where the closed loops intersect the path from $x$ to $y$. We now introduce the set $\mathcal{N}^l_{(x \to x'),(y' \to y)}$ of non-backtracking walks of length $l$, starting on the directed edge $(x \to x')$ and terminating on the edge $(y' \to y)$. For a non-backtracking walk $w$, we denote by $\mathcal{E}(w)$ the list of edges crossed by $w$, where each edge appears with a multiplicity equal to the number of times $w$ crosses this edge. Since any path in $\mathcal{P}_{xy}$ is a non-backtracking walk (though the reverse is again, in the presence of loops, not true), it holds that

$$\mathbb{E}(\sigma_x \sigma_y) \leqslant \sum\limits_{l \geqslant 1} \sum\limits_{\substack{x' \in \partial x \\ y' \in \partial y}} \sum\limits_{w \in \mathcal{N}^l_{(x \to x'),(y' \to y)}} \prod\limits_{(ij) \in \mathcal{E}(w)} b_{ij} \qquad (9.45)$$

where $\partial x$ denotes the set of neighbors of $x$ in the graph $G$. In order to write this last expression in terms of the non-backtracking operator, we introduce the vector $u_{x \to x'} \in \mathbb{R}^{|\vec{E}(G)|}$ with entries all equal to $0$ except for the $(x \to x')$ entry, which is equal to 1. Similarly, we introduce the vector $v_{y' \to y} \in \mathbb{R}^{|\vec{E}(G)|}$ whose only non-zero entry is equal to $b_{yy'}$, in position $(y \to y')$. Then we have

$$\sum\limits_{g \in \mathcal{N}^l_{(x \to x'),(y' \to y)}} \prod\limits_{(ij) \in E(g)} b_{ij} = v_{y' \to y}^\top B^{l-1} u_{x \to x'}, \qquad (9.46)$$

so that

$$\mathbb{E}(\sigma_x \sigma_y) \leqslant \left( \sum\limits_{y' \in \partial y} v_{y' \to y} \right)^\top \sum\limits_{l \geqslant 1} B^{l-1} \left( \sum\limits_{x' \in \partial x} u_{x \to x'} \right) \qquad (9.47)$$

$$\leqslant \left( \sum\limits_{y' \in \partial y} v_{y' \to y} \right)^\top (I - B)^{-1} \left( \sum\limits_{x' \in \partial x} u_{x \to x'} \right) \qquad (9.48)$$

where we have used the assumption $\rho(B) < 1$, so that the series of powers of $B$ is summable. To write this last equation in a more compact way, recall the definition of the *susceptibility matrix* $\chi \in \mathbb{R}^{n \times n}$, with elements $\chi_{xy} = \mathbb{E}(\sigma_x \sigma_y)$. Then it holds element-wise that

$$\chi \leqslant P(I - B)^{-1}Q^{\mathsf{T}} + I_{n \times n} \tag{9.49}$$

where $P, Q \in \mathbb{R}^{n \times |\vec{E}(G)|}$ are defined in equation (9.13). Note that the addition of an identity matrix ensures that the inequality also holds on the diagonal of $\chi$. This completes the proof of theorem 9.2.2.

## 9.4 CONCLUSION

In this chapter, we have considered the ferromagnetic Ising model, and used the non-backtracking operator and Bethe Hessian to derive rigorous bounds on the partition function, the magnetizations, and the correlations, valid on arbitrary graphs, in a certain high temperature region specified by the spectral radius of the non-backtracking operator. Our approach builds upon previous work by explicitly computing an upper bound which was known, but for which a tractable expression existed only in very special cases. By contrast, our bounds can be easily implemented and efficiently computed on arbitrary finite graphs, with arbitrary (positive) couplings. As a by-product, we were able to show that, in the same high temperature region, the popular susceptibility propagation algorithm converges to approximate correlations that admit an analytical expression, and that these correlations are an upper bound on the true correlations. In the next, and final chapter of this dissertation, we hint at a possible application of some of the previous ideas to a *learning* problem.

# OUTLOOK: LEARNING AND THE NATURAL GRADIENT

Throughout the dissertation, we have been interested exclusively in *inference*, i. e. the task of computing the marginals of a given pairwise MRF. In this last, short chapter, we hint at the possible applicability of some of the ideas developed previously in the context of *learning*, where the aim is to recover a pairwise MRF from a dataset assumed to be sampled from it. After presenting succinctly the learning problem in exponential families, we introduce the concept of natural gradient, which modifies the usual maximum likelihood learning rule to take into account information geometry considerations. We show how to give a mean-field approximation to the natural gradient, and illustrate our computation on a binary restricted Boltzmann Machine (RBM). Our very preliminary results still lack an empirical validation, but may hopefully serve as a basis for future investigations.

## 10.1 LEARNING IN EXPONENTIAL MODELS

We consider here an exponential model with so-called *canonical* parameters $\theta = (\theta_\alpha)_{\alpha \in \mathcal{I}} \in \mathbb{R}^{|\mathcal{I}|}$, for some index set $\mathcal{I}$, defined by the joint probability distribution

$$\mathbb{P}_\theta(\sigma) = \exp\left(\langle \theta, f(\sigma) \rangle + \mathcal{F}(\theta)\right) \tag{10.1}$$

*Exponential model*

where $\sigma \in \mathcal{X}^n$ is a collection of random variables, $f(\sigma) = (f_\alpha(\sigma))_{\alpha \in \mathcal{I}} \in \mathbb{R}^{|\mathcal{I}|}$ is a vector-valued function of $\sigma$ called *sufficient statistics*, and the angular brackets denote the canonical inner product in $\mathbb{R}^{|\mathcal{I}|}$. Recall that, as seen in section 1.3.2, the exponential model (10.1) is the maximum entropy distribution with constrained expectations

$$\mathbb{E}_{\mathbb{P}_\theta}[f_\alpha(\sigma)], \quad \text{for } \alpha \in \mathcal{I}. \tag{10.2}$$

In particular, when the sufficient statistics $f(\sigma)$ contain only functions of single variables $\sigma_i$ or pairs of variables $(\sigma_i, \sigma_j)$, the model (10.1) defines a pairwise MRF in its exponential representation (1.13). The quantity $\mathcal{F}(\theta)$ enforces the normalization of the probability distribution and equals the Helmholtz free energy of the model

$$\mathcal{F}(\theta) = -\log\left(\sum_{\sigma \in \mathcal{X}^n} \exp\langle \theta, f(\sigma) \rangle\right) \tag{10.3}$$

*Helmholtz free energy*

We assume that we are given $p$ i.i.d samples $\hat{\sigma} = \left(\hat{\sigma}^{(i)}\right)_{i \in [p]}$ drawn from the distribution (10.1), and we are looking to *learn* the param-eter $\theta$ from these samples. It can be shown [152] that the optimal estimator $\hat{\theta}$ of $\theta$ that minimizes the mean square error is given by

*Optimal estimator of θ*

$$\hat{\theta} = \text{argmax}_\theta \, \ell\,(\theta, \hat{\sigma}) \,, \qquad (10.4)$$

where $\ell\,(\theta, \hat{\sigma})$ is the *log-likelihood* of the observed samples, given by

*Log-likelihood*

$$\ell\,(\theta, \hat{\sigma}) = \frac{1}{p} \sum_{i=1}^{p} \log \mathbb{P}_\theta \left( \hat{\sigma}^{(i)} \right) \,, \qquad (10.5)$$

$$= \langle \theta, \hat{\mu} \rangle + \mathcal{F}(\theta) \,, \qquad (10.6)$$

where $\hat{\mu} = (\hat{\mu}_\alpha)_{\alpha \in \mathcal{I}} \in \mathbb{R}^{|\mathcal{I}|}$ is a vector of empirical means

$$\hat{\mu}_\alpha = \frac{1}{p} \sum_{i=1}^{p} f_\alpha \left( \hat{\sigma}^{(i)} \right) \,. \qquad (10.7)$$

*Gradient ascent*

To maximize the log-likelihood in (10.5), a popular algorithm is (stochastic [1]) gradient ascent. With a small, time-dependent *learning rate* $\eta_t$, we iteratively update $\theta^{t+1} \leftarrow \theta^t + \Delta\theta^t$ where for $\alpha \in \mathcal{I}$,

*Classical learning rule*

$$\Delta\theta_\alpha^t = \eta_t \, (\nabla \ell\,(\theta, \hat{\sigma}))_\alpha \,,$$
$$= \eta_t \left( \hat{\mu}_\alpha - \mathbb{E}_{\mathbb{P}_{\theta^t}} \left[ f_\alpha(\sigma) \right] \right) \,, \qquad (10.8)$$

where we have used that

$$\frac{\partial \mathcal{F}(\theta)}{\partial \theta_\alpha} = -\mathbb{E}_{\mathbb{P}_{\theta^t}} \left[ f_\alpha(\sigma) \right] \,. \qquad (10.9)$$

In particular, the algorithm has converged when the moments of the exponential distribution (10.1) match the empirical means (10.7).

## 10.2 NATURAL GRADIENT

*Rationale behind the classical learning rule*

It is easy, but instructive, to re-derive heuristically the previous learning rule. At each iteration $t$, we are looking for a *small* move $\Delta\theta^t$ in the Euclidean space $\mathbb{R}^{|\mathcal{I}|}$ that maximally increases the log-likelihood. Therefore, we are solving

$$\Delta\theta^t = \underset{\substack{\Delta\theta \in \mathbb{R}^{|\mathcal{I}|} \\ \|\Delta\theta\| = \epsilon}}{\text{argmax}} \, \ell\left( \theta^t + \Delta\theta, \hat{\sigma} \right) \,, \qquad (10.10)$$

where $\|\Delta\theta\|$ is the Euclidean norm of $\Delta\theta$. We are therefore solving a constrained optimization problem with Lagrangian

$$\mathcal{L} = \ell\left( \theta^t + \Delta\theta, \hat{\sigma} \right) - \lambda \left( \|\Delta\theta\|^2 - \epsilon^2 \right) \,,$$
$$\approx \ell(\theta^t) + \Delta\theta \, \nabla \ell\,(\theta, \hat{\sigma}) - \lambda \left( \|\Delta\theta\|^2 - \epsilon^2 \right) \,. \qquad (10.11)$$

---

1. Stochastic gradient ascent considers, at each iteration, a random subset of the samples, called a mini-batch.

where $\lambda$ is a Lagrange multiplier, and we have used that $\epsilon$ is small to expand, to first order, the log-likelihood. Note that we neglect here the second order term in the expansion of $\ell$, which would give us a (regularized) Newton algorithm. Equating the gradient of this Lagrangian to $0$, we obtain the classical learning rule

$$\Delta\theta \propto \nabla\ell(\theta, \hat{\sigma}) . \tag{10.12}$$

*The natural gradient approach*

When deriving this learning rule, we have identified our probability distribution with its vector $\theta$ of canonical parameters, i. e. with an element of an Euclidean space. A fruitful approach, pioneered by Amari [9], is to consider instead that the optimization takes place on the manifold of exponential models of the form (10.1). In other words, instead of considering that we optimize over $\theta$, we optimize directly over the probability distribution $\mathbb{P}_\theta$. The geometry of this manifold is non-trivial, as there is no clear notion of distance between probability distributions (recall that the KL-divergence introduced in section 1.5.2 is not symmetric). Still, we can endow the manifold of exponential models of the form (10.1) with a local Riemannian structure by showing [9, 123] that, locally, the KL-divergence defines a metric $F$ given in matrix form by

$$D_{KL}(\mathbb{P}_\theta \,\|\, \mathbb{P}_{\theta+\Delta\theta}) \underset{\Delta\theta\to 0}{\sim} D_{KL}(\mathbb{P}_{\theta+\Delta\theta} \,\|\, \mathbb{P}_\theta) \underset{\Delta\theta\to 0}{\sim} \frac{1}{2}\Delta\theta^\intercal F\,\Delta\theta , \tag{10.13}$$

where $F$ is the *Fisher information matrix* (hereafter the Fisher)

$$F = \mathbb{E}_{\mathbb{P}_\theta}\left[-\frac{\partial^2 \log \mathbb{P}_\theta(\sigma)}{\partial\theta^2}\right] . \tag{10.14}$$

The natural gradient approach seeks to find a small update $\Delta\theta$ that minimizes $\ell(\theta + \Delta\theta, \hat{\sigma})$ under the constraint that $D_{KL}(\mathbb{P}_\theta \,\|\, \mathbb{P}_{\theta+\Delta\theta})$ (rather than the euclidean norm of $\Delta\theta$) is fixed. The Lagrangian (10.11) then becomes

$$\mathcal{L} \approx \ell(\theta^t) + \Delta\theta\,\nabla\ell(\theta, \hat{\sigma}) - \lambda\left(\frac{1}{2}\Delta\theta^\intercal F\,\Delta\theta - \epsilon^2\right) , \tag{10.15}$$

so that the update rule, with a time-dependent learning rate, becomes

*Natural gradient learning rule*

$$\Delta\theta^t = \eta_t\,F^{-1}\nabla\ell(\theta) . \tag{10.16}$$

The natural gradient approach enjoys desirable properties such as invariance with respect to the parametrization of the exponential model (10.1) [9], and it has proven to speed up learning in various problems [9, 123, 134]. One major drawback, however, of this method is that it requires to store and, more importantly, to invert the very large matrix $F$. Several approximation schemes have therefore been proposed. Notably, the authors of [36] tune the architecture of a deep neural network in such a way that the natural gradient is approximately equal to the classical gradient. As a consequence, the usual (stochastic) gradient ascent learning rule (10.8) is approximately identical to the natural gradient learning rule. The authors report an impressive improvement in training time on various popular datasets. In the rest of this chapter, we would like to argue that we can compute directly an approximation of the *inverse* Fisher $F^{-1}$ using mean-field approximations, eliminating the need to invert it. Additionally, our approximation turns out to be sparse, making its implementation efficient both in terms of time and memory complexities.

*Equivalence with Newton's method on exponential models*

It is worth noting that in the case of exponential models defined by (10.1), the natural gradient approach is equivalent to Newton's method. Indeed,

$$F = \mathbb{E}_{\mathbb{P}_\theta}\left[-\frac{\partial^2 \log \mathbb{P}_\theta(\sigma)}{\partial \theta^2}\right] \tag{10.17}$$

$$= -\frac{\partial^2 \mathcal{F}(\theta)}{\partial \theta^2} \tag{10.18}$$

$$= -\frac{\partial^2 \ell(\theta, X)}{\partial \theta^2} \tag{10.19}$$

so that $-F$ is in fact the Hessian of the objective function. We note that this fact is not general [134] and stems from the log-linear nature of model (10.1).

## 10.3   MEAN-FIELD APPROXIMATION TO THE NATURAL GRADIENT

*Legendre transform and entropy*

Using the variational approach developed in section 1.5.2, it can be shown (see chapter 3 of [151] for the details) that the Helmholtz free energy $\mathcal{F}(\theta)$ is related to entropy by the following Legendre transform

$$\mathcal{F}(\theta) = \inf_{\mu \in \mathcal{M}} \{-\langle \theta, \mu \rangle - S(\mu)\} \tag{10.20}$$

where, with an abuse of notations, we write $S(\mu)$ for the entropy of the distribution $\mathbb{P}_{\theta(\mu)}$, where $\theta(\mu)$ is defined by the conditions

$$\mathbb{E}_{\mathbb{P}_{\theta(\mu)}}[f_\alpha(\sigma)] = \mu_\alpha, \qquad \forall \alpha \in \mathcal{I}. \tag{10.21}$$

In equation (10.20), $\mathcal{M}$ is the set of *realizable* mean parameters, i.e. the set of mean parameters $\mu$ that can be written as the moments of a certain exponential model

$$\mathcal{M} = \{\mu \in \mathbb{R}^{|\mathcal{I}|} \mid \exists \theta \in \mathbb{R}^{|\mathcal{I}|} \text{ such that } \forall \alpha \in \mathcal{I}, \ \mathbb{E}_{\mathbb{P}_\theta}[f_\alpha(\sigma)] = \mu_\alpha\} \tag{10.22}$$

The main ingredient of our approach is the following standard result in conjugate duality

*Inverse of the Fisher and Hessian of the entropy*

$$F^{-1} = -\left(\frac{\partial^2 \mathcal{F}(\theta)}{\partial \theta^2}\right)^{-1} = -\frac{\partial^2 S(\mu)}{\partial \mu^2}\bigg|_{\mu = \mu^\star}, \tag{10.23}$$

where the last derivative has to be evaluated at $\mu^\star$ such that

$$\nabla \mathcal{F}(\theta) = -\mu^\star \quad \Longleftrightarrow \quad \frac{\partial S(\mu)}{\partial \mu}\bigg|_{\mu = \mu^\star} = -\theta. \tag{10.24}$$

We therefore have an explicit formula for the inverse of the Fisher. Unfortunately, computing this entropy is still intractable in any interesting problem. However, we have seen various mean-field approximations to the entropy in chapter 1, and we illustrate here how they can be used to approximate the inverse Fisher. For instance, within the Bethe approximation [2], we may try to approximate equation (10.23) as

$$F^{-1} \approx -\frac{\partial^2 S^{\text{Bethe}}(\mu)}{\partial \mu^2}, \tag{10.25}$$

where the derivative is evaluated at $\mu^\star$ given by equation (10.24), which translates, in our approximation, as

$$\left.\frac{\partial S^{\text{Bethe}}(\mu)}{\partial \mu}\right|_{\mu=\mu^\star} = -\theta. \tag{10.26}$$

In other words, $\mu^\star$ is a stationary point of the Bethe free energy, since the previous relation can be written as

$$\left.\frac{\partial \mathcal{F}^{\text{Bethe}}(\theta, \mu)}{\partial \mu}\right|_{\mu=\mu^\star} = 0, \tag{10.27}$$

where by its definition (1.64), the Bethe free energy reads

$$\mathcal{F}^{\text{Bethe}}(\theta, \mu) = -\langle \theta, \mu \rangle - S^{\text{Bethe}}(\mu). \tag{10.28}$$

Note that we have made here the dependence of $\mathcal{F}^{\text{Bethe}}$ on the canonical parameters explicit. Finally, since the internal energy $-\langle \theta, \mu \rangle$ is linear in $\theta$, we may rewrite equation (10.25) as

$$F^{-1} \approx \left.\frac{\partial^2 \mathcal{F}^{\text{Bethe}}(\theta, \mu)}{\partial \mu^2}\right|_{\mu=\mu^\star}, \tag{10.29}$$

where $\mu^\star$ is a local minimum [3] of the Bethe free energy (over a corresponding set $\mathcal{M}$ of realizable parameters translating the consistency constraints (1.63) on the beliefs, which we write explicitly on an example in the following). We have therefore expressed the natural gradient in terms of the Hessian of the Bethe free energy, a matrix we have become well-acquainted with. Note that contrary to our past approaches, we evaluate here the Hessian at a non-trivial value of $\mu$, which is always a local minimum of the Bethe free energy, so that this Hessian should always be positive definite.

---

2. These arguments can be adapted to other mean-field approximations.

3. The "true" free energy $-\langle \theta, \mu \rangle - S(\mu)$ where $\mu$ is defined by (10.21) is convex, and has a unique stationary point specified by the condition (10.24), which is also a global minimum. In our mean-field approximation, we have to deal with the non-convex Bethe free energy, which may have multiple saddle points and local minima that all verify the condition (10.27), so that it is not obvious how to make sense of this definition of $\mu^\star$. It is reasonable to require, in addition to the stationarity condition (10.27), that $\mu^\star$ be a local minimum (not a saddle) of $\mathcal{F}^{\text{Bethe}}$, since the Fisher should be positive definite in well-defined problems.

Let us illustrate our approach on a simple binary example. We consider a Boltzmann machine (i. e. an Ising model) on a graph $G = ([n], E)$, where $E$ is the set of edges of the corresponding pairwise MRF. The Boltzmann distribution is then given by

*Boltzmann machine*

$$\mathbb{P}_\theta(\sigma) = \exp\left(\sum_{i=1}^{n} \theta_i \sigma_i + \sum_{(ij) \in E} \theta_{ij} \sigma_i \sigma_j + \mathcal{F}(\theta)\right). \quad (10.30)$$

where the sufficient statistics are here $f_i(x) = \sigma_i, f_{ij}(\sigma) = \sigma_i \sigma_j$ for $i \in [n], (ij) \in E$. As is usual in the machine learning literature, we assume here that $\sigma_i \in \{0, 1\}$ for $i \in [n]$. The canonical parameters are here the $\theta_i, \theta_{ij}$ for $i \in [n], (ij) \in E$. The Bethe free energy can be written in terms of mean parameters $\mu_i, \mu_{ij}$, for $i \in [n], (ij) \in E$ as[4] ([154])

*$\sigma_i \in \{0, 1\}$*

*Bethe free energy of the Boltzmann machine*

$$\mathcal{F}^{\text{Bethe}}(\theta, \mu) = \sum_{i=1}^{n} \theta_i \mu_i + \sum_{(ij) \in E} \theta_{ij} \mu_{ij} + \sum_{i=1}^{n} (1 - |\partial_i|) \left[\eta(\mu_i) + \eta(1 - \mu_i)\right]$$
$$+ \sum_{(ij) \in E} \eta(\mu_{ij}) + \eta(\mu_{ij} + 1 - \mu_i - \mu_j) + \eta(\mu_i - \mu_{ij}) + \eta(\mu_j - \mu_{ij}). \quad (10.31)$$

where $|\partial_i|$ is the degree of vertex $i$ in the graph $G$, and $\eta(x) = x \log x$. The set of realizable mean parameters is here ([154])

*Realizable mean parameters*

$$\mathcal{M} = \Big\{(\mu_i, \mu_{ij}) \in [0, 1]^{n + |E|} \mid \forall (i, j) \in E,$$
$$\max(0, \mu_i + \mu_j - 1) \leqslant \mu_{ij} \leqslant \min(\mu_i, \mu_j)\Big\}. \quad (10.32)$$

Computing the Hessian of the Bethe free energy is then a simple matter of algebra. We find

$$\frac{\partial^2 \mathcal{F}^{\text{Bethe}}(\theta, \mu)}{\partial \mu_i \mu_j} = \mathbf{1}(i = j)\frac{1 - d_i}{\mu_i(1 - \mu_i)} + \frac{\mathbf{1}(j \in \partial i)}{\mu_{ij} + 1 - \mu_i - \mu_j}, \quad (10.33)$$

*Hessian of the Bethe free energy*

$$\frac{\partial^2 \mathcal{F}^{\text{Bethe}}(\theta, \mu)}{\partial \mu_{ij} \mu_k} = -\mathbf{1}(i = k)\frac{1 - \mu_j}{(\mu_i - \mu_{ij})(\mu_{ij} + 1 - \mu_i - \mu_j)},$$
$$-\mathbf{1}(j = k)\frac{1 - \mu_i}{(\mu_j - \mu_{ij})(\mu_{ij} + 1 - \mu_i - \mu_j)} \quad (10.34)$$

$$\frac{\partial^2 \mathcal{F}^{\text{Bethe}}(\theta, \mu)}{\partial \mu_{ij} \mu_{kl}} = \frac{\mathbf{1}(i = k)\mathbf{1}(j = l)}{\mu_{ij}(\mu_i - \mu_{ij})(\mu_j - \mu_{ij})(\mu_{ij} + 1 - \mu_i - \mu_j)}. \quad (10.35)$$

To compute the natural gradient, one only needs to find a local minimum $\mu^\star$ of the Bethe free energy, and evaluate the previous Hessian at $\mu^\star$. This local minimum can be computed in various ways, from the BP algorithm (which however may output a saddle point of the Bethe free energy), to the direct (constrained) optimization of the

---

4. Note that this expression is slightly different from the one used in section 2.3.2 because we consider here "spins" $\sigma_i \in \{0, 1\}$.

Bethe free energy on the set $\mathcal{M}$ [154]. A simple and efficient approach is to use the TAP approximation of [47], which approximates (to good accuracy when the couplings $\theta_{ij}$ for $(ij) \in E$ are small enough [5]) the stationary points of the Bethe free energy [154].

Finally, we note that although the Hessian (10.33) is of size $(n + |E|)^2$, *Sparsity of the* it is very sparse (even when $G$ is a dense graph), and contains only *Hessian* $n + 7|E|$ non-zero elements. For example, for a RBM [6] trained on the MNIST dataset of handwritten digits, with $n_v = 784$ visible units and $n_h = 500$ hidden units [7], the Hessian is of size $(500 + 784 + 500 \times 784)^2 = 393284^2$, but it contains only $2745284$ non-zero entries, i.e. less than $0.002\%$ of the total number of entries. Since the natural gradient learning rule requires multiplying the gradient of the log-likelihood by this matrix at each iteration, the sparsity of the Hessian is a very important feature in practice.

## 10.4 CONCLUSION

In this chapter, we have outlined a possible application of the ideas developed previously in this dissertation to a learning problem. In particular, we have argued that the Hessian of the Bethe free energy may allow to approximate the natural gradient, thus providing a more efficient learning rule. We ran some preliminary numerical tests of the MNIST dataset of handwritten digits, and found that this mean-field natural gradient approach allows to train a RBM to good accuracy. It is however still unclear whether this approach improves the speed of convergence, compared to the classical contrastive divergence approach [64] (which uses a short Markov chain to sample from the Boltzmann distribution), or the TAP approach of [47]. The problem we found is that, as is well-known, the training speed largely depends on the learning rate $\eta$ in all of these methods, which should be chosen in different ways depending on the precise learning rule. Before being able to perform a consistent comparison between these methods, it is important to find a principled way to fix this learning rate. This problem is interesting both in theory and in practice, and should be the focus of future work.

---

5. It is good practice to regularize the log-likelihood by adding a penalty term favoring small couplings.

6. A RBM is defined on the complete bipartite graph with $n_v$ visible spins (or neurons) and $n_h$ hidden ones.

7. When dealing with hidden units $\sigma = (v, h)$, various approaches to the natural gradient are possible [123]. We could ascend the manifold of probability distributions on the visible units by integrating out the hidden units, i.e. $\mathbb{P}_\theta(v) = \sum_h \mathbb{P}_\theta(v, h)$. We choose instead to consider the manifold of joint probability distributions $\mathbb{P}_\theta(v, h)$.

## CONCLUDING REMARKS

In this dissertation, we have considered the problem of estimating the marginals of a pairwise MRF using spectral methods derived from mean-field approximations. We have identified a class of factorized and symmetric models for which BP possesses a trivial, uninformative fixed point. By analyzing the stability of this fixed point, we have uncovered a phase transition expressed in terms of the spectral radius of a certain operator, called the non-backtracking operator. We have linked this operator to another, smaller and symmetric matrix which we called the Bethe Hessian. This latter operator controls the stability of the paramagnetic phase of an Ising model associated with the original pairwise MRF, in the Bethe approximation. By studying the spectral properties of these two operators on random graphs, as well as their relationship to each other, we have argued that both allowed to design accurate and efficient spectral algorithms on sparse graphs. We have finally illustrated the performance of these algorithms on various problems, in the fields of community detection, similarity-based unsupervised and semi-supervised clustering, and matrix completion. In some cases, we have been able to produce rigorous guarantees for the performance of the algorithms based on the non-backtracking operator.

There are various natural directions for further investigations. On the theoretical side, although we have presented several indirect arguments for the optimality of the Bethe Hessian, we still lack rigorous guarantees. In addition to its impact on the inference problems where the Bethe Hessian is applicable, a rigorous result on the negative eigenvalues of this operator might also have consequences on (part of) the phase diagram of sparse Ising spin glasses, which are notoriously hard to obtain. For the non-backtracking operator, the situation is better. Relying heavily on the techniques developed in [19] in the case of the SBM, we were able to sketch rigorous arguments concerning the spectrum of the (weighted) non-backtracking operator of the lSBM, in the special case of $q = 2$ clusters, with $\alpha_{in} = \alpha_{out}$, i. e. in the case where the graph is a homogeneous weighted Erdős-Rényi random graph. Adapting these results to a general instance of the lSBM (i. e. proving the conjectures of section 3.1.3) is still an interesting open problem, which should be attainable by adapting the techniques developed in [19] and extended in this dissertation.

On the practical side, one of the most pressing point is to understand how to adapt the methods presented here to more realistic graphs. In particular, it was shown in [68], perhaps unsurprisingly, that the methods presented here, which rely on the Bethe approxi-mation, are very sensitive to the addition of loops in the graph. A possible way out, based on more complex Kikuchi approximations, was hinted at in section 2.3.3. Another possibility is to systematically improve the mean-field approximation used with adaptive methods such as those presented by [120] in the context of the TAP approximation. Note that when we get to *choose* the graph, *e. g.* in the context of randomly subsampled clustering as in chapters 6 and 7, the issue raised by [68] does not apply. Finally, it would be interesting to find more applications of the ideas developed here. One idea, sketched as an outlook in chapter 10, is to use free energy Hessians to approximate the natural gradient and speed up learning. This direction is attractive, since the use of ideas closely related to the natural gradient has recently been shown to provide important improvements in deep learning applications [36]. More generally, the ideas developed in this dissertation can be though of as an approximate optimization scheme. As argued in section 1.7, when given a cost function (or energy) $E$, we may define a probabilistic model of the form

$$\mathbb{P}(\sigma) \propto \exp\left(-\beta\, E(\sigma)\right) .$$

It is reasonable to expect that, for a carefully chosen range of values of $\beta$, the distribution thus defined will concentrate around configurations that are good approximations to the minimizers of $E$. We may therefore minimize $E$ by computing approximate marginals using the methods introduced in this dissertation. Finding interesting use cases, with accurate corresponding mean-field approximations suited to the structure of the energy $E$, may provide a promising family of extensions of the present work.

# APPENDIX: JACOBIAN OF BELIEF PROPAGATION AT THE TRIVIAL FIXED POINT OF FACTORIZED MODELS

We compute the Jacobian $\mathcal{J}$ of the function $F$, whose components are defined in equation (2.15), at the fixed point $b^\star$ of equation (2.11). We use throughout the computation the symmetry property of pairwise potentials, equation (1.12), i.e. $\psi_{lk}(\sigma, \sigma') = \psi_{kl}(\sigma', \sigma)$ for any $(kl) \in E, \sigma, \sigma' \in \mathcal{X}$. We have for $(i \to j), (k \to l) \in \vec{E}, \sigma, \sigma' \in \mathcal{X}$,

$$\mathcal{J}_{i \to j, k \to l}^{(\sigma, \sigma')} = \frac{\partial f_{i \to j}^{(\sigma)}(b)}{\partial b_{k \to l}(\sigma')}\Bigg|_{b = b^\star}$$

By the definition of $f_{i \to j}^{(\sigma)}$ in equation (2.15), this quantity is non-zero only if $l = i$ and $k \neq j$. In this case we have, for all $k \in \partial i \backslash j, \sigma, \sigma' \in \mathcal{X}$,

$$\frac{\partial f_{i \to j}^{(\sigma)}(b)}{\partial b_{k \to i}(\sigma')} = \frac{1}{\mathcal{Z}_{i \to j}} \phi_i(\sigma) \psi_{ik}(\sigma, \sigma') \prod_{l \in \partial i \backslash \{j, k\}} \sum_{\sigma_l \in \mathcal{X}} \psi_{il}(\sigma, \sigma_l) b_{l \to i}(\sigma_l)$$
$$- \frac{1}{\mathcal{Z}_{i \to j}^2} \frac{\partial \mathcal{Z}_{i \to j}}{\partial b_{k \to i}(\sigma')} \phi_i(\sigma) \prod_{l \in \partial i \backslash j} \sum_{\sigma_l \in \mathcal{X}} \psi_{il}(\sigma, \sigma_l) b_{l \to i}(\sigma_l),$$

and the derivative of the normalization is given by

$$\frac{\partial \mathcal{Z}_{i \to j}}{\partial b_{k \to i}(\sigma')} = \frac{\partial}{\partial b_{k \to i}(\sigma')} \sum_{\sigma_i \in \mathcal{X}} \phi_i(\sigma_i) \prod_{l \in \partial i \backslash j} \sum_{\sigma_l \in \mathcal{X}} \psi_{il}(\sigma_i, \sigma_l) b_{l \to i}(\sigma_l)$$
$$= \sum_{\sigma_i \in \mathcal{X}} \phi_i(\sigma_i) \psi_{ik}(\sigma_i, \sigma') \prod_{l \in \partial i \backslash \{j, k\}} \sum_{\sigma_l \in \mathcal{X}} \psi_{il}(\sigma_i, \sigma_l) b_{l \to i}(\sigma_l).$$

We recall that the factorized condition imposes that the following quantities are independent of $\sigma$

$$\sum_{\sigma_l \in \mathcal{X}} \psi_{il}(\sigma, \sigma_l) \phi_l(\sigma_l) = r_{i \to l} \qquad \forall l \in \partial i, \sigma \in \mathcal{X}$$

When evaluating the previous quantities at the fixed point $b^\star$, we have, recalling that we assume $\forall i \in [n], \sum_{\sigma \in \mathcal{X}} \phi_i(\sigma) = 1$,

$$\mathcal{Z}_{i \to j}\big|_{b = b^\star} = \prod_{l \in \partial i \backslash j} r_{i \to l}.$$

$$\frac{\partial \mathcal{Z}_{i \to j}}{\partial b_{k \to i}(\sigma')}\bigg|_{b = b^\star} = r_{k \to i} \prod_{l \in \partial i \backslash \{j, k\}} r_{i \to l} = \frac{r_{k \to i}}{r_{i \to k}} \mathcal{Z}_{i \to j}\big|_{b = b^\star}.$$

Evaluating the non-zero elements of the Jacobian at the fixed point, we therefore get

$$
\begin{aligned}
\frac{\partial f_{i \to j}^{(\sigma)}(b)}{\partial b_{k \to i}(\sigma')}\bigg|_{b=b^*} ={}& \phi_i(\sigma)\psi_{ik}(\sigma, \sigma')\frac{\prod_{l \in \partial i \setminus \{j,k\}} r_{i \to l}}{\prod_{l \in \partial i \setminus j} r_{i \to l}} \\
& - \phi_i(\sigma)\frac{r_{k \to i}}{r_{i \to k}}\frac{\prod_{l \in \partial i \setminus j} r_{i \to l}}{\prod_{l \in \partial i \setminus j} r_{i \to l}}, \\
={}& \phi_i(\sigma)\left(\frac{\psi_{ik}(\sigma, \sigma')}{r_{i \to k}} - \frac{r_{k \to i}}{r_{i \to k}}\right).
\end{aligned}
$$

The full Jacobian can therefore finally be written as

$$
\mathcal{J}_{i \to j, k \to l}^{(\sigma, \sigma')} = \phi_l(\sigma)\left(\frac{\psi_{kl}(\sigma', \sigma)}{r_{l \to k}} - \frac{r_{k \to l}}{r_{l \to k}}\right)\mathbf{1}\left(l = i\right)\mathbf{1}\left(k \neq j\right).
$$

## Résumé

Face au déluge actuel de données principalement non structurées, les graphes ont démontré, dans une variété de domaines scientifiques, leur importance croissante comme language abstrait pour décrire des interactions complexes entre des objets complexes. L'un des principaux défis posés par l'étude de ces réseaux est l'inférence de propriétés macroscopiques à grande échelle, affectant un grand nombre d'objets ou d'agents, sur la seule base des interactions microscopiques qu'entretiennent leurs constituants élémentaires. La physique statistique, créée précisément dans le but d'obtenir les lois macroscopiques de la thermodynamique à partir d'un modèle idéal de particules en interaction, fournit une intuition décisive dans l'étude des réseaux complexes.

Dans cette thèse, nous utilisons des méthodes issues de la physique statistique des systèmes désordonnés pour mettre au point et analyser de nouveaux algorithmes d'inférence sur les graphes. Nous nous concentrons sur les méthodes spectrales, utilisant certains vecteurs propres de matrices bien choisies, et sur les graphes parcimonieux, qui contiennent une faible quantité d'information. Nous développons une théorie originale de l'inférence spectrale, fondée sur une relaxation de l'optimisation de certaines énergies libres en champ moyen. Notre approche est donc entièrement probabiliste, et diffère considérablement des motivations plus classiques fondées sur l'optimisation d'une fonction de coût. Nous illustrons l'efficacité de notre approche sur différents problèmes, dont la détection de communautés, la classification non supervisée à partir de similarités mesurées aléatoirement, et la complétion de matrices.

## Abstract

In an era of unprecedented deluge of (mostly unstructured) data, graphs are proving more and more useful, across the sciences, as a flexible abstraction to capture complex relationships between complex objects. One of the main challenges arising in the study of such networks is the inference of macroscopic, large-scale properties affecting a large number of objects, based solely on the microscopic interactions between their elementary constituents. Statistical physics, precisely created to recover the macroscopic laws of thermodynamics from an idealized model of interacting particles, provides significant insight to tackle such complex networks.

In this dissertation, we use methods derived from the statistical physics of disordered systems to design and study new algorithms for inference on graphs. Our focus is on spectral methods, based on certain eigenvectors of carefully chosen matrices, and sparse graphs, containing only a small amount of information. We develop an original theory of spectral inference based on a relaxation of various mean-field free energy optimizations. Our approach is therefore fully probabilistic, and contrasts with more traditional motivations based on the optimization of a cost function. We illustrate the efficiency of our approach on various problems, including community detection, randomized similarity-based clustering, and matrix completion.

## Mots Clés

## Keywords